\documentclass[12pt, fleqn, a4paper]{article} % A4 size
\usepackage{hyperref}
\hypersetup{pdfstartview  = Fit,
            pdfpagelayout = SinglePage,
            pdfborder  = {0 0 0.5},
            colorlinks = false
}
\usepackage{epsfig, graphics, graphicx, subcaption}
\usepackage{clshan-math, clshan-dm-dd, clshan-dm-ddd}
\usepackage{multirow}
\def \lsim {\:\raisebox{-0.7 ex}{$\stackrel{\textstyle<}{\sim}$}\:}
\def \gsim {\:\raisebox{-0.7 ex}{$\stackrel{\textstyle>}{\sim}$}\:}
 \def \PeriodAa     {0      -- 365}
 \def \PlotNumberAa {00000}
 \def \PeriodBa     {49.0   -- 109.0}
 \def \PeriodBb     {140.25 -- 200.25}
 \def \PeriodBc     {231.50 -- 291.50}
 \def \PeriodBd     {322.75 -- 382.75}
 \def \PlotNumberBa {07900}
 \def \PlotNumberBb {17025}
 \def \PlotNumberBc {26150}
 \def \PlotNumberBd {35275}
 \def \PeriodCa     {19.49  --  79.49}
 \def \PeriodCb     {110.74 -- 170.74}
 \def \PeriodCc     {201.99 -- 261.99}
 \def \PeriodCd     {293.24 -- 353.24}
 \def \PlotNumberCa {04949}
 \def \PlotNumberCb {14074}
 \def \PlotNumberCc {23199}
 \def \PlotNumberCd {32324}
 \def \PeriodDa     {177.66 -- 237.66}
 \def \PeriodDb     {360.16 -- 420.16}
 \def \PlotNumberDa {20766}
 \def \PlotNumberDb {39016}
\newcommand{\Dunderline} [1] {\underline{\underline{#1}}}
\newcommand{\InsertPlotNphitheta} [1] {
\begin{figure} [t!]
\begin{center}
 \begin{subfigure} [c] {8.25 cm}
  \includegraphics [width = 8.25 cm]
   {N_phi_theta-\ShortFrame-\EventNumber-\PlotNumbera}
 \caption{\Perioda\ day}
 \end{subfigure}
 \hspace{0.1 cm}
 \begin{subfigure} [c] {8.25 cm}
  \includegraphics [width = 8.25 cm]
   {N_phi_theta-\ShortFrame-\EventNumber-\PlotNumberb}
 \caption{\Periodb\ day}
 \end{subfigure}
 \\ \vspace{0.6 cm}
 \begin{subfigure} [c] {8.25 cm}
  \includegraphics [width = 8.25 cm]
   {N_phi_theta-\ShortFrame-\EventNumber-\PlotNumberc}
 \caption{\Periodc\ day}
 \end{subfigure}
 \hspace{0.1 cm}
 \begin{subfigure} [c] {8.25 cm}
  \includegraphics [width = 8.25 cm]
   {N_phi_theta-\ShortFrame-\EventNumber-\PlotNumberd}
 \caption{\Periodd\ day}
 \end{subfigure}
\end{center}
\caption{
 #1
}
\label{fig:N_phi_theta-\ShortFrame-\EventNumber-\PlotNumbera}
\end{figure}
}
\newcommand{\InsertPlotNphithetaLab} [1] {
\begin{figure} [t!]
\begin{center}
 \begin{subfigure} [c] {8.25 cm}
  \includegraphics [width = 8.25 cm]
   {N_phi_theta-\ShortFrame-\EventNumber-\PlotNumberA-\LabName}
 \caption{\PeriodA\ day}
 \end{subfigure}
 \\ \vspace{0.6 cm}
 \begin{subfigure} [c] {8.25 cm}
  \includegraphics [width = 8.25 cm]
   {N_phi_theta-\ShortFrame-\EventNumber-\PlotNumbera-\LabName}
 \caption{\Perioda\ day}
 \end{subfigure}
 \hspace{0.1 cm}
 \begin{subfigure} [c] {8.25 cm}
  \includegraphics [width = 8.25 cm]
   {N_phi_theta-\ShortFrame-\EventNumber-\PlotNumberb-\LabName}
 \caption{\Periodb\ day}
 \end{subfigure}
 \\ \vspace{0.6 cm}
 \begin{subfigure} [c] {8.25 cm}
  \includegraphics [width = 8.25 cm]
   {N_phi_theta-\ShortFrame-\EventNumber-\PlotNumberc-\LabName}
 \caption{\Periodc\ day}
 \end{subfigure}
 \hspace{0.1 cm}
 \begin{subfigure} [c] {8.25 cm}
  \includegraphics [width = 8.25 cm]
   {N_phi_theta-\ShortFrame-\EventNumber-\PlotNumberd-\LabName}
 \caption{\Periodd\ day}
 \end{subfigure}
\end{center}
\caption{
 #1
}
\label{fig:N_phi_theta-\ShortFrame-\EventNumber-\PlotNumbera-\LabName}
\end{figure}
}
\newcommand{\InsertPlotNphithetaDiurnal} [1] {
\begin{figure} [t!]
\begin{center}
 \begin{subfigure} [c] {8.25 cm}
  \includegraphics [width = 8.25 cm]
   {N_phi_theta-Lab-\EventNumber-\PlotNumberD-00-\LabName}
 \caption{0 -- 2 + 22 -- 24 hour}
 \end{subfigure}
 \hspace{0.1 cm}
 \begin{subfigure} [c] {8.25 cm}
  \includegraphics [width = 8.25 cm]
   {N_phi_theta-Lab-\EventNumber-\PlotNumberD-06-\LabName}
 \caption{4 -- 8 hour}
 \end{subfigure}
 \\ \vspace{0.6 cm}
 \begin{subfigure} [c] {8.25 cm}
  \includegraphics [width = 8.25 cm]
   {N_phi_theta-Lab-\EventNumber-\PlotNumberD-12-\LabName}
 \caption{10 -- 14 hour}
 \end{subfigure}
 \hspace{0.1 cm}
 \begin{subfigure} [c] {8.25 cm}
  \includegraphics [width = 8.25 cm]
   {N_phi_theta-Lab-\EventNumber-\PlotNumberD-18-\LabName}
 \caption{16 -- 20 hour}
 \end{subfigure}
\end{center}
\caption{
 #1
}
\label{fig:N_phi_theta-Lab-\EventNumber-\PlotNumberD-00-\LabName}
\end{figure}
}
\newcommand{\InsertPlotNphithetaDiurnalLab} [1] {
\begin{figure} [t!]
\vspace{-0.65 cm}
\begin{center}
 \begin{subfigure} [c] {7 cm}
  \includegraphics [width = 7 cm]
   {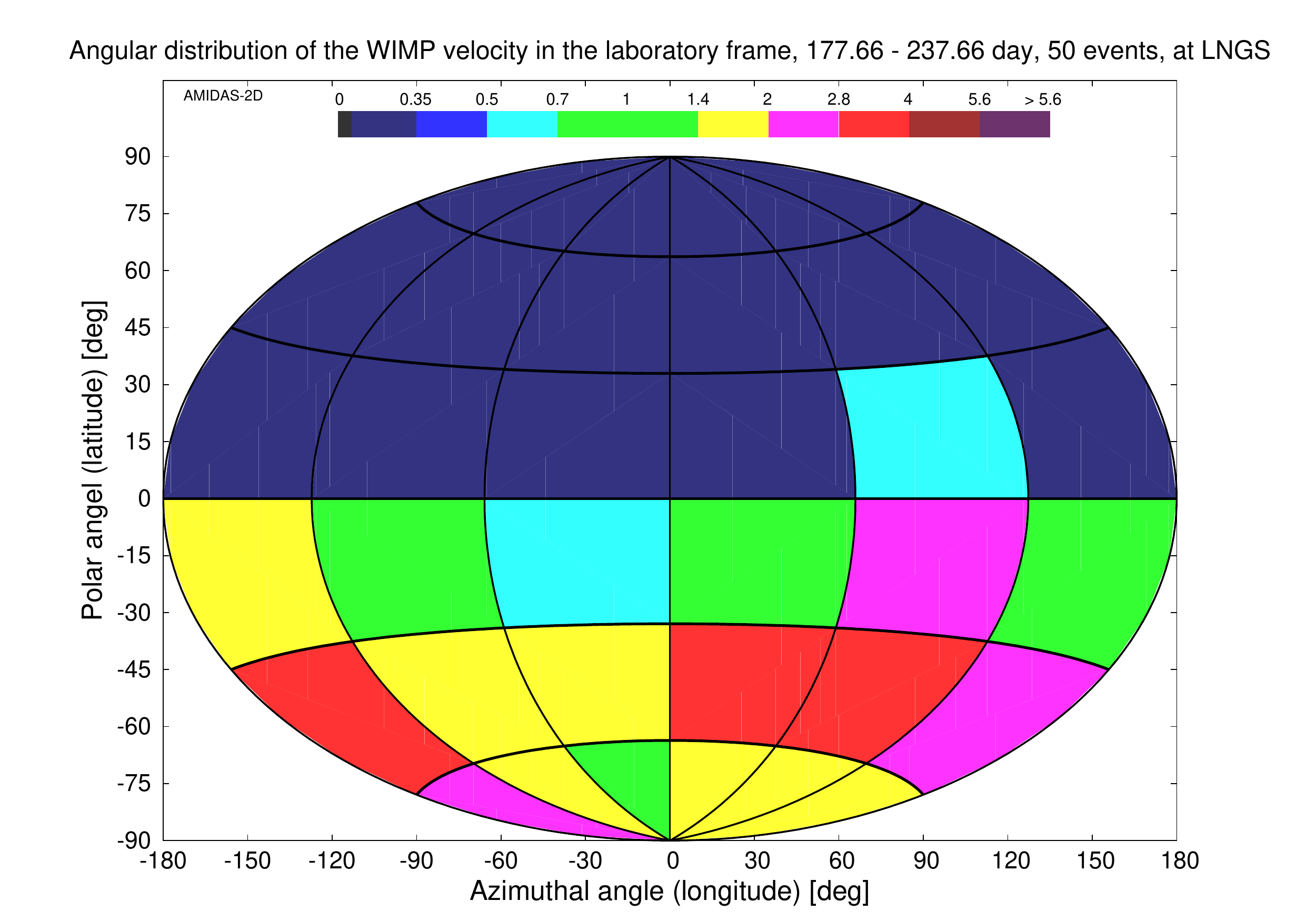}
 \caption{\PeriodDa\ day, 22 --  2 hour}
 \end{subfigure}
 \hspace{0.5 cm}
 \begin{subfigure} [c] {7 cm}
  \includegraphics [width = 7 cm]
   {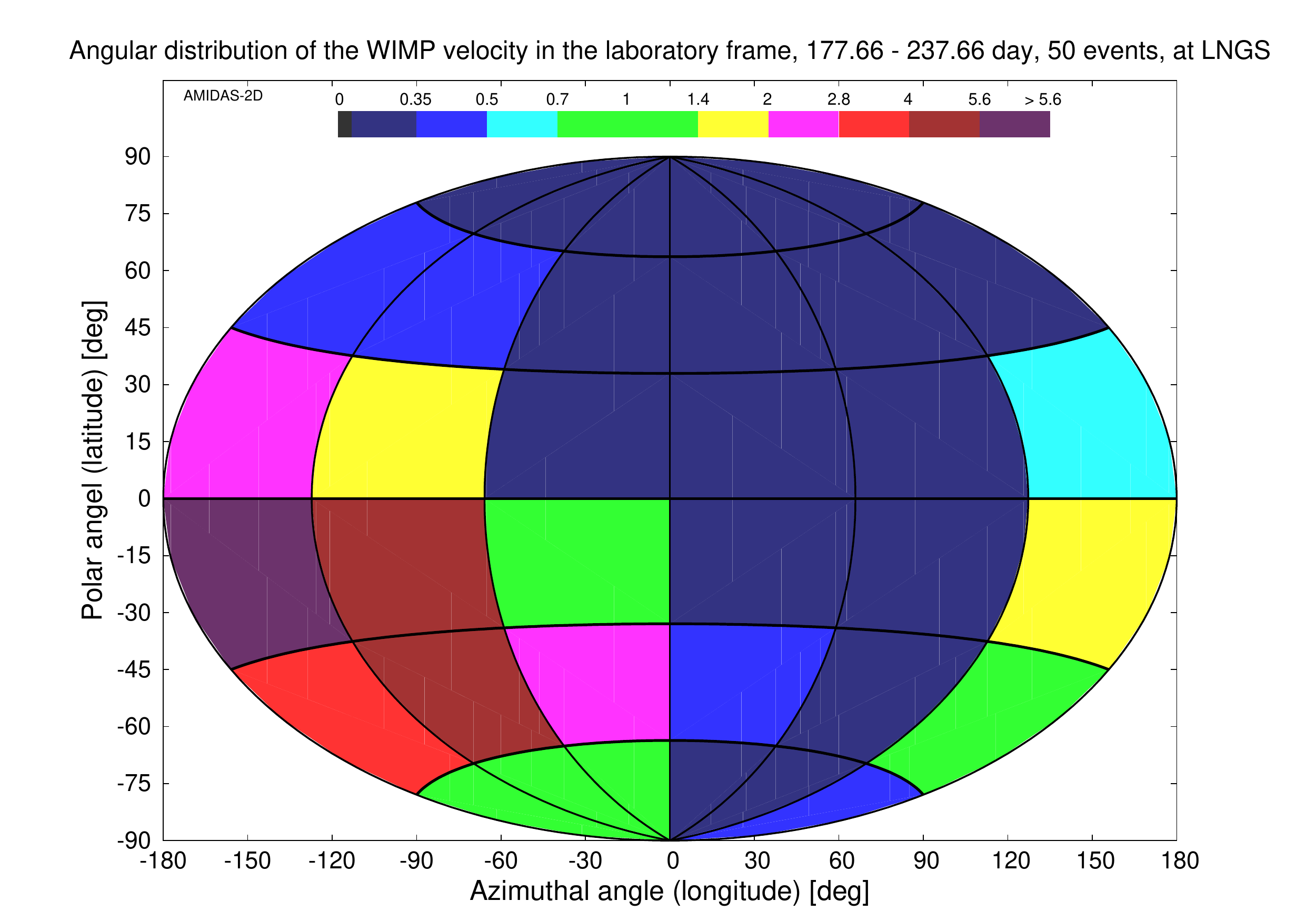}
 \caption{\PeriodDa\ day,  4 --  8 hour}
 \end{subfigure}
 \\ \vspace{0.1 cm}
 \begin{subfigure} [c] {7 cm}
  \includegraphics [width = 7 cm]
   {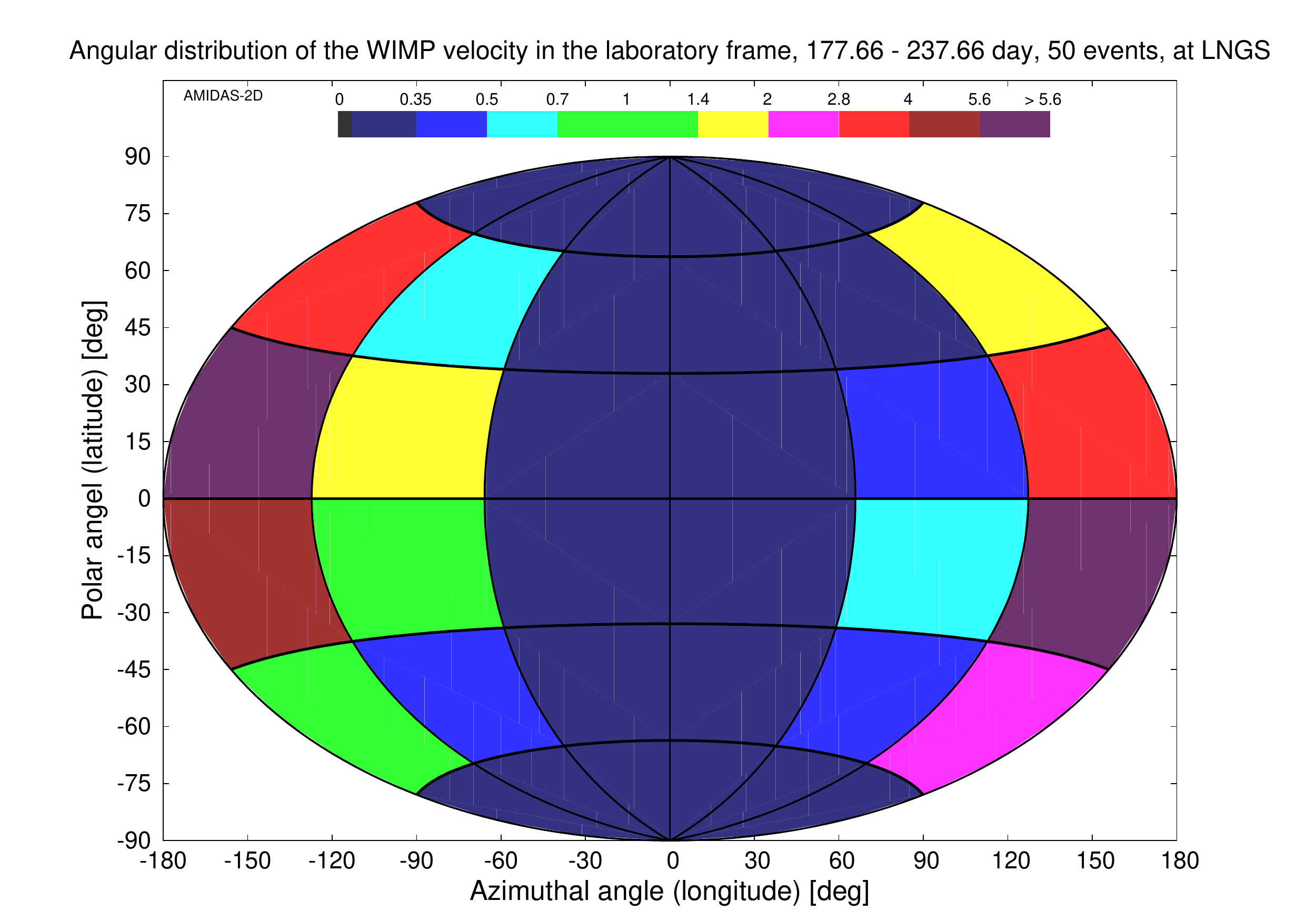}
 \caption{\PeriodDa\ day, 10 -- 14 hour}
 \end{subfigure}
 \hspace{0.5 cm}
 \begin{subfigure} [c] {7 cm}
  \includegraphics [width = 7 cm]
   {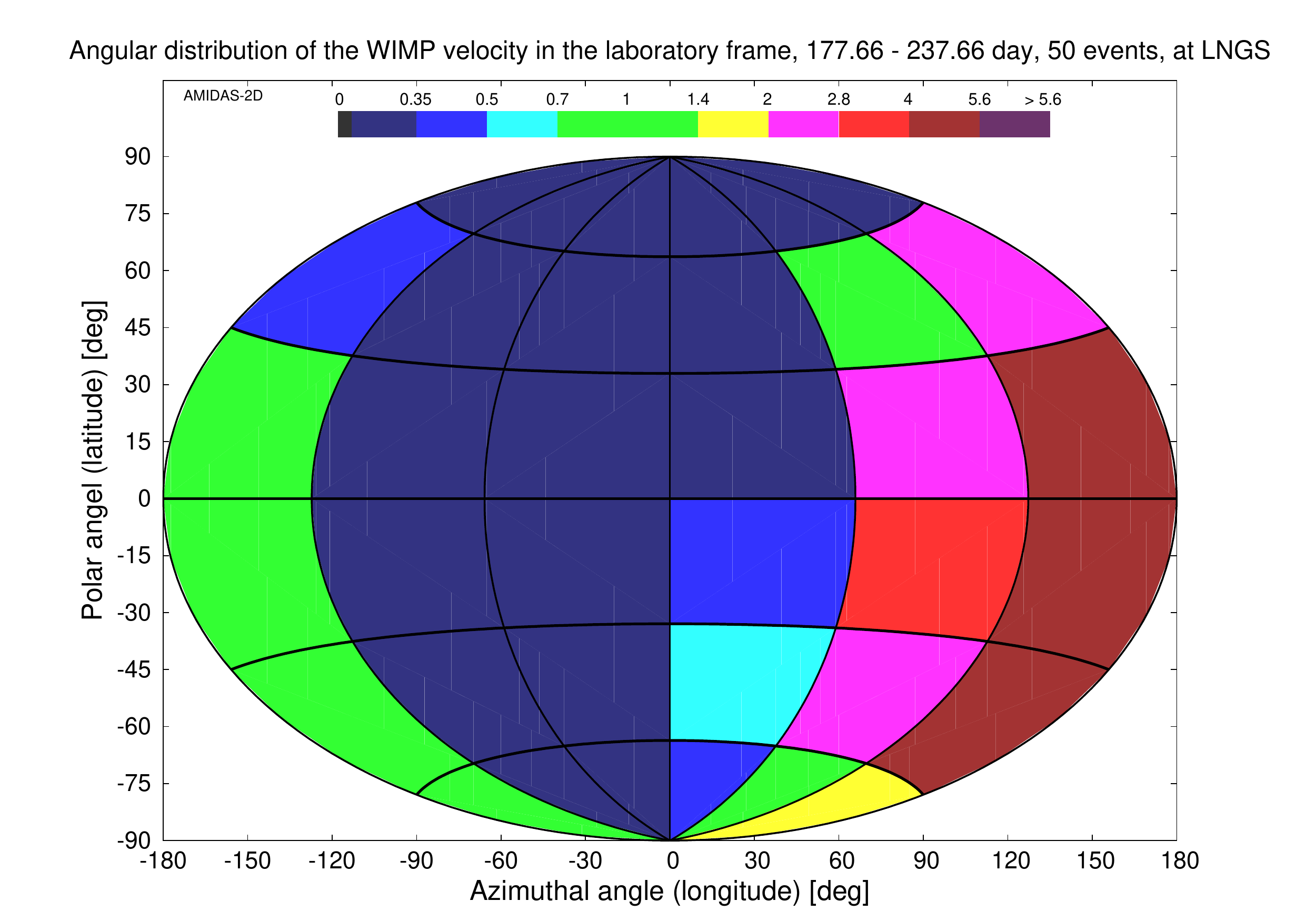}
 \caption{\PeriodDa\ day, 16 -- 20 hour}
 \end{subfigure}
 \\ \vspace{0.6 cm}
 \begin{subfigure} [c] {7 cm}
  \includegraphics [width = 7 cm]
   {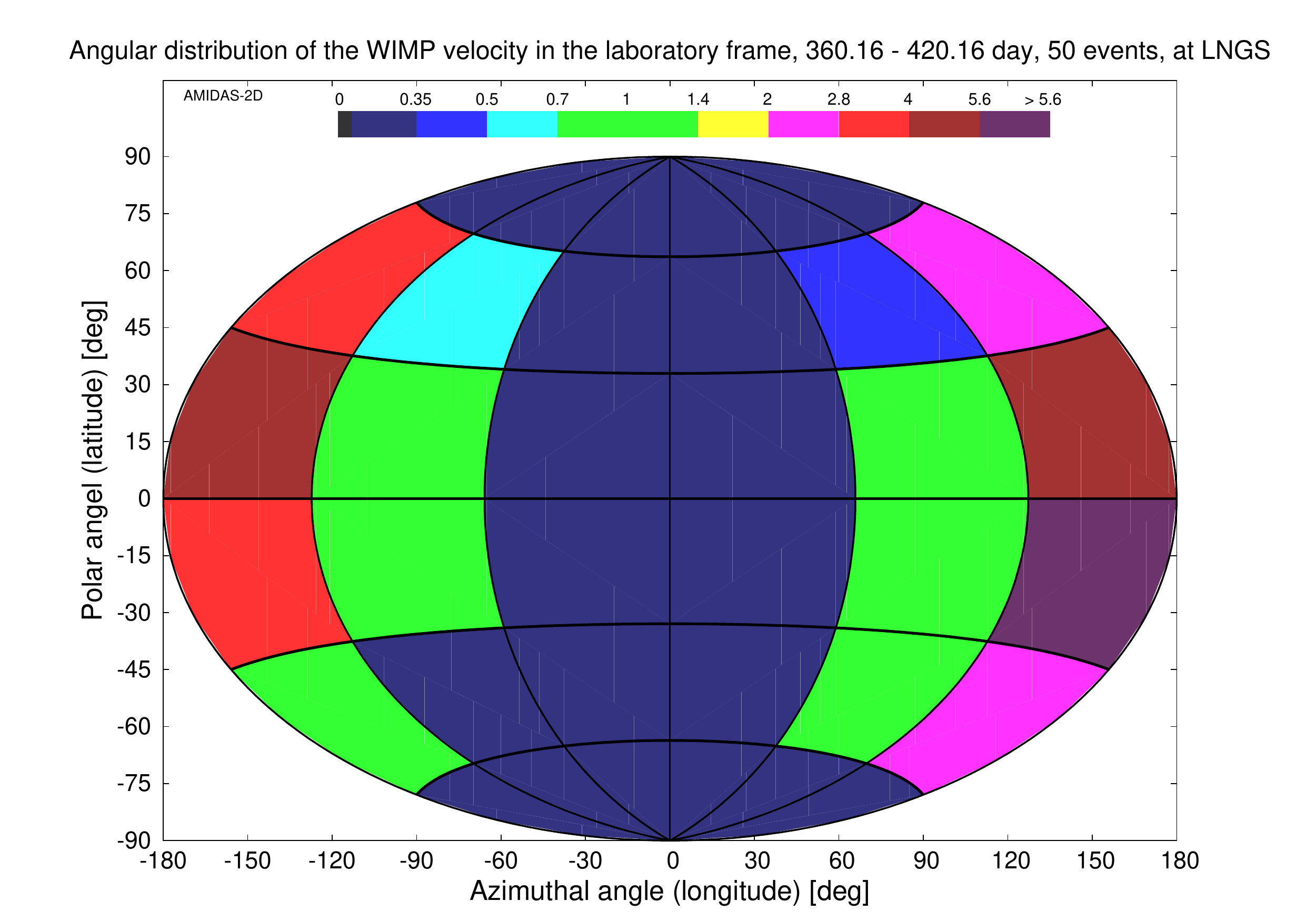}
 \caption{\PeriodDb\ day, 22 --  2 hour}
 \end{subfigure}
 \hspace{0.5 cm}
 \begin{subfigure} [c] {7 cm}
  \includegraphics [width = 7 cm]
   {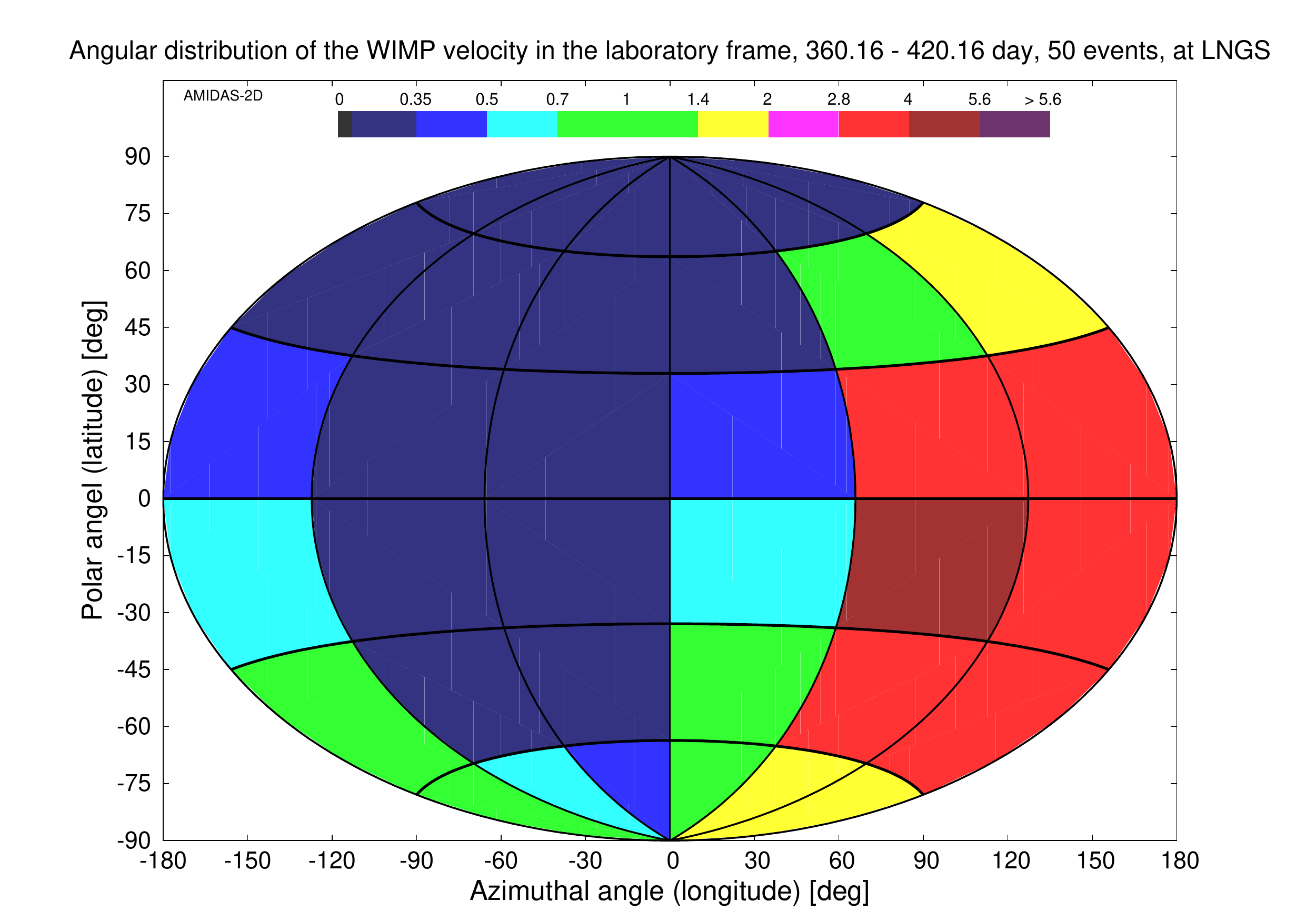}
 \caption{\PeriodDb\ day,  4 --  8 hour}
 \end{subfigure}
 \\ \vspace{0.1 cm}
 \begin{subfigure} [c] {7 cm}
  \includegraphics [width = 7 cm]
   {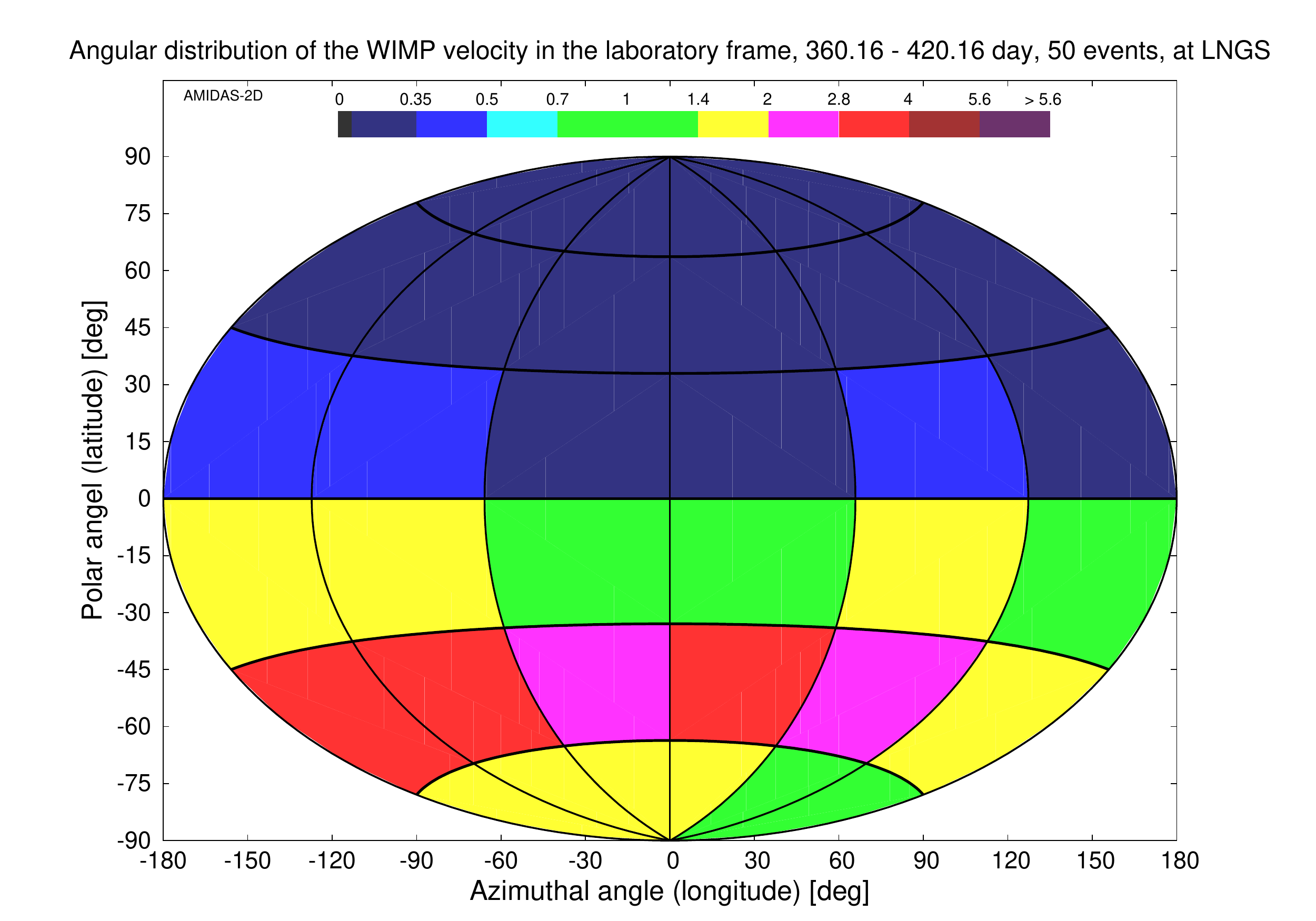}
 \caption{\PeriodDb\ day, 10 -- 14 hour}
 \end{subfigure}
 \hspace{0.5 cm}
 \begin{subfigure} [c] {7 cm}
  \includegraphics [width = 7 cm]
   {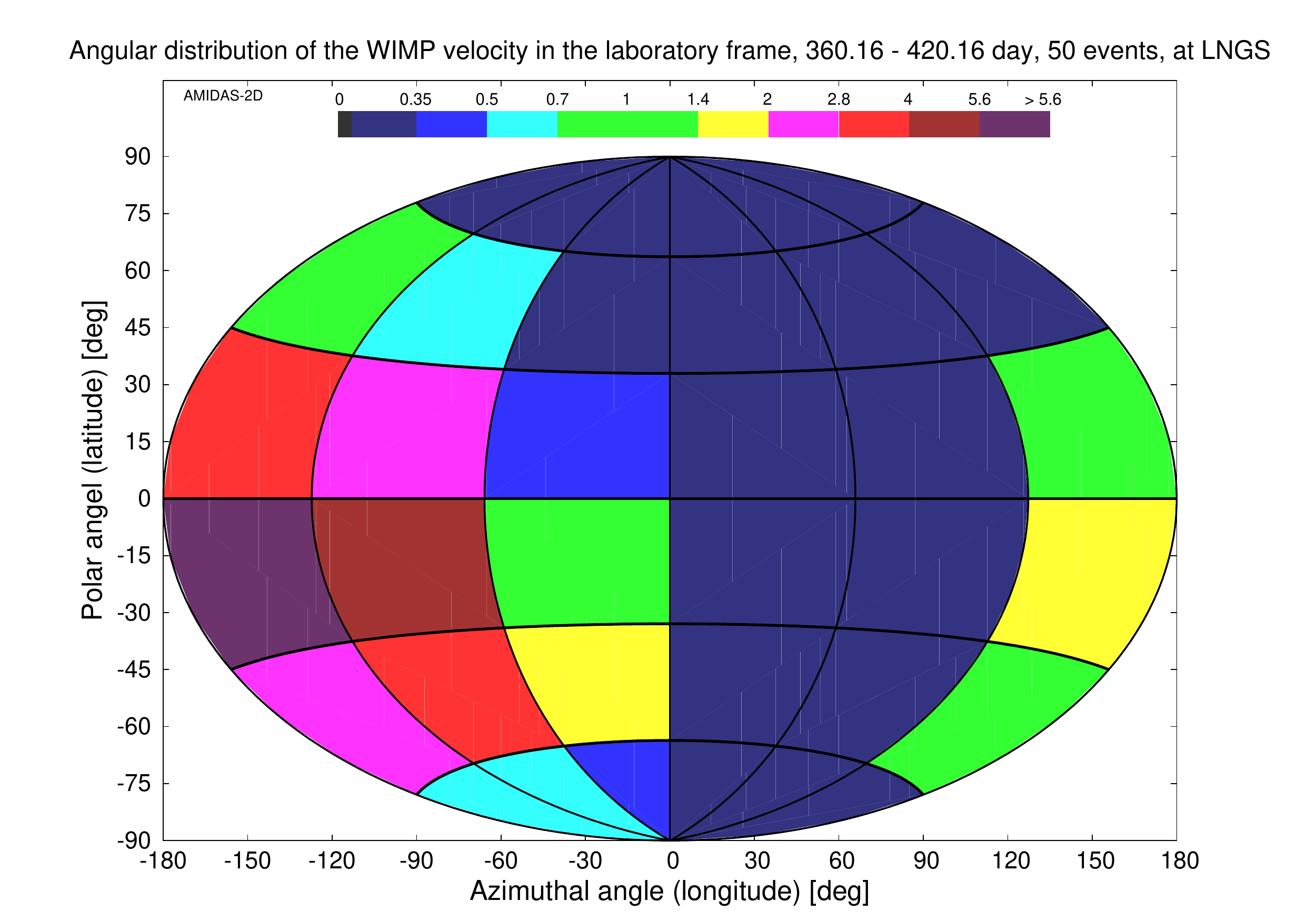}
 \caption{\PeriodDb\ day, 16 -- 20 hour}
 \end{subfigure}
 \vspace{-0.3 cm}
\end{center}
\caption{
 #1
}
\label{fig:N_phi_theta-Lab-\EventNumber-\PlotNumberDa-\PlotNumberDb-00-\LabName}
\end{figure}
}
\newcommand{\InsertPlotNphithetaULab} {
 \def \ShortFrame        {H}
 \def \EventNumber       {050}
 \InsertPlotNphithetaLab
  {The angular distributions of the 3-D WIMP velocity
   in the horizontal coordinate system
   at the location of the \LabName\ laboratory
   \LabLocation.
   50 total events on average
   in one entire year (a)
   and in each 60-day observation period of four advanced seasons
   (b -- e)
   have been simulated.
   }
 \def \EventNumber       {500}
 \InsertPlotNphithetaLab
  {As in Figs.~\ref{fig:N_phi_theta-H-050-\PlotNumbera-\LabName},
   except that
   500 total events on average
   in one entire year (a)
   and in each 60-day observation period of four advanced seasons
   (b -- e)
   have been simulated.
   \vspace{1 cm}
   }
 \def \ShortFrame        {Lab}
 \def \EventNumber       {050}
 \InsertPlotNphithetaLab
  {The angular distributions of the 3-D WIMP velocity
   transformed from events
   shown in Fig.~\ref{fig:N_phi_theta-H-050-04949-\LabName}
   to the laboratory coordinate system
   at the location of the \LabName\ laboratory.
   All simulation setup and notations are the same
   as in Fig.~\ref{fig:N_phi_theta-H-050-04949-\LabName}.
   \vspace{0.5 cm}
   }
 \def \EventNumber       {500}
 \InsertPlotNphithetaLab
  {As in Figs.~\ref{fig:N_phi_theta-Lab-050-\PlotNumbera-\LabName},
   except that
   500 total events on average
   in one entire year (a)
   and in each 60-day observation period of four advanced seasons
   (b -- e)
   have been simulated.
   \vspace{1 cm}
   }
 \def \EventNumber              {050}
 \InsertPlotNphithetaDiurnalLab
  {The angular distributions of the 3-D WIMP velocity
   observed at the location of the \LabName\ laboratory.
   50 total events on average
   in each 4-hour daily shift
   in the 60-day observation periods of
       \PeriodDa\           day (a -- d)
   and \PeriodDb\ (= 55.16) day (e -- h)
   have been simulated.
   \vspace{-1.45 cm}
   }
 \def \EventNumber              {500}
 \InsertPlotNphithetaDiurnalLab
  {As in Figs.~\ref{fig:N_phi_theta-Lab-050-20766-39016-00-\LabName},
   except that
   500 total events on average
   in each 4-hour daily shift
   in the 60-day observation periods
   have been simulated.
   \vspace{-0.85 cm}
   }
}
\newcommand{\InsertSKPPlotS} [3] [10.5] {
\begin{figure} [t!]
\begin{center}
 \includegraphics [width = #1 cm] {skp-#2}
\end{center}
\caption{
 #3
}
\label{fig:#2}
\end{figure}
}
\newcommand{\InsertSKPPlotD} [5] [t!] {
\begin{figure} [#1]
\begin{center}
 \begin{subfigure} [c] {8.25 cm}
  \includegraphics [width = 8.25 cm] {skp-#2}
 \caption{}
 \end{subfigure}
 \hspace{0.1 cm}
 \begin{subfigure} [c] {8.25 cm}
  \includegraphics [width = 8.25 cm] {skp-#3}
 \caption{}
 \end{subfigure}
\end{center}
\caption{
 #5
}
\label{#4}
\end{figure}
}
\newcommand{\InsertPlotNvBayesianD} [3] [t!] {
\begin{figure} [#1]
\begin{center}
 \begin{subfigure} [c] {8.25 cm}
  \includegraphics [width = 8.25 cm]
   {N_v-Bayesian-\ShortFrame-\Fittingfv-\EventNumber-\PlotNumberA}
 \caption{}
 \end{subfigure}
 \hspace{0.1 cm}
 \begin{subfigure} [c] {8.25 cm}
  \includegraphics [width = 8.25 cm]
   {dis-\ShortFrame-\Fittingfv-#2-\EventNumber-\PlotNumberA}
 \caption{}
 \end{subfigure}
\end{center}
\caption{
 #3
}
\label{fig:N_v-Bayesian-\ShortFrame-\Fittingfv-\EventNumber-\PlotNumberA}
\end{figure}
}
\newcommand{\InsertPlotNvBayesianQ} [3] {
\begin{figure} [t!]
\begin{center}
 \begin{subfigure} [c] {8.25 cm}
  \includegraphics [width = 8.25 cm]
   {N_v-Bayesian-\ShortFrame-\Fittingfv-\EventNumber-\PlotNumberA}
 \caption{}
 \end{subfigure}
 \hspace{0.1 cm}
 \begin{subfigure} [c] {8.25 cm}
  \includegraphics [width = 8.25 cm]
   {dis-\ShortFrame-\Fittingfv-#1-#2-\EventNumber-\PlotNumberA}
 \caption{}
 \end{subfigure}
 \\ \vspace{0.6 cm}
 \begin{subfigure} [c] {8.25 cm}
  \includegraphics [width = 8.25 cm]
   {dis-\ShortFrame-\Fittingfv-#1-\EventNumber-\PlotNumberA}
 \caption{}
 \end{subfigure}
 \hspace{0.1 cm}
 \begin{subfigure} [c] {8.25 cm}
  \includegraphics [width = 8.25 cm]
   {dis-\ShortFrame-\Fittingfv-#2-\EventNumber-\PlotNumberA}
 \caption{}
 \end{subfigure}
\end{center}
\caption{
 #3
}
\label{fig:N_v-Bayesian-\ShortFrame-\Fittingfv-\EventNumber-\PlotNumberA}
\end{figure}
}
\newcommand{\InsertPlotNvBayesianAnnual} [1] {
\begin{figure} [p!]
\begin{center}
\vspace{-0.85 cm}
 \begin{subfigure} [c] {15.5 cm}
  \includegraphics [width = 7.5 cm]
   {N_v-Bayesian-\ShortFrame-\Fittingfv-\EventNumber-\PlotNumbera}        \hspace{ 0.5 cm}
  \includegraphics [width = 7.5 cm]
   {dis-\ShortFrame-\Fittingfv-\FittingPara-\EventNumber-\PlotNumbera} \\ \vspace{-0.7 cm}
 \caption{\Perioda\ day}
 \end{subfigure}
 \\ \vspace{0.25 cm}
 \begin{subfigure} [c] {15.5 cm}
  \includegraphics [width = 7.5 cm]
   {N_v-Bayesian-\ShortFrame-\Fittingfv-\EventNumber-\PlotNumberb}        \hspace{ 0.5 cm}
  \includegraphics [width = 7.5 cm]
   {dis-\ShortFrame-\Fittingfv-\FittingPara-\EventNumber-\PlotNumberb} \\ \vspace{-0.7 cm}
 \caption{\Periodb\ day}
 \end{subfigure}
 \\ \vspace{0.25 cm}
 \begin{subfigure} [c] {15.5 cm}
  \includegraphics [width = 7.5 cm]
   {N_v-Bayesian-\ShortFrame-\Fittingfv-\EventNumber-\PlotNumberc}        \hspace{ 0.5 cm}
  \includegraphics [width = 7.5 cm]
   {dis-\ShortFrame-\Fittingfv-\FittingPara-\EventNumber-\PlotNumberc} \\ \vspace{-0.7 cm}
 \caption{\Periodc\ day}
 \end{subfigure}
 \\ \vspace{0.25 cm}
 \begin{subfigure} [c] {15.5 cm}
  \includegraphics [width = 7.5 cm]
   {N_v-Bayesian-\ShortFrame-\Fittingfv-\EventNumber-\PlotNumberd}        \hspace{ 0.5 cm}
  \includegraphics [width = 7.5 cm]
   {dis-\ShortFrame-\Fittingfv-\FittingPara-\EventNumber-\PlotNumberd} \\ \vspace{-0.7 cm}
 \caption{\Periodd\ day}
 \end{subfigure}
 \vspace{-0.5 cm}
\end{center}
\caption{
 #1
}
\label{fig:N_v-Bayesian-\ShortFrame-\Fittingfv-\EventNumber-\PlotNumbera}
\end{figure}
}
\newcommand{\InsertResultsTableNvBayesian} [4] [t!] {
\begin{table} [#1]
\begin{center}
\renewcommand{\arraystretch}{2}
\setlength{\tabcolsep}{3 pt}
{\footnotesize
 \begin{tabular}{|| c || c | c | c | c | c ||}
  \hline
  \hline
  \makebox[2.25 cm] [c] {Fitting dist.}               &
  \makebox[2.25 cm] [c] {Parameter}                   &
  \makebox[2.5  cm] [c] {Max.~${\rm P}_{\rm median}$} &
  \makebox[3.65 cm] [c] {Median}                      &
  \makebox[2.55 cm] [c] {1$\sigma$ range}             &
  \makebox[2.55 cm] [c] {2$\sigma$ range}             \\
  \hline
  \hline
   #2
  \hline
  \hline
   \multicolumn {6} {|| l ||} {~#3}                   \\
  \hline
  \hline
 \end{tabular}
 }
\end{center}
\caption{
 #4
}
\label{tab:N_v-Bayesian-\ShortFrame-\EventNumber-\PlotNumber}
\end{table}
}
\newcommand{\InsertResultsTableNvBayesianAnnual} [6] [t!] {
\begin{table} [#1]
\begin{center}
\renewcommand{\arraystretch}{2}
\setlength{\tabcolsep}{3 pt}
{\footnotesize
 \begin{tabular}{|| c || c | c | c | c | c ||}
  \hline
  \hline
  \multicolumn{6}{|| c ||}
  {\bf\boldmath
   Fitting distribution: #2}                          \\
  \hline
  \hline
  \makebox[3.25 cm] [c]
   {\begin{minipage} {3.25 cm}
     \begin{center}
        ~                   \\ \vspace{1.1  ex}
        Central date        \\ \vspace{0.65 ex}
       (observation period) \\ \vspace{0.65 ex}
       (day)                \\ \vspace{1.1  ex}
     \end{center}
    \end{minipage}}                                   &
  \makebox[2    cm] [c] {Parameter}                   &
  \makebox[2.25 cm] [c] {Max.~${\rm P}_{\rm median}$} &
  \makebox[3.25 cm] [c] {Median}                      &
  \makebox[2.5  cm] [c] {1$\sigma$ range}             &
  \makebox[2.5  cm] [c] {2$\sigma$ range}             \\
  \hline
  \hline
   #3
  \hline
  \hline
   \multicolumn {6} {|| l ||}
   {~#4 total events on average in one observation period of 60 days} \\
  \hline
  \hline
 \end{tabular}
 }
\vspace{-0.25 cm}
\end{center}
\caption{
 The summary of
 the reconstructed results of the fitting
 #5
 their 1(2)$\sigma$ statistical uncertainty ranges
 of the median values
 by using the #2
 with #4 total events on average
 in each 60--day observation period of the #6 seasons.
}
\label{tab:N_v-Bayesian-\ShortFrame-\Fittingfv-\EventNumber-\PlotNumber}
\end{table}
}
\newcommand{\InsertResultsTableNvBayesianAnnualD} [7] {
\begin{table} [p!]
\begin{center}
\renewcommand{\arraystretch}{2}
\setlength{\tabcolsep}{3 pt}
{\footnotesize
 \begin{tabular}{|| c || c | c | c | c | c ||}
  \hline
  \hline
  \multicolumn{6}{|| c ||}
  {\bf\boldmath
   Fitting distribution: #1}                          \\
  \hline
  \hline
  \makebox[3.25 cm] [c]
   {\begin{minipage} {3.25 cm}
     \begin{center}
        ~                   \\ \vspace{1.1  ex}
        Central date        \\ \vspace{0.65 ex}
       (observation period) \\ \vspace{0.65 ex}
       (day)                \\ \vspace{1.1  ex}
     \end{center}
    \end{minipage}}                                   &
  \makebox[2    cm] [c] {Parameter}                   &
  \makebox[2.25 cm] [c] {Max.~${\rm P}_{\rm median}$} &
  \makebox[3.25 cm] [c] {Median}                      &
  \makebox[2.5  cm] [c] {1$\sigma$ range}             &
  \makebox[2.5  cm] [c] {2$\sigma$ range}             \\
  \hline
  \hline
   #2
  \hline
  \hline
   \multicolumn {6} {|| l ||}
   {~#3 total events on average in one observation period of 60 days} \\
  \hline
  \hline
   \multicolumn {6} {   l   }
   {~} \\
  \hline
  \hline
  \makebox[3.25 cm] [c]
   {\begin{minipage} {3.25 cm}
     \begin{center}
        ~                   \\ \vspace{1.1  ex}
        Central date        \\ \vspace{0.65 ex}
       (observation period) \\ \vspace{0.65 ex}
       (day)                \\ \vspace{1.1  ex}
     \end{center}
    \end{minipage}}                                   &
  \makebox[2    cm] [c] {Parameter}                   &
  \makebox[2.25 cm] [c] {Max.~${\rm P}_{\rm median}$} &
  \makebox[3.25 cm] [c] {Median}                      &
  \makebox[2.5  cm] [c] {1$\sigma$ range}             &
  \makebox[2.5  cm] [c] {2$\sigma$ range}             \\
  \hline
  \hline
   #4
  \hline
  \hline
   \multicolumn {6} {|| l ||}
   {~#5 total events on average in one observation period of 60 days} \\
  \hline
  \hline
 \end{tabular}
 }
\vspace{-0.25 cm}
\end{center}
\caption{
 The summary of
 the reconstructed results of the fitting
 #6
 their 1(2)$\sigma$ statistical uncertainty ranges
 of the median values
 by using the #1
 with #3 (upper) and #5 (lower) total events on average
 in each 60--day observation period of the #7 seasons.
}
\label{tab:N_v-Bayesian-\ShortFrame-\Fittingfv-\PlotNumber}
\end{table}
}
\newcommand{\InsertResultsTableNphithetaLab} {
\begin{table} [h!]
\begin{center}
\renewcommand{\arraystretch}{2}
\setlength{\tabcolsep}{3 pt}
{\footnotesize
 \begin{tabular}{|| c || c | c ||}
  \hline
  \hline
  \multirow {2} {*}
   {\makebox[ 4.5 cm] [c]
    {\begin{minipage} {4.5 cm}
      \begin{center}
         Central date               \\ \vspace{0.2  ex}
        (observation period)        \\ \vspace{0.2  ex}
        (day)
      \end{center}
     \end{minipage}}}                               &
  \multicolumn {2} { c ||}
   {\makebox[12   cm] [c] {Most--event directions}} \\
  \cline{2-3}
                                                    &
   {\makebox[ 6   cm] [c] {50  events}}             &
   {\makebox[ 6   cm] [c] {500 events}}             \\
  \hline
  \hline
  \begin{minipage} {4.5 cm}
   \begin{center}
      ~          \\ \vspace{3.1  ex}
      \PeriodAa  \\ \vspace{3.9  ex}
   \end{center}
  \end{minipage}            &
  \begin{minipage} {6   cm}
   \begin{center}
    \DirectionAaTens
   \end{center}
  \end{minipage}            &
  \begin{minipage} {6   cm}
   \begin{center}
    \DirectionAaHundreds
   \end{center}
  \end{minipage}            \\
  \hline
  \hline
  \begin{minipage} {4.5 cm}
   \begin{center}
      ~          \\ \vspace{1.2  ex}
       49.49     \\ \vspace{0.65 ex}
     (\PeriodCa) \\ \vspace{1.5  ex}
   \end{center}
  \end{minipage}            &
  \begin{minipage} {6   cm}
   \begin{center}
    \DirectionCaTens
   \end{center}
  \end{minipage}            &
  \begin{minipage} {6   cm}
   \begin{center}
    \DirectionCaHundreds
   \end{center}
  \end{minipage}            \\
  \hline
  \begin{minipage} {4.5 cm}
   \begin{center}
      ~          \\ \vspace{1.2  ex}
      140.74     \\ \vspace{0.65 ex}
     (\PeriodCb) \\ \vspace{1.5  ex}
   \end{center}
  \end{minipage}            &
  \begin{minipage} {6   cm}
   \begin{center}
    \DirectionCbTens
   \end{center}
  \end{minipage}            &
  \begin{minipage} {6   cm}
   \begin{center}
    \DirectionCbHundreds
   \end{center}
  \end{minipage}            \\
  \hline
  \begin{minipage} {4.5 cm}
   \begin{center}
      ~          \\ \vspace{1.2  ex}
      231.99     \\ \vspace{0.65 ex}
     (\PeriodCc) \\ \vspace{1.5  ex}
   \end{center}
  \end{minipage}            &
  \begin{minipage} {6   cm}
   \begin{center}
    \DirectionCcTens
   \end{center}
  \end{minipage}            &
  \begin{minipage} {6   cm}
   \begin{center}
    \DirectionCcHundreds
   \end{center}
  \end{minipage}            \\
  \hline
  \begin{minipage} {4.5 cm}
   \begin{center}
      ~          \\ \vspace{1.2  ex}
      323.24     \\ \vspace{0.65 ex}
     (\PeriodCd) \\ \vspace{1.5  ex}
   \end{center}
  \end{minipage}            &
  \begin{minipage} {6   cm}
   \begin{center}
    \DirectionCdTens
   \end{center}
  \end{minipage}            &
  \begin{minipage} {6   cm}
   \begin{center}
    \DirectionCdHundreds
   \end{center}
  \end{minipage}            \\
  \hline
  \hline
 \end{tabular}
 }
\end{center}
\caption{
 The summary of
 the directions of the simulated 3-D WIMP velocity
 with the highest event numbers
 ($>$ 4 times of the all--sky average value)
 in the horizontal coordinate system
 in one entire year and four advanced seasons
 at the location of the \LabName\ laboratory
 \LabLocation.
 50 and 500 total events on average
 in each observation period of 60 days
 have been simulated.
}
\label{tab:N_phi_thetan-H-\PlotNumberCa-\LabName}
\end{table}
}
\begin{document}
\thispagestyle{empty}
\begin{flushright}
 May 2019
\end{flushright}
\begin{center}
{\Large\bf
 Simulations of
 the 3-Dimensional Velocity Distribution of     \\
 Halo Weakly Interacting Massive Particles for  \\ \vspace{0.2 cm}
 Directional Dark Matter Detection Experiments} \\
\vspace*{0.7 cm}
 {\sc Chung-Lin Shan}                           \\~\\
\vspace{0.5 cm}
 {\it Physics Division,
      National Center for Theoretical Sciences  \\
      No.~101, Sec.~2, Kuang-Fu Road,
      Hsinchu City 30013, Taiwan, R.O.C.}       \\~\\
 {\it E-mail:} {\tt clshan@phys.nthu.edu.tw}
\end{center}
\vspace{2 cm}
\begin{abstract}

 In this paper,
 as a preparation of developing data analysis procedures
 for using 3-dimensional information
 offered by directional Dark Matter (DM) detection experiments,
 we study the patterns of
 the angular distribution of
 the Monte Carlo--generated 3-D velocity of
 halo Weakly Interacting Massive Particles (WIMPs)
 as well as
 apply the Bayesian fitting technique to reconstruct
 the radial distribution of the 3-D WIMP velocity.
 Besides
 the diurnal modulation of
 the angular WIMP velocity distribution,
 the so--called ``directionality'' of DM signals
 proposed in literature,
 we will also demonstrate
 possible ``annual'' modulations of
 both of the angular and the radial distributions
 of the 3-D WIMP velocity.
 Our Bayesian reconstruction results of
 (the annual modulation of)
 the radial WIMP velocity distribution
 will also be discussed in detail.
 For readers' reference,
 the angular distribution patterns of the 3-D WIMP velocity
 in the ``laboratory (location)--dependent'' reference frames
 of several underground laboratories
 are given in the Appendix.

\end{abstract}
\clearpage
\tableofcontents
\addtocontents{toc}{}
\clearpage
\section{Introduction}

 So far
 Weakly Interacting Massive Particles (WIMPs) $\chi$
 arising in several extensions of
 the Standard Model of particle physics
 are still one of the most favorite
 candidates for cosmological Dark Matter (DM).
 In the last (more than) three decades,
 a large number of experiments has been built
 and is being planned
 to search for different WIMP candidates
 by direct detection of
 the scattering recoil energy
 of ambient WIMPs off target nuclei
 in low--background underground laboratory detectors
 (see Refs.~%
  \cite{SUSYDM96,
        Gaitskell04,
        Baudis12c, Baudis15,
        Drees18a,
        Schumann19}
  for reviews).

 Among these direct detection experiments
 measuring recoil energies deposited in detectors,
 the ``directional'' detection of Galatic DM particles
 has been proposed more than one decade
 to be a promising experimental strategy
 for discriminating signals from backgrounds
 by using additional 3-dimensional information
 (the recoil tracks and/or the head--tail senses)
 of (elastic) WIMP--nucleus scattering events
 (see Refs.~%
  \cite{Ahlen09,
        Mayet11, Vahsen14, Phan15,
        Mayet16, Hochberg16, Battat16b}).
 Several experimental collaborations
 investigate different detector materials and techniques
 and have achieved recently great progress
 \cite{Mayet16, Battat16b}.

 For instance,
 as the first directional detection experiment,
 the DRIFT Collab.~demonstrated
 the reconstruction ability of the range component signature
 by using their DRIFT--IId detector
 with a gas mixture of CS$_2$ + CF$_4$ + O$_2$
 \cite{DRIFT-III,
       Burgos07, Burgos08b,
       Majewski09,
       Battat14, Battat16a, Battat17a, Battat17b}.
 Meanwhile,
 the MIMAC experiment
 started with $\rmXA{He}{3}$ as detector material
 for searching for light WIMPs
 \cite{MIMAC,
       Moulin05, Santos07},
 but turned later to develop micromegas $\mu$TPC detetor
 with a gas mixture of CF$_4$ + CHF$_3$ + C$_4$H$_{10}$
 and also demonstrated the ability
 of determining 3-D positions of primary ionization electrons
 for reconstructing in turn
 the track of the recoiled nucleus
 \cite{Mayet09a,
       Santos10,
       Billard12a,
       Riffard16,
       YTao19}.

 In contrast to the use of gas mixtures,
 pure CF$_4$ gas has also been studied
 by the NEWAGE Collab.~with also the $\mu$TPC technique
 \cite{Miuchi10, Miuchi11, Nakamura15b}
 as well as
 by the DMTPC Collab.~%
 using an optical (CCD pixel image) readout
 combined with a transient charge readout
 \cite{DMTPC,
       Dujmic08,
       Sciolla08c,
       Sciolla09b,
       Ahlen10,
       Deaconu15, Deaconu17}.
 Recently,
 while
 the D3 Collab.~considers $\rm S F_6$ gas
 with sulfur as target nucleus
 \cite{Vahsen11,
       Ross14},
 the NEWSdm Collab.~investigates
 the nuclear emulsion technique
 with AgBr(I) as detector material
 \cite{NaKa11,
       Aleksandrov16, Agafonova17}.
 Furthermore,
 a larger CYGNUS Collab.~has been built
 for combining members, efforts and achievements
 from different experimental collaborations
 \cite{CYGNUS}.

\InsertSKPPlotD
 {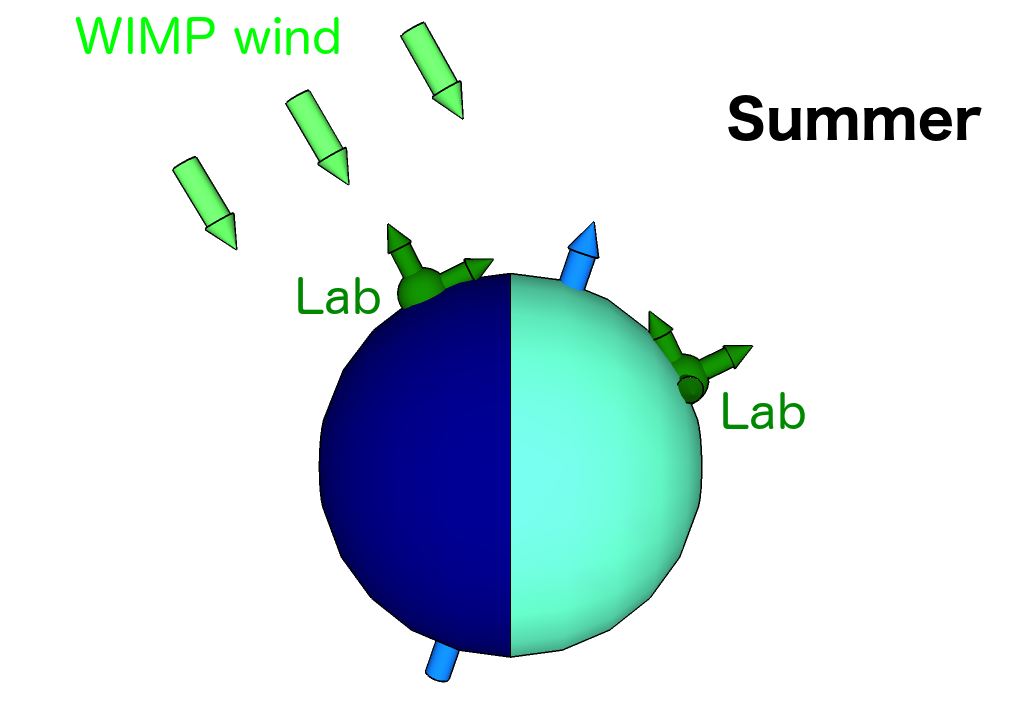}
 {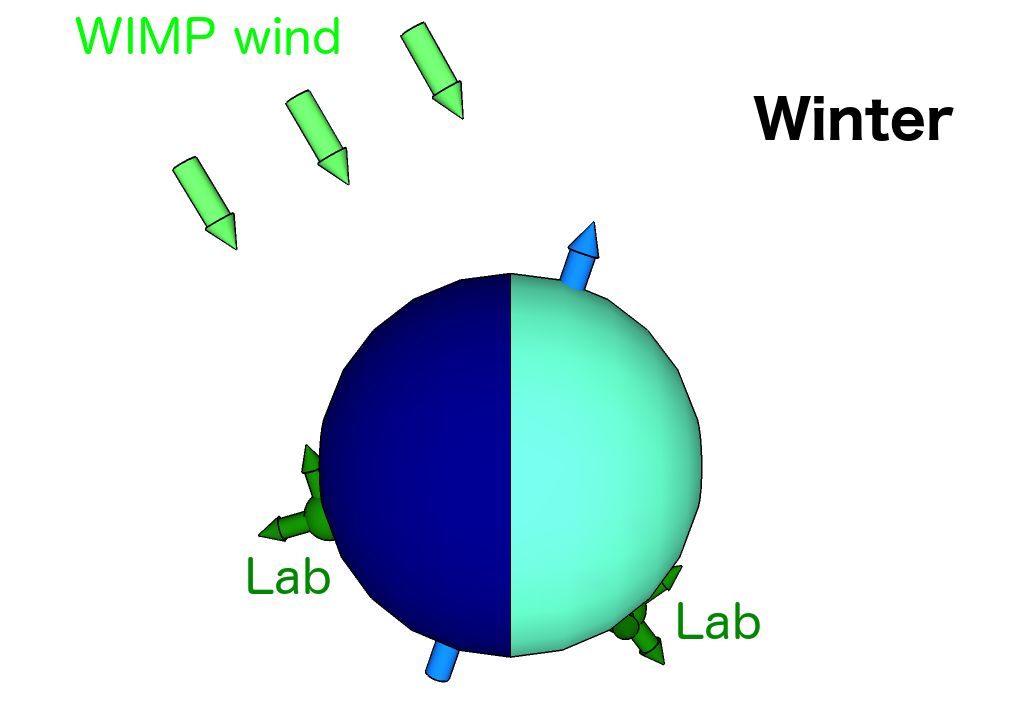}
 {fig:directional-1293-summer}
 {Sketches of
  the basic concepts of
  directional Dark Matter detection:
  (a)
  the diurnal modulation of
  the main direction of incident WIMPs,
  the so--called ``directionality'' of the WIMP wind,
  for a laboratory located in the Northern Hemisphere
  (in Summer);
  (b)
  except of the directionality of the WIMP wind,
  the event number of WIMP signals
  observed at a laboratory located in the Southern Hemisphere
  (in Winter)
  could also have the diurnal modulation
  caused by the Earth's shielding of the WIMP flux.
  The darkened/lightened (left/right--hand) spheres
  indicate that
  the laboratory of interest is in the night/day.
  See Figs.~\ref{fig:directional-1293-winter}
  and \ref{fig:v_Earth_chi_S-20766}
  as well as
  the text there
  for more details.
  }

 The basic concept of
 directional Dark Matter detection
 is based on the rotation of the Earth.
 As sketched in Figs.~\ref{fig:directional-1293-summer},
 there are two kinds of
 possible ``diurnal'' modulation of WIMP signals
 to observe:
 the diurnal modulation of the main incident direction of WIMP events,
 the so--called ``directionality'' of the WIMP wind,
 as well as
 that of the number (scattering rate) of WIMP events
 caused by Earth's shielding of the WIMP flux.
 Directional DM detection experiments
 aim hence,
 as the first step,
 to identify positive {\em modulated anisotropic} WIMP signals
 and discriminate them
 from theoretically (approximately) isotropic background events.

 Once positive WIMP signals could be observed
 and more and more events could be accumulated,
 reconstructions of particle and/or
 astronomical properties of halo WIMPs
 would be considered as the next stage.
 For this purpose,
 different methods have been developed.
 In literature,
 a large number of earlier studies is focused on
 (the use of) the distribution patterns of
 the nuclear recoil energy and the WIMP flux,
 e.g.,
 works done by N.~Bozorgnia, G.~B.~Gelmini and P.~Gondolo
 \cite{Gondolo02, Alenazi07, Bozorgnia11, Bozorgnia12}
 as well as
 by B.~Morgan, A.~M.~Green, C.~A.~J.~O'Hare and B.~J.~Kavanagh
 \cite{Morgan04, Morgan05,
       OHare14, Kavanagh15, OHare15b, Kavanagh16, OHare17}.
 Meanwhile,
 J.~Billard, F.~Mayet and D.~Santos developed their Bayesian analysis,
 which considers
 not only the 3-D velocity dispersion of
 the local WIMP velocity distribution
 and the main direction of the recoiled nucleus,
 but also
 the WIMP mass and the WIMP--nucleon cross section
 as fitting parameters simultaneously
 \cite{Billard09,
       Billard10b,
       Billard10d}.
 Moreover,
 some authors focused on theoretically analyzing of
 the (local) Dark Matter velocity distribution,
 e.g.,
 works done by
 S.~K.~Lee and A.~H.~G.~Peter
 \cite{SKLee12, SKLee14}.

 Besides these works,
 in this paper,
 as a preparation for our future study
 on the development of data analysis procedures
 for using and/or combining 3-D information
 offered by directional detection experiments,
 to,
 e.g.,
 reconstruct the 3-dimensional WIMP velocity distribution,
 we investigate the patterns of
 the angular distribution of
 the Monte Carlo--generated 3-dimensional WIMP velocity
 as well as
 apply the Bayesian fitting technique to reconstruct
 the radial distribution (magnitude)
 of the 3-D WIMP velocity.
 Not only
 the diurnal modulation of
 the angular WIMP velocity distribution
 mentioned above,
 we will also demonstrate
 possible ``annual'' modulations of
 both of the angular and the radial distributions
 of the 3-D WIMP velocity.

 The remainder of this paper is organized as follows.
 In Sec.~2,
 we describe all needed tools
 for our simulations and analysis procedures
 presented in this paper:
 our definitions of different celestial coordinate systems,
 the event generation process
 based on the Monte Carlo method,
 as well as
 the Bayesian fitting procedure
 used for reconstructing
 the radial component of
 the 3-dimensional WIMP velocity distribution.
 Then
 we present at first
 the (annual/diurnal modulated) patterns of
 the angular WIMP velocity distribution
 in different celestial coordinate systems
 one by one in Sec.~3
 and the reconstructions of
 (the annual modulation of)
 the radial WIMP velocity distribution
 in both of the Equatorial and the Galactic coordinate systems
 in Sec.~4.
 In Sec.~5,
 we raise the total number
 of generated WIMP events
 and demonstrate
 the simulation results
 with an improved analysis resolution.
 We conclude in Sec.~6.
 Some technical details for our analyses
 and
 a summary of the laboratory--dependent simulation results
 will be given in Appendix.

%
% 1/10
 %
%
\section{Toolbox}
\label{sec:toolbox}

 In this section,
 we describe all tools
 needed in our simulations and data analyses
 presented in this paper.
 At first
 we give our definitions of
 different celestial coordinate systems
 considered for demonstrating
 the 3-dimensional WIMP velocity distribution.
 Then
 we describe
 the event generation process
 based on the Monte Carlo method
 as well as
 the Bayesian fitting procedure
 used for reconstructing
 the radial distribution of the 3-D WIMP velocity.

\subsection{Our definitions of
            different celestial coordinate systems}
\label{sec:XYZ}

 In this subsection,
 we give our definitions of
 four ``laboratory--independent''
 (Galactic,
  Ecliptic,
  Equatorial
  and Earth)
 and two ``laboratory--dependent''
 (horizontal and laboratory)
 coordinate systems%
\footnote{
 Note that
 some of our definitions are different
 from their astronomical conventions,
 but would be more convenient
 for our data analyses,
 in particular,
 for comparing results
 obtained by using data
 from different underground laboratories.
}.
 The analytic and/or the numerical forms
 of all transformation matrices
 between these coordinate systems
 will be derived in detail
 in Appendix \ref{appx:XYZ}.

\subsubsection{Definition of the Galactic coordinate system}
\label{sec:XYZ_G}
\InsertSKPPlotS
 {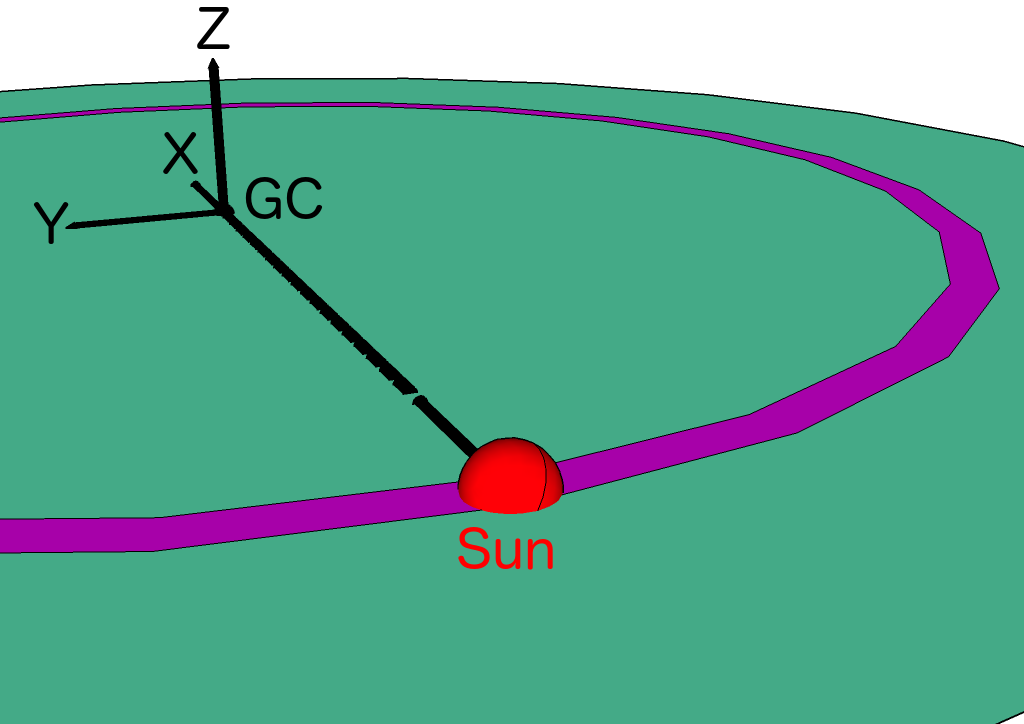}
 {The sketch of
  the definition of
  the (black) Galactic coordinate system:
  the origin is at the Galactic Center (GC)
  (not the center of the Sun),
  the primary direction (the $\xG$--axis)
  points from the Solar center to
  the approximate center of our Galaxy,
  the $\zG$--axis
  points to the Galactic North Pole (GNP),
  and the $\yG$--axis
  is then defined by
  the right--handed convention.
  The fundamental ($\xG - \yG$) plane
  is the approximate Galactic plane
  and the magenta circular band
  indicates an approximate path of
  the orbital motion of the Solar system
  in the Galaxy.
  }

 As shown in Fig.~\ref{fig:G-rotated},
 the Galactic coordinate system
 in our simulations presented in this paper
 is defined as follows:
  the origin is at the Galactic Center (GC),
 the primary direction (the $\xG$--axis)
 points from the Solar center to
 the approximate center of our Galaxy,
 and
 the $\zG$--axis
 points to the Galactic North Pole (GNP).
 As usual,
 the right--handed convention
 is used for defining the $\yG$--axis
 and the fundamental ($\xG - \yG$) plane
 is the approximate Galactic plane
 \cite{Wiki-Galactic}.
 Note that,
 as discussed in detail
 in Appendix \ref{appx:v_Sun_G}
 and sketched in Fig.~\ref{fig:v_Sun_Eq-S-G-rotated},
 the direction of the Solar movement
 towards the CYGNUS constellation
 is {\em not parallel to},
 but only {\em approximately along}
 the $\yG$--axis of the Galactic coordinate system.

\subsubsection{Definitions of the Ecliptic and the Equatorial coordinate systems}
\label{sec:XYZ_S-Eq}
\InsertSKPPlotD
 {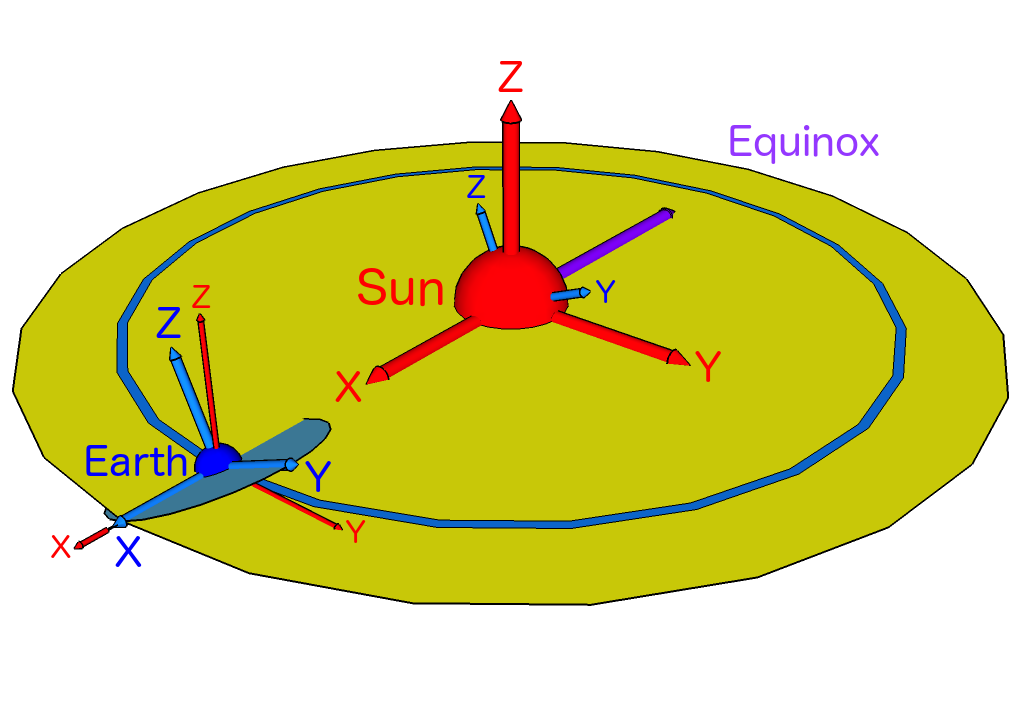}
 {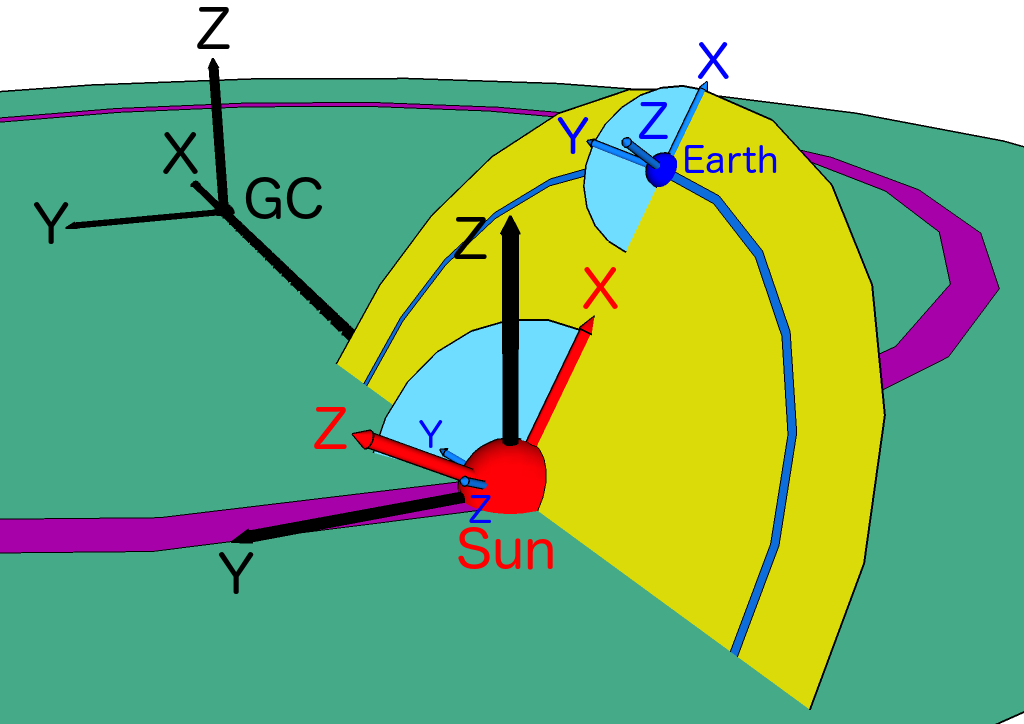}
 {fig:Eq-S}
 {(a)
  The sketch of
  the definitions of
  the (red) Ecliptic and
  the (blue) Equatorial coordinate systems:
  their origins are located at
  the center of the Sun
  and that of the Earth,
  respectively,
  the common primary direction (the $\xS$/$\xEq$--axis)
  is the direction
  pointing from the Solar center to that of the Earth
  at 12 midnight (the end) of
  the date of the vernal equinox,
  the $\zS$-- and the $\zEq$--axes
  are perpendicular to the (yellow) Ecliptic
  or the (blue) Equatorial plane,
  respectively,
  and their $\yS$-- and $\yEq$--axes
  are then defined as usual by
  the right--handed convention.
  The blue circular band indicates
  the Earth's orbit around the Sun.
  The purple arrow
  points to the celestial Equinox,
  which is the conventional (common) primary direction of
  the Ecliptic and the Equatorial coordinate systems.
  (b)
  The sketch of
  the relative orientations between
  the (black) Galactic,
  the (red) Ecliptic and the (blue) Equatorial coordinate systems
  (on the date of the vernal equinox).
  }

 As shown in Fig.~\ref{fig:Eq-S}(a),
 the Ecliptic and
 the Equatorial coordinate systems
 are defined as follows:
 their origins are located at
 the center of the Sun
 and that of the Earth,
 respectively,
 the common primary direction (the $\xS$/$\xEq$--axis)
 is the direction
 pointing from the Solar center to that of the Earth
 at 12 midnight (the {\em end}) of
 the date of the vernal equinox%
\footnote{
 Note that,
 in our simulations presented in this paper,
 this date has been fixed exactly as
 the 79th day (the May 20th) of a 365-day year.
},
 the $\zS$-- and the $\zEq$--axes
 are perpendicular to the Ecliptic
 or the Equatorial plane,
 respectively,
 and their $\yS$-- and $\yEq$--axes
 are then defined as usual by
 the right--handed convention.

 Additionally,
 in Fig.~\ref{fig:Eq-S}(b),
 we sketch
 the relative orientations between
 the Galactic,
 the Ecliptic and the Equatorial coordinate systems
 (on the date of the vernal equinox).
 Note that,
 the Ecliptic and the Equatorial coordinate systems
 only move,
 but do not rotate with the Sun nor with the Earth.

\subsubsection{Definition of the Earth coordinate system}
\label{sec:XYZ_E}
\InsertSKPPlotS
 {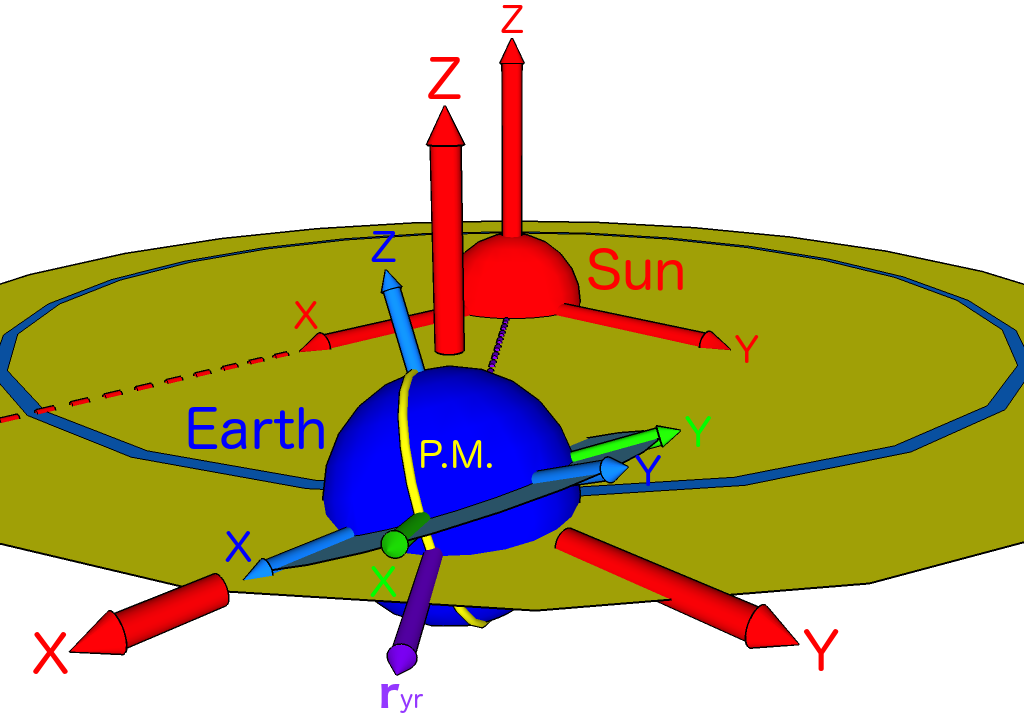}
 {The sketch of
  the definition of
  the (light--green) Earth coordinate system:
  while
  the origin is also
  located at the Earth's center
  and
  the $\zE$--axis is still
  the Earth's north polar axis,
  the primary direction (the $\xE$--axis)
  points now from the Earth's center
  to the (yellow) Prime Meridian
  (the longitude 0$^{\circ}$)
  at 12 midnight
  (i.e.,
   the moment
   when the Prime Meridian passes the purple arrow
   pointing from the Solar center to that of the Earth,
   $\ryear$)
  of ``each single'' day.
  Finally,
  the fundamental ($\xE - \yE$) plane is also
  the (blue) Equatorial plane
  and the right--hand convention is used
  to define the $\yE$--axis
  (see also Figs.~\ref{fig:E-H-north} and \ref{fig:E-Lab}).
  The (red) Ecliptic and the (blue) Equatorial coordinate systems
  as well as
  the (blue) Earth's orbit around the Sun
  are also given here.
  }

 As shown in Fig.~\ref{fig:E-Eq-S},
 we define the Earth coordinate system
 as follows:
 while
 the origin is also
 located at the Earth's center
 and
 the $\zE$--axis is still
 the Earth's north polar axis,
 the primary direction (the $\xE$--axis)
 points now from the Earth's center to the Prime Meridian
 (the longitude 0$^{\circ}$)
 at 12 midnight
 (i.e.,
  the moment
  when the Prime Meridian passes the direction
  pointing from the Solar center to that of the Earth)
 of ``each single'' day.
 Finally,
 the fundamental ($\xE - \yE$) plane is also
 the Equatorial plane
 and the right--hand convention is used
 to define the $\yE$--axis
 (see also Figs.~\ref{fig:E-H-north} and \ref{fig:E-Lab}).

 Note that,
 the Earth coordinate system
 is fixed but {\em rotates} with the Earth,
 during the Earth's orbital motion around the Sun.
 Hence,
 in our simulations presented in this paper,
 we consider the position of the Earth coordinate system
 at 12 midnight (the {\em beginning}) of
 (the Coordinated Universal Time (UTC) of)
 each single ``Solar'' day%
\footnote{
 Originally,
 the Earth coordinate system
 was defined for connecting
 the horizontal and the laboratory coordinate systems
 (defined in Secs.~\ref{sec:XYZ_H} and \ref{sec:XYZ_Lab})
 with the Equatorial and the Ecliptic coordinate systems
 \cite{Bandyopadhyay10}.
}.
 This means then that
 our Earth coordinate system
 changes daily and {\em discretely}
 (see Table \ref{tab:CSs}
  for the summary of
  the styles of the movement and the rotation
  of different coordinate systems).

\subsubsection{Definition of the horizontal coordinate system}
\label{sec:XYZ_H}
\InsertSKPPlotS
 {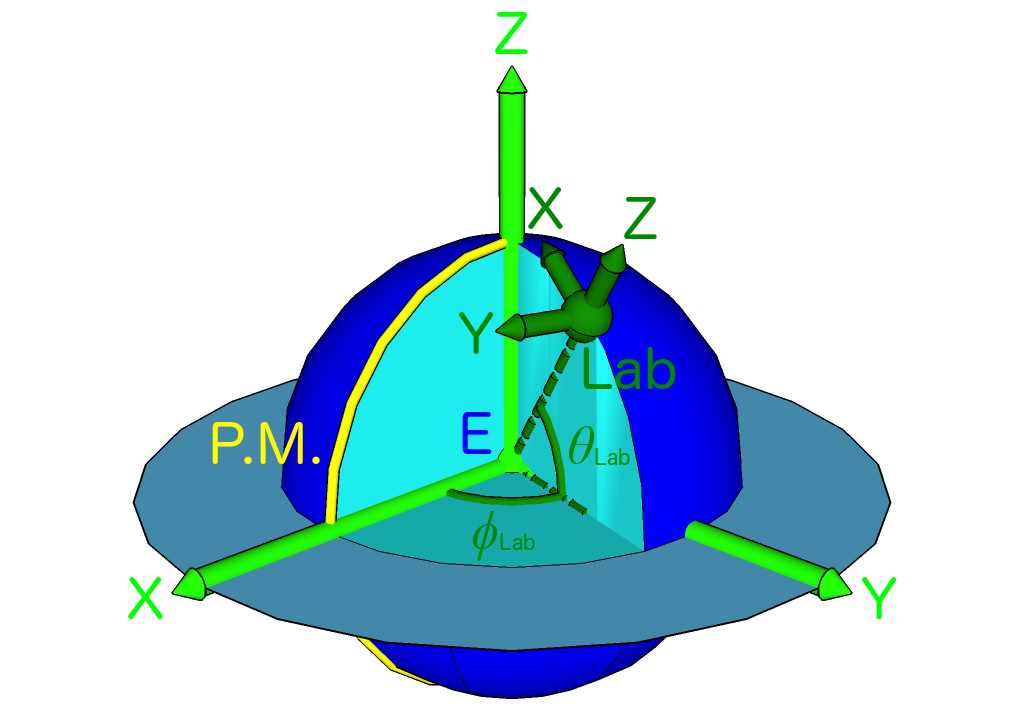}
 {The sketch of
  the definition of
  the (dark--green) horizontal coordinate system:
  the origin is the geographic location
  of the laboratory of interest
  at 12 midnight (the beginning) of
  (the UTC time of)
  each single day,
  the primary direction (the $\xH$--axis)
  and the $\zH$--axis
  point towards north
  and the zenith,
  respectively,
  and the right--handed convention is used
  for defining the $\yH$--axis.
  Our (light--green)
  Earth coordinate system
  is also given here
  and
  $\phiLab$ and $\thetaLab$ indicate
  the longitude and the latitude of
  the location of the laboratory,
  respectively.
  }

 Moreover,
 for demonstrating
 (the annual modulation of) the angular distribution
 of the 3-D WIMP velocity
 observed in a specified underground laboratory,
 we define first
 the horizontal coordinate system with
 the origin at the geographic location
 of the laboratory of interest
 at 12 midnight (the {\em beginning}) of
 (the UTC time of)
 each single day,
 the primary direction (the $\xH$--axis)
 and the $\zH$--axis
 pointing towards north
 and the zenith,
 respectively,
 and the right--handed convention
 for defining the $\yH$--axis
 (see Fig.~\ref{fig:E-H-north}).
 Note that,
 as the Earth coordinate system,
 for each single Solar day,
 our horizontal coordinate system is fixed
 at 12 midnight (the beginning) of the day
 and thus
 changes daily and discretely
 (see Table \ref{tab:CSs}
  for the summary of
  the movement and the rotation styles
  of different coordinate systems).

\subsubsection{Definition of the laboratory coordinate system}
\label{sec:XYZ_Lab}
\InsertSKPPlotS
 {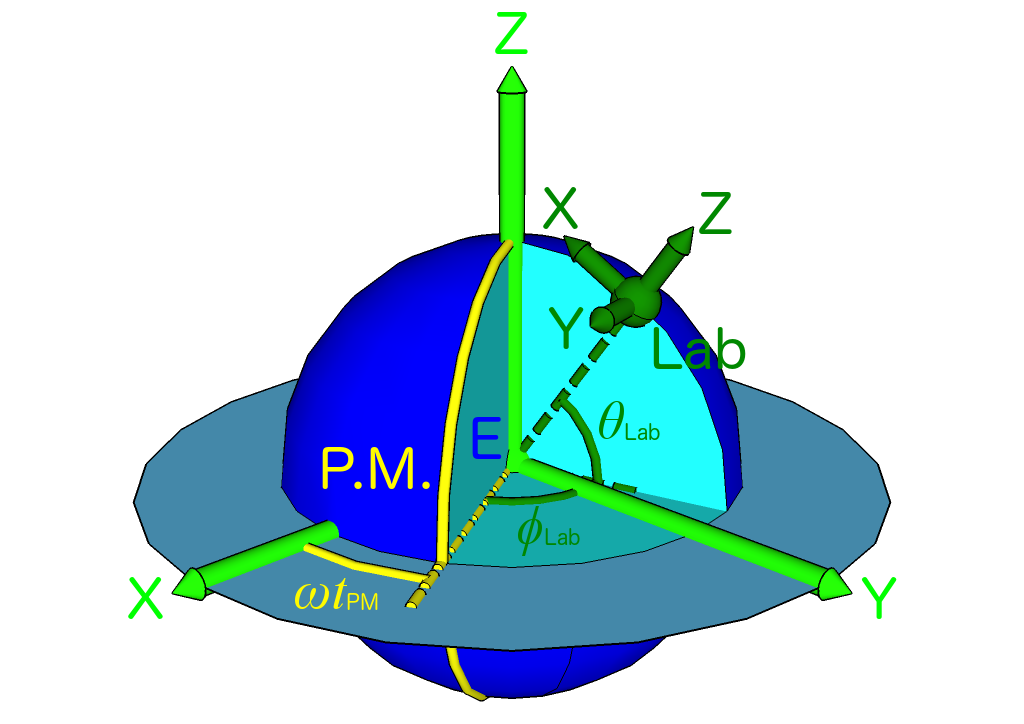}
 {The sketch of
  the definition of
  the (dark--green) laboratory coordinate system:
  the same as the horizontal coordinate system
  in Fig.~\ref{fig:E-H-north},
  but rotates with the laboratory of interest
  around the Earth's north polar ($\zEq/\zE$--)axis
  {\em instantaneously}.
  $\omega \tPM$ indicates
  the rotated angle of the (yellow) Prime Meridian
  from 12 midnight (the beginning) of
  (the UTC time of)
  each single day.
  Our (light--green)
  Earth coordinate system
  is also given here.
  }

 Finally,
 considering the long running time of
 (directional) direct Dark Matter detection experiments
 as well as
 for identifying the diurnal modulation of
 the angular WIMP velocity distribution,
 we define,
 besides the discretely variated horizontal coordinate system,
 the laboratory coordinate system
 by taking into account
 the {\em instantaneous} measuring time of
 each recorded WIMP scattering event.
 This means that
 our laboratory coordinate system is defined
 basically the same as the horizontal coordinate system,
 but rotates with the laboratory of interest
 around the Earth's north polar ($\zEq/\zE$--)axis instantaneously
 by an angle of $\omega \tPM$
 (see Fig.~\ref{fig:E-Lab}),
 where
 we define
\beq
         \omega
 \equiv  \frac{2 \pi}{1~{\rm day}}
\~,
\label{eqn:omega}
\eeq
 and
 $\tPM$ indicates
 the fractional part of
 the measuring UTC time $t$ of each recorded WIMP event
 in unit of day.
 Note that,
 our laboratory coordinate system changes
 (rotates around the Earth's north polar axis)
 {\em event by event}.

\subsubsection*{}
\begin{table} [t!]
\small
\begin{center}
\renewcommand{\arraystretch}{1.5}
 \begin{tabular}{|| c || c | c || c ||}
\hline
\hline
 \makebox[4   cm][c]{Coordinate system} &
 \makebox[2.5 cm][c]{Movement}          &
 \makebox[2.5 cm][c]{Rotation}          &
 \makebox[6   cm][c]{Style}             \\
\hline
\hline
 Galactic   & $\times$          & $\times^\dagger$ & --- \\
\hline
 Ecliptic   & $\surd$           & $\times$         & Orbital $\to$ approximately linear \\
 Equatorial & $\surd$           & $\times$         & Linear + orbital $\to$ spiral      \\
\hline
 Earth      & $\times^\ddagger$ & $\surd$          & Daily and discrete           \\
 Horizontal & $\times^\ddagger$ & $\surd$          & Daily and discrete           \\
 Laboratory & $\times^\ddagger$ & $\surd$          & Instantaneous and continuous \\
\hline
\hline
\end{tabular}
\end{center}
\caption{
 The summary of
 the styles of the movement and the rotation
 of all six celestial coordinate systems
 defined in our simulations presented in this paper.
\\
 $^\dagger$:
 The tiny angle
 swept by the connection
 between the Solar and the Galactic centers
 during the orbital motion of the Solar system
 in the Galaxy
 is ignored here.
\\
 $^\ddagger$:
 Fixed on the Earth
 and combined additionally with
 the ``linear + orbital $\to$ spiral'' movement of
 the Equatorial coordinate system.
}
\label{tab:CSs}
\end{table}

 In Table \ref{tab:CSs},
 we summarize
 the styles of the movement and the rotation
 of all six celestial coordinate systems
 defined in our simulations presented in this paper.
 Note that,
 except of
 the ``linear + orbital $\to$ spiral'' motion of
 the Equatorial coordinate system,
 the Earth,
 the horizontal and the laboratory coordinate systems
 are fixed on the Earth
 and only rotate around
 the Earth's north polar
 (i.e., $\zEq/\zE$--)axis.
 Hence,
 the radial component (magnitude)
 of the 3-D WIMP velocity distribution
 in these three coordinate systems
 are equal to that
 in the Equatorial coordinate system.

\subsection{Event generation in the Galactic coordinate system}
\label{sec:N_v_phi_theta-G}

 In this subsection,
 we describe
 the generation procedure of
 the 3-dimensional information
 on the WIMP velocity
 (the magnitude and the direction
  as well as
  the measuring time)
 in the Galactic coordinate system
 based on the Monte Carlo method.
 These generated 3-D WIMP velocities
 will be transformed
 into different celestial coordinate systems
 for data analyses and distribution reconstructions
 presented in Secs.~\ref{sec:N_phi_theta-050},
 \ref{sec:N_v-Bayesian-050},
 and \ref{sec:N_v_phi_theta-500}
 as well as
 in Appendices \ref{appx:N_phi_theta-ULabs}
 and \ref{appx:N_v-Bayesian-Eq-07900}.

\subsubsection{Radial component of the 3-D WIMP velocity distribution}
\label{sec:N_v-G}

 For generating
 the radial component (magnitude) of the 3-D WIMP velocity
 in the Galactic coordinate system,
 we consider
 the simple Maxwellian velocity distribution
 truncated at the Galactic escape velocity
 \cite{SUSYDM96}:
\beq
     f_{1, \Gau}(v)
  =  \cleft{\renewcommand{\arraystretch}{1.75}
            \begin{array}{l c l}
             \D
             N_{\Gau}
             \afrac{v^2}{v_0^3}
             e^{-v^2 / v_0^2}      \~, & ~~~~ ~~~~ & % 8
             {\rm for}~v \le \vesc \~, \\
             0                     \~, &           &
             {\rm for}~v >   \vesc \~,
            \end{array}}
\label{eqn:f1v_Gau_vesc}
\eeq
 with the normalization constant
\beq
     N_{\Gau}
  =  \bbrac{  \afrac{\sqrt{\pi}}{4} \erf\afrac{\vesc}{v_0}
            - \afrac{\vesc}{2 v_0}  e^{-\vesc^2 / v_0^2}   }^{-1}
\~,
\label{eqn:N_Gau_vesc}
\eeq
 where
 $v_0 \simeq 220$ km/s
 is the Solar orbital speed around the Galactic center
 and $498~{\rm km/s} < \vesc < 608~{\rm km/s}$
 is the escape velocity from our Galaxy
 at the position of the Solar system
 \cite{RPP18AP}.
\begin{figure} [t!]
\begin{center}
 \includegraphics [width = 15 cm] {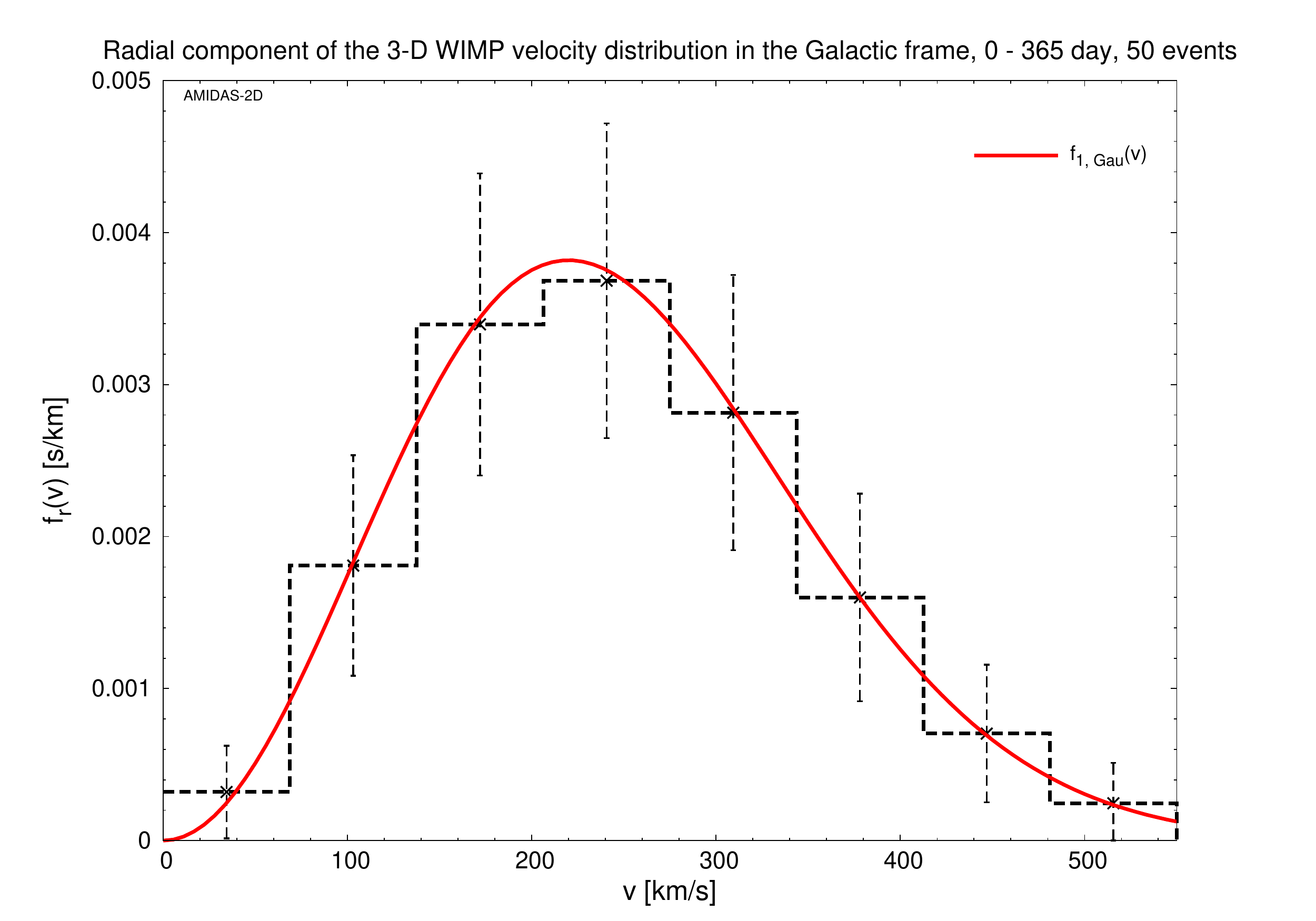}
\end{center}
\caption{
 The radial distribution
 of the 3-dimensional WIMP velocity
 in the Galactic coordinate system
 generated by Eqs.~(\ref{eqn:f1v_Gau_vesc}) to (\ref{eqn:f1v_t_G}).
 50 total events on average in one experiment
 (in one entire year)
 have been generated
 and binned into 8 bins.
 The solid red curve
 is the generating
 simple Maxwellian velocity distribution
 $f_{1, \Gau}(v)$
 given in Eq.~(\ref{eqn:f1v_Gau_vesc})
 with the Solar Galactic velocity $v_0 = 220$ km/s,
 while
 the dashed black histogram shows
 the generated WIMP velocities
 and the thin vertical dashed black lines
 indicate the 1$\sigma$ Poisson statistical uncertainties
 on the recorded event numbers in the $v$--bins.
 The Galactic escape velocity
 has been set as
 $\vesc = 550$ km/s.
 See the text for further details.
}
\label{fig:N_v-G-050-00000}
\end{figure}

 In Fig.~\ref{fig:N_v-G-050-00000},
 we show
 the radial distribution
 of the 3-dimensional WIMP velocity
 in the Galactic coordinate system
 generated by Eqs.~(\ref{eqn:f1v_Gau_vesc}) to (\ref{eqn:f1v_t_G}).
 One entire year (0 to 365 day)
 for the measuring time of WIMP events
 has been considered.
 50 total events on average in one experiment
 have been generated
 and binned into 8 bins.
 5,000 experiments have been simulated.
 The solid red curve
 is the generating
 simple Maxwellian velocity distribution
 $f_{1, \Gau}(v)$
 given in Eq.~(\ref{eqn:f1v_Gau_vesc})
 with the Solar Galactic velocity $v_0 = 220$ km/s,
 while
 the dashed black histogram shows
 the generated WIMP velocities
 and the thin vertical dashed black lines
 indicate the 1$\sigma$ Poisson statistical uncertainties
 on the recorded event numbers in the $v$--bins.
 The Galactic escape velocity
 has been set as
 $\vesc = 550$ km/s.
 Not surprisingly,
 the recorded event numbers
 (averaged by all simulated experiments)
 in all $v$--bins match
 our generating simple Maxwellian velocity distribution
 perfectly.

\subsubsection{Angular component of the 3-D WIMP velocity distribution}
\label{sec:N_phi_theta-G}

 Since
 the simplest model of the Galactic Dark Matter halo
 is assumed to be isothermal, spherical and isotropic,
 the angular distribution of the 3-D WIMP velocity
 in the Galactic coordinate system
 has been considered to be isotropic
 and thus
 the velocity directions
 (i.e.,
  the $\phi$-- and $\theta$--angles
  in the longitude and the latitude directions,
  respectively)
 are generated with a constant probability
 in our simulations:
\beq
     f_{\phi, {\rm G}}(\phi)
  =  1
\~,
     ~~~~ ~~~~ ~~ % 10
     \phi \in (-\pi,~\pi]
\~,
\label{eqn:f1v_phi_G}
\eeq
 and
\beq
     f_{\theta, {\rm G}}(\theta)
  =  1
\~,
     ~~~~ ~~~~ ~~ % 10
    \theta \in [-\pi / 2,~\pi / 2]
\~.
\label{eqn:f1v_theta_G}
\eeq
\begin{figure} [t!]
\begin{center}
 \includegraphics [width = 15 cm] {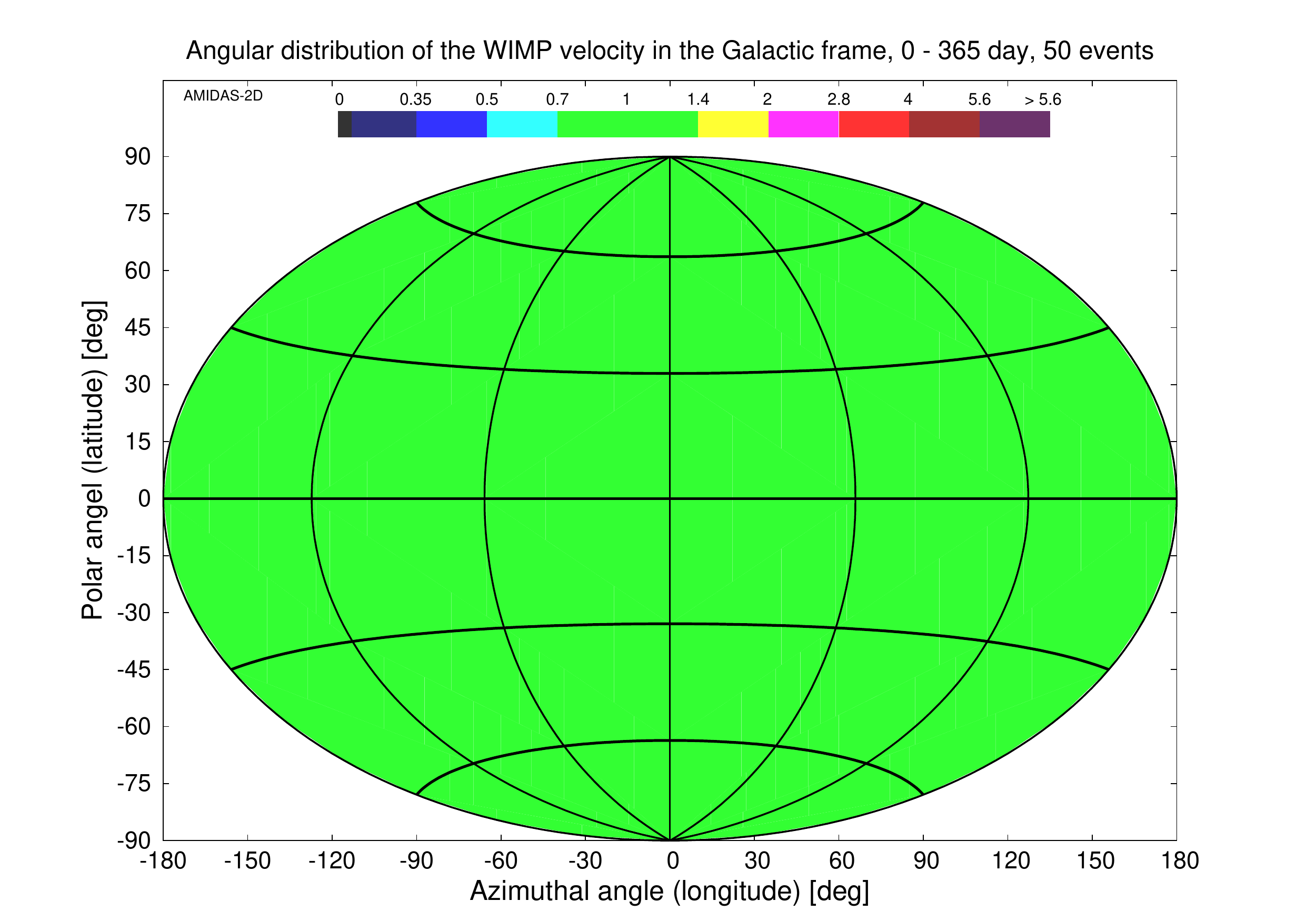}
\end{center}
\caption{
 The angular distribution
 of the 3-dimensional WIMP velocity
 in the Galactic coordinate system
 generated by Eqs.~(\ref{eqn:f1v_Gau_vesc}) to (\ref{eqn:f1v_t_G}).
 50 total events on average in one experiment
 (in one entire year)
 have been generated
 and binned into 6 $\times$ 6 bins
 in the longitude and the latitude directions,
 respectively.
 The horizontal color bar on the top of the plot
 indicates
 the mean value of the recorded event number
 (averaged over all simulated experiments)
 in each angular bin
 in unit of the all--sky average value
 (50 events / 36 bins $\cong$ 1.39 events/bin here).
 See the text for further details.
}
\label{fig:N_phi_theta-G-050-00000}
\end{figure}

 In Fig.~\ref{fig:N_phi_theta-G-050-00000},
 we show
 the angular distribution
 of the 3-dimensional WIMP velocity
 in the Galactic coordinate system
 generated by Eqs.~(\ref{eqn:f1v_Gau_vesc}) to (\ref{eqn:f1v_t_G}).
 50 total events on average in one experiment
 have been binned into 6 $\times$ 6 bins
 in the longitude and the latitude directions,
 respectively.
 The horizontal color bar on the top of the plot
 indicates
 the mean value of the recorded event number
 (averaged over all simulated experiments)
 in each angular bin
 in unit of the all--sky average value
 (50 events / 36 bins $\cong$ 1.39 events/bin here).

 As expected,
 the event numbers
 in all angular bins
 are between 70\% and 1.4 times
 of the all--sky average value
 (0.97 events/bin to 1.94 events/bin).
 More exactly,
 with $\cal O$(50) total WIMP events,
 the maximal perturbation of
 the angular distribution of
 the simulated isotropic WIMP halo
 would be less than $\pm 2.5\%$
 of the all--sky average.

\subsubsection{Measuring time and observation periods/shifts of WIMP events}
\label{sec:N_t-G}

 Although due to
 the orbital motion of the Earth around the Sun
 and the Earth's rotation around its axis,
 the theoretically predicted
 radial and angular WIMP velocity distributions
 as well as
 the event rate for WIMP--nucleus scattering
 should be time--dependent
 (the so--called ``annual modulation'')
 \cite{Freese88},
 in the Galactic point of view,
 WIMP--nucleus scattering events
 should be observed randomly and constantly.
 Hence,
 in our simulations
 we considered a constant probability
 for generating the measuring UTC time of
 the recorded WIMP scattering events:
\beq
     f_{t, {\rm G}}(t)
  =  1
\~,
     ~~~~ ~~~~ ~~ % 10
     t \in [t_{\rm start},~t_{\rm end}]
\~.
\label{eqn:f1v_t_G}
\eeq
 For generating
 the WIMP events shown
 in Figs.~\ref{fig:N_v-G-050-00000}
 and \ref{fig:N_phi_theta-G-050-00000},
 the observation period has been set as
 $[t_{\rm start},~t_{\rm end}] = [0, 365~{\rm day}]$.

\paragraph{Annual modulation of the 3-D WIMP velocity}
 ~\\

\InsertSKPPlotD
 {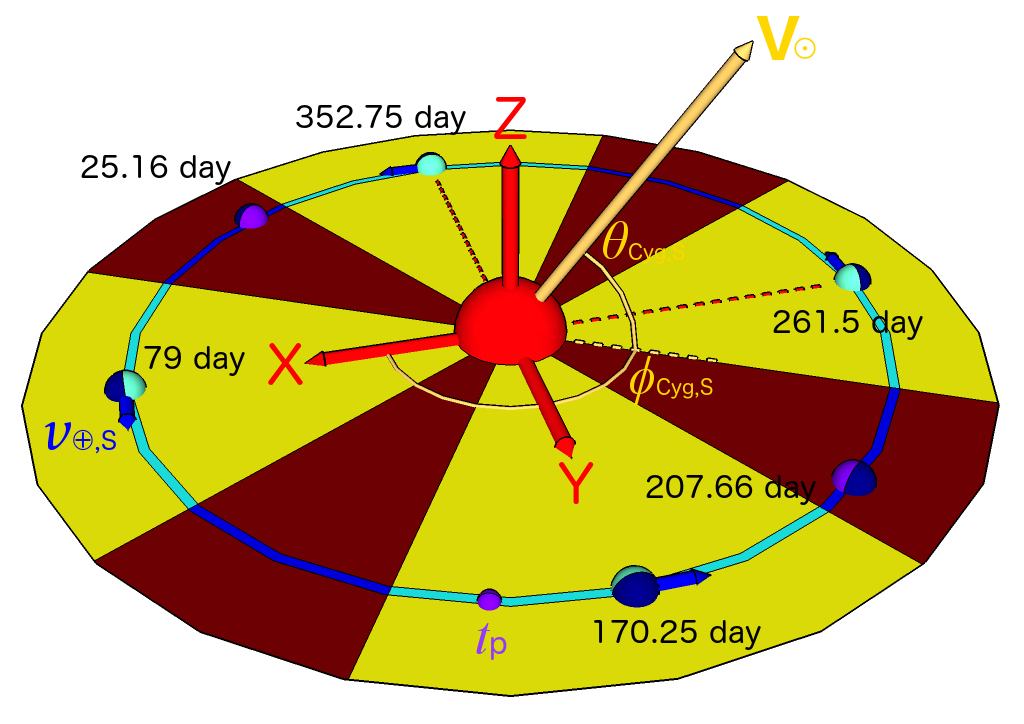}
 {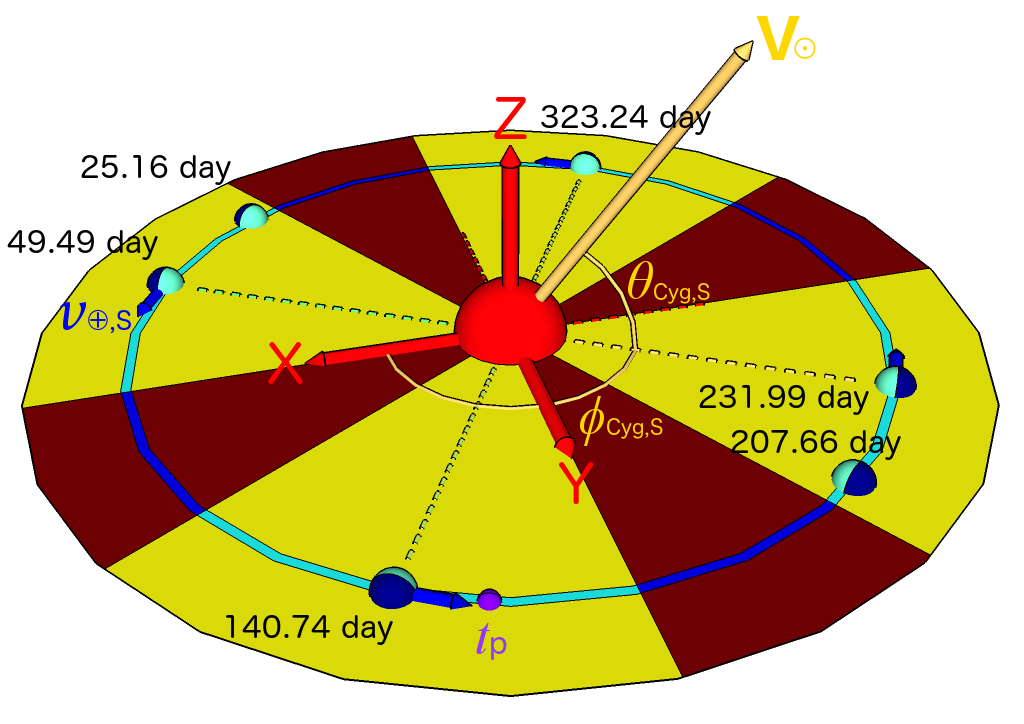}
 {fig:v_Earth_chi_S}
 {The sketches of
  two options for the observation periods
  (lightened areas)
  considered in our simulations presented in this paper
  (see also Table \ref{tab:period_year}).
  (a)
  Four normal seasons
  with a common 60-day period
  on the central dates of
   79.0  day,
  170.25 day,
  261.50 day,
  and
  352.75 day,
  respectively.
  (b)
  Four ``advanced'' seasons
  with a 60-day period
  on the central dates of
   49.49 day,
  140.74 day,
  231.99 day,
  and
  323.24 day,
  respectively.
  The golden arrow indicates
  the moving direction of the Solar system
  towards the CYGNUS constellation
  with the velocity of \mbox{$v_0 \simeq 220$ km/s},
  and
  the four short dark--blue arrows
  (in front of the Earths)
  indicate the (average) orbital velocity of the Earth,
  $|\VEarthS| \simeq 30$ km/s,
  on the central dates of
  four normal or four advanced seasons.
  While
  the small purple point at the bottom of each sketch
  indicates the location,
  where the Earth's orbital speed is maximal
  on the date around June 2nd ($t_{\rm p} = 152.5$ day)
  \cite{Freese88},
  two Earths without the velocity arrow
  indicate the locations,
  where the theoretical main direction of the WIMP wind
  (the opposite direction of the Solar movement)
  points straightly to the Prime Meridian
  in the night (207.66 day)
  or the day (25.16 day),
  corresponding (approximately) to the locations
  drawn in Figs.~\ref{fig:directional-1293-summer},
  \ref{fig:directional-1293-winter},
  and \ref{fig:v_Earth_chi_S-20766}.
  Other notations are the same as
  in Fig.~\ref{fig:Eq-S}(a).
  See the text
  as well as
  Appendices \ref{appx:v_Earth_chi_S}
  and \ref{appx:v_Earth_chi_S-20766}
  for further details.
  }
\begin{table} [t!]
\small
\begin{center}
\renewcommand{\arraystretch}{1.5}
 \begin{tabular}{|| c || c | c ||}
\hline
\hline
 \makebox[4.5 cm][c]{Option}             &
 \makebox[5   cm][c]{Central date (day)} &
 \makebox[5   cm][c]{Period       (day)} \\
\hline
\hline
 One entire year
 & ---    & \PeriodAa \\
\hline
 \multirow{4}{*}{Four normal seasons}
 &  79.0  & \PeriodBa            \\
 & 170.25 & \PeriodBb            \\
 & 261.50 & \PeriodBc            \\
 & 352.75 & \PeriodBd\ (= 17.75) \\
\hline
 \multirow{4}{*}{Four advanced seasons}
 &  49.49 & \PeriodCa \\
 & 140.74 & \PeriodCb \\
 & 231.99 & \PeriodCc \\
 & 323.24 & \PeriodCd \\
\hline
 \multirow{2}{*}{For diurnal modulation}
 & 207.66           & \PeriodDa            \\
 & 390.16 (= 25.16) & \PeriodDb\ (= 55.16) \\
\hline
\hline
\end{tabular}
\end{center}
\caption{
 The list of
 four options for the observation periods
 in a 365-day year
 considered in our simulations presented in this paper.
 Note that,
 first,
 in each of the normal and advanced season,
 a period of 60 days ($\pm 30$ days)
 has been set.
 This means that
 each pair of the corresponding season
 has an overlap of around 30 days
 (see the lightened and darkened areas
  in two sketches of Figs.~\ref{fig:v_Earth_chi_S}).
 Second,
 the last option is considered
 only for demonstrating the diurnal modulation of
 the angular distribution of
 the 3-D WIMP velocity
 (see Appendix \ref{appx:v_Earth_chi_S-20766}
  for the detailed calculation).
}
\label{tab:period_year}
\end{table}

 As shown in Fig.~\ref{fig:v_Earth_chi_S}(a),
 on the central dates of four normal seasons,
 the orbital velocity of the Earth's rotation around the Sun
 in the Ecliptic coordinate system
 is along the $\xS$-- or the $\yS$--axis.
 This would thus be the most natural choice
 for demonstrating
 the annual modulation of the Earth's velocity
 relative to the Dark Matter halo.
 However,
 as shown in Fig.~\ref{fig:v_Earth_chi_S}(b)
 and discussed in detail
 in Appendix \ref{appx:v_Earth_chi_S},
 the relative velocity of the Earth to the DM halo
 should be the maximum (minimum),
 when its orbital velocity is (anti--)parallel to
 the projection of the direction of the Solar movement
 on the Ecliptic plane,
 namely,
 on the date around the 21st of May (140.74 day)
 (the 20th of November, 323.24 day).
 Hence,
 for demonstrating
 the annual modulation of
 the radial and angular distributions
 of the 3-D WIMP velocity,
 besides the 60-day ($\pm$ 30 days)
 observation periods of four normal seasons
 on the central dates of
  79.0  day,
 170.25 day,
 261.50 day,
 and
 352.75 day,
 respectively,
 we considered also
 the periods of four ``advanced''
 ($\sim$ 30 days earlier) seasons
 on the central dates of
  49.49 day,
 140.74 day,
 231.99 day,
 and
 323.24 day,
 respectively.
 In Figs.~\ref{fig:v_Earth_chi_S},
 we sketch
 these two simulation options
 as the lightened areas.

\InsertSKPPlotD [b!]
 {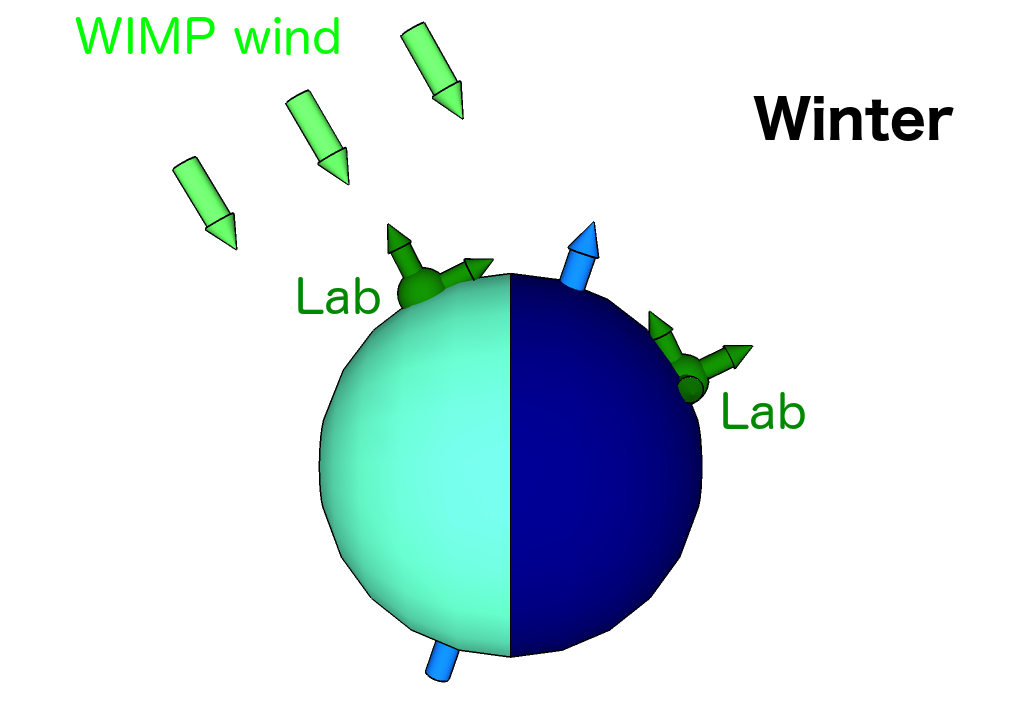}
 {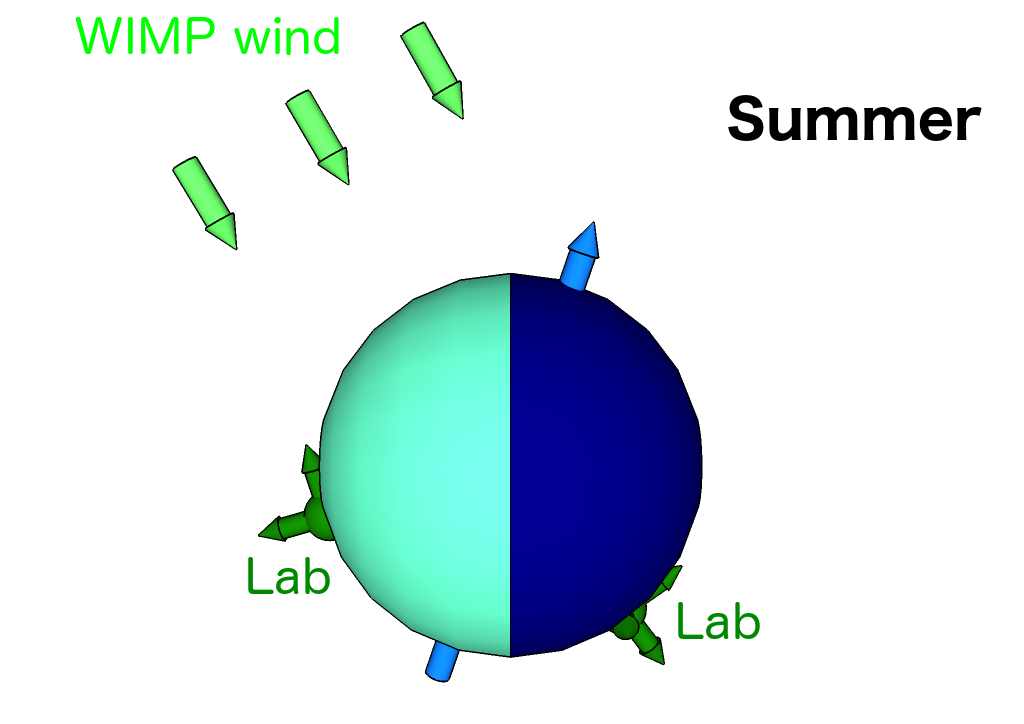}
 {fig:directional-1293-winter}
 {As in Figs.~\ref{fig:directional-1293-summer},
  except that
  the laboratory
  in the Northern Hemisphere
  is now in Winter (a),
  while
  that
  in the Southern Hemisphere
  is in Summer (b).
  }

 Moreover,
 in Table \ref{tab:period_year}
 we list,
 including the normal and the advanced seasons,
 four different options for the observation periods
 considered in our simulations presented in this paper.
 Note that,
 since
 in each of the normal and the advanced season,
 a period of 60 days ($\pm 30$ days)
 has been set,
 each pair of the corresponding season
 has an overlap of around 30 days
 (compare two sketches in Figs.~\ref{fig:v_Earth_chi_S}).

\paragraph{Diurnal modulation of the 3-D WIMP velocity}
 ~\\

\InsertSKPPlotS
 {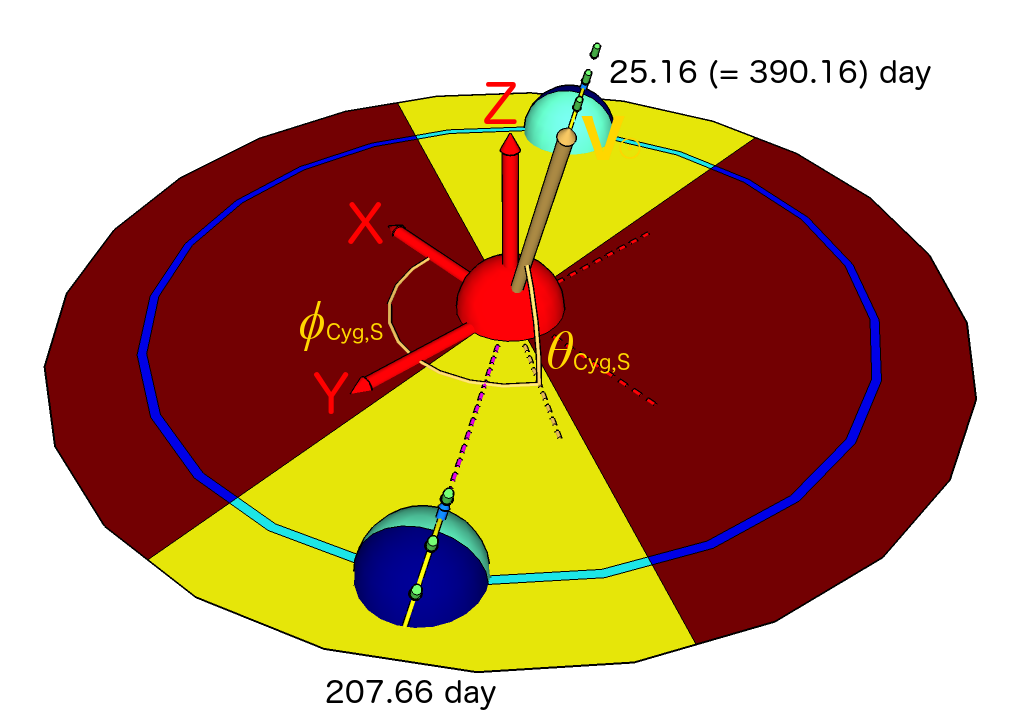}
 {The sketch of
  two options for the observation periods
  (lightened areas)
  considered for demonstrating the diurnal modulation of
  the angular distribution of
  the 3-D WIMP velocity
  presented in this paper
  (see also Table \ref{tab:period_year}):
  two 60-day ($\pm$ 30 days) observation period
  on the central date of
  207.66 day and 25.16 (= 390.16) day,
  respectively,
  on which
  the (light--green) WIMP wind
  points straightly to the (yellow) Prime Meridian
  in the night (207.66 day)
  or the day (25.16 day),
  corresponding (approximately) to the locations
  drawn in Figs.~\ref{fig:directional-1293-summer}
  and \ref{fig:directional-1293-winter},
  respectively.
  Other notations are the same as
  in Figs.~\ref{fig:v_Earth_chi_S}.
  Note that
  the angle of view in this sketch
  turns $\sim 63^{\circ}$ counterclockwise
  from the sketches in Figs.~\ref{fig:v_Earth_chi_S}.
  See the text
  and Appendix \ref{appx:v_Earth_chi_S-20766}
  for further details.
  }

 Due to the very small unexcluded
 WIMP--nucleus cross section,
 nowadays it is required to run
 a direct Dark Matter detection experiment
 for a (very) long period,
 especially for directional low--mass gas detectors.
 Then,
 by comparing Figs.~\ref{fig:directional-1293-summer}
 with Figs.~\ref{fig:directional-1293-winter}
 and
 as shown in detail
 in Figs.~\ref{fig:v_Earth_chi_S}
 and \ref{fig:v_Earth_chi_S-20766},
 one can find that
 the orignial motivation of
 directional DM detection experiments
 --- identifying the directionality of incident WIMPs ---
 could be (strongly) reduced or even vanish.
 For demonstrating however
 the orignial motivation of
 directional detection experiments,
 we considered four observation intervals of
 4 hours ($\pm 2$ hours)
 at the central ({\em local}, not the UTC) times of
  0 o'clock,
  6 o'clock,
 12 o'clock, and
 18 o'clock,
 respectively
 (see Table \ref{tab:interval_day}),
 in the 60-day periods
 centered on the
 207.66 day
 and
 390.16 (= 25.16) day,
 respectively
 (see Table \ref{tab:period_year} and
  Fig.~\ref{fig:v_Earth_chi_S-20766}),
 on which
 the theoretical main direction of
 the WIMP wind
 points straightly to the Prime Meridian
 in the night (207.66 day)
 or the day (25.16 day)%
\footnote{
 Note that
 these pure theoretically estimated dates
 (under some simplified assumptions)
 have in fact an $\sim 8$-hour difference
 from the midnight or the noon,
 respectively,
 which could nevertheless be neglected
 in the 60-day observation periods.
},
 corresponding (approximately) to the locations
 drawn in Figs.~\ref{fig:directional-1293-summer}
 and \ref{fig:directional-1293-winter},
 respectively
 (see Appendix \ref{appx:v_Earth_chi_S-20766}
  for the detailed calculation).

 In Figs.~\ref{fig:v_Earth_chi_S-20766},
 we sketch
 these two specified observation periods
 as the lightened areas
 (see also Table \ref{tab:period_year}).
 And
 in Table \ref{tab:interval_day},
 we list
 two options for the observation intervals
 in a (Solar) day
 considered for demonstrating (the diurnal modulation of)
 the angular distribution of
 the 3-D WIMP velocity
 presented in this paper.

\begin{table} [t!]
\small
\begin{center}
\renewcommand{\arraystretch}{1.5}
 \begin{tabular}{|| c || c | c ||}
\hline
\hline
 \makebox[4.5 cm][c]{Option}              &
 \makebox[5   cm][c]{Central time (hour)} &
 \makebox[5   cm][c]{Interval     (hour)} \\
\hline
\hline
 One entire day
 & --- &  0 -- 24 \\
\hline
 \multirow{4}{*}{Four daily shifts}
 &  0  &  0 --  2, 22 -- 24 \\
 &  6  &  4 --  8 \\
 & 12  & 10 -- 14 \\
 & 18  & 16 -- 20 \\
\hline
\hline
\end{tabular}
\end{center}
\caption{
 The list of
 two options for the observation intervals
 in a (Solar) day
 considered for demonstrating the diurnal modulation of
 the angular distribution of
 the 3-D WIMP velocity
 presented in this paper.
 Note that,
 in each shift of the second option,
 an interval of 4 hours ($\pm 2$ hours)
 has been set.
}
\label{tab:interval_day}
\end{table}
\subsection{Bayesian reconstruction of
            the radial distribution of the 3-D WIMP velocity}
\label{sec:N_v-Bayesian}

 In this subsection,
 we describe briefly
 the Bayesian fitting procedure
 applied for reconstructing the radial component (magnitude)
 of the 3-D WIMP velocity distribution
 presented in Secs.~\ref{sec:N_v-Bayesian-050}
 and \ref{sec:N_v-Bayesian-500}
 as well as
 in Appendix \ref{appx:N_v-Bayesian-Eq-07900}.

\subsubsection{Bayesian analysis}

 We start with the basic formula
 for Bayesian analysis
 \cite{DAgostini95}:
\beq
     {\rm p}(\Theta | {\rm data})
  =  \frac{{\rm p}({\rm data} | \Theta)}
          {{\rm p}({\rm data})} \cdot {\rm p}(\Theta)
\~.
\label{eqn:Bayesian_analysis}
\eeq
 Here
 $\Theta = \cbig{a_1, a_2, \cdots, a_{N_{\rm Bayesian}}}$
 denotes a specified (combination of the) value(s)
 of the fitting parameter(s);
 ${\rm p}(\Theta)$,
 called the ``prior probability'',
 represents our degree of belief about $\Theta$
 being the true value(s) of fitting parameter(s),
 which is often given
 in form of (the multiplication of)
 the probability distribution(s) of the fitting parameter(s).
 ${\rm p}({\rm data})$,
 called the ``evidence'',
 is the total probability of obtaining the particular set of data,
 which is in practice
 irrespective of the actual value(s) of the
 parameter(s) and
 can be treated as a normalization constant;
 it will not be of interest
 in our analysis
 presented in this paper.
 ${\rm p}({\rm data} | \Theta)$
 denotes the probability of the observed result,
 once the specified (combination of the) value(s)
 of the fitting parameter(s)
 happens,
 which can usually be described
 by the likelihood function of $\Theta$,
 ${\cal L}(\Theta)$.
 Finally,
 ${\rm p}(\Theta | {\rm data})$,
 called the ``posterior probability density function''
 for $\Theta$,
 representes
 the probability of that
 the specified (combination of the) value(s)
 of the fitting parameter(s)
 happens,
 given the observed result.

\subsubsection[Bayesian reconstruction of          $f_{\rm r}(v)$]
              {Bayesian reconstruction of \boldmath$f_{\rm r}(v)$}

 Below
 we describe
 the procedure of our Bayesian reconstruction of
 the radial distribution of the 3-D WIMP velocity
 in detail.

 First,
 as shown in Figs.~\ref{fig:N_v-G-050-00000}
 and \ref{fig:N_v-Eq-050-00000},
 the magnitudes of the (transformed) 3-D velocity of
 the recorded WIMP scattering events
 are binned into $B$ bins as
\beq
          v_{n, {\rm min}}
  \equiv  v_n - \frac{b_v}{2}
  \le     v_{n, i}
  \le     v_n + \frac{b_v}{2}
  \equiv  v_{n, {\rm max}}
\~,
     ~~~~ ~~
     i
  =  1,~2,~\cdots,~N_n,~
     n
  =  1,~2,~\cdots,~B.
\label{eqn:N_v}
\eeq
 Here
 the entire velocity range below the maximal cut--off
 ($\vesc$ in the Galactic coordinate system and
  $\vmax$ in the Equatorial coordinate system)
 has been divided into $B$ bins
 with central points $v_n$ and
 a common width $b_v$.
 In the $n$th bin,
 $N_n$ events are recorded and
\beq
     N_{\rm tot}
  =  \sum_{n = 1}^{B} N_n
\label{eqn:N_tot}
\eeq
 is the number of total WIMP events
 in the dataset to be analyzed.
 This means that
 in the $n$th $v$--bin $\bbrac{v_{n, {\rm min}}, v_{n, {\rm max}}}$,
 the {\em normalized} event number is
\beq
     f_{\rm r, expt}(v_n)
  =  \frac{1}{N_{\rm tot}} \afrac{N_n}{b_v}
\~.
\label{eqn:f1v_expt_n}
\eeq
 Choosing a theoretical prediction of
 the one--dimensional WIMP velocity distribution: \\
 $f_{\rm r, th} (v; a_1, a_2, \cdots, a_{N_{\rm Bayesian}})$,
 where $\abig{a_1, a_2, \cdots, a_{N_{\rm Bayesian}}}$
 are the $N_{\rm Bayesian}$ fitting parameters,
 and
 since
 the recorded event number in each $v$--bin
 should be Poisson--distributed
 around the theoretical predictions
 $f_{\rm r, th} (v_n; a_1, a_2, \cdots, a_{N_{\rm Bayesian}})$,
 the likelihood function for ${\rm p}({\rm data} | \Theta)$
 can be defined by
\beqn
 \conti
     {\cal L} \aBig{f_{\rm r, expt}(v_n),~n = 1,~2,~\cdots,~B;~
                    a_i,~i = 1,~2,~\cdots,~N_{\rm Bayesian}}
     \non\\
 \eqnequiv
     \prod_{n = 1}^{B}
     {\rm Poi} \aBig{v_n,
                     f_{\rm r, expt}(v_n);
                     a_1, a_2, \cdots, a_{N_{\rm Bayesian}}}
\~,
\label{eqn:calL}
\eeqn
 where
\beqn
 \conti
     {\rm Poi} \aBig{v_n,
                     f_{\rm r, expt}(v_n);
                     a_1, a_2, \cdots, a_{N_{\rm Bayesian}}}
     \non\\
 \eqnequiv
     \frac{f_{\rm r, th}^{f_{\rm r, expt}(v_n)} \abrac{v_n; a_1, a_2, \cdots, a_{N_{\rm Bayesian}}}
           e^{-f_{\rm r, th} \abrac{v_n; a_1, a_2, \cdots, a_{N_{\rm Bayesian}}}} }
          {\Gamma\abrac{f_{\rm r, expt}(v_n) + 1}}
\~,
\label{eqn:Bayesian_DF_Gau}
\eeqn
 and
\beq
     \Gamma(x)
  =  \intz t^{x - 1} e^{-t} \~ dt
\label{eqn:Gamma}
\eeq
 is the gamma function.
 Then
 the posterior probability density
 on the left--hand side of Eq.~(\ref{eqn:Bayesian_analysis})
 can be given as
\beqn
 \conti
     {\rm p} \aBig{a_i,~i = 1,~2,~\cdots,~N_{\rm Bayesian}~\Big|~
                   f_{\rm r, expt}(v_n),~n = 1,~2,~\cdots,~B}
     \non\\
 \eqnpropto
     {\cal L} \aBig{f_{\rm r, expt}(v_n),~n = 1,~2,~\cdots,~B;~
                    a_i,~i = 1,~2,~\cdots,~N_{\rm Bayesian}}
     \prod_{i = 1}^{N_{\rm Bayesian}} {\rm p}_i(a_i)
\~.
\label{eqn:P_Bayesian}
\eeqn
 Regarding our degree of belief about
 each fitting parameter $a_i$,
 i.e.~${\rm p}_i(a_i)$ in Eq.~(\ref{eqn:P_Bayesian}),
 two probability distribution functions
 have been considered.
 The simplest one is the flat--distribution:
\beq
     {\rm p}_i(a_i)
  =  1
\~,
    ~~~~ ~~~~ ~~~~ ~~~~ % 16
    {\rm for~} a_{i, \rm min} \le a_i \le a_{i, \rm max},
\label{eqn:Bayesian_DF_a_flat}
\eeq
 where $a_{i, \rm (min, max)}$ denote
 the minimal and maximal bounds of the scanning interval of
 the fitting parameter $a_i$.
 On the other hand,
 for the case that
 we have already prior knowledge about
 one fitting parameter,
 a Gaussian--distribution:
\beq
     {\rm p}_i(a_i; \mu_{a, i}, \sigma_{a, i})
  =  \frac{1}{\sqrt{2 \pi} \~ \sigma_{a, i}} \~
     e^{-(a_i - \mu_{a, i})^2 / 2 \sigma_{a, i}^2}
\label{eqn:Bayesian_DF_a_Gau}
\eeq
 with the expectation value $\mu_{a, i}$ of and
 the 1$\sigma$ uncertainty $\sigma_{a, i}$ on
 the fitting parameter $a_i$
 is used.
 Note that,
 in one simulated experiment,
 we scan the parameter space
 $(a_1, a_2, \cdots, a_{N_{\rm Bayesian}})$
 in the volume
 $a_i \in [a_{i, {\rm min}}, a_{i, {\rm max}}]$,
 $i = 1, 2, \cdots, N_{\rm Bayesian}$,
 to find a particular point
 $(a_1^{\ast}, a_2^{\ast}, \cdots, a_{N_{\rm Bayesian}}^{\ast})$,
 which maximizes (the numerator of)
 the posterior probability density: \\
 ${\rm p} \aBig{a_i,~i = 1,~2,~\cdots,~N_{\rm Bayesian}~\Big|~
                f_{\rm r, expt}(v_n),~n = 1,~2,~\cdots,~B}$.
 And after that
 all simulations have been done,
 we determine the mean value of
 the event number in each $v$--bin
 from all simulated experiments,
 denoted as $N_{n, {\rm mean}}$,
 and define then
\beq
     f_{\rm r, mean}(v_n)
  =  \frac{1}{N_{\rm tot, ave}} \afrac{N_{n, {\rm mean}}}{b_v}
\~,
\label{eqn:f1v_expt_n}
\eeq
 where
 $N_{\rm tot, ave}$ is the expectation value of the event number
 in one simulated experiment%
\footnote{
 Note that
 in our numerical simulations
 presented in this paper,
 the total number of the generated WIMP events
 in each experiment,
 $N_{\rm tot}$,
 is Poisson--distributed around
 the expectation value $N_{\rm tot, ave}$.
}.
 Hence,
 we can define further
\beqn
 \conti
     {\rm P}_{\rm mean} \abrac{a_i,~i = 1,~2,~\cdots,~N_{\rm Bayesian}}
     \non\\
 \eqnequiv
     {\rm p } \abrac{a_i,~i = 1,~2,~\cdots,~N_{\rm Bayesian}~\Big|~
                     f_{\rm r, mean}(v_n),~n = 1,~2,~\cdots,~B}
\~,
\label{eqn:P_Bayesian_mean}
\eeqn
 and scan the points
 $(a_1^{\ast}, a_2^{\ast}, \cdots, a_{N_{\rm Bayesian}}^{\ast})$
 obtained from all simulated experiments
 one--by--one
 to find the ``best--fit'' point
 $(a_{1, {\rm Pmax}}, a_{2, {\rm Pmax}}, \cdots, a_{N_{\rm Bayesian}, {\rm Pmax}})$,
 which maximizes
 ${\rm P}_{\rm mean} (a_i,~i = 1,~2,~\cdots,~N_{\rm Bayesian})$.

\subsubsection[Fitting velocity distribution          $f_{\rm r, th}(v)$]
              {Fitting velocity distribution \boldmath$f_{\rm r, th}(v)$}
\label{sec:N_v-Bayesian-f1v}

 By taking into account
 the orbital motion of the Solar system around our Galaxy
 as well as
 that of the Earth around the Sun,
 the shifted Maxwellian velocity distribution
 has been derived in detail
 by Lewin and Smith
 \cite{Lewin96}:
\beqn
     f_{1, \sh, {\rm vesc}}(v)
 \=  \cleft{\renewcommand{\arraystretch}{0.75}
            \begin{array}{l l}
             \D
             N_{\sh, {\rm vesc}}
             \afrac{v}{v_0 \ve}
             \bBig{  e^{-(v - \ve)^2 / v_0^2}
                   - e^{-(v + \ve)^2 / v_0^2} } \~, &
             {\rm for}~v \le \vesc - \ve        \~, \\
             ~ & ~ \\
             \D
             N_{\sh, {\rm vesc}}
             \afrac{v}{v_0 \ve}
             \bBig{  e^{-(v - \ve)^2 / v_0^2}
                   - e^{-\vesc^2     / v_0^2} }          \~, &
             {\rm for}~\vesc - \ve \le v \le \vesc + \ve \~, \\
             ~ & ~ \\
             0                                           \~, &
             {\rm for}~v \ge \vesc + \ve \~ \equiv \vmax \~,
            \end{array}}
     \non\\
\label{eqn:f1v_sh_vesc}
\eeqn
 with the normalization constant
\beq
     N_{\sh, {\rm vesc}}
  =  \bbrac{  \sqrt{\pi} \~
              \erf{\D \afrac{\vesc}{v_0}}
            - \afrac{2 \vesc}{v_0}
              e^{-\vesc^2 / v_0^2}        }^{-1}
\~.
\label{eqn:N_sh_vesc}
\eeq
 Here
 $\ve$ is the time--dependent Earth's speed
 in the Galactic frame,
 which has been given in Ref.~\cite{Freese88}
 by
\beq
     \ve(t)
  =  v_0 \bbrac{1.05 + 0.07 \cos\afrac{2 \pi (t - t_{\rm p})}{1~{\rm yr}}}
\~,
\label{eqn:ve}
\eeq
 with $t_{\rm p} \simeq$ June 2nd
 (the purple points
  sketched in Figs.~\ref{fig:v_Earth_chi_S}),
 and,
 since the Galactic escape velocity
 has been set as $\vesc = 550$ km/s
 in our simulations,
 the maximal cut--off
 on the radial distribution of the 3-D WIMP velocity
 in the Equatorial coordinate system
 is given as $\vmax = 781$ km/s.
 Meanwhile,
 a simplified expression for
 the shifted Maxwellian velocity distribution
 $f_{1, \sh, {\rm vesc}}(v)$
 given in Eq.~(\ref{eqn:f1v_sh_vesc})
 is often adopted in literature
 \cite{SUSYDM96}:
\beq
     f_{1, \sh}(v)
  =  \cleft{\renewcommand{\arraystretch}{1.75}
            \begin{array}{l c l}
             \D
             N_{\sh}
             \afrac{v}{v_0 \ve}
             \bBig{  e^{-(v - \ve)^2 / v_0^2}
                   - e^{-(v + \ve)^2 / v_0^2} } \~, & ~~~~ ~~~~ & % 8
             {\rm for}~v \le \vmax              \~, \\
             0                                  \~, &           &
             {\rm for}~v >   \vmax              \~,
            \end{array}}
\label{eqn:f1v_sh_vmax}
\eeq
 with the normalization constant
\beqn
     N_{\sh}
 \=  \cleft{   \afrac{\sqrt{\pi}}{2}
               \bbrac{  \erf{\D \afrac{\vmax + \ve}{v_0}}
                      + \erf{\D \afrac{\vmax - \ve}{v_0}} }  }
     \non\\
 \conti ~~~~ ~~~~ ~~~~ ~~~~ ~~~~ % 20
     \cright{+ \afrac{v_0}{2 \ve}
               \bbigg{  e^{-(\vmax + \ve)^2 / v_0^2}
                      - e^{-(\vmax - \ve)^2 / v_0^2}      }  }^{-1}
\~.
\label{eqn:N_sh_vmax}
\eeqn
 In Appendix \ref{appx:f1v_sh_vesc},
 we will show that,
 with the Galactic escape velocity of $\vesc = 500$ km/s,
 the shape difference between
 the exact analytic expression
 $f_{1, \sh, {\rm vesc}}(v)$
 given in Eq.~(\ref{eqn:f1v_sh_vesc})
 and the simplified form
 $f_{1, \sh}(v)$
 in Eq.~(\ref{eqn:f1v_sh_vmax})
 is pretty tiny and negligible
 compared to
 the annual variation of two expressions
 as well as
 the much larger statistical uncertainties
 on the recorded WIMP events
 (shown in e.g., Fig.~\ref{fig:N_v-Eq-050-00000});
 this tiny shape difference
 could even vanish
 once the Galactic escape velocity
 would be as large as $\vesc = 600$ km/s.
 Hence,
 for our Bayesian reconstruction of
 the radial distribution of the 3-D WIMP velocity
 presented in this paper,
 we took the simplified expression (\ref{eqn:f1v_sh_vmax})
 and considered the following
 four fitting distribution functions.

 The first one is
 the ``one--parameter''
 shifted Maxwellian velocity distribution
 with the Solar Galactic velocity $v_0$
 as the unique fitting parameter
 \cite{DMDDf1v-Bayesian}:
\beq
     f_{1, \sh, v_0}(v; v_0)
  =  N_{\sh}(v_0)
     \afrac{v}{v_0 \ve}
     \bBig{  e^{-(v - \ve)^2 / v_0^2}
           - e^{-(v + \ve)^2 / v_0^2}  }
\~,
\label{eqn:f1v_sh_v0}
\eeq
 and the constraint on $\ve$:%
\footnote{
 Although
 in Appendix \ref{appx:v_Earth_chi_S}
 we will show that
 the first--order term of
 the time--dependence between $v_0$ and $\ve$
 would only be 1.0078,
 we took still the commonly used value of 1.05
 for our reconstructions
 presented in this paper.
}
\beq
     \ve
  =  1.05 \~ v_0
\~.
\label{eqn:ve_ave}
\eeq
 Considering the (annual) variation of
 the Earth's Galactic velocity
 in different observation periods,
 we introduce a ``$v_0$--fixed''
 shifted Maxwellian velocity distribution
 with now the Earth's Galactic velocity $\ve$
 as the fitting parameter:
\beq
     f_{1, \sh, \ve}(v; \ve)
  =  N_{\sh}(\ve)
     \afrac{v}{v_0 \ve}
     \bBig{  e^{-(v - \ve)^2 / v_0^2}
           - e^{-(v + \ve)^2 / v_0^2}  }
\~,
\label{eqn:f1v_sh_ve}
\eeq
 and the constraint of $v_0 = 220$ km/s.
 Moreover,
 in order to obtain reconstruction results
 which can match the recorded radial WIMP velocity distribution
 as well as possible,
 the simplified shifted Maxwellian velocity distribution
 with $v_0$ and $v_e$
 as two {\em independent} fitting parameters
 should certainly be considered:
\beq
     f_{1, \sh}(v; v_0, \ve)
  =  N_{\sh}(v_0, \ve)
     \afrac{v}{v_0 \ve}
     \bBig{  e^{-(v - \ve)^2 / v_0^2}
           - e^{-(v + \ve)^2 / v_0^2}  }
\~.
\label{eqn:f1v_sh}
\eeq
 And,
 finally,
 as the auxiliary fitting function
 for confirming and/or improving our reconstruction results,
 the ``modified'' shifted Maxwellian velocity distribution
 with $v_0$ and $\Delta v \equiv \ve - v_0$
 as two independent fitting parameters
 introduced in our earlier work
 \cite{DMDDf1v-Bayesian}
 has also be used:
\beq
     f_{1, \sh, \Delta v}(v; v_0, \Delta v)
  =  N_{\sh}(v_0, \Delta v)
     \bfrac{v}{v_0 \abrac{v_0 + \Delta v}}
     \cbigg{  e^{-\bbrac{v - \abrac{v_0 + \Delta v}}^2 / v_0^2}
            - e^{-\bbrac{v + \abrac{v_0 + \Delta v}}^2 / v_0^2}  }
\~.
\label{eqn:f1v_sh_Dv}
\eeq
%

%

%
% 2/10
 %
%
\section{Angular distributions of the 3-D WIMP velocity
         in different coordinate systems}
\label{sec:N_phi_theta-050}

 In this section,
 we present
 the angular distributions of
 the (transformed) 3-D WIMP velocity
 in three ``laboratory--independent''
 (Ecliptic,
  Equatorial,
  and Earth)
 coordinate systems
 and then
 in two ``laboratory--dependent''
 (horizontal
  and laboratory)
 coordinate systems
 one by one.
 As readers' reference,
 a summary of
 the angular distributions
 observed in the horizontal and laboratory coordinate systems
 of different underground laboratories
 will be given
 in Appendix \ref{appx:N_phi_theta-ULabs}.

\subsection{Angular WIMP velocity distribution in the Ecliptic frame}
\label{sec:N_phi_theta-S-050}

 In this subsection,
 we present at first
 the angular distribution of the 3-D WIMP velocity
 in the Ecliptic coordinate system.
 Remind that
 the Ecliptic coordinate system
 only moves (approximately) linearly
 with the Solar Galactic orbital velocity $v_0 \simeq 220$ km/s
 and its tiny rotation is imperceptible.

 In our simulations presented in this paper,
 the $i$th 3-D WIMP velocity
 transformed from the $i$th velocity
 generated in the Galactic coordinate system
 to the Ecliptic coordinate system
 can be given by
\cheqna
\beq
     {\bf v}_{\chi, i, {\rm S}}
  =  \MaGS \abrac{{\bf v}_{\chi, i, {\rm G}} - \VSunG}
\~,
\label{eqn:V_chi_i_S-V_chi_i_G}
\eeq
 where $\MaGS$ is the transformation matrix
 given in Eq.~(\ref{eqn:Ma_G_S})
 and $\VSunG$ is the moving velocity of the Solar system
 in the Galactic coordinate system
 given in Eq.~(\ref{eqn:V_Sun_G}).
 Conversely,
 for analyzing real data
 provided by experimental collaborations
 (hopefully) in the future,
 the $i$th 3-D velocity
 in the Galactic coordinate system
 can by transformed from that
 in the Ecliptic coordinate system by
\cheqnb
\beq
     {\bf v}_{\chi, i, {\rm G}}
  =  \MaSG {\bf v}_{\chi, i, {\rm S}} + \VSunG
\~,
\label{eqn:V_chi_i_G-V_chi_i_S}
\eeq
\cheqn
 where the transformation matrix $\MaSG$
 is given in Eq.~(\ref{eqn:Ma_S_G}).

\begin{figure} [t!]
\begin{center}
 \includegraphics [width = 15 cm] {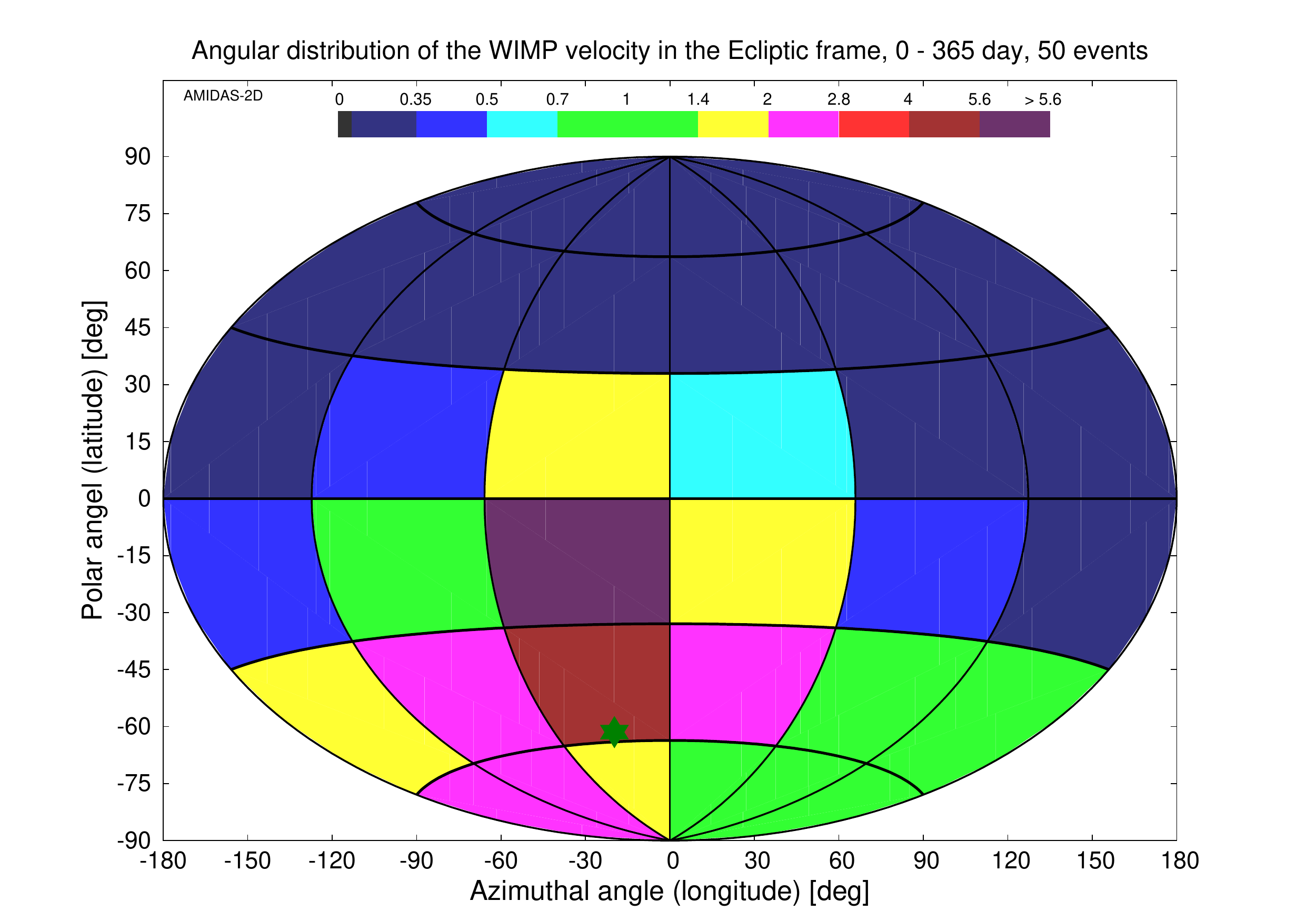}
\end{center}
\caption{
 The angular distribution of the 3-D WIMP velocity
 transformed from events
 shown in Fig.~\ref{fig:N_phi_theta-G-050-00000}
 to the Ecliptic coordinate system
 by Eq.~(\ref{eqn:V_chi_i_S-V_chi_i_G}).
 The dark--green star indicates
 the theoretical main direction of incident WIMPs
 (the opposite direction of the Solar Galactic movement)
 in the Ecliptic coordinate system:
 57.40$^{\circ}$S, 29.10$^{\circ}$W.
 All simulation setup and other notations are the same
 as in Fig.~\ref{fig:N_phi_theta-G-050-00000}.
 See the text for further details.
}
\label{fig:N_phi_theta-S-050-00000}
\end{figure}

 In Fig.~\ref{fig:N_phi_theta-S-050-00000},
 we show
 the angular distribution of the 3-D WIMP velocity
 transformed from events
 shown in Fig.~\ref{fig:N_phi_theta-G-050-00000}
 to the Ecliptic coordinate system
 by Eq.~(\ref{eqn:V_chi_i_S-V_chi_i_G}).
 One entire year (0 to 365 day)
 and 50 total events on average in one experiment
 have been considered.
 It can be found that,
 while
 in the most part of the northern sky
 and a small part of the southern sky,
 the numbers of WIMP events
 would be less than 35\% (0.49 events/bin)
 of the all--sky average value (1.39 events/bin),
 the event numbers
 in the bins
 from the center
 (i.e.,
  the direction of the $\xS$--axis)
 to southwest
 would be larger than 1.4 times
 (1.94 events/bin)
 of the all--sky average value;
 in particular,
 in the dark--red and dark--purple bins
 in the southern sky,
 the numbers of WIMP events
 are at least 4 times
 (5.56 events/bin)
 or even larger than 5.6 times
 (7.78 events/bin)
 of the all--sky average value,
 respectively.
 This means that
 the average event numbers
 from the center to the southwest part
 could be at least 4 times
 or even 16 times larger than
 the rest part of the sky
 and
 would hence be a clear identification
 of the anisotropy
 of the incident direction of Galactic WIMPs.

 However,
 comparing with
 the dark--green star on the map,
 which indicates
 the theoretical main direction of incident WIMPs
 (the opposite direction of the Solar Galactic movement)
 in the Ecliptic coordinate system:
 57.40$^{\circ}$S, 29.10$^{\circ}$W,%
\footnote{
 See Appendix \ref{appx:v_Sun_G}
 for the detailed calculation.
}
 the (dark--purple) bin with the most WIMP events
 would not match
 the theoretically predicted direction of the WIMP wind
 and has a 30$^{\circ}$ to 40$^{\circ}$
 northward deviation.
 In Sec.~\ref{sec:N_phi_theta-S-500},
 we will see that,
 with $\cal O$(500) total WIMP events
 and a higher analysis resolution,
 this systematic bias could be strongly reduced,
 but a small amount of deviation would still exist.

\subsection{Angular WIMP velocity distribution in the Equatorial frame}
\label{sec:N_phi_theta-Eq-050}

 In this subsection,
 we present
 the angular distribution of the 3-D WIMP velocity
 in the Equatorial coordinate system.
 Remind that
 the Equatorial coordinate system
 moves orbitally around
 (and also linearly with) the Sun,
 but doesn't rotate.
 Thus
 its axes are fixed.

 Similar to Eqs.~(\ref{eqn:V_chi_i_S-V_chi_i_G})
 and (\ref{eqn:V_chi_i_G-V_chi_i_S}),
 in our simulations presented in this paper,
 the $i$th 3-D WIMP velocity
 transformed from the $i$th velocity
 in (transformed to) the Ecliptic coordinate system
 to the Equatorial coordinate system
 can be given by
\cheqna
\beq
     {\bf v}_{\chi, i, {\rm Eq}}
  =  \MaSEq \bBig{{\bf v}_{\chi, i, {\rm S}} - \VEarthS(t_i)}
\~,
\label{eqn:V_chi_i_Eq-V_chi_i_S}
\eeq
 where $t_i$ is the generated observation time
 of the $i$th WIMP event,
 $\MaSEq$ is the transformation matrix
 given in Eq.~(\ref{eqn:Ma_S_Eq})
 and $\VEarthS(t)$ is the time--dependent
 Earth's orbital velocity around the Sun
 in the Ecliptic coordinate system
 given in Eq.~(\ref{eqn:V_Earth_S}).
 Conversely,
 for analyzing real data
 provided by experimental collaborations in the future,
 the $i$th 3-D velocity
 in the Ecliptic coordinate system
 can by transformed from the $i$th velocity
 measured at time $t_i$
 in the Equatorial coordinate system by
\cheqnb
\beq
     {\bf v}_{\chi, i, {\rm S}}
  =  \MaEqS {\bf v}_{\chi, i, {\rm Eq}} + \VEarthS(t_i)
\~,
\label{eqn:V_chi_i_S-V_chi_i_Eq}
\eeq
\cheqn
 where the transformation matrix $\MaEqS$
 is given in Eq.~(\ref{eqn:Ma_Eq_S}).

\begin{figure} [t!]
\begin{center}
 \includegraphics [width = 15 cm] {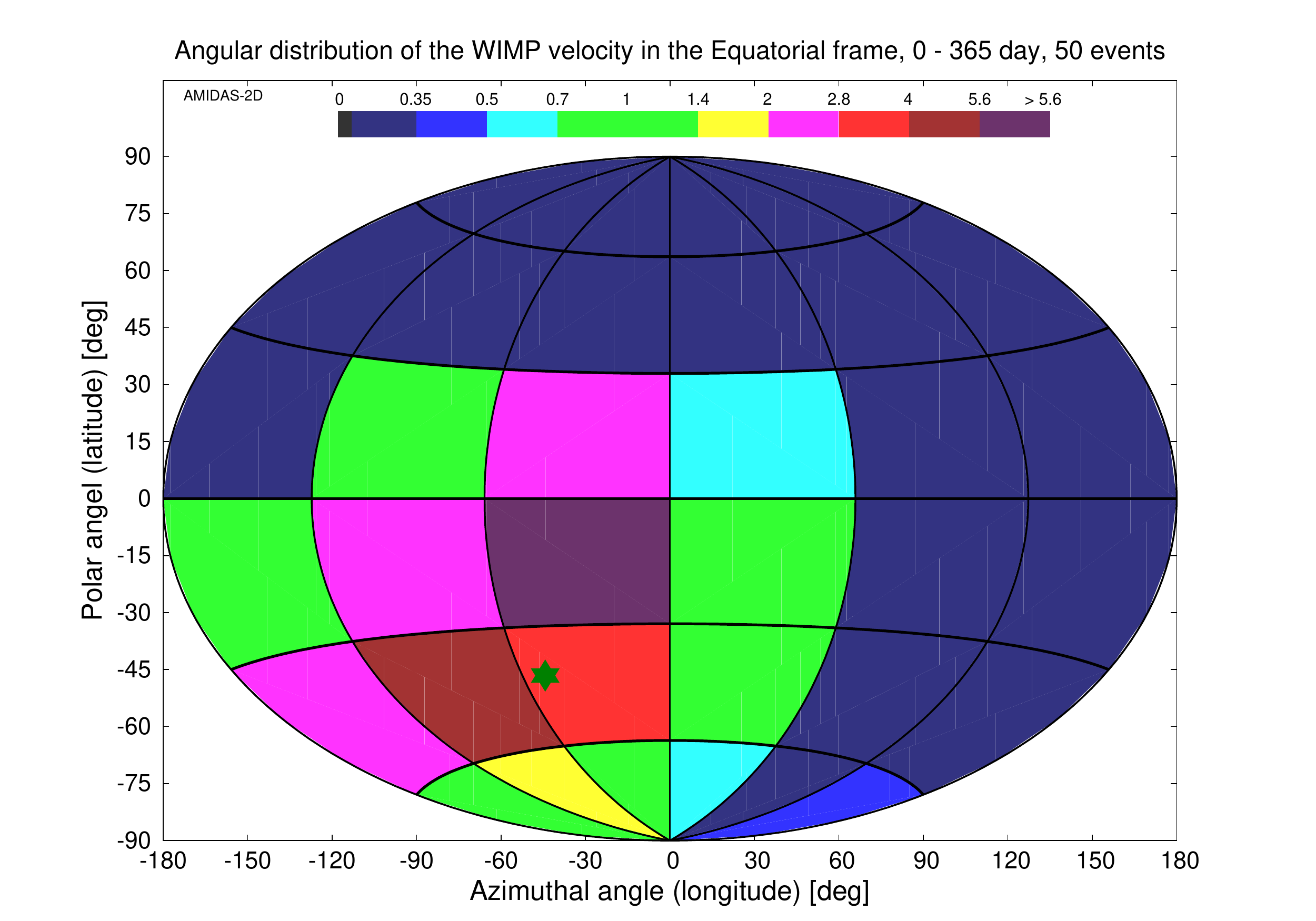}
\end{center}
\caption{
 The angular distribution of the 3-D WIMP velocity
 transformed from events
 shown in Fig.~\ref{fig:N_phi_theta-S-050-00000}
 to the Equatorial coordinate system
 by Eq.~(\ref{eqn:V_chi_i_Eq-V_chi_i_S}).
 The dark--green star indicates now
 the theoretical main direction of incident WIMPs
 in the Equatorial coordinate system
 \cite{Bandyopadhyay10}:
 42.00$^{\circ}$S, 50.70$^{\circ}$W.
 All simulation setup and other notations are the same
 as in Fig.~\ref{fig:N_phi_theta-S-050-00000}.
}
\label{fig:N_phi_theta-Eq-050-00000}
\end{figure}

 In Fig.~\ref{fig:N_phi_theta-Eq-050-00000},
 we show
 the angular distribution of the 3-D WIMP velocity
 transformed from events
 shown in Figs.~\ref{fig:N_phi_theta-G-050-00000}
 and \ref{fig:N_phi_theta-S-050-00000}
 to the Equatorial coordinate system
 by Eq.~(\ref{eqn:V_chi_i_Eq-V_chi_i_S}).
 One entire year (0 to 365 day)
 and 50 total events on average in one experiment
 have been considered.
 As in Fig.~\ref{fig:N_phi_theta-S-050-00000},
 it can also be seen clearly here that,
 while
 in the most part of the northern
 and the eastern skies,
 the numbers of WIMP events
 would be less than 35\% (0.49 events/bin)
 of the all--sky average value (1.39 events/bin),
 the event numbers
 in the southwest part of the sky
 would be at least 2 times
 (2.78 events/bin)
 or even more than 5.6 times
 (7.78 events/bin)
 of the all--sky average value,
 respectively.
 This means that
 the average event numbers
 from the center
 (i.e.,
  the direction of the $\xEq$--axis)
 to the southwest part
 could be at least 5.7 times
 or even 16 times larger than
 the rest part of the sky
 and
 would hence also be a clear identification
 of the anisotropy
 of the incident direction of Galactic WIMPs.

 In Fig.~\ref{fig:N_phi_theta-Eq-050-00000},
 we also put
 a dark--green star
 to indicate
 the theoretical main direction of incident WIMPs
 in the Equatorial coordinate system
 \cite{Bandyopadhyay10}:
 42.00$^{\circ}$S, 50.70$^{\circ}$W.
 Now
 the deviation between
 the (dark--purple) bin with the most WIMP events
 and
 the theoretically predicted direction of the WIMP wind
 would only be $\sim$ 10$^{\circ}$
 northwestward,
 but they can still not match each other.
 In Sec.~\ref{sec:N_phi_theta-Eq-500},
 we will show that,
 with $\cal O$(500) total WIMP events
 and a higher analysis resolution,
 a small deviation might still exist.

 \def \ShortFrame     {Eq}
 \def \EventNumber    {050}
 \def \Perioda        {\PeriodBa}
 \def \Periodb        {\PeriodBb}
 \def \Periodc        {\PeriodBc}
 \def \Periodd        {\PeriodBd}
 \def \PlotNumbera    {\PlotNumberBa}
 \def \PlotNumberb    {\PlotNumberBb}
 \def \PlotNumberc    {\PlotNumberBc}
 \def \PlotNumberd    {\PlotNumberBd}
 \InsertPlotNphitheta
  {The angular distributions of the 3-D WIMP velocity
   in the Equatorial coordinate system.
   Four observation periods of the normal seasons
   listed in Table \ref{tab:period_year}
   and sketched in Fig.~\ref{fig:v_Earth_chi_S}(a)
   as well as
   50 total events on average
   in each 60-day observation period
   have been considered.
   Besides the dark--green star
   indicating
   the theoretical main direction of incident WIMPs,
   the blue--yellow point in each plot indicates
   the opposite direction of
   the Earth's velocity relative to the Dark Matter halo
   on the central date of the observation period
   (listed in Table \ref{tab:vEarthchiEqT}).
   All other simulation setup and notations are the same
   as in Fig.~\ref{fig:N_phi_theta-Eq-050-00000}.
   }
 \def \Perioda        {\PeriodCa}
 \def \Periodb        {\PeriodCb}
 \def \Periodc        {\PeriodCc}
 \def \Periodd        {\PeriodCd}
 \def \PlotNumbera    {\PlotNumberCa}
 \def \PlotNumberb    {\PlotNumberCb}
 \def \PlotNumberc    {\PlotNumberCc}
 \def \PlotNumberd    {\PlotNumberCd}
 \InsertPlotNphitheta
  {As in Figs.~\ref{fig:N_phi_theta-Eq-050-07900},
   except that
   four observation periods of the advanced seasons
   listed in Table \ref{tab:period_year}
   and sketched in Fig.~\ref{fig:v_Earth_chi_S}(b)
   have been considered.
   }
\subsubsection{Annual modulation of the angular velocity distribution
               in the Equatorial frame}
\label{sec:N_phi_theta-Eq-050-04949}

 Moreover,
 in order to demonstrate
 the annual modulation of
 the angular distribution pattern of
 the 3-D WIMP velocity,
 we show
 in Figs.~\ref{fig:N_phi_theta-Eq-050-07900}
 and \ref{fig:N_phi_theta-Eq-050-04949}
 the angular distributions
 in the Equatorial coordinate system
 in four observation periods of
 the normal and the advanced seasons
 listed in Table \ref{tab:period_year}
 and sketched in Figs.~\ref{fig:v_Earth_chi_S},
 respectively.
 Note that
 50 total events on average
 in each 60-day observation period
 have been simulated.
 This means although that
 $\sim$ 300 total events in one year
 would be required,
 considering the laboratory--independence of
 the Equatorial coordinate system,
 we could in practice collect
 the WIMP events
 observed in several different underground laboratories
 for such an analysis.

 In each plot of Figs.~\ref{fig:N_phi_theta-Eq-050-07900}
 and \ref{fig:N_phi_theta-Eq-050-04949},
 besides the dark--green star
 indicating
 the theoretical main direction of incident WIMPs,
 we also put
 a blue--yellow point to indicate
 the opposite direction of
 the Earth's velocity relative to the Dark Matter halo
 on the central date of the observation period%
\footnote{
 See Appendix \ref{appx:v_Earth_chi_S}
 for the detailed calculations.
}.
 By comparing the plots
 in four normal and four advanced seasons
 carefully,
 one could find that
 the event numbers indeed variate
 (become more and then fewer) slightly
 and
 this variation follows also
 the circular clockwise movement of the blue--yellow point.
 This would be,
 besides the pure anisotropy of the WIMP velocity,
 a second (important) characteristic
 for identifying directional WIMP signals
 and discriminating from any (unexpected) backgrounds
 with some specified incoming directions.

\subsection{Angular WIMP velocity distribution in the Earth frame}
\label{sec:N_phi_theta-E-050}

 In this subsection,
 we present
 the angular distribution of the 3-D WIMP velocity
 in the Earth coordinate system.
 Remind that
 the Earth coordinate system
 not only moves orbitally around
 (and also linearly with) the Sun,
 but also rotates daily and discretely.

 As mentioned at the end of Sec.~\ref{sec:XYZ},
 the transformations of
 the $i$th 3-D WIMP velocity
 measured at time $t_i$
 between the Equatorial and the Earth coordinate systems
 are pure rotations,
 which can be given by
\cheqna
\beq
     {\bf v}_{\chi, i, {\rm E}}
  =  \MaEqE(t_i) \~ {\bf v}_{\chi, i, {\rm Eq}}
\~,
\label{eqn:V_chi_i_E-V_chi_i_Eq}
\eeq
 and,
 conversely,
\cheqnb
\beq
     {\bf v}_{\chi, i, {\rm Eq}}
  =  \MaEEq(t_i) \~ {\bf v}_{\chi, i, {\rm E}}
\~,
\label{eqn:V_chi_i_Eq-V_chi_i_E}
\eeq
\cheqn
 where the transformation matrices $\MaEqE(t)$ and $\MaEEq(t)$
 are given in Eqs.~(\ref{eqn:Ma_Eq_E}) and (\ref{eqn:Ma_E_Eq}).

\begin{figure} [t!]
\begin{center}
 \includegraphics [width = 15 cm] {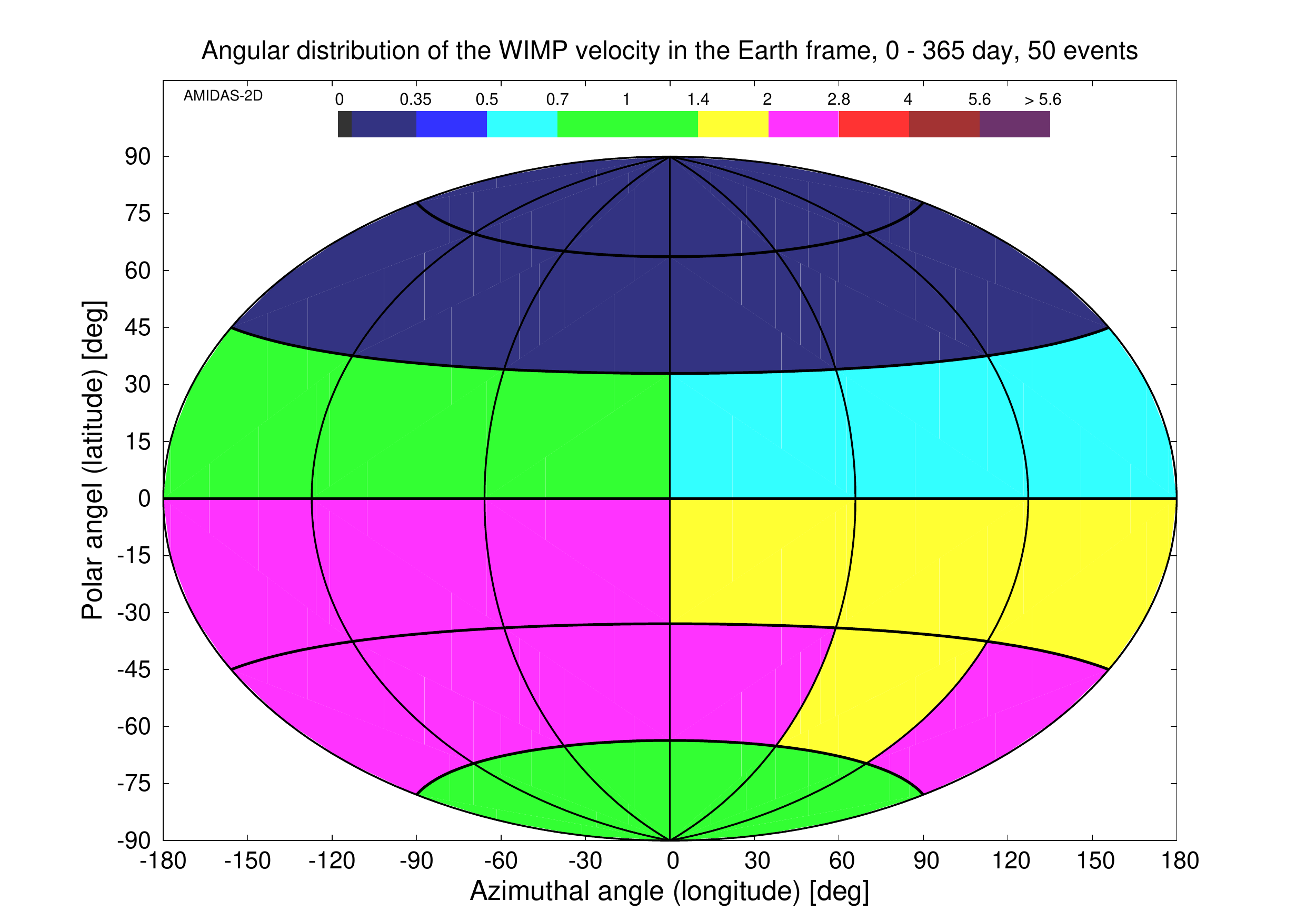}
\end{center}
\caption{
 The angular distribution of the 3-D WIMP velocity
 transformed from events
 shown in Fig.~\ref{fig:N_phi_theta-Eq-050-00000}
 to the Earth coordinate system
 by Eq.~(\ref{eqn:V_chi_i_E-V_chi_i_Eq}).
 All simulation setup and notations are the same
 as in Fig.~\ref{fig:N_phi_theta-Eq-050-00000}.
}
\label{fig:N_phi_theta-E-050-00000}
\end{figure}

 In Fig.~\ref{fig:N_phi_theta-E-050-00000},
 we show
 the angular distribution of the 3-D WIMP velocity
 transformed from events
 shown in Figs.~\ref{fig:N_phi_theta-G-050-00000},
 \ref{fig:N_phi_theta-S-050-00000},
 and \ref{fig:N_phi_theta-Eq-050-00000}
 to the Earth coordinate system
 by Eq.~(\ref{eqn:V_chi_i_E-V_chi_i_Eq}).
 One entire year (0 to 365 day)
 with 50 total events on average
 have been considered.
 It could be seen that,
 first,
 due to the Earth's orbital motion around the Sun
 and thus the Earth coordinate system
 rotates daily,
 (the anisotropy of)
 the angular distribution of the 3-D WIMP velocity
 spreads out latitudinally.
 However,
 one could still find that
 the event numbers are
 at fewest
 (less than 35\% of the all--sky average value,
  $<$ 0.49 events/bin among $\cal O$(50) total events)
 in the bins more northern than 60$^{\circ}$N,
 and
 at least 2 times of the all--sky average
 ($>$ 2.78 events/bin)
 in the southwest part of the sky.

\subsubsection{Annual modulation of the angular velocity distribution
               in the Earth frame}
\label{sec:N_phi_theta-E-050-04949}
 \def \ShortFrame     {E}
 \def \Perioda        {\PeriodBa}
 \def \Periodb        {\PeriodBb}
 \def \Periodc        {\PeriodBc}
 \def \Periodd        {\PeriodBd}
 \def \PlotNumbera    {\PlotNumberBa}
 \def \PlotNumberb    {\PlotNumberBb}
 \def \PlotNumberc    {\PlotNumberBc}
 \def \PlotNumberd    {\PlotNumberBd}
 \InsertPlotNphitheta
  {The angular distributions of the 3-D WIMP velocity
   in the Earth coordinate system.
   Four observation periods of the normal seasons
   as well as
   50 total events on average
   in each 60-day observation period
   have been considered.
   All other simulation setup and notations are the same
   as in Fig.~\ref{fig:N_phi_theta-E-050-00000}.
   }
 \def \Perioda        {\PeriodCa}
 \def \Periodb        {\PeriodCb}
 \def \Periodc        {\PeriodCc}
 \def \Periodd        {\PeriodCd}
 \def \PlotNumbera    {\PlotNumberCa}
 \def \PlotNumberb    {\PlotNumberCb}
 \def \PlotNumberc    {\PlotNumberCc}
 \def \PlotNumberd    {\PlotNumberCd}
 \InsertPlotNphitheta
  {As in Figs.~\ref{fig:N_phi_theta-E-050-07900},
   except that
   four observation periods of the advanced seasons
   have been considered.
   }

 In Figs.~\ref{fig:N_phi_theta-E-050-07900}
 and \ref{fig:N_phi_theta-E-050-04949},
 we show
 the angular distributions of the 3-D WIMP velocity
 in the Earth coordinate system
 in
 four observation periods of
 the normal and the advanced seasons,
 respectively.
 Remind that,
 as in Figs.~\ref{fig:N_phi_theta-Eq-050-07900}
 and \ref{fig:N_phi_theta-Eq-050-04949},
 50 total events on average
 in each 60-day observation period
 have been simulated.

 By comparing with
 Figs.~\ref{fig:N_phi_theta-Eq-050-07900}
 and \ref{fig:N_phi_theta-Eq-050-04949},
 one can see clearly
 the similarity between
 the angular distribution patterns of the 3-D WIMP velocity
 observed in each 60-day observation period.
 The reason is understandable:
 since
 by definition
 the Earth coordinate system
 rotates daily with a one--year period,
 the directions of the $\xE$--axis
 on the central dates of the observation period
 in four plots shown
 in Fig.~\ref{fig:N_phi_theta-E-050-07900}
 or \ref{fig:N_phi_theta-E-050-04949}
 rotates 90$^{\circ}$ eastwards
 on the celestial Equator.
 Hence,
 four distribution patterns
 in the normal and the advanced seasons
 rotates 90$^{\circ}$ westwards around the $\zE$--axis.
 Due to the same reason,
 the distribution patterns
 in each pair of the normal and the advanced seasons
 would also rotate $\sim$ 30$^{\circ}$ westwards.
 In general,
 the angular distribution patterns
 in the Earth coordinate system
 should approximately be
 those in the Equatorial coordinate system
 combined with a westward rotation
 of a one--year period.
 This periodic variation of
 the anisotropic angular distribution of the 3-D WIMP velocity
 could be observed more clearly
 in our simulations
 with $\cal O$(500) total WIMP events
 and a higher analysis resolution
 given in Sec.~\ref{sec:N_phi_theta-E-500}.

\subsection{Angular WIMP velocity distribution in the horizontal frame}
\label{sec:N_phi_theta-H-050}

 In this subsection,
 we present
 the angular distribution of the 3-D WIMP velocity
 in the laboratory--dependent
 horizontal coordinate system.
 Remind that,
 as the Earth coordinate system,
 the horizontal coordinate system
 not only moves orbitally around
 (and also linearly with) the Sun,
 but also rotates daily and discretely.

 Similar to Eqs.~(\ref{eqn:V_chi_i_E-V_chi_i_Eq})
 and (\ref{eqn:V_chi_i_Eq-V_chi_i_E}),
 the transformations (pure rotations) of
 the $i$th 3-D WIMP velocity
 between the Earth and the horizontal coordinate systems
 can be given by
\cheqna
\beq
     {\bf v}_{\chi, i, {\rm H}}
  =  \MaEH(\phiLab, \thetaLab) \~ {\bf v}_{\chi, i, {\rm E}}
\~,
\label{eqn:V_chi_i_H-V_chi_i_E}
\eeq
 and,
 conversely,
\cheqnb
\beq
     {\bf v}_{\chi, i, {\rm E}}
  =  \MaHE(\phiLab, \thetaLab) \~ {\bf v}_{\chi, i, {\rm H}}
\~,
\label{eqn:V_chi_i_E-V_chi_i_H}
\eeq
\cheqn
 where the transformation matrices
 $\MaEH(\phiLab, \thetaLab)$ and $\MaHE(\phiLab, \thetaLab)$
 depending on the longitude and the latitude of the laboratory
 $(\phiLab, \thetaLab)$
 are given in Eqs.~(\ref{eqn:Ma_E_H}) and (\ref{eqn:Ma_H_E}).

\subsubsection{Annual modulation of the angular velocity distribution
               in the horizontal frame}
\label{sec:N_phi_theta-H-050-04949}
 \def \ShortFrame        {H}
 \def \PeriodA           {\PeriodAa}
 \def \Perioda           {\PeriodCa}
 \def \Periodb           {\PeriodCb}
 \def \Periodc           {\PeriodCc}
 \def \Periodd           {\PeriodCd}
 \def \PlotNumberA       {\PlotNumberAa}
 \def \PlotNumbera       {\PlotNumberCa}
 \def \PlotNumberb       {\PlotNumberCb}
 \def \PlotNumberc       {\PlotNumberCc}
 \def \PlotNumberd       {\PlotNumberCd}
 \def \LabName           {LNGS}
 \def \LabLocation       {(42.45$^{\circ}$N, 13.58$^{\circ}$E)}
 \InsertPlotNphithetaLab
  {The angular distributions of the 3-D WIMP velocity
   in the horizontal coordinate system
   at the location of the \LabName\ laboratory
   \LabLocation.
   50 total events on average
   in one entire year (a)
   and in each of four {\em advanced} seasons (b -- e)
   have been considered.
   All simulation setup and notations are the same
   as in Figs.~\ref{fig:N_phi_theta-E-050-00000}
   or \ref{fig:N_phi_theta-E-050-04949},
   respectively.
   \label{fig:N_phi_theta-H-050-04949-LNGS'}
   }
 \def \LabName           {SUPL}
 \def \LabLocation       {(37.07$^{\circ}$S, 142.81$^{\circ}$E)}
 \InsertPlotNphithetaLab
  {As in Figs.~\ref{fig:N_phi_theta-H-050-04949-LNGS'},
   except that
   the \LabName\ laboratory
   \LabLocation,
   as the so far unique functionable underground laboratory
   in the Southern Hemisphere,
   has been considered.
   \vspace{0.75 cm}
   \label{fig:N_phi_theta-H-050-04949-SUPL'}
   }

 In Figs.~\ref{fig:N_phi_theta-H-050-04949-LNGS'},
 we show
 the angular distributions of
 the 3-D WIMP velocity
 transformed from events
 shown in Figs.~\ref{fig:N_phi_theta-E-050-00000}
 and \ref{fig:N_phi_theta-E-050-04949}
 to the horizontal coordinate system
 at the location of the LNGS laboratory
 (42.45$^{\circ}$N, 13.58$^{\circ}$E)
 as a demonstration
 for an underground laboratory
 located in the Northern Hemisphere.
 50 total events on average in one experiment
 in one entire year (a)
 and in each of four 60-day observation periods
 of the {\em advanced} seasons
 (b -- e)
 have been considered.

 From the sketch of an underground laboratory
 located in the Northern Hemisphere in Winter
 in Fig.~\ref{fig:directional-1293-winter}(a)
 combined with the sketches
 of the Earth's positions
 in Figs.~\ref{fig:v_Earth_chi_S}(b)
 and \ref{fig:v_Earth_chi_S-20766},
 one can find that,
 first,
 during the observation period of the advanced Spring
 (see Fig.~\ref{fig:N_phi_theta-H-050-04949-LNGS'}(b)),
 the main direction of incident WIMPs
 should be approximately towards
 the $-{\xH}_{\rm , \~ LNGS}$--axis.
 Thus
 the highest--WIMP--flux bins
 is between 30$^{\circ}$N and 30$^{\circ}$S,
 from 150$^{\circ}$E to 180$^{\circ}$.
 Note here that
 there is a difference of $\sim$ 24 days
 between the central date of the advanced Spring (49.49 day)
 and the date sketched
 in Fig.~\ref{fig:directional-1293-winter}(a) (25.16 day).
 This corresponds to a shift of
 the angular distribution pattern of
 $\sim$ 24$^{\circ}$ westwards.
 Additionally,
 by definition of the horizontal coordinate system,
 the geographical difference
 between the LNGS laboratory and the Prime Meridian
 enlarge the shift of the angular distribution pattern
 by 13.58$^{\circ}$.
 Totally,
 the overall shift of
 the angular distribution observed at the LNGS laboratory
 in the advanced Spring
 is $\sim$ 38$^{\circ}$ westwards.

 Similar reason can explain
 the angular distribution of the advanced Winter
 (Fig.~\ref{fig:N_phi_theta-H-050-04949-LNGS'}(e)).
 A difference of $\sim$ 67 days
 between the central date of the advanced Winter (323.24 day)
 and the date sketched
 in Fig.~\ref{fig:directional-1293-winter}(a) (390.16 day)
 corresponds to a shift of
 the angular distribution pattern of
 $\sim$ 66$^{\circ}$ eastwards now.
 In addition,
 in this observation period,
 more events should {\em come from}
 the direction {\em above} the horizon
 and result in
 the angular distribution
 with slightly higher event numbers
 in the {\em southern} sky.

 On the other hand,
 from the sketch of a laboratory
 in the Northern Hemisphere in Summer
 (Fig.~\ref{fig:directional-1293-summer}(a))
 combined with the sketches
 in Figs.~\ref{fig:v_Earth_chi_S}(b)
 and \ref{fig:v_Earth_chi_S-20766},
 one can also conclude that,
 from the (advanced) Summer to the (advanced) Autumn,
 most WIMPs should come from the zenith
 and the most part in the northern sky
 in Figs.~\ref{fig:N_phi_theta-H-050-04949-LNGS'}(c) and (d)
 would hence have the lowest event numbers.

 Finally,
 the angular distribution
 shown in Fig.~\ref{fig:N_phi_theta-H-050-04949-LNGS'}(a)
 in one entire year
 is approximately a combination of
 the plots of four advanced seasons.
 Hence,
 as the angular distribution
 shown in Fig.~\ref{fig:N_phi_theta-E-050-00000},
 due to the Earth's orbital motion around the Sun,
 the anisotropy of the incident direction of Galactic WIMPs
 would be averaged out,
 except of the lowest--WIMP--flux bins
 from the center
 (i.e., the north at the laboratory location)
 to the north pole
 (the zenith of the laboratory).
 From the sketches
 in Figs.~\ref{fig:directional-1293-summer}(a)
 and \ref{fig:directional-1293-winter}(a)
 combined with the sketches
 in Figs.~\ref{fig:v_Earth_chi_S}(b)
 and \ref{fig:v_Earth_chi_S-20766},
 one can find that
 these bins correspond exactly to
 the sky around the Earth's North Pole
 (see Fig.~\ref{fig:N_phi_theta-E-050-00000}).

 Moreover,
 in order to demonstrate the {\em location}--dependence of
 (the annual modulation of)
 the angular distribution pattern,
 in Figs.~\ref{fig:N_phi_theta-H-050-04949-SUPL'}
 we show also
 the angular distributions of the 3-D WIMP velocity
 in the horizontal coordinate system
 at the location of the SUPL laboratory
 (37.07$^{\circ}$S, 142.81$^{\circ}$E),
 which is so far
 the unique functionable underground laboratory
 in the Southern Hemisphere.

 From Figs.~\ref{fig:directional-1293-summer}(b)
 and \ref{fig:directional-1293-winter}(b)
 and the detailed discussions above,
 one would expect that,
 from the (advanced) Spring to the (advanced) Summer
 (Figs.~\ref{fig:N_phi_theta-H-050-04949-SUPL'}(b) and (c)),
 the main direction of incident WIMPs
 should be towards the $-{\xH}_{\rm , \~ SUPL}$--axis.
 This implies that
 the high--WIMP--flux bins
 should be close to the 180$^{\circ}$ longitude.
 Moreover,
 from the (advanced) Autumn to the (advanced) Winter,
 most WIMPs should come from the ground
 and thus
 the most part in the southern sky
 in Figs.~\ref{fig:N_phi_theta-H-050-04949-SUPL'}(d) and (e)
 have the lowest event numbers.
 On the other hand,
 in contrast to Fig.~\ref{fig:N_phi_theta-H-050-04949-LNGS'}(a),
 the angular distribution in one entire year
 in Fig.~\ref{fig:N_phi_theta-H-050-04949-SUPL'}(a)
 shows the pattern
 with the lowest--WIMP--flux bins
 from the center to the south pole.

\subsection{Angular WIMP velocity distribution in the laboratory frame}
\label{sec:N_phi_theta-Lab-050}

 In this subsection,
 we present
 the angular distribution of the 3-D WIMP velocity
 in the laboratory coordinate system.
 Remind that
 the laboratory coordinate system
 not only moves orbitally around
 (and also linearly with) the Sun,
 but also rotates continuously.
 Thus
 the angular distribution
 in the laboratory coordinate system
 would also be averaged out,
 when a long observation period
 for accumulating enough WIMP events
 is required.
 Hence,
 in this subsection
 we only discuss
 the diurnal modulation of
 the angular distribution patterns.
 The angular distributions
 in one entire year
 and in four {\em advanced} seasons
 at the locations of several underground laboratories
 will be summarized
 in Appendix \ref{appx:N_phi_theta-ULabs}
 as readers' reference.

 Similar to Eqs.~(\ref{eqn:V_chi_i_H-V_chi_i_E})
 and (\ref{eqn:V_chi_i_E-V_chi_i_H}),
 the transformations (pure rotations) of
 the $i$th 3-D WIMP velocity
 between the horizontal and the laboratory coordinate systems
 can be given by
\cheqna
\beq
     {\bf v}_{\chi, i, {\rm Lab}}
  =  \MaHLab(t_i, \phiLab, \thetaLab) \~ {\bf v}_{\chi, i, {\rm H}}
\~,
\label{eqn:V_chi_i_Lab-V_chi_i_H}
\eeq
 and,
 conversely,
\cheqnb
\beq
     {\bf v}_{\chi, i, {\rm H}}
  =  \MaLabH(t_i, \phiLab, \thetaLab) \~ {\bf v}_{\chi, i, {\rm Lab}}
\~,
\label{eqn:V_chi_i_H-V_chi_i_Lab}
\eeq
\cheqn
 where the transformation matrices
 $\MaHLab(t, \phiLab, \thetaLab)$ and $\MaLabH(t, \phiLab, \thetaLab)$
 depending not only on the longitude and the latitude of the laboratory
 $(\phiLab, \thetaLab)$
 but also on the measuring time $t$ ($\tPM$) of each recorded WIMP event
 are given in Eqs.~(\ref{eqn:Ma_H_Lab}) and (\ref{eqn:Ma_Lab_H}).

\subsubsection{Diurnal modulation of
               the angular velocity distribution
               in the laboratory frame}
\label{sec:N_phi_theta-Lab-050-20766}
 \def \PeriodD               {\PeriodDa}
 \def \PlotNumberD           {\PlotNumberDa}
 \def \LabName               {LNGS}
 \def \LabLocation           {(42.45$^{\circ}$N, 13.58$^{\circ}$E)}
 \InsertPlotNphithetaDiurnal
  {The angular distributions of the 3-D WIMP velocity
   observed at the location of the \LabName\ laboratory
   \LabLocation\
   in four daily shifts
   listed in Table \ref{tab:interval_day}
   in the observation period of \PeriodD\ day.
   50 total events on average
   in each 4-hour daily shift
   in the 60-day observation period
   have been considered.
   All other simulation setup and notations are the same
   as in Figs.~\ref{fig:N_phi_theta-H-050-04949-LNGS'}.
   }
 \def \PeriodD               {\PeriodDb}
 \def \PlotNumberD           {\PlotNumberDb}
 \InsertPlotNphithetaDiurnal
  {As in Figs.~\ref{fig:N_phi_theta-Lab-\EventNumber-\PlotNumberDa-00-\LabName},
   except that
   the observation period is now
   \PeriodDb\ (= 55.16) day.
   }

 In Figs.~\ref{fig:N_phi_theta-Lab-050-20766-00-LNGS}
 and \ref{fig:N_phi_theta-Lab-050-39016-00-LNGS},
 we show
 the angular distributions of the 3-D WIMP velocity
 observed at the location of the LNGS laboratory
 (42.45$^{\circ}$N, 13.58$^{\circ}$E)
 in four daily shifts
 listed in Table \ref{tab:interval_day}%
\footnote{
 Note that,
 in the simulations of
 the diurnal modulation of
 the angular WIMP velocity distribution
 presented in this paper,
 we have re--calculated the local time
 of each simulated event
 of the considered laboratory
 from the generated measuring UTC time.
}
 in the observation period of \PeriodDa\ day
 and \PeriodDb\ (= 55.16) day,
 respectively.
 As usual,
 50 total events on average in one experiment
 in each 4-hour daily shift
 in the 60-day observation period
 have been considered.

 Not surprisingly,
 by comparing these two figures with each other,
 we can find first that,
 in two observation periods
 with a half--year time difference,
 the angular distribution patterns
 show indeed a 12-hour shift.
 Moreover,
 the diurnal modulations
 (from the midnight and morning shifts
  to the noon and evening shifts)
 in both of two periods
 look also similar to the annual modulation
 shown in Figs.~\ref{fig:N_phi_theta-H-050-04949-LNGS'}
 (b -- e).

 \def \PeriodD               {\PeriodDa}
 \def \PlotNumberD           {\PlotNumberDa}
 \def \LabName               {SUPL}
 \def \LabLocation           {(37.07$^{\circ}$S, 142.81$^{\circ}$E)}
 \InsertPlotNphithetaDiurnal
  {As in Figs.~\ref{fig:N_phi_theta-Lab-\EventNumber-\PlotNumberDa-00-LNGS},
   except that
   the \LabName\ laboratory
   \LabLocation,
   as the so far unique functionable underground laboratory
   in the Southern Hemisphere,
   has been considered.
   }
 \def \PeriodD               {\PeriodDb}
 \def \PlotNumberD           {\PlotNumberDb}
 \InsertPlotNphithetaDiurnal
  {As in Figs.~\ref{fig:N_phi_theta-Lab-\EventNumber-\PlotNumberDa-00-\LabName},
   except that
   the observation period is now
   \PeriodDb\ (= 55.16) day.
   }

 As in Sec.~\ref{sec:N_phi_theta-H-050},
 in order to demonstrate the location--dependence of
 the diurnal modulation of the angular distribution
 of the 3-D WIMP velocity,
 in Figs.~\ref{fig:N_phi_theta-Lab-050-20766-00-SUPL}
 and \ref{fig:N_phi_theta-Lab-050-39016-00-SUPL}
 we show also
 the angular distribution patterns
 observed at the location of the SUPL laboratory
 (37.07$^{\circ}$S, 142.81$^{\circ}$E)
 in four daily shifts
 in two observation periods.
 As expected,
 by comparing these two figures with each other
 and with Figs.~\ref{fig:N_phi_theta-H-050-04949-SUPL'},
 not only two observations described above,
 but also
 the location--dependence of
 (the annual modulation of)
 the angular distribution patterns
 observed in the horizontal coordinate system
 can be clearly confirmed.

 In Appendix \ref{appx:N_phi_theta-ULabs},
 one can find
 more simulation results and discussions
 for other underground laboratories.
 Note only that,
 although
 the diurnal modulation
 in the laboratory coordinate system
 (or,
  in fact,
  in any laboratory's own reference frame)
 can be pretty clearly observed,
 a few tens of WIMP events
 accumulated in each a--few--hour daily shift
 in one observation period of a few tens of days
 would be required.
 This means in turn that
 a couple of thousands of WIMP events
 would be needed to observe in one or two years,
 which should be
 a very hard challenge
 for (directional) direct Dark Matter detection experiments
 in the near future.

%

%
% 3/10
 %
%
\section{Bayesian reconstruction of
         the radial distribution of the 3-D WIMP velocity}
\label{sec:N_v-Bayesian-050}

 In this section,
 we turn to analyze
 the radial distribution (magnitude) of
 the (transformed) 3-D WIMP velocity
 in the Equatorial coordinate system%
\footnote{
 Remind that
 the transformations between the Equatorial,
 the Earth,
 the horizontal,
 and the laboratory coordinate systems
 are only rotations.
 These vary thus only
 the directions (angular components) of
 the generated 3-D WIMP velocities,
 but keep their magnitudes (radial components) unchanged.
}.
 We discuss at first
 the radial distribution of the 3-D WIMP velocity
 transformed from events
 shown in Fig.~\ref{fig:N_v-G-050-00000}
 to the Equatorial coordinate system.
 Then we present
 the reconstruction results of
 (the annual modulation of)
 the radial WIMP velocity distribution
 by using the Bayesian fitting procedure
 described in Sec.~\ref{sec:N_v-Bayesian}.

\subsection{Radial WIMP velocity distribution in the Equatorial frame}
\label{sec:N_v-Eq-050}
\begin{figure} [t!]
\begin{center}
 \includegraphics [width = 15 cm] {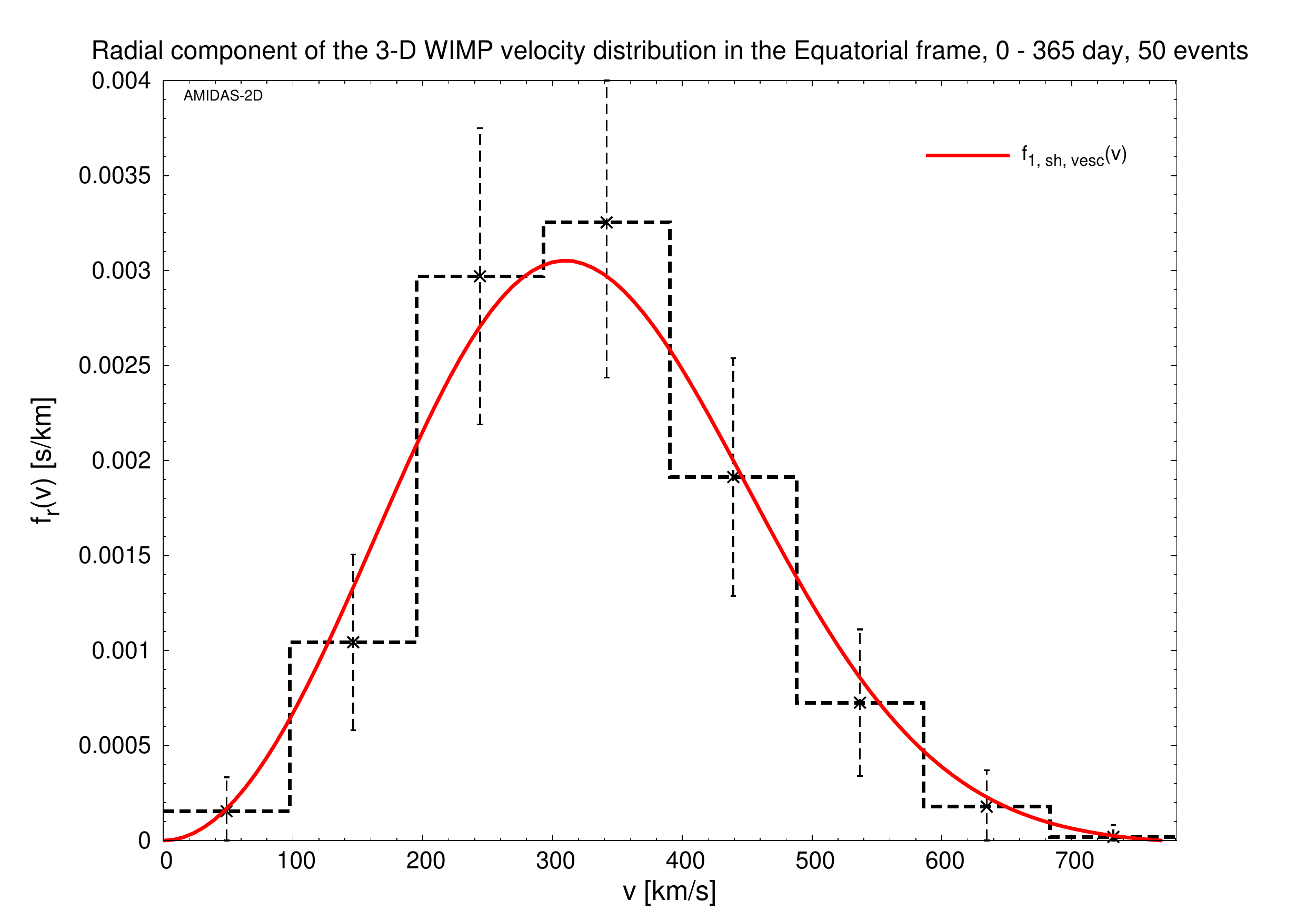}
\end{center}
\caption{
 The radial distribution of the 3-D WIMP velocity
 transformed from events
 shown in Fig.~\ref{fig:N_v-G-050-00000}
 to the Equatorial coordinate system
 by using Eqs.~(\ref{eqn:V_chi_i_S-V_chi_i_G})
 and (\ref{eqn:V_chi_i_Eq-V_chi_i_S}).
 The solid red curve is
 the shifted Maxwellian velocity distribution
 $f_{1, \sh, \rm vesc}(v)$
 given in Eq.~(\ref{eqn:f1v_sh_vesc})
 with the input value of $v_0 = 220$ km/s
 and $\ve = 1.05 \~ v_0$,
 while
 the dashed black histogram shows
 the binned radial component of
 the transformed 3-D WIMP velocities
 and the thin vertical dashed black lines
 indicate the 1$\sigma$ Poisson statistical uncertainties
 on the recorded event numbers in the $v$--bins.
 One entire year (0 to 365 day)
 and 50 total events on average
 have been considered.
 Remind that
 the maximal cut--off velocity here is given by
 $\vmax = 781$ km/s.
}
\label{fig:N_v-Eq-050-00000}
\end{figure}

 In Fig.~\ref{fig:N_v-Eq-050-00000}
 we show the radial distribution of the 3-D WIMP velocity
 transformed from events
 shown in Fig.~\ref{fig:N_v-G-050-00000}
 to the Equatorial coordinate system
 by using Eqs.~(\ref{eqn:V_chi_i_S-V_chi_i_G})
 and (\ref{eqn:V_chi_i_Eq-V_chi_i_S}).
 The solid red curve is
 the shifted Maxwellian velocity distribution
 $f_{1, \sh, \rm vesc}(v)$
 given in Eq.~(\ref{eqn:f1v_sh_vesc})
 with the input value of $v_0 = 220$ km/s
 and $\ve = 1.05 \~ v_0$,
 while
 the dashed black histogram shows
 the binned radial component of
 the transformed 3-D WIMP velocities
 and the thin vertical dashed black lines
 indicate the 1$\sigma$ Poisson statistical uncertainties
 on the recorded event numbers in the $v$--bins.
 One entire year (0 to 365 day)
 and 50 total events on average in one experiment
 have been considered.
 Remind that
 the maximal cut--off velocity here is given by
 $\vmax = 781$ km/s.

 It can be found that,
 firstly,
 although
 the theoretical (solid red) curve falls
 inside the 1$\sigma$ statistical uncertainty range of
 the radial distribution of the 3-D WIMP velocity,
 the theoretical distribution
 seems to be slightly higher than
 the histogram in the velocity ranges of
 100 km/s $\lsim \~ v \~ \lsim$ 200 km/s
 and 400 km/s $\lsim \~ v \~ \lsim$ 700 km/s,
 while
 in the velocity range around the peak
 (close to the average value)
 of the theoretical distribution
 (200 km/s $\lsim \~ v \~ \lsim$ 400 km/s),
 the radial distribution of
 the simulated WIMP velocity
 is obviously higher than
 the theoretical prediction.
 This indicates that
 there would be more WIMPs
 with middle--value velocities,
 but fewer WIMPs
 with relatively lower or higher velocities
 than theoretically predicted.
 In Sec.~\ref{sec:N_v-Bayesian-500},
 we will see that,
 with $\cal O$(500) total WIMP events
 and a higher analysis resolution,
 the difference between
 the simulated radial WIMP velocity distribution
 and the theoretical prediction
 would become more obviously.

\subsection{Fitted radial WIMP velocity distributions in the Equatorial frame}
\label{sec:N_v-Bayesian-Eq-050-00000}

 Since
 we found {\em unexpectedly} that
 the simulated radial WIMP velocity distribution
 in the Equatorial coordinate system
 could not match the theoretical prediction well,
 we considered then to apply
 the Bayesian fitting technique
 to reconstruct the radial component of
 the transformed 3-D WIMP velocity
 shown in Fig.~\ref{fig:N_v-Eq-050-00000},
 in order to find out the most suitable
 radial WIMP velocity distribution
 as well as
 to obtain the best--fit values of
 the Solar and Earth's Galactic velocities,
 $v_0$ and $\ve$.
 In this and the next subsections,
 we discuss our reconstruction results
 by using the Bayesian fitting procedure
 with four fitting velocity distributions
 given in Sec.~\ref{sec:N_v-Bayesian-f1v}.

\subsubsection[With the one--parameter velocity distribution
               $f_{1, \sh, v_0}(v; v_0)$]
              {\boldmath
               With the one--parameter velocity distribution
               $f_{1, \sh, v_0}(v; v_0)$}
\label{sec:N_v-Bayesian-Eq-v0-050-00000}

 We consider at first the one--parameter
 shifted Maxwellian velocity distribution
 $f_{1, \sh, v_0}(v; v_0)$
 given by Eq.~(\ref{eqn:f1v_sh_v0})
 with the constraint that
 $\ve = 1.05 \~ v_0$.
 The fitting parameter $v_0$
 has been scanned in the range of
 160 km/s $\le v_0 \le$ 270 km/s.

 \def \ShortFrame       {Eq}
 \def \Fittingfv        {v0}
 \InsertPlotNvBayesianD {v_0}
  {(a)
   The reconstructed radial distributions of
   the 3-D WIMP velocity
   and the 1(2)$\sigma$ statistical uncertainty bands
   by using the one--parameter
   shifted Maxwellian velocity distribution
   $f_{1, \sh, v_0}(v; v_0)$
   given by Eq.~(\ref{eqn:f1v_sh_v0})
   to fit data shown in Fig.~\ref{fig:N_v-Eq-050-00000}.
   The solid red curve
   is the shifted Maxwellian velocity distribution
   $f_{1, \sh, \rm vesc}(v)$
   given in Eq.~(\ref{eqn:f1v_sh_vesc})
   with the input value of $v_0 = 220$ km/s
   and $\ve = 1.05 \~ v_0$.
   While
   the dashed green curve
   labeled with the subscript ``median''
   indicates the reconstructed velocity distribution
   with the fitting parameter $v_0$
   given by the median value of all simulated experiments,
   the dash--dotted blue curve
   labeled with the subscript ``Pmax''
   indicates the reconstructed velocity distribution with $v_0$
   maximizing ${\rm P}_{\rm mean}\abrac{a_i,~i = 1,~2,~\cdots,~N_{\rm Bayesian}}$
   defined in Eq.~(\ref{eqn:P_Bayesian_mean}).
   (b)
   The distribution of the fitting parameter $v_0$
   in all simulated experiments.
   The red vertical line labeled with the subscript ``input''
   indicates the input value of $v_0$,
   whereas
   the green vertical line labeled with the subscript ``median''
   and the blue one labeled with the subscript ``Pmax''
   indicate the median value of the simulated results
   and the value which maximizes
   ${\rm P}_{\rm mean}$,
   respectively.
   In addition,
   the horizontal thick (thin) green bars show
   the 1(2)$\sigma$ ranges of the reconstructed results.
   See the text for further details.
   }

 In Fig.~\ref{fig:N_v-Bayesian-Eq-v0-050-00000}(a),
 we show the reconstructed radial distributions of
 the 3-D WIMP velocity
 and the 1(2)$\sigma$ statistical uncertainty bands
 by using $f_{1, \sh, v_0}(v; v_0)$
 to fit data shown in Fig.~\ref{fig:N_v-Eq-050-00000}.
 The solid red curve
 is the shifted Maxwellian velocity distribution
 $f_{1, \sh, \rm vesc}(v)$
 given in Eq.~(\ref{eqn:f1v_sh_vesc})
 with the input value of $v_0 = 220$ km/s
 and $\ve = 1.05 \~ v_0$.
 While
 the dashed green curve
 labeled with the subscript ``median''
 indicates the reconstructed velocity distribution
 with the fitting parameter $v_0$
 given by the median value of all simulated experiments,
 the dash--dotted blue curve
 labeled with the subscript ``Pmax''
 indicates the reconstructed velocity distribution with $v_0$
 maximizing ${\rm P}_{\rm mean}\abrac{a_i,~i = 1,~2,~\cdots,~N_{\rm Bayesian}}$
 defined in Eq.~(\ref{eqn:P_Bayesian_mean}).

 Additionally,
 the light--green (light--blue) area shown here
 indicate the 1(2)$\sigma$ statistical uncertainty bands
 of the Bayesian reconstructed velocity distribution,
 which has been determined as follows.
 After scanning
 the reconstructed fitting parameter $v_0$
 obtained from all simulated experiments
 and ordering
 according to their ${\rm P}_{\rm mean}$ values
 defined in Eq.~(\ref{eqn:P_Bayesian_mean})
 descendingly,
 we can not only
 determine the point
 which maximizes ${\rm P}_{\rm mean}$
 (labeled with the subscript ``Pmax'' in our plots hereafter),
 but also
 the smallest and largest values of
 the first 68.27\% (95.45\%) of
 all reconstructed $v_0$'s.
 We then use the smallest (largest) value of
 the first 68.27\% (95.45\%) reconstructed $v_0$'s
 to give the 1(2)$\sigma$ lower (upper) boundaries
 of the Bayesian reconstructed velocity distribution.
 This means that
 all of the velocity distributions with $v_0$'s
 which give the largest 68.27\% (95.45\%) ${\rm P}_{\rm mean}$ values
 should be in the 1(2)$\sigma$ light--green (light--blue) areas.

 Meanwhile,
 Fig.~\ref{fig:N_v-Bayesian-Eq-v0-050-00000}(b) shows
 the distribution of the fitting parameter $v_0$
 in all simulated experiments.
 The red vertical line labeled with the subscript ``input''
 indicates the input value of $v_0$,
 whereas
 the green vertical line labeled with the subscript ``median''
 and the blue one labeled with the subscript ``Pmax''
 indicate the median value of the simulated results
 and the value which maximizes
 ${\rm P}_{\rm mean}$,
 respectively.
 In addition,
 the horizontal thick (thin) green bars show
 the 1(2)$\sigma$ ranges of the reconstructed results.

 It can be seen that
 the best--fit value of $v_0 = 215.0$ km/s
 is only a little bit (0.45$\sigma$,
 see Table \ref{tab:N_v-Bayesian-Eq-050-00000})
 smaller than
 the input value of 220 km/s
 and thus
 the best--fit distribution curve
 would be very close to
 the theoretically derived
 shifted Maxwellian velocity distribution
 $f_{1, \sh, \vesc}(v)$.
 However,
 as found in Sec.~\ref{sec:N_v-Eq-050},
 the best--fit velocity distribution shown here
 differs clearly from
 the simulated radial distribution
 of the 3-D WIMP velocity.

\subsubsection[With the $v_0$--fixed velocity distribution
               $f_{1, \sh, \ve}(v; \ve)$]
              {\boldmath
               With the $v_0$--fixed velocity distribution
               $f_{1, \sh, \ve}(v; \ve)$}
\label{sec:N_v-Bayesian-Eq-ve-050-00000}

 As shown in Sec.~\ref{sec:N_v-Bayesian-Eq-v0-050-00000},
 by using $v_0$ as the unique fitting parameter
 with the fixed relation between $v_0$ and $\ve$,
 the reconstructed velocity distribution can {\em not} fit
 the simulated radial distribution of the 3-D WIMP velocity
 as well as we hoped.
 Hence,
 we consider here the $v_0$--fixed
 shifted Maxwellian velocity distribution
 $f_{1, \sh, \ve}(v; \ve)$
 given by Eq.~(\ref{eqn:f1v_sh_ve})
 with the input condition that
 $v_0 = 220$ km/s
 (although
  the best--fit value of $v_0$
  obtained in Sec.~\ref{sec:N_v-Bayesian-Eq-v0-050-00000}
  is 215.0 km/s)
 and scan the Earth's Galactic velocity $\ve$
 in the range of 90 km/s $\le \ve \le$ 330 km/s.

 \def \Fittingfv        {ve}
 \InsertPlotNvBayesianD {v_e}
  {(a)
   The reconstructed radial distributions of
   the 3-D WIMP velocity
   and the 1(2)$\sigma$ statistical uncertainty bands
   by using the $v_0$--fixed
   shifted Maxwellian velocity distribution
   $f_{1, \sh, \ve}(v; \ve)$
   given by Eq.~(\ref{eqn:f1v_sh_ve})
   to fit data
   shown in Fig.~\ref{fig:N_v-Eq-050-00000}.
   (b)
   The distribution of the fitting parameter $\ve$
   in all simulated experiments.
   The red vertical line
   labeled with the subscript ``input''
   indicates
   the theoretical value of
   $\ve = 1.05 \~ v_0 = 231$ km/s.
   Other notations are the same as
   in Figs.~\ref{fig:N_v-Bayesian-Eq-v0-050-00000}.
   }

 In Fig.~\ref{fig:N_v-Bayesian-Eq-ve-050-00000}(a),
 we show the reconstructed radial distributions of
 the 3-D WIMP velocity
 and the 1(2)$\sigma$ statistical uncertainty bands
 by using $f_{1, \sh, \ve}(v; \ve)$
 to fit data
 shown in Fig.~\ref{fig:N_v-Eq-050-00000}.
 Also,
 in Fig.~\ref{fig:N_v-Bayesian-Eq-ve-050-00000}(b),
 we give
 the distribution of the fitting parameter $\ve$
 in all simulated experiments.
 The red vertical line
 labeled with the subscript ``input''
 indicates
 the theoretical value of
 \mbox{$\ve = 1.05 \~ v_0 = 231$ km/s}.
 Unfortunately and unexpectedly,
 we have obtained a similar result as
 in Sec.~\ref{sec:N_v-Bayesian-Eq-v0-050-00000}:
 the best--fit value of $\ve = 224.4$ km/s
 is only a little bit (0.25$\sigma$,
 see Table \ref{tab:N_v-Bayesian-Eq-050-00000})
 smaller than
 the theoretical value of 231 km/s,
 and thus
 the best--fit distribution curve
 would be very close to
 the theoretical derivation
 $f_{1, \sh, \vesc}(v)$,
 but {\em not} be improved to fit
 the simulated radial WIMP velocity distribution
 very much
 (as we hoped).

\subsubsection[With the simplified velocity distribution
               $f_{1, \sh}(v; v_0, \ve)$]
              {\boldmath
               With the simplified velocity distribution
               $f_{1, \sh}(v; v_0, \ve)$}
\label{sec:N_v-Bayesian-Eq-sh-050-00000}

 Our fitting results
 shown in Secs.~\ref{sec:N_v-Bayesian-Eq-v0-050-00000}
 and \ref{sec:N_v-Bayesian-Eq-ve-050-00000}
 indicate that,
 although
 the best--fit values of $v_0 = 215.0$ km/s
 and $\ve = 224.4$ km/s
 are only a little bit smaller than
 the theoretical values
 and thus the best--fit distribution curves
 would be very close to
 the theoretically predicted
 shifted Maxwellian velocity distribution
 $f_{1, \sh, \vesc}(v)$,
 there would obviously be some systematic bias
 between the theoretical derivation
 and our (Monte Carlo--simulated) radial velocity distribution.
 Hence,
 we release now
 the constraints on both of
 the fitting parameters $v_0$ and $\ve$,
 use the simplified shifted Maxwellian velocity distribution
 $f_{1, \sh}(v; v_0, \ve)$
 given by Eq.~(\ref{eqn:f1v_sh})
 and scan the parameter plane of
 80 km/s $\le v_0 \le$ 340 km/s
 and 0 $\le \ve \le$ 380 km/s
 in order to obtain
 a better--fitted WIMP velocity distribution%
\footnote{
 Note that
 the scanning ranges of both of $v_0$ and $\ve$
 are extended from those
 used for the one--parameter and the $v_0$--fixed
 velocity distributions
 in Secs.~\ref{sec:N_v-Bayesian-Eq-v0-050-00000}
 and \ref{sec:N_v-Bayesian-Eq-ve-050-00000}.
}.

 In Fig.~\ref{fig:N_v-Bayesian-Eq-sh-050-00000}(a),
 we show the reconstructed radial distributions of
 the 3-D WIMP velocity
 and the 1(2)$\sigma$ statistical uncertainty bands
 by using $f_{1, \sh}(v; v_0, \ve)$
 to fit data
 shown in Fig.~\ref{fig:N_v-Eq-050-00000}.
 It can be seen that,
 while the red theoretically derived velocity distribution
 (\ref{eqn:f1v_sh_vesc})
 could still be covered by
 the 1$\sigma$ statistical uncertainty band,
 except of in the ranges around two turning points:
 $v \sim 270$ km/s and $v \sim 550$ km/s,
 the best--fit velocity distributions
 would now clearly differ from the theoretical derivation,
 but indeed fit to our simulated radial distribution
 of the 3-D WIMP velocity much better.

 Meanwhile,
 Fig.~\ref{fig:N_v-Bayesian-Eq-sh-050-00000}(b)
 shows the distribution of the fitting parameters $v_0$ and $\ve$
 in all simulated experiments on the $v_0 - \ve$ plane.
 The light--green (light--blue, golden) square points
 indicate the 1(2, $> 2$)$\sigma$ areas
 of the reconstructed combination of $v_0$ and $\ve$.
 While
 the red upward--triangle
 labeled with the subscript ``input'' indicates
 the theoretical values of \mbox{$v_0 = 220$ km/s}
 and $\ve = 231$ km/s,
 the green disk
 labeled with the subscript ``median''
 and the blue downward--triangle
 labeled with the subscript ``Pmax'' indicate
 the median values of the simulated results
 and the point which maximizes ${\rm P}_{\rm mean}$,
 respectively.
 Additionally,
 Figs.~\ref{fig:N_v-Bayesian-Eq-sh-050-00000}(c) and (d)
 show the distributions of the fitting parameters $v_0$ and $\ve$
 in all simulated experiments separately.

 \def \Fittingfv        {sh}
 \InsertPlotNvBayesianQ {v_0} {v_e}
  {(a)
   The reconstructed radial distributions of
   the 3-D WIMP velocity
   and the 1(2)$\sigma$ statistical uncertainty bands
   by using the simplified
   shifted Maxwellian velocity distribution
   $f_{1, \sh}(v; v_0, \ve)$
   given by Eq.~(\ref{eqn:f1v_sh})
   to fit data
   shown in Fig.~\ref{fig:N_v-Eq-050-00000}.
   Notations are the same as
   in Fig.~\ref{fig:N_v-Bayesian-Eq-v0-050-00000}(a).
   (b)
   The distribution of the fitting parameters $v_0$ and $\ve$
   in all simulated experiments on the $v_0 - \ve$ plane.
   The light--green (light--blue, golden) square points
   indicate the 1(2, $> 2$)$\sigma$ areas
   of the reconstructed combination of $v_0$ and $\ve$.
   While
   the red upward--triangle
   labeled with the subscript ``input'' indicates
   the theoretical values of $v_0 = 220$ km/s and $\ve = 231$ km/s,
   the green disk
   labeled with the subscript ``median''
   and the blue downward--triangle
   labeled with the subscript ``Pmax'' indicate
   the median values of the simulated results
   and the point which maximizes ${\rm P}_{\rm mean}$,
   respectively.
   The meaning of
   the horizontal and vertical thick (thin) green bars
   are the same as
   in Figs.~\ref{fig:N_v-Bayesian-Eq-v0-050-00000}(b)
   and \ref{fig:N_v-Bayesian-Eq-ve-050-00000}(b).
   (c)
   As in Figs.~\ref{fig:N_v-Bayesian-Eq-v0-050-00000}(b).
   (d)
   As in Figs.~\ref{fig:N_v-Bayesian-Eq-ve-050-00000}(b).
   }

 Corresponding to the reconstructed radial distributions,
 from these three plots
 one can find that,
 while
 the best--fit values of $v_0 \simeq 190$ km/s
 are $\sim 14\%$ smaller than the input value,
 those of $\ve = 258.4$ km/s
 are now $\sim 12\%$ {\em larger} than its theoretical value
 (see also Table \ref{tab:N_v-Bayesian-Eq-050-00000}),
 although
 (the combination of) the theoretical values
 of our fitting parameters
 could still fall inside
 their 1$\sigma$ statistical uncertainty (green) area/ranges.
 Consequently,
 the ratio between the best--fit values of $v_0$ and $\ve$
 is $\sim$ 1.4,
 much larger than the commonly adopted theoretical value of 1.05
 \cite{Freese88}
 and our estimate of 1.0078.%
\footnote{
 See Appendix \ref{appx:v_Earth_chi_S}
 for the detailed calculation.
}
 Note also that,
 due to the pretty small ($\cal O$(50))
 simulated total event number
 and thus a large statistical fluctuation,
 there is a long distribution tail
 at the lower scanning boundary of $\ve$
 in Fig.~\ref{fig:N_v-Bayesian-Eq-sh-050-00000}(b).
 In Sec.~\ref{sec:N_v-Bayesian-500},
 we will see that,
 once the total event number
 is raised to $\cal O$(500)
 and thus the statistics is improved,
 this tail would disappear.

\subsubsection[With the modified velocity distribution
               $f_{1, \sh, \Delta v}(v; v_0, \Delta v)$]
              {\boldmath
               With the modified velocity distribution
               $f_{1, \sh, \Delta v}(v; v_0, \Delta v)$}
\label{sec:N_v-Bayesian-Eq-Dv-050-00000}

 In Ref.~\cite{DMDDf1v-Bayesian},
 we introduced the modified
 shifted Maxwellian velocity distribution
 $f_{1, \sh, \Delta v}(v; v_0, \Delta v)$
 given by Eq.~(\ref{eqn:f1v_sh_Dv})
 for reducing the systematic bais
 of the best--fit results
 obtained by using the simplified velocity distribution
 $f_{1, \sh}(v; v_0, \ve)$.
 For checking and potentially also improving
 our reconstruction results
 shown in Sec.~\ref{sec:N_v-Bayesian-Eq-sh-050-00000},
 we apply now the modified velocity distribution
 $f_{1, \sh, \Delta v}(v; v_0, \Delta v)$
 and scan the parameter plane of
 80 km/s $\le v_0 \le$ 340 km/s
 and $-190$ km/s $\le \Delta v \le$ 230 km/s.

 \def \Fittingfv        {Dv}
 \InsertPlotNvBayesianQ {v_0} {D_v}
  {(a)
   The reconstructed radial distributions of
   the 3-D WIMP velocity
   and the 1(2)$\sigma$ statistical uncertainty bands
   by using the modified
   shifted Maxwellian velocity distribution
   $f_{1, \sh, \Delta v}(v; v_0, \Delta v)$
   given by Eq.~(\ref{eqn:f1v_sh_Dv})
   to fit data
   shown in Fig.~\ref{fig:N_v-Eq-050-00000}.
   Notations are the same as
   in Fig.~\ref{fig:N_v-Bayesian-Eq-v0-050-00000}(a).
   (b)
   The distribution of the fitting parameters $v_0$ and $\Delta v$
   in all simulated experiments on the $v_0 - \Delta v$ plane.
   The light--green (light--blue, golden) square points
   indicate the 1(2, $> 2$)$\sigma$ areas
   of the reconstructed combination of $v_0$ and $\Delta v$.
   The red upward--triangle
   labeled with the subscript ``input''   indicates
   the theoretical values of $v_0 = 220$ km/s
   and \mbox{$\Delta v = 11$ km/s}.
   Other notations are the same as
   in Fig.~\ref{fig:N_v-Bayesian-Eq-sh-050-00000}(b).
   (c)
   As in Figs.~\ref{fig:N_v-Bayesian-Eq-v0-050-00000}(b).
   (d)
   As in Figs.~\ref{fig:N_v-Bayesian-Eq-ve-050-00000}(b).
   }

 In Fig.~\ref{fig:N_v-Bayesian-Eq-Dv-050-00000}(a)
 we show the reconstructed radial distributions of
 the 3-D WIMP velocity
 and the 1(2)$\sigma$ statistical uncertainty bands
 by using $f_{1, \sh, \Delta v}(v; v_0, \Delta v)$
 to fit data
 shown in Fig.~\ref{fig:N_v-Eq-050-00000}.
 Not surprisingly,
 the reconstructed velocity distributions
 as well as
 the 1(2)$\sigma$ statistical uncertainty bands
 have similar shapes as those shown
 in Fig.~\ref{fig:N_v-Bayesian-Eq-sh-050-00000}(a).

 Meanwhile,
 Fig.~\ref{fig:N_v-Bayesian-Eq-Dv-050-00000}(b)
 shows the distribution of the fitting parameters $v_0$ and $\Delta v$
 in all simulated experiments on the $v_0 - \Delta v$ plane.
 The light--green (light--blue, golden) square points
 indicate the 1(2, $> 2$)$\sigma$ areas
 of the reconstructed combination of $v_0$ and $\Delta v$.
 While
 the red upward--triangle
 labeled with the subscript ``input'' indicates
 the theoretical values of \mbox{$v_0 = 220$ km/s}
 and $\Delta v = 11$ km/s,
 the green disk
 labeled with the subscript ``median''
 and the blue downward--triangle
 labeled with the subscript ``Pmax'' indicate
 the median values of the simulated results
 and the point which maximizes ${\rm P}_{\rm mean}$,
 respectively.
 As shown in \ref{fig:N_v-Bayesian-Eq-sh-050-00000}(b),
 one can also see here
 a long distribution tail
 at the lower scanning boundary of $\Delta v$.
 Additionally,
 Figs.~\ref{fig:N_v-Bayesian-Eq-Dv-050-00000}(c) and (d)
 show the distribution of the fitting parameters $v_0$ and $\Delta v$
 in all simulated experiments separately.

 From these three plots
 one can confirm that,
 first,
 the best--fit values of $v_0 \simeq 190$ km/s
 are the same as the results
 obtained by using the simplified velocity distribution
 (see also Table \ref{tab:N_v-Bayesian-Eq-050-00000}).
 Second,
 the best--fit values of $\Delta v \simeq 70$ km/s
 are also equal to
 the difference between $v_0$ and $\ve$
 obtained previously.
 Note however that,
 the 1(2)$\sigma$ statistical uncertainties on $\Delta v$
 would be at least 1.5 times larger than those on $\ve$
 (see Table \ref{tab:N_v-Bayesian-Eq-050-00000}).

\subsubsection*{}
 \def \PlotNumber             {\PlotNumberAa}
\InsertResultsTableNvBayesian
{{$f_{1, \sh, v_0}(v)$}      &
  $v_0$ [km/s]               &
  215.0 & $215.0    \pm 11.0         \~ (^{+ 21.2}_{- 22.0})$ & [204.0, 226.0] & [193.0, 236.2] \\
 \hline
 \hline
 {$f_{1, \sh, \ve}(v)$}      &
  $\ve$ [km/s]               &
  224.4 & $224.4 \~ ^{+26.4}_{-28.8} \~ (^{+ 50.4}_{- 60.0})$ & [195.6, 250.8] & [164.4, 274.8] \\
 \hline
 \hline
 \multirow{2}{*}
 {$f_{1, \sh}(v)$}           &
  $v_0$ [km/s]               &
  191.8 & $186.6 \~ ^{+36.4}_{-26.0} \~ (^{+101.4}_{-~44.2})$ & [160.6, 223.0] & [142.4, 288.0] \\
 \cline{2-6}
                             &
  $\ve$ [km/s]               &
  258.4 & $258.4 \~ ^{+30.4}_{-45.6} \~ (^{+~53.2}_{-235.6})$ & [212.8, 288.8] &  [22.8, 311.6] \\
 \hline
 \hline
 \multirow{2}{*}
 {$f_{1, \sh, \Delta v}(v)$} &
  $v_0$ [km/s]               &
  191.8 & $186.6 \~ ^{+36.4}_{-26.0} \~ (^{+ 93.6}_{- 44.2})$ & [160.6, 223.0] & [142.4, 280.2] \\
 \cline{2-6}
                             &
  $\Delta v$ [km/s]          &
   66.2 & $ 74.6 \~ ^{+46.2}_{-79.8} \~ (^{+~84.0}_{-264.6})$ & [$-5.2$, 120.8] & [$-190.0$, 158.6] \\
 }
{50 total events on average in the observation period of \PeriodA\ day
 }
{The summary of
 the reconstructed results of the fitting parameters
 and their 1(2)$\sigma$ statistical uncertainty ranges
 of the median values
 by using four considered fitting velocity distributions
 with 50 total events on average in one entire year
 shown in Figs.~\ref{fig:N_v-Bayesian-Eq-v0-050-00000},
 \ref{fig:N_v-Bayesian-Eq-ve-050-00000},
 \ref{fig:N_v-Bayesian-Eq-sh-050-00000},
 and \ref{fig:N_v-Bayesian-Eq-Dv-050-00000}.
 }

 In Table \ref{tab:N_v-Bayesian-Eq-050-00000},
 we summarize
 the reconstructed results of the fitting parameters
 and their 1(2)$\sigma$ statistical uncertainty ranges
 of the median values
 with 50 total events on average in one entire year
 for all four considered fitting velocity distributions
 shown in Figs.~\ref{fig:N_v-Bayesian-Eq-v0-050-00000},
 \ref{fig:N_v-Bayesian-Eq-ve-050-00000},
 \ref{fig:N_v-Bayesian-Eq-sh-050-00000},
 and \ref{fig:N_v-Bayesian-Eq-Dv-050-00000}.

\subsection{Annual modulation of
            the radial distribution of the 3-D WIMP velocity}
\label{sec:N_v-Bayesian-Eq-050-04949}

 In this subsection,
 we discuss the annual modulation of
 the radial distribution of
 the 3-D WIMP velocity
 and present the reconstructed radial WIMP velocity distributions
 in the observation periods of four {\em advanced} seasons%
\footnote{
 For the sake of completeness and readers' reference,
 the reconstructed radial WIMP velocity distributions
 for four normal seasons
 will be given in Appendix \ref{appx:N_v-Bayesian-Eq-07900}.
}.
\subsubsection[With the one--parameter velocity distribution
               $f_{1, \sh, v_0}(v; v_0)$]
              {\boldmath
               With the one--parameter velocity distribution
               $f_{1, \sh, v_0}(v; v_0)$}
\label{sec:N_v-Bayesian-Eq-v0-050-04949}
 \def \Perioda         {\PeriodCa}
 \def \Periodb         {\PeriodCb}
 \def \Periodc         {\PeriodCc}
 \def \Periodd         {\PeriodCd}
 \def \PlotNumbera     {\PlotNumberCa}
 \def \PlotNumberb     {\PlotNumberCb}
 \def \PlotNumberc     {\PlotNumberCc}
 \def \PlotNumberd     {\PlotNumberCd}
 \def \Fittingfv       {v0}
 \def \FittingPara     {v_0}
 \InsertPlotNvBayesianAnnual
  {As in Figs.~\ref{fig:N_v-Bayesian-Eq-v0-050-00000},
   reconstructed with the one--parameter velocity distribution
   $f_{1, \sh, v_0}(v; v_0)$,
   except that
   four 60-day observation periods of the advanced seasons
   have been considered.
   }
 \def \PlotNumber                   {\PlotNumberCa}
\InsertResultsTableNvBayesianAnnual
{one--parameter velocity distribution
 $f_{1, \sh, v_0}(v; v_0)$}
{\begin{minipage} {3.25 cm}
  \begin{center}
     ~          \\ \vspace{1.1  ex}
      49.49     \\ \vspace{0.65 ex}
    (\PeriodCa) \\ \vspace{1.1  ex}
  \end{center}
 \end{minipage}             &
 $v_0$ [km/s]               &
 215.0 & $215.0    \pm 11.0         \~ (^{+22.0}_{-20.9})$ & [204.0, 226.0] & [194.1, 237.0] \\
 \hline
 \begin{minipage} {3.25 cm}
  \begin{center}
     ~          \\ \vspace{1.1  ex}
     140.74     \\ \vspace{0.65 ex}
    (\PeriodCb) \\ \vspace{1.1  ex}
  \end{center}
 \end{minipage}             &
 $v_0$ [km/s]               &
 220.5 & $220.5    \pm 11.0         \~ (^{+23.1}_{-20.9})$ & [209.5, 231.5] & [199.6, 243.6] \\
 \hline
 \begin{minipage} {3.25 cm}
  \begin{center}
     ~          \\ \vspace{1.1  ex}
     231.99     \\ \vspace{0.65 ex}
    (\PeriodCc) \\ \vspace{1.1  ex}
  \end{center}
 \end{minipage}             &
 $v_0$ [km/s]               &
 215.0 & $215.0    \pm 11.0         \~ (^{+22.0}_{-20.9})$ & [204.0, 226.0] & [194.1, 237.0] \\
 \hline
 \begin{minipage} {3.25 cm}
  \begin{center}
     ~          \\ \vspace{1.1  ex}
     323.24     \\ \vspace{0.65 ex}
    (\PeriodCd) \\ \vspace{1.1  ex}
  \end{center}
 \end{minipage}             &
 $v_0$ [km/s]               &
 209.5 & $208.4 \~ ^{+11.0}_{-~9.9} \~ (^{+22.0}_{-19.8})$ & [198.5, 219.4] & [188.6, 230.4] \\
 }
{50}
{parameter $v_0$ and}
{advanced}

 As in Sec.~\ref{sec:N_v-Bayesian-Eq-050-00000},
 we consider at first
 the one--parameter velocity distribution
 $f_{1, \sh, v_0}(v; v_0)$
 with $v_0$ as the fitting parameter
 and the constraint that
 $\ve = 1.05 \~ v_0$.
 Note however that,
 although
 the fitting parameter $v_0$
 is in fact the Solar Galactic velocity
 and should be fixed
 in the whole year,
 it has been used here
 as a normal fitting parameter%
\footnote{
 In Secs.~\ref{sec:N_v-Bayesian-Eq-sh-050-04949}
 and \ref{sec:N_v-Bayesian-Eq-Dv-050-04949},
 we will see that,
 once our constraints on $v_0$ and $\ve$ are released
 and two fitting parameters,
 $v_0$ and $\ve$ (or $\Delta v$),
 are used,
 the best--fit values of $v_0$
 would indeed be a constant.
}.
 \def \Fittingfv       {ve}
 \def \FittingPara     {v_e}
 \InsertPlotNvBayesianAnnual
  {As in Figs.~\ref{fig:N_v-Bayesian-Eq-ve-050-00000},
   reconstructed with the $v_0$--fixed velocity distribution
   $f_{1, \sh, \ve}(v; \ve)$,
   except that
   four 60-day observation periods of the advanced seasons
   have been considered.
   \\~
   }

 In Figs.~\ref{fig:N_v-Bayesian-Eq-v0-050-04949},
 we show the reconstructed radial WIMP velocity distributions
 and the 1(2)$\sigma$ statistical uncertainty bands
 by using $f_{1, \sh, v_0}(v; v_0)$
 as well as
 the distributions of the fitting parameter $v_0$
 in all simulated experiments
 for four advanced seasons.
 Firstly,
 from the plots in the left column
 of Figs.~\ref{fig:N_v-Bayesian-Eq-v0-050-04949},
 one can see
 the variation of the radial distributions
 of the 3-D WIMP velocity
 in different seasons:
 the event numbers in the high--velocity bins
 (\mbox{$v \~ \gsim \~ 290$ km/s})
 vary (from top to bottom) at first more and then fewer;
 in contrast,
 the event numbers in low--velocity bins
 ($v \~ \lsim \~ 290$ km/s)
 at first fewer and then more.
 This can be observed in particularly clearly
 from the variation of the difference
 between the third and the forth $v$--bins
 around the distribution peak.
 This implies that
 the (average) velocity of the simulated WIMP events
 indeed increases at first to the maximum
 (in the advanced Summer)
 and then reduces to the minimum
 (in the advanced Winter).
 Secondly and more importantly,
 the reconstructed velocity distributions
 and their 1(2)$\sigma$ statistical uncertainty bands
 follow the variation of the histogram
 in different seasons as tightly as possible.
 Note that
 the red theoretically derived curve
 shown in the left column
 of Figs.~\ref{fig:N_v-Bayesian-Eq-v0-050-04949}
 is time--{\em independent}.

 Moreover,
 from the plots in the right column
 of Figs.~\ref{fig:N_v-Bayesian-Eq-v0-050-04949},
 the {\em symmetric periodic} variation of
 the fitting parameter $v_0$
 of 215.0 ($\pm \sim 5.5$) km/s
 can also be seen clearly.
 It would be worth to notice here that
 the $\sim \pm 5.5$ km/s annual variation of
 the best--fit value of $v_0$
 is $\sim$ 0.5$\sigma$ of
 the statistical uncertainties on $v_0$
 (see the summary
  given in Table \ref{tab:N_v-Bayesian-Eq-050-00000})
 and approximately equal to
 the difference between
 the annual average value of $v_0 = 215.0$ km/s
 (see Table \ref{tab:N_v-Bayesian-Eq-050-00000})
 and its theoretical value.
 In Sec.~\ref{sec:N_v-Bayesian-Eq-500-04949},
 we will show that,
 with $\cal O$(500) total WIMP events
 and thus \mbox{$\sim$ 1/3} reduced statistical uncertainties on $v_0$,
 the annual variation of
 the fitting parameter $v_0$
 and in turn
 that of the radial WIMP velocity distribution
 could be identified with a confidence level
 of 2$\sigma$ to 3$\sigma$.

 In Table \ref{tab:N_v-Bayesian-Eq-v0-050-04949},
 we summarize
 the reconstructed results of the fitting parameter $v_0$
 and their 1(2)$\sigma$ statistical uncertainty ranges
 of the median values
 by using the one--parameter velocity distribution
 $f_{1, \sh, v_0}(v; v_0)$
 with 50 total events on average
 in each 60-day observation period.

\subsubsection[With the $v_0$--fixed velocity distribution
               $f_{1, \sh, \ve}(v; \ve)$]
              {\boldmath
               With the $v_0$--fixed velocity distribution
               $f_{1, \sh, \ve}(v; \ve)$}
\label{sec:N_v-Bayesian-Eq-ve-050-04949}
\InsertResultsTableNvBayesianAnnual
{$v_0$--fixed velocity distribution
 $f_{1, \sh, \ve}(v; \ve)$}
{\begin{minipage} {3.25 cm}
  \begin{center}
     ~          \\ \vspace{1.1  ex}
      49.49     \\ \vspace{0.65 ex}
    (\PeriodCa) \\ \vspace{1.1  ex}
  \end{center}
 \end{minipage}             &
 $\ve$ [km/s]               &
 224.4 & $224.4 \~ ^{+26.4}_{-28.8} \~ (^{+50.4}_{-58.2})$ & [195.6, 250.8] & [166.2, 274.8] \\
 \hline
 \begin{minipage} {3.25 cm}
  \begin{center}
     ~          \\ \vspace{1.1  ex}
     140.74     \\ \vspace{0.65 ex}
    (\PeriodCb) \\ \vspace{1.1  ex}
  \end{center}
 \end{minipage}             &
 $\ve$ [km/s]               &
 238.8 & $238.8    \pm 26.4         \~ (^{+50.4}_{-52.8})$ & [212.4, 265.2] & [186.0, 289.2] \\
 \hline
 \begin{minipage} {3.25 cm}
  \begin{center}
     ~          \\ \vspace{1.1  ex}
     231.99     \\ \vspace{0.65 ex}
    (\PeriodCc) \\ \vspace{1.1  ex}
  \end{center}
 \end{minipage}             &
 $\ve$ [km/s]               &
 224.4 & $224.4    \pm 26.4         \~ (^{+50.4}_{-57.6})$ & [198.0, 250.8] & [166.8, 274.8] \\
 \hline
 \begin{minipage} {3.25 cm}
  \begin{center}
     ~          \\ \vspace{1.1  ex}
     323.24     \\ \vspace{0.65 ex}
    (\PeriodCd) \\ \vspace{1.1  ex}
  \end{center}
 \end{minipage}             &
 $\ve$ [km/s]               &
 207.6 & $207.6    \pm 28.8         \~ (^{+52.8}_{-60.0})$ & [178.8, 236.4] & [147.6, 260.4] \\
 }
{50}
{parameter $\ve$ and}
{advanced}

 In order to check
 the identification possibility of
 the variation of the Earth's Galactic velocity $\ve$,
 we consider then
 the $v_0$--fixed velocity distribution
 $f_{1, \sh, \ve}(v; \ve)$
 with $\ve$ as the fitting parameter
 and the input condition that
 $v_0 = 220$ km/s.

 In Figs.~\ref{fig:N_v-Bayesian-Eq-ve-050-04949},
 we show the reconstructed radial WIMP velocity distributions
 and the 1(2)$\sigma$ statistical uncertainty bands
 by using $f_{1, \sh, \ve}(v; \ve)$
 as well as
 the distributions of the fitting parameter $\ve$
 in all simulated experiments
 for four advanced seasons.
 Firstly,
 from the plots in the left column
 of Figs.~\ref{fig:N_v-Bayesian-Eq-ve-050-04949},
 one can find that
 the reconstructed velocity distributions
 and their 1(2)$\sigma$ statistical uncertainty bands
 also follow
 the variation of the simulated radial velocity distribution
 in different seasons.
 Meanwhile,
 from the plots in the right column
 of Figs.~\ref{fig:N_v-Bayesian-Eq-ve-050-04949},
 one can also see
 the (approximately) symmetric periodic variation of
 the fitting parameter $\ve$
 of \mbox{224.4 ($^{+14.4}_{-16.8}$) km/s} clearly;
 this $^{+14.4}_{-16.8}$ km/s annual variation of
 the best--fit value of $\ve$
 is also \mbox{$\sim$ 0.5$\sigma$} of
 the statistical uncertainties on $\ve$
 (see the summary
  given in Table \ref{tab:N_v-Bayesian-Eq-050-00000}).
 In Sec.~\ref{sec:N_v-Bayesian-Eq-500-04949},
 we will show that,
 with $\cal O$(500) total WIMP events
 and thus $\sim$ 1/3 reduced statistical uncertainties on $\ve$,
 its annual variation
 could be identified with a confidence level of $> 2 \sigma$.

 In Table \ref{tab:N_v-Bayesian-Eq-ve-050-04949},
 we summarize
 the reconstructed results of the fitting parameter $\ve$
 and their 1(2)$\sigma$ statistical uncertainty ranges
 of the median values
 by using the $v_0$--fixed velocity distribution
 $f_{1, \sh, \ve}(v; \ve)$
 with 50 total events on average
 in each 60-day observation period.

\subsubsection[With the simplified velocity distribution
               $f_{1, \sh}(v; v_0, \ve)$]
              {\boldmath
               With the simplified velocity distribution
               $f_{1, \sh}(v; v_0, \ve)$}
\label{sec:N_v-Bayesian-Eq-sh-050-04949}

 Now
 we release the constraints on
 the fitting parameters $v_0$ and $\ve$
 and consider the simplified velocity distribution
 $f_{1, \sh}(v; v_0, \ve)$,
 in order to identify
 the variation of the Earth's Galactic velocity $\ve$
 as well as
 to check the {\em invariability} of
 the Solar Galactic velocity $v_0$
 with the better--fitted radial WIMP velocity distributions
 in the advanced seasons.

 \def \Fittingfv       {sh}
 \def \FittingPara     {v_0-v_e}
 \InsertPlotNvBayesianAnnual
  {As in Figs.~\ref{fig:N_v-Bayesian-Eq-sh-050-00000},
   reconstructed with the simplified velocity distribution
   $f_{1, \sh}(v; v_0, \ve)$,
   except that
   four 60-day observation periods of the advanced seasons
   have been considered.
   \\~
   }

 In Figs.~\ref{fig:N_v-Bayesian-Eq-sh-050-04949},
 we show the reconstructed radial WIMP velocity distributions
 and the 1(2)$\sigma$ statistical uncertainty bands
 by using $f_{1, \sh}(v; v_0, \ve)$
 as well as
 the distributions of the fitting parameters $v_0$ and $\ve$
 in all simulated experiments on the $v_0 - \ve$ plane
 for four advanced seasons.
 Firstly,
 as found in Sec.~\ref{sec:N_v-Bayesian-Eq-sh-050-00000},
 from the plots in the left column
 of Figs.~\ref{fig:N_v-Bayesian-Eq-sh-050-04949},
 one can observe clearly that
 the reconstructed velocity distributions
 and their 1(2)$\sigma$ statistical uncertainty bands
 indeed follow
 the variation of the simulated radial velocity distribution
 in different seasons
 (much tighter than those shown
  in Figs.~\ref{fig:N_v-Bayesian-Eq-v0-050-04949}
  and \ref{fig:N_v-Bayesian-Eq-ve-050-04949}
  and)
 almost perfectly.
 Meanwhile,
 from the plots in the right column
 of Figs.~\ref{fig:N_v-Bayesian-Eq-sh-050-04949}
 as well as
 our summary given
 in Table \ref{tab:N_v-Bayesian-Eq-sh-050-04949},
 we can find that,
 although the first fitting parameter $v_0$ is unconstraint,
 its best--fit values
 in four advanced seasons are indeed a constant
 equal to its annual average value
 with also the similar 1(2)$\sigma$ statistical uncertainties.
 Remind however that,
 as its annual average value
 obtained in Secs.~\ref{sec:N_v-Bayesian-Eq-sh-050-00000}
 and \ref{sec:N_v-Bayesian-Eq-Dv-050-00000},
 the invariant best--fit value of $v_0 \simeq 190$ km/s
 is $\sim 14\%$ smaller than the input value.

 In contrast,
 in four advanced seasons
 the best--fit values of the second fitting parameter $\ve$
 show clearly an (approximately) symmetric periodic variation
 of 258.4 ($^{+11.4}_{-15.2}$) km/s,
 also around its annual average value
 obtained in Sec.~\ref{sec:N_v-Bayesian-Eq-sh-050-00000}.
 As mentioned in Sec.~\ref{sec:N_v-Bayesian-Eq-ve-050-04949},
 this $^{+11.4}_{-15.2}$ km/s annual variation of
 the best--fit value of $\ve$
 is $\sim$ 1/3 of the statistical uncertainties on $\ve$
 (see the summary
  given in Table \ref{tab:N_v-Bayesian-Eq-050-00000}).
 This indicates that,
 with $\cal O$(500) total WIMP events
 and thus $\sim$ 1/3 reduced statistical uncertainties on $\ve$,
 its annual variation
 could be identified with a confidence level of $\sim 1 \sigma$
 (see Sec.~\ref{sec:N_v-Bayesian-Eq-500-04949}).

\InsertResultsTableNvBayesianAnnual
{simplified velocity distribution
 $f_{1, \sh}(v; v_0, \ve)$}
{\multirow {2} {*}
  {\begin{minipage} {3.25 cm}
    \begin{center}
        49.49       \\ \vspace{0.65 ex}
      (\PeriodCa)
    \end{center}
   \end{minipage}}          &
 $v_0$ [km/s]               &
 189.2 & $186.6 \~ ^{+33.8}_{-26.0} \~ (^{+ 96.2}_{- 46.8})$ & [160.6, 220.4] & [139.8, 282.8] \\
 \cline{2-6}
                            &
 $\ve$ [km/s]               &
 258.4 & $258.4 \~ ^{+30.4}_{-41.8} \~ (^{+~54.2}_{-231.8})$ & [216.6, 288.8] &  [26.6, 312.6] \\
 \hline
 \multirow {2} {*}
  {\begin{minipage} {3.25 cm}
    \begin{center}
       140.74       \\ \vspace{0.65 ex}
      (\PeriodCb)
    \end{center}
   \end{minipage}}          &
 $v_0$ [km/s]               &
 191.8 & $186.6 \~ ^{+33.8}_{-23.4} \~ (^{+101.4}_{-~46.8})$ & [163.2, 220.4] & [139.8, 288.0] \\
 \cline{2-6}
                            &
 $\ve$ [km/s]               &
 269.8 & $273.6 \~ ^{+26.6}_{-41.8} \~ (^{+~49.4}_{-239.4})$ & [231.8, 300.2] &  [34.2, 323.0] \\
 \hline
 \multirow {2} {*}
  {\begin{minipage} {3.25 cm}
    \begin{center}
       231.99       \\ \vspace{0.65 ex}
      (\PeriodCc)
    \end{center}
   \end{minipage}}          &
 $v_0$ [km/s]               &
 189.2 & $186.6 \~ ^{+33.8}_{-23.4} \~ (^{+ 98.8}_{- 46.8})$ & [163.2, 220.4] & [139.8, 285.4] \\
 \cline{2-6}
                            &
 $\ve$ [km/s]               &
 258.4 & $258.4 \~ ^{+30.4}_{-41.8} \~ (^{+~53.2}_{-228.0})$ & [216.6, 288.8] &  [30.4, 311.6] \\
 \hline
 \multirow {2} {*}
  {\begin{minipage} {3.25 cm}
    \begin{center}
       323.24       \\ \vspace{0.65 ex}
      (\PeriodCd)
    \end{center}
   \end{minipage}}          &
 $v_0$ [km/s]               &
 191.8 & $186.6 \~ ^{+36.4}_{-26.0} \~ (^{+ 93.6}_{- 49.4})$ & [160.6, 223.0] & [137.2, 280.2] \\
 \cline{2-6}
                            &
 $\ve$ [km/s]               &
 243.2 & $247.0 \~ ^{+30.4}_{-49.4} \~ (^{+~53.2}_{-228.0})$ & [197.6, 277.4] &  [19.0, 300.2] \\
 }
{50}
{parameters $v_0$ and $\ve$ as well as}
{advanced}

 In Table \ref{tab:N_v-Bayesian-Eq-sh-050-04949},
 we summarize
 the reconstructed results of the fitting parameters $v_0$ and $\ve$
 as well as
 their 1(2)$\sigma$ statistical uncertainty ranges
 of the median values
 by using the simplified velocity distribution
 $f_{1, \sh}(v; v_0, \ve)$
 with 50 total events on average
 in each 60-day observation period.

\subsubsection[With the modified velocity distribution
               $f_{1, \sh, \Delta v}(v; v_0, \Delta v)$]
              {\boldmath
               With the modified velocity distribution
               $f_{1, \sh, \Delta v}(v; v_0, \Delta v)$}
\label{sec:N_v-Bayesian-Eq-Dv-050-04949}

 Finally,
 as the confirmation of the invariability of
 the Solar Galactic velocity $v_0$
 and that of the variation of the Earth's Galactic velocity $\ve$
 shown in Sec.~\ref{sec:N_v-Bayesian-Eq-sh-050-04949},
 we consider here the modified velocity distribution
 $f_{1, \sh, \Delta v}(v; v_0, \Delta v)$
 with $v_0$ and $\Delta v$
 as two free fitting parameters.

 In Figs.~\ref{fig:N_v-Bayesian-Eq-Dv-050-04949},
 we show the reconstructed radial WIMP velocity distributions
 and the 1(2)$\sigma$ statistical uncertainty bands
 by using $f_{1, \sh, \Delta v}(v; v_0, \Delta v)$
 as well as
 the distributions of the fitting parameters $v_0$ and $\Delta v$
 in all simulated experiments on the $v_0 - \Delta v$ plane
 for four advanced seasons.
 Firstly,
 by comparing the plots in the left column
 of Figs.~\ref{fig:N_v-Bayesian-Eq-Dv-050-04949}
 with the plots in the left column
 of Figs.~\ref{fig:N_v-Bayesian-Eq-sh-050-04949},
 one could see that,
 except of the 2$\sigma$ statistical uncertainty bands%
\footnote{
 Note that,
 due to the wide--extended distribution tails
 at the lower scanning boundaries of $\ve$ and $\Delta v$,
 the lower bounds of
 the 2$\sigma$ statistical uncertainties on $\ve$ and $\Delta v$
 and in turn the shapes of
 the 2$\sigma$ uncertainty bands
 are actually limited by
 our simulation setup for their scanning ranges.
 In Sec.~\ref{sec:N_v-Bayesian-Eq-500-04949}
 we will see that,
 with $\cal O$(500) total WIMP events
 the 1(2)$\sigma$ uncertainty contours become closed
 and the lower bounds of
 the 2$\sigma$ uncertainties on $\ve$ and $\Delta v$
 can be determined accurately.
 Then
 the shape differences between
 the 2$\sigma$ uncertainty bands
 of the WIMP velocity distributions
 reconstructed with the simplified and
 the modified velocity distributions
 becomes very tiny.
},
 the velocity distributions
 and their 1$\sigma$ statistical uncertainty bands
 reconstructed with the simplified and
 the modified velocity distributions
 would be identical.

 \def \Fittingfv       {Dv}
 \def \FittingPara     {v_0-D_v}
 \InsertPlotNvBayesianAnnual
  {As in Figs.~\ref{fig:N_v-Bayesian-Eq-Dv-050-00000},
   reconstructed with the modified velocity distribution
   $f_{1, \sh, \Delta v}(v; v_0, \Delta v)$,
   except that
   four 60-day observation periods of the advanced seasons
   have been considered.
   \\~
   }
\InsertResultsTableNvBayesianAnnual
{modified velocity distribution
 $f_{1, \sh, \Delta v}(v; v_0, \Delta v)$}
{\multirow {2} {*}
  {\begin{minipage} {3.25 cm}
    \begin{center}
        49.49       \\ \vspace{0.65 ex}
      (\PeriodCa)
    \end{center}
   \end{minipage}}          &
 $v_0$ [km/s]               &
 189.2 & $186.6 \~ ^{+36.4}_{-26.0} \~ (^{+ 88.4}_{- 46.8})$ & [160.6, 223.0] & [139.8, 275.0] \\
 \cline{2-6}
                            &
 $\Delta v$ [km/s]          &
  70.4 & $ 74.6 \~ ^{+46.2}_{-75.6} \~ (^{+~88.2}_{-264.6})$ & [$-1.0$, 120.8] & [$-190.0$, 162.8] \\
 \hline
 \multirow {2} {*}
  {\begin{minipage} {3.25 cm}
    \begin{center}
       140.74       \\ \vspace{0.65 ex}
      (\PeriodCb)
    \end{center}
   \end{minipage}}          &
 $v_0$ [km/s]               &
 191.8 & $186.6 \~ ^{+33.8}_{-23.4} \~ (^{+ 93.6}_{- 46.8})$ & [163.2, 220.4] & [139.8, 280.2] \\
 \cline{2-6}
                            &
 $\Delta v$ [km/s]          &
  78.8 & $ 87.2 \~ ^{+46.2}_{-71.4} \~ (^{+~84.0}_{-273.0})$ &  [15.8, 133.4] & [$-185.8$, 171.2] \\
 \hline
 \multirow {2} {*}
  {\begin{minipage} {3.25 cm}
    \begin{center}
       231.99       \\ \vspace{0.65 ex}
      (\PeriodCc)
    \end{center}
   \end{minipage}}          &
 $v_0$ [km/s]               &
 189.2 & $186.6 \~ ^{+33.8}_{-23.4} \~ (^{+ 91.0}_{- 46.8})$ & [163.2, 220.4] & [139.8, 277.6] \\
 \cline{2-6}
                            &
 $\Delta v$ [km/s]          &
  70.4 & $ 74.6 \~ ^{+50.4}_{-75.6} \~ (^{+~88.2}_{-264.6})$ & [$-1.0$, 125.0] & [$-190.0$, 162.8] \\
 \hline
 \multirow {2} {*}
  {\begin{minipage} {3.25 cm}
    \begin{center}
       323.24       \\ \vspace{0.65 ex}
      (\PeriodCd)
    \end{center}
   \end{minipage}}          &
 $v_0$ [km/s]               &
 189.2 & $186.6 \~ ^{+36.4}_{-26.0} \~ (^{+ 88.4}_{- 49.4})$ & [160.6, 223.0] & [137.2, 275.0] \\
 \cline{2-6}
                            &
 $\Delta v$ [km/s]          &
  57.8 & $ 62.0 \~ ^{+50.4}_{-88.2} \~ (^{+~88.2}_{-252.0})$ & [$-26.2$, 112.4] & [$-190.0$, 150.2] \\
 }
{50}
{parameters $v_0$ and $\Delta v$ as well as}
{advanced}

 Additionally,
 from the plots in the right column
 of Figs.~\ref{fig:N_v-Bayesian-Eq-Dv-050-04949}
 as well as
 our summary given
 in Table \ref{tab:N_v-Bayesian-Eq-Dv-050-04949},
 one could confirm that
 the best--fit values of the first fitting parameter $v_0$
 in four advanced seasons
 are fixed around
 the annual average value of $v_0 \simeq 190$ km/s
 with the similar 1(2)$\sigma$ statistical uncertainties.
 Meanwhile,
 the best--fit values of the second fitting parameter $\Delta v$
 show also clearly
 an (approximately) symmetric periodic variation
 of 70.4 ($^{+~8.4}_{-12.6}$) km/s
 in four advanced seasons,
 which is approximately equal to both of
 the annual average value of \mbox{$\Delta v \simeq 70$ km/s}
 (see Table \ref{tab:N_v-Bayesian-Eq-050-00000})
 and the difference between
 the best--fit values of $v_0$ and $\ve$
 obtained in Sec.~\ref{sec:N_v-Bayesian-Eq-sh-050-04949}
 (see Table \ref{tab:N_v-Bayesian-Eq-sh-050-04949}).
 It would be worth to notice that,
 while
 the $^{+~8.4}_{-12.6}$ km/s annual variation of
 the best--fit value of $\Delta v$
 is smaller than
 the $^{+11.4}_{-15.2}$ km/s annual variation of $\ve$,
 the 1$\sigma$ statistical uncertainties on $\Delta v$
 are in contrast much ($\gsim \~ 1.5$ times) larger than
 those on $\ve$.

 In Table \ref{tab:N_v-Bayesian-Eq-Dv-050-04949},
 we summarize
 the reconstructed results of the fitting parameters $v_0$ and $\Delta v$
 as well as
 their 1(2)$\sigma$ statistical uncertainty ranges
 of the median values
 by using the modified velocity distribution
 $f_{1, \sh, \Delta v}(v; v_0, \Delta v)$
 with 50 total events on average
 in each 60-day observation period.

\subsection{Fitted radial WIMP velocity distribution
            in the Galactic frame}
\label{sec:N_v-Bayesian-G-050}

 As preparation
 for analyzing (future) real experimental data
 to understand the 3-dimensional velocity distribution
 of halo WIMPs,
 we have also tried to transform
 the saved 3-D velocity information of the simulated WIMP events
 (including measuring times)
 in the laboratory coordinate system back
 by using Eqs.~(\ref{eqn:V_chi_i_H-V_chi_i_Lab}),
 (\ref{eqn:V_chi_i_E-V_chi_i_H}),
 (\ref{eqn:V_chi_i_Eq-V_chi_i_E}),
 (\ref{eqn:V_chi_i_S-V_chi_i_Eq}),
 and (\ref{eqn:V_chi_i_G-V_chi_i_S})
 and reconstruct the radial WIMP velocity distribution
 in the Galactic coordinate system.

 In Figs.~\ref{fig:N_v-Bayesian-G-v0-050-00000},
 we show the reconstructed radial distributions of
 the 3-D WIMP velocity
 and the 1(2)$\sigma$ statistical uncertainty bands
 by using the {\em simple} Maxwellian velocity distribution
 $f_{1, \Gau}(v)$
 given by Eq.~(\ref{eqn:f1v_Gau_vesc})
 to fit data shown in Fig.~\ref{fig:N_v-G-050-00000}
 as well as
 the distribution of the fitting parameter $v_0$
 in all simulated experiments.
 Not surprisingly,
 our best--fit reconstruction distributions
 could match the input
 simple Maxwellian velocity distribution perfectly
 and the fitting parameter $v_0$
 could also be pinned down very precisely%
\footnote{
 The tiny difference of 0.2 km/s
 between the best--fit and the input values of $v_0$
 is only due to the scanning resolution.
}
 (see the summary
  given in Table \ref{tab:N_v-Bayesian-G-050-00000}).

 \def \ShortFrame       {G}
 \def \Fittingfv        {v0}
 \InsertPlotNvBayesianD {v_0}
  {(a)
   The reconstructed radial distributions of
   the 3-D WIMP velocity
   and the 1(2)$\sigma$ statistical uncertainty bands
   by using the {\em simple} Maxwellian velocity distribution
   $f_{1, \Gau}(v)$
   given by Eq.~(\ref{eqn:f1v_Gau_vesc})
   to fit data shown in Fig.~\ref{fig:N_v-G-050-00000}.
   The solid red curve
   is the input simple Maxwellian velocity distribution
   with the input value of $v_0 = 220$ km/s.
   (b)
   The distribution of the fitting parameter $v_0$
   in all simulated experiments.
   Other notations are the same as
   in Figs.~\ref{fig:N_v-Bayesian-Eq-v0-050-00000}.
   }
 \def \PlotNumber             {\PlotNumberAa}
\InsertResultsTableNvBayesian [b!]
{{$f_{1, \Gau}(v)$}          &
  $v_0$ [km/s]               &
  220.2 & $220.2   \pm 14           \~ (^{+29.4}_{-26.6})$ & [206.2, 234.2] & [193.6, 249.6] \\
 }
{50 total events on average in the observation period of \PeriodA\ day
 }
{The summary of
 the reconstructed results of the fitting parameter $v_0$
 and their 1(2)$\sigma$ statistical uncertainty ranges
 of the median values
 by using the simple Maxwellian velocity distribution
 with 50 total events on average
 in one entire year
 shown in Figs.~\ref{fig:N_v-Bayesian-G-v0-050-00000}.
 }
\section{With a raised total event number}
\label{sec:N_v_phi_theta-500}

 In this section,
 we raise the simulated total event number
 in one observation period
 (365 days/year,
  60 days/season,
  or 4 hours/shift $\times$ 60 days)
 to 500 events on average in one experiment.
 We present at first
 the angular distribution patterns
 of the 3-dimensional WIMP velocity
 and then discuss
 the Bayesian reconstructions of
 its radial distributions.

%
% 5/10
 %
%
\subsection{Angular distributions of the 3-D WIMP velocity
            in the laboratory--independent frames}
\label{sec:N_phi_theta-500}

 In this subsection,
 we present only
 the angular distributions of
 the (transformed) 3-D WIMP velocity
 in three laboratory--independent
 (Ecliptic,
  Equatorial,
  and Earth)
 coordinate systems.
 The angular distributions
 in two laboratory--dependent
 (horizontal and laboratory)
 coordinate systems
 observed in different underground laboratories
 can be found
 in Appendix \ref{appx:N_phi_theta-ULabs}.

\subsubsection{Angular WIMP velocity distribution in the Ecliptic frame}
\label{sec:N_phi_theta-S-500}
\begin{figure} [t!]
\begin{center}
 \includegraphics [width = 15 cm] {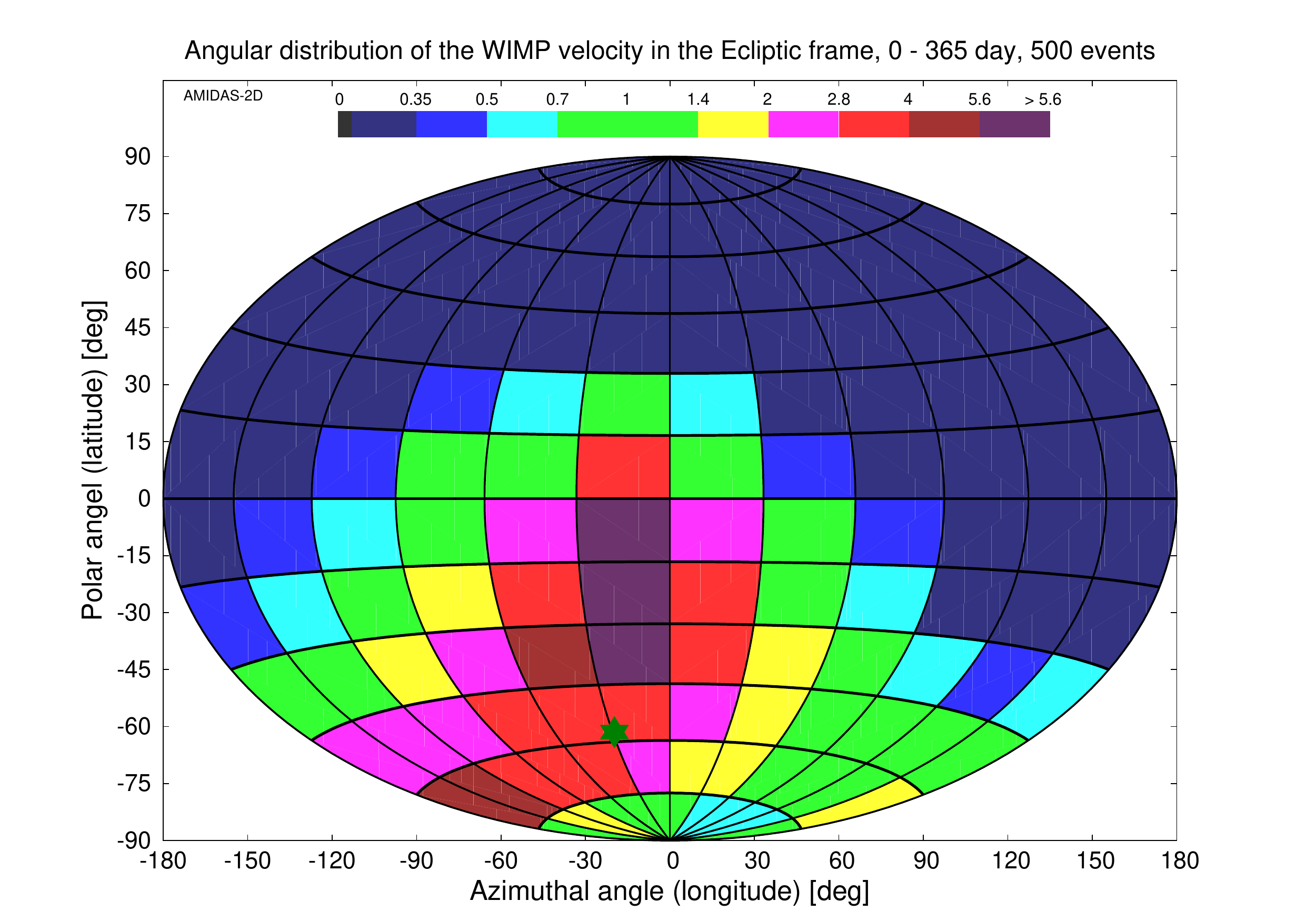}
\end{center}
\caption{
 As in Fig.~\ref{fig:N_phi_theta-S-050-00000},
 the angular distribution of the 3-D WIMP velocity
 transformed to the Ecliptic coordinate system,
 except that
 500 total events on average in one entire year
 have been generated
 and binned into 12 $\times$ 12 bins
 for the longitude and latitude directions,
 respectively.
}
\label{fig:N_phi_theta-S-500-00000}
\end{figure}

 In Fig.~\ref{fig:N_phi_theta-S-500-00000},
 we show
 the angular distribution of the 3-D WIMP velocity
 transformed to the Ecliptic coordinate system.
 500 total events on average in one experiment
 in one entire year
 have been generated
 and binned into 12 $\times$ 12 bins
 for the longitude and latitude directions,
 respectively.

 It can be found that,
 first,
 with a higher analysis resolution,
 the bins with higher event numbers
 spreading from the center to the southwest part
 can be seen more obviously,
 including
 the bins with the highest event number
 ($>$ 5.6 times of the all--sky average value of
  500 events / 144 bins = 3.47 events/bin,
  i.e.,
  $>$ 19.44 events/bin)
 between 45$^{\circ}$S and 0$^{\circ}$ latitude,
 30$^{\circ}$W and 0$^{\circ}$ longitude.
 Second,
 the deviation
 between the center of the band of the high--WIMP--flux bins
 and the theoretical main direction of incident WIMPs
 in the Ecliptic coordinate system
 (57.40$^{\circ}$S, 29.10$^{\circ}$W)
 (the dark--green star)
 would now be $\lsim$ 15$^{\circ}$ and $\lsim$ 30$^{\circ}$
 in the latitude and the longitude direction,
 respectively.

\subsubsection{Angular WIMP velocity distribution in the Equatorial frame}
\label{sec:N_phi_theta-Eq-500}
\begin{figure} [t!]
\begin{center}
 \includegraphics [width = 15 cm] {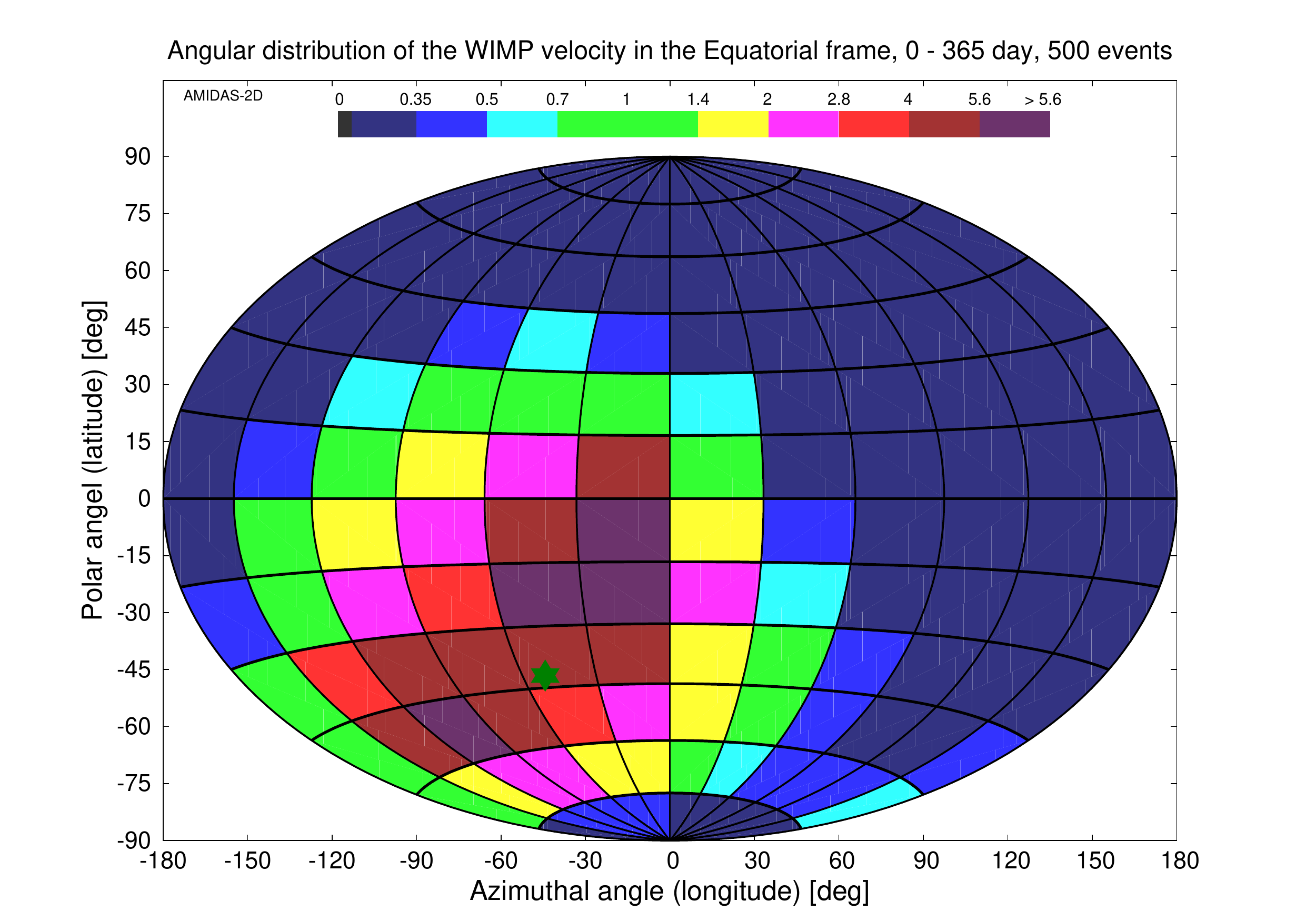}
\end{center}
\caption{
 As in Fig.~\ref{fig:N_phi_theta-Eq-050-00000},
 the angular distribution of the 3-D WIMP velocity
 transformed to the Equatorial coordinate system,
 except that
 500 total events on average in one entire year
 have been simulated.
}
\label{fig:N_phi_theta-Eq-500-00000}
\end{figure}

 In Fig.~\ref{fig:N_phi_theta-Eq-500-00000},
 we show at first
 the angular distribution of the 3-D WIMP velocity
 transformed to the Equatorial coordinate system
 with 500 total events on average
 in one entire year.
 With a higher analysis resolution,
 one can also find that,
 first,
 as in the Ecliptic coordinate system,
 the bins with higher event numbers
 spreading from the center to the southwest part
 can be seen more obviously
 and
 there seems still to be a small deviation
 between the center of the band of the high--WIMP--flux bins
 and the theoretical main direction of incident WIMPs
 (in the Equatorial coordinate system,
  42.00$^{\circ}$S, 50.70$^{\circ}$W)
 (the dark--green star).
 Second,
 not only the bins around the north pole
 (more northern than 45$^{\circ}$N)
 and in the most part of the eastern sky,
 a few bins around the south pole
 (more southern than 75$^{\circ}$S)
 would also be a ``WIMP hole''
 with event numbers
 less than 35\% of the all--sky average
 (i.e.,
  $<$ 1.22 events/bin
  among $\cal O$(500) total events).

 \def \ShortFrame     {Eq}
 \def \EventNumber    {500}
 \def \Perioda        {\PeriodBa}
 \def \Periodb        {\PeriodBb}
 \def \Periodc        {\PeriodBc}
 \def \Periodd        {\PeriodBd}
 \def \PlotNumbera    {\PlotNumberBa}
 \def \PlotNumberb    {\PlotNumberBb}
 \def \PlotNumberc    {\PlotNumberBc}
 \def \PlotNumberd    {\PlotNumberBd}
 \InsertPlotNphitheta
  {As in Figs.~\ref{fig:N_phi_theta-Eq-050-07900},
   the angular distributions of the 3-D WIMP velocity
   in the Equatorial coordinate system,
   except that
   500 total events on average
   in each 60-day observation period of four normal seasons
   have been considered.
   }
 \def \Perioda        {\PeriodCa}
 \def \Periodb        {\PeriodCb}
 \def \Periodc        {\PeriodCc}
 \def \Periodd        {\PeriodCd}
 \def \PlotNumbera    {\PlotNumberCa}
 \def \PlotNumberb    {\PlotNumberCb}
 \def \PlotNumberc    {\PlotNumberCc}
 \def \PlotNumberd    {\PlotNumberCd}
 \InsertPlotNphitheta
  {As in Figs.~\ref{fig:N_phi_theta-Eq-050-04949},
   the angular distributions of the 3-D WIMP velocity
   in the Equatorial coordinate system,
   except that
   500 total events on average
   in each 60-day observation period of four advanced seasons
   have been considered.
   }

 Moreover,
 in order to demonstrate
 the annual modulation of
 the angular distribution pattern of
 the 3-D WIMP velocity,
 in Figs.~\ref{fig:N_phi_theta-Eq-500-07900}
 and \ref{fig:N_phi_theta-Eq-500-04949}
 we show
 the angular distributions
 in the Equatorial coordinate system
 in four normal and four advanced seasons,
 respectively.
 Remind that
 500 total events on average in each 60-day observation period
 have been simulated.
 Now,
 with a higher analysis resolution,
 the clockwise circular variation of
 the angular distribution patterns
 following the blue--yellow point
 around the theoretical main direction of incident WIMPs
 could be observed more clearly.
 In addition,
 the difference of
 the angular distribution patterns
 between the corresponding observation periods
 of four normal and four advanced seasons
 can already be observed.
 In fact,
 one could even combine all eight plots
 of the normal and the advanced seasons
 to build a more--detailed rotated pattern
 of the angular WIMP velocity distribution.

\subsubsection{Angular WIMP velocity distribution in the Earth frame}
\label{sec:N_phi_theta-E-500}
\begin{figure} [t!]
\begin{center}
 \includegraphics [width = 15 cm] {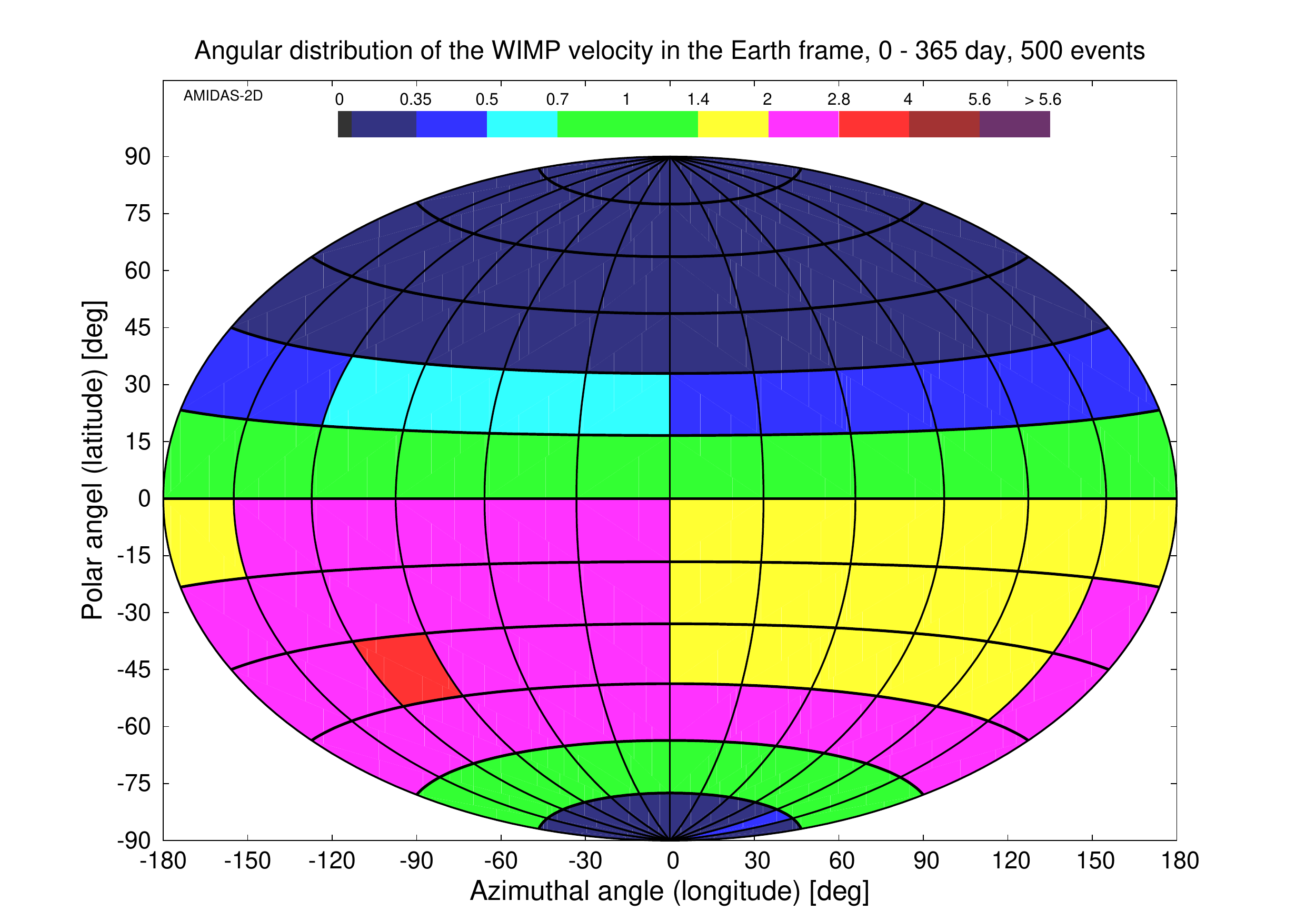}
\end{center}
\caption{
 As in Fig.~\ref{fig:N_phi_theta-E-050-00000},
 the angular distribution of the 3-D WIMP velocity
 transformed to the Earth coordinate system,
 except that
 500 total events on average in one entire year
 have been simulated.
}
\label{fig:N_phi_theta-E-500-00000}
\end{figure}

 In Fig.~\ref{fig:N_phi_theta-E-500-00000},
 we show
 the angular distribution of the 3-D WIMP velocity
 transformed to the Earth coordinate system
 with 500 total events on average in one entire year.
 With a higher analysis resolution,
 as in the Equatorial coordinate system,
 not only the bins around the north pole
 (more northern than 30$^{\circ}$N),
 a few bins around the south pole
 (more southern than 75$^{\circ}$S)
 would also be the ``WIMP hole''
 with event numbers
 less than 35\% of the all--sky average
 (i.e.,
  \mbox{$<$ 1.22 events/bin}
  among $\cal O$(500) total events).

 \def \ShortFrame     {E}
 \def \Perioda        {\PeriodBa}
 \def \Periodb        {\PeriodBb}
 \def \Periodc        {\PeriodBc}
 \def \Periodd        {\PeriodBd}
 \def \PlotNumbera    {\PlotNumberBa}
 \def \PlotNumberb    {\PlotNumberBb}
 \def \PlotNumberc    {\PlotNumberBc}
 \def \PlotNumberd    {\PlotNumberBd}
 \InsertPlotNphitheta
  {As in Figs.~\ref{fig:N_phi_theta-E-050-07900},
   the angular distributions of the 3-D WIMP velocity
   in the Earth coordinate system,
   except that
   500 total events on average
   in each 60-day observation period of four normal seasons
   have been considered.
   }
 \def \Perioda        {\PeriodCa}
 \def \Periodb        {\PeriodCb}
 \def \Periodc        {\PeriodCc}
 \def \Periodd        {\PeriodCd}
 \def \PlotNumbera    {\PlotNumberCa}
 \def \PlotNumberb    {\PlotNumberCb}
 \def \PlotNumberc    {\PlotNumberCc}
 \def \PlotNumberd    {\PlotNumberCd}
 \InsertPlotNphitheta
  {As in Figs.~\ref{fig:N_phi_theta-E-050-04949},
   the angular distributions of the 3-D WIMP velocity
   in the Earth coordinate system,
   except that
   500 total events on average
   in each 60-day observation period of four advanced seasons
   have been considered.
   }

 Moreover,
 in order to demonstrate
 the annual modulation of
 the angular distribution pattern of
 the 3-D WIMP velocity,
 in Figs.~\ref{fig:N_phi_theta-E-500-07900}
 and \ref{fig:N_phi_theta-E-500-04949}
 we show
 the angular distributions
 in the Earth coordinate system
 in four normal and four advanced seasons,
 respectively.
 Remind that
 500 total events on average in each 60-day observation period
 have been simulated.
 Now,
 with a higher analysis resolution,
 the rotation of the angular distribution patterns
 around the $\zE$--axis
 can be clearly observed,
 not only in four normal and four advanced seasons,
 but also between each pair of the corresponding seasons.

%

%
% 6/10
%
 %
%
\subsection{Bayesian reconstruction of
            the radial distribution of the 3-D WIMP velocity}
\label{sec:N_v-Bayesian-500}

 In this subsection,
 we present our Bayesian reconstructions of
 the radial distribution of
 the (transformed) 3-D WIMP velocity
 in the Equatorial coordinate system.
 As in Sec.~\ref{sec:N_v-Bayesian-050},
 we discuss at first the one--year simulations
 and then the results of the annual modulation
 in the observation periods of four advanced seasons%
\footnote{
 The reconstructed radial WIMP velocity distributions
 in the observation periods of four normal seasons
 will be given
 in Appendix \ref{appx:N_v-Bayesian-Eq-07900}
 for readers' reference.
}.
 All four fitting velocity distributions
 given in Sec.~\ref{sec:N_v-Bayesian-f1v}
 will be considered.
 At the end of this subsection,
 we will also show
 the reconstruction result of the WIMP velocity distribution
 in the Galactic coordinate system.

\subsubsection{Fitted radial WIMP velocity distributions
               in the Equatorial frame}
\label{sec:N_v-Bayesian-Eq-500-00000}

 We discuss at first our reconstruction results
 with 500 total events on average in one experiment
 recorded in one entire year.

\paragraph[With the one--parameter and the $v_0$--fixed velocity distributions
           $f_{1, \sh, v_0}(v; v_0)$ and $f_{1, \sh, \ve}(v; \ve)$]
          {\boldmath
           With the one--parameter and the $v_0$--fixed velocity distributions
           $f_{1, \sh, v_0}(v; v_0)$ and $f_{1, \sh, \ve}(v; \ve)$}
 ~\\

 As in Secs.~\ref{sec:N_v-Bayesian-Eq-v0-050-00000}
 and \ref{sec:N_v-Bayesian-Eq-ve-050-00000},
 we consider at first
 the one--parameter and the $v_0$--fixed
 velocity distribution functions,
 $f_{1, \sh, v_0}(v; v_0)$ and $f_{1, \sh, \ve}(v; \ve)$,
 with the constraints that
 $\ve = 1.05 \~ v_0$ and
 $v_0 = 220$ km/s,
 respectively.
 Note that
 the scanning ranges of the fitting parameters
 have been shrunk to
 190 km/s $\le v_0 \le$ 240 km/s or
 180 km/s $\le \ve \le$ 270 km/s,
 respectively.

 \def \ShortFrame       {Eq}
 \def \PeriodA          {\PeriodAa}
 \def \PlotNumberA      {\PlotNumberAa}
 \def \Fittingfv        {v0}
 \InsertPlotNvBayesianD {v_0}
  {As in Figs.~\ref{fig:N_v-Bayesian-Eq-v0-050-00000},
   reconstructed with the one--parameter velocity distribution
   $f_{1, \sh, v_0}(v; v_0)$,
   except that
   500 total events on average in one entire year
   have been simulated.
   Note that
   the scanning range of the fitting parameter $v_0$
   is shrunk to between 190 km/s and \mbox{240 km/s}.
   }
 \def \Fittingfv        {ve}
 \InsertPlotNvBayesianD [b!] {v_e}
  {As in Figs.~\ref{fig:N_v-Bayesian-Eq-ve-050-00000},
   reconstructed with the $v_0$--fixed velocity distribution
   $f_{1, \sh, \ve}(v; \ve)$,
   except that
   500 total events on average in one entire year
   have been simulated.
   Note that
   the scanning range of the fitting parameter $\ve$
   is shrunk to between 180 km/s and 270 km/s.
   }
 \def \Fittingfv        {sh}
 \InsertPlotNvBayesianQ {v_0} {v_e}
  {As in Figs.~\ref{fig:N_v-Bayesian-Eq-sh-050-00000},
   reconstructed with the simplified velocity distribution
   $f_{1, \sh}(v; v_0, \ve)$,
   except that
   500 total events on average in one entire year
   have been simulated.
   Note that
   the scanning ranges of the fitting parameters $v_0$ and $\ve$
   are shrunk to between 140 km/s and \mbox{240 km/s}
   and between 200 km/s and 310 km/s,
   respectively.
   }
 \def \Fittingfv        {Dv}
 \InsertPlotNvBayesianQ {v_0} {D_v}
  {As in Figs.~\ref{fig:N_v-Bayesian-Eq-Dv-050-00000},
   reconstructed with the modified velocity distribution
   $f_{1, \sh, \Delta v}(v; v_0, \Delta v)$,
   except that
   500 total events on average in one entire year
   have been simulated.
   Note that
   the scanning ranges of the fitting parameters $v_0$ and $\Delta v$
   are shrunk to between 140 km/s and \mbox{240 km/s}
   and between $-20$ km/s and 150 km/s,
   respectively.
   }
 \def \PlotNumber             {\PlotNumberAa}
\InsertResultsTableNvBayesian
{{$f_{1, \sh, v_0}(v)$}      &
  $v_0$ [km/s]               &
  214.5 & $214.5    \pm  3 .5        \~ (\pm \~ 6.5)$       & [211.0, 218.0] & [208.0, 221.0] \\
 \hline
 \hline
 {$f_{1, \sh, \ve}(v)$}      &
  $\ve$ [km/s]               &
  225.0 & $225.0 \~ ^{+ 8.1}_{- 9.0} \~ (^{+16.2}_{-17.1})$ & [216.0, 233.1] & [207.9, 241.2] \\
 \hline
 \hline
 \multirow{2}{*}
 {$f_{1, \sh}(v)$}           &
  $v_0$ [km/s]               &
  186.0 & $185.0 \~ ^{+10.0}_{-~8.0} \~ (^{+20.0}_{-16.0})$ & [177.0, 195.0] & [169.0, 205.0] \\
 \cline{2-6}
                             &
  $\ve$ [km/s]               &
  261.6 & $262.7 \~ ^{+~8.8}_{-11.0} \~ (^{+17.6}_{-24.2})$ & [251.7, 271.5] & [238.5, 280.3] \\
 \hline
 \hline
 \multirow{2}{*}
 {$f_{1, \sh, \Delta v}(v)$} &
  $v_0$ [km/s]               &
  186.0 & $185.0 \~ ^{+10.0}_{-~8.0} \~ (^{+20.0}_{-16.0})$ & [177.0, 195.0] & [169.0, 205.0] \\
 \cline{2-6}
                             &
  $\Delta v$ [km/s]          &
   75.2 & $ 76.9 \~ ^{+17.0}_{-18.7} \~ (^{+32.3}_{-40.8})$ &  [58.2,  93.9] &  [36.1, 109.2] \\
 }
{500 total events on average in the observation period of \PeriodA\ day
 }
{The summary of
 the reconstructed results of the fitting parameters
 and their 1(2)$\sigma$ statistical uncertainty ranges
 of the median values
 by using four considered fitting velocity distributions
 with 500 total events on average in one entire year
 shown in Figs.~\ref{fig:N_v-Bayesian-Eq-v0-500-00000},
 \ref{fig:N_v-Bayesian-Eq-ve-500-00000},
 \ref{fig:N_v-Bayesian-Eq-sh-500-00000},
 and \ref{fig:N_v-Bayesian-Eq-Dv-500-00000}.
 }

 In Figs.~\ref{fig:N_v-Bayesian-Eq-v0-500-00000}
 and \ref{fig:N_v-Bayesian-Eq-ve-500-00000},
 we show the reconstructed radial distributions of
 the 3-D WIMP velocity
 and the 1(2)$\sigma$ statistical uncertainty bands
 by using the one--parameter and the $v_0$--fixed
 velocity distributions,
 respectively.
 One can find that,
 although
 two 1$\sigma$ statistical uncertainty bands
 could still cover the theoretical (solid red) velocity distribution,
 the difference between
 the best--fit and the input values of $v_0$
 is already $\sim 1.6 \sigma$,
 while
 the difference between these two values of $\ve$
 is also $\sim 0.7 \sigma$
 (see the summary
  given in Table \ref{tab:N_v-Bayesian-Eq-500-00000}).
 More importantly,
 both of the theoretical velocity distribution
 and those reconstructed with two fitting distribution functions
 are now clearly $\sim 1 \sigma$ deviated from
 the radial distribution of the simulated 3-D WIMP velocity
 in most of the $v$--bins.
 These would imply that
 the radial distribution of the 3-D WIMP velocity
 in the Equatorial as well as in the laboratory coordinate systems
 would differ from the theoretically derived analytic form
 $f_{1, \sh, {\rm vesc}}(v)$
 given in Eq.~(\ref{eqn:f1v_sh_vesc})
 \cite{Lewin96}.

\paragraph[With the simplified and the modified velocity distributions
           $f_{1, \sh}(v; v_0, \ve)$ and \\ $f_{1, \sh, \Delta v}(v; v_0, \Delta v)$]
          {\boldmath
           With the simplified and the modified velocity distributions
           $f_{1, \sh}(v; v_0, \ve)$ and \\ $f_{1, \sh, \Delta v}(v; v_0, \Delta v)$}
 ~\\

 As in Secs.~\ref{sec:N_v-Bayesian-Eq-sh-050-00000}
 and \ref{sec:N_v-Bayesian-Eq-Dv-050-00000},
 we release now the constraints that
 $\ve = 1.05 \~ v_0$ and \mbox{$v_0 = 220$ km/s}
 and consider
 the simplified and the modified
 velocity distribution functions,
 $f_{1, \sh}(v; v_0, \ve)$ and $f_{1, \sh, \Delta v}(v; v_0, \Delta v)$.
 Note that
 the scanning ranges of the fitting parameters
 have been shrunk to
 140 km/s $\le v_0 \le$ 240 km/s,
 200 km/s $\le \ve \le$ 310 km/s,
 and $-20$ km/s $\le \Delta v \le$ 150 km/s.

 In Figs.~\ref{fig:N_v-Bayesian-Eq-sh-500-00000}
 and \ref{fig:N_v-Bayesian-Eq-Dv-500-00000},
 we show the reconstructed radial distributions of
 the 3-D WIMP velocity
 and the 1(2)$\sigma$ statistical uncertainty bands
 by using the simplified and the modified velocity distributions,
 respectively.
 One can find firstly that,
 although
 the long distribution tails at the lower scanning boundaries
 on the $v_0 - \ve$ and the $v_0 - \Delta v$ planes
 disappear now and
 the 1(2)$\sigma$ statistical uncertainty contours become closed,
 two 1(2)$\sigma$ uncertainty bands could not cover
 the theoretical (solid red) velocity distribution any more.
 Additionally,
 two reconstructed radial distributions
 as well as
 their 1(2)$\sigma$ uncertainty bands
 seem not to fit
 the radial distribution of the simulated 3-D WIMP velocity
 in the velocity range of 200 km/s $\lsim \~ v \~ \lsim$ 300 km/s
 very well.
 Moreover,
 in Figs.~\ref{fig:N_v-Bayesian-Eq-sh-500-00000}(b)
 and \ref{fig:N_v-Bayesian-Eq-Dv-500-00000}(b),
 one can see that
 the theoretical values of the fitting parameters
 (the red upward--triangle)
 are now clearly out of
 the (light--blue) 2$\sigma$ statistical uncertainty contours;
 the difference between
 the best--fit and the input values of $v_0$
 becomes now $\sim 3.4 \sigma$,
 and the difference between these two values of $\ve$
 is also enlarged to $\sim 2.8 \sigma$
 (see the summary
  given in Table \ref{tab:N_v-Bayesian-Eq-500-00000}).
 These would imply not only that
 the radial distribution of the 3-D WIMP velocity
 in the Equatorial as well as in the laboratory coordinate systems
 would differ from the theoretically derived analytic form
 (\ref{eqn:f1v_sh_vesc})
 of $f_{1, \sh, {\rm vesc}}(v)$,
 but also
 a requirement of
 the modification of the analytic form of
 the fitting velocity distribution function.

 In Table \ref{tab:N_v-Bayesian-Eq-500-00000},
 we summarize
 the reconstructed results of the fitting parameters
 and their 1(2)$\sigma$ statistical uncertainty ranges
 of the median values
 with 500 total events on average in one entire year
 for all four considered fitting velocity distributions
 shown in Figs.~\ref{fig:N_v-Bayesian-Eq-v0-050-00000},
 \ref{fig:N_v-Bayesian-Eq-ve-050-00000},
 \ref{fig:N_v-Bayesian-Eq-sh-050-00000},
 and \ref{fig:N_v-Bayesian-Eq-Dv-050-00000}.

\subsubsection{Annual modulation of the radial WIMP velocity distribution}
\label{sec:N_v-Bayesian-Eq-500-04949}

 Now we come to present the reconstruction results of
 the annual modulation of
 the radial WIMP velocity distribution
 in the observation periods of four advanced seasons.

\paragraph[With the one--parameter and the $v_0$--fixed velocity distributions
           $f_{1, \sh, v_0}(v; v_0)$ and $f_{1, \sh, \ve}(v; \ve)$]
          {\boldmath
           With the one--parameter and the $v_0$--fixed velocity distributions
           $f_{1, \sh, v_0}(v; v_0)$ and $f_{1, \sh, \ve}(v; \ve)$}
 ~\\

 As in Secs.~\ref{sec:N_v-Bayesian-Eq-v0-050-04949}
 and \ref{sec:N_v-Bayesian-Eq-ve-050-04949},
 we consider at first
 the one--parameter and the $v_0$--fixed
 velocity distribution functions,
 $f_{1, \sh, v_0}(v; v_0)$ and $f_{1, \sh, \ve}(v; \ve)$,
 with the constraints that
 $\ve = 1.05 \~ v_0$ and $v_0 = 220$ km/s,
 respectively.

 In Figs.~\ref{fig:N_v-Bayesian-Eq-v0-500-04949}
 and Figs.~\ref{fig:N_v-Bayesian-Eq-ve-500-04949},
 we show the reconstructed radial distributions of
 the 3-D WIMP velocity
 and the 1(2)$\sigma$ statistical uncertainty bands
 as well as
 the distributions of the fitting parameters $v_0$ and $\ve$
 in all simulated experiments
 by using the one--parameter and the $v_0$--fixed velocity distributions
 in four observation periods of the advanced seasons,
 respectively.
 It can be find that
 the (approximately) symmetric periodic variations of
 the fitting parameters $v_0$ and $\ve$
 observed in Secs.~\ref{sec:N_v-Bayesian-Eq-v0-050-04949}
 and \ref{sec:N_v-Bayesian-Eq-ve-050-04949}
 can now be seen more obviously
 and pinned down more precisely as
 $214.5 \pm 6.0$ km/s
 and 225.0 ($^{+15.3}_{-17.1}$) km/s,
 respectively;
 the $\pm 6.0$ km/s and the $^{+15.3}_{-17.1}$ km/s
 annual variations of
 the best--fit values of $v_0$ and $\ve$
 are almost $\sim 2 \sigma$ of
 their statistical uncertainties
 (see Table \ref{tab:N_v-Bayesian-Eq-500-00000}).
 Additionally,
 it would be worth to also notice that,
 in (the advanced) Winter
 the deviations of the best--fit values of $v_0$ and $\ve$
 from their theoretical (annual--average) values of 220 km/s and 231 km/s
 could be as large as $\sim 3.3 \sigma$ and $\sim 2.6 \sigma$
 of their statistical uncertainties,
 respectively.

 In Tables \ref{tab:N_v-Bayesian-Eq-v0-500-04949}
 and \ref{tab:N_v-Bayesian-Eq-ve-500-04949},
 we summarize
 the reconstructed results of the fitting parameters $v_0$ and $\ve$
 as well as
 their 1(2)$\sigma$ statistical uncertainty ranges
 of the median values
 by using the one--parameter and the $v_0$--fixed velocity distribution
 with 500 total events on average
 in each 60-day observation period of the advanced seasons,
 respectively.

\paragraph[With the simplified and the modified velocity distributions
           $f_{1, \sh}(v; v_0, \ve)$ and \\ $f_{1, \sh, \Delta v}(v; v_0, \Delta v)$]
          {\boldmath
           With the simplified and the modified velocity distributions
           $f_{1, \sh}(v; v_0, \ve)$ and \\ $f_{1, \sh, \Delta v}(v; v_0, \Delta v)$}
 ~\\

 As in Secs.~\ref{sec:N_v-Bayesian-Eq-sh-050-04949}
 and \ref{sec:N_v-Bayesian-Eq-Dv-050-04949},
 we release now the constraints that
 $\ve = 1.05 \~ v_0$ and $v_0 = 220$ km/s
 and consider the simplified and the modified
 velocity distribution functions,
 $f_{1, \sh}(v; v_0, \ve)$ and
 $f_{1, \sh, \Delta v}(v; v_0, \Delta v)$,
 respectively,
 in order to check the identification possibility
 of the invariability of the Solar Galactic velocity $v_0$
 as well as
 that of the variation of the Earth's Galactic velocity $\ve$
 observed in Secs.~\ref{sec:N_v-Bayesian-Eq-sh-050-04949}
 and \ref{sec:N_v-Bayesian-Eq-Dv-050-04949}.

 In Figs.~\ref{fig:N_v-Bayesian-Eq-sh-500-04949}
 and Figs.~\ref{fig:N_v-Bayesian-Eq-Dv-500-04949},
 we show the reconstructed radial distributions of
 the 3-D WIMP velocity
 and the 1(2)$\sigma$ statistical uncertainty bands
 as well as
 the distributions of the fitting parameters $v_0$, $\ve$ and $\Delta v$
 in all simulated experiments
 on the $v_0 - \ve$ and $v_0 - \Delta v$ planes
 by using the simplified and the modified
 velocity distributions
 in four observation periods of the advanced seasons,
 respectively.
 As observed in Secs.~\ref{sec:N_v-Bayesian-Eq-sh-050-04949}
 and \ref{sec:N_v-Bayesian-Eq-Dv-050-04949},
 although the first fitting parameter $v_0$ is unconstraint,
 its best--fit values in four advanced seasons
 would be fixed as
 its annual average value of $\simeq 185$ km/s,
 with the similar 1(2)$\sigma$ statistical uncertainties of
 \mbox{$^{+10.0}_{-~8.0} \~ (^{+20.0}_{-16.0})$ km/s}.
 In contrast,
 in four advanced seasons
 the best--fit values of the second fitting parameter $\ve$ and $\Delta v$
 show clearly the (approximately) symmetric periodic variations
 of \mbox{262.7 ($\pm \sim 13$) km/s}
 and \mbox{78.6 ($\pm \sim 12$) km/s},
 respectively.
 Additionally,
 these annual variations
 would now be comparable with
 (or even larger than)
 their strongly reduced 1$\sigma$ statistical uncertainties.
 Remind that,
 as shown in Figs.~\ref{fig:N_v-Bayesian-Eq-sh-500-00000}(b)
 and \ref{fig:N_v-Bayesian-Eq-Dv-500-00000}(b),
 the theoretical values of the fitting parameters
 (the red upward--triangle)
 are now clearly out of
 the (light--blue) 2$\sigma$ statistical uncertainty contours.

 In Tables \ref{tab:N_v-Bayesian-Eq-sh-500-04949}
 and \ref{tab:N_v-Bayesian-Eq-Dv-500-04949},
 we summarize
 the reconstructed results of the fitting parameters $v_0$, $\ve$ and $\Delta v$
 as well as
 their 1(2)$\sigma$ statistical uncertainty ranges
 of the median values
 by using the simplified and the modified velocity distribution
 with 500 total events on average
 in each 60-day observation period of the advanced seasons,
 respectively.

 \def \Perioda     {\PeriodCa}
 \def \Periodb     {\PeriodCb}
 \def \Periodc     {\PeriodCc}
 \def \Periodd     {\PeriodCd}
 \def \PlotNumbera {\PlotNumberCa}
 \def \PlotNumberb {\PlotNumberCb}
 \def \PlotNumberc {\PlotNumberCc}
 \def \PlotNumberd {\PlotNumberCd}
 \def \Fittingfv             {v0}
 \def \FittingPara           {v_0}
 \InsertPlotNvBayesianAnnual
  {As in Figs.~\ref{fig:N_v-Bayesian-Eq-v0-050-04949},
   reconstructed with the one--parameter velocity distribution
   $f_{1, \sh, v_0}(v; v_0)$,
   except that
   500 total events on average
   in each 60-day observation period of the advanced seasons
   have been simulated.
   }
 \def \PlotNumber                   {\PlotNumberCa}
\InsertResultsTableNvBayesianAnnual
{one--parameter velocity distribution
 $f_{1, \sh, v_0}(v; v_0)$}
{\begin{minipage} {3.25 cm}
  \begin{center}
     ~          \\ \vspace{1.1  ex}
      49.49     \\ \vspace{0.65 ex}
    (\PeriodCa) \\ \vspace{1.1  ex}
  \end{center}
 \end{minipage}             &
 $v_0$ [km/s]               &
 214.5 & $214.5 \~ ^{+3.0}_{-3.5}   \~ (\pm \~ 6.5)$       & [211.0, 217.5] & [208.0, 221.0] \\
 \hline
 \begin{minipage} {3.25 cm}
  \begin{center}
     ~          \\ \vspace{1.1  ex}
     140.74     \\ \vspace{0.65 ex}
    (\PeriodCb) \\ \vspace{1.1  ex}
  \end{center}
 \end{minipage}             &
 $v_0$ [km/s]               &
 220.5 & $220.5    \pm  3 .5        \~ (\pm \~ 6.5)$       & [217.0, 224.0] & [214.0, 227.0] \\
 \hline
 \begin{minipage} {3.25 cm}
  \begin{center}
     ~          \\ \vspace{1.1  ex}
     231.99     \\ \vspace{0.65 ex}
    (\PeriodCc) \\ \vspace{1.1  ex}
  \end{center}
 \end{minipage}             &
 $v_0$ [km/s]               &
 214.5 & $214.5 \~ ^{+3.5}_{-3.0}   \~ (\pm \~ 6.5)$       & [211.5, 218.0] & [208.0, 221.0] \\
 \hline
 \begin{minipage} {3.25 cm}
  \begin{center}
     ~          \\ \vspace{1.1  ex}
     323.24     \\ \vspace{0.65 ex}
    (\PeriodCd) \\ \vspace{1.1  ex}
  \end{center}
 \end{minipage}             &
 $v_0$ [km/s]               &
 208.5 & $208.5 \~ ^{+3.5}_{-3.0}   \~ (\pm \~ 6.5)$       & [205.5, 212.0] & [202.0, 215.0] \\
 }
{500}
{parameter $v_0$ and}
{advanced}
 \def \Fittingfv             {ve}
 \def \FittingPara           {v_e}
 \InsertPlotNvBayesianAnnual
  {As in Figs.~\ref{fig:N_v-Bayesian-Eq-ve-050-04949},
   reconstructed with the $v_0$--fixed velocity distribution
   $f_{1, \sh, \ve}(v; \ve)$,
   except that
   500 total events on average
   in each 60-day observation period of the advanced seasons
   have been simulated.
   }
\InsertResultsTableNvBayesianAnnual
{$v_0$--fixed velocity distribution
 $f_{1, \sh, \ve}(v; \ve)$}
{\begin{minipage} {3.25 cm}
  \begin{center}
     ~          \\ \vspace{1.1  ex}
      49.49     \\ \vspace{0.65 ex}
    (\PeriodCa) \\ \vspace{1.1  ex}
  \end{center}
 \end{minipage}             &
 $\ve$ [km/s]               &
 225.0 & $225.0 \~ ^{+7.2}_{-9.0} \~ (^{+15.3}_{-17.1})$ & [216.0, 232.2] & [207.9, 240.3] \\
 \hline
 \begin{minipage} {3.25 cm}
  \begin{center}
     ~          \\ \vspace{1.1  ex}
     140.74     \\ \vspace{0.65 ex}
    (\PeriodCb) \\ \vspace{1.1  ex}
  \end{center}
 \end{minipage}             &
 $\ve$ [km/s]               &
 240.3 & $240.3 \~ ^{+7.2}_{-8.1} \~ (^{+15.3}_{-17.1})$ & [232.2, 247.5] & [223.2, 255.6] \\
 \hline
 \begin{minipage} {3.25 cm}
  \begin{center}
     ~          \\ \vspace{1.1  ex}
     231.99     \\ \vspace{0.65 ex}
    (\PeriodCc) \\ \vspace{1.1  ex}
  \end{center}
 \end{minipage}             &
 $\ve$ [km/s]               &
 225.0 & $225.0    \pm 8.1        \~ (^{+16.2}_{-17.1})$ & [216.9, 233.1] & [207.9, 241.2] \\
 \hline
 \begin{minipage} {3.25 cm}
  \begin{center}
     ~          \\ \vspace{1.1  ex}
     323.24     \\ \vspace{0.65 ex}
    (\PeriodCd) \\ \vspace{1.1  ex}
  \end{center}
 \end{minipage}             &
 $\ve$ [km/s]               &
 207.9 & $207.9 \~ ^{+9.0}_{-8.1} \~ (\pm \~ 17.1)$      & [199.8, 216.9] & [190.8, 225.0] \\
 }
{500}
{parameter $\ve$ and}
{advanced}
 \def \Fittingfv             {sh}
 \def \FittingPara           {v_0-v_e}
 \InsertPlotNvBayesianAnnual
  {As in Figs.~\ref{fig:N_v-Bayesian-Eq-sh-050-04949},
   reconstructed with the simplified velocity distribution
   $f_{1, \sh}(v; v_0, \ve)$,
   except that
   500 total events on average
   in each 60-day observation period of the advanced seasons
   have been simulated.
   }
\InsertResultsTableNvBayesianAnnual
{simplified velocity distribution
 $f_{1, \sh}(v; v_0, \ve)$}
{\multirow {2} {*}
  {\begin{minipage} {3.25 cm}
    \begin{center}
        49.49       \\ \vspace{0.65 ex}
      (\PeriodCa)
    \end{center}
   \end{minipage}}          &
 $v_0$ [km/s]               &
 185.0 & $185.0    \pm  9.0         \~ (^{+20.0}_{-16.0})$ & [176.0, 194.0] & [169.0, 205.0] \\
 \cline{2-6}
                            &
 $\ve$ [km/s]               &
 262.7 & $262.7 \~ ^{+~9.9}_{-11.0} \~ (^{+18.7}_{-24.2})$ & [251.7, 272.6] & [238.5, 281.4] \\
 \hline
 \multirow {2} {*}
  {\begin{minipage} {3.25 cm}
    \begin{center}
       140.74       \\ \vspace{0.65 ex}
      (\PeriodCb)
    \end{center}
   \end{minipage}}          &
 $v_0$ [km/s]               &
 186.0 & $185.0 \~ ^{+ 9.0}_{- 8.0} \~ (^{+19.0}_{-16.0})$ & [177.0, 194.0] & [169.0, 204.0] \\
 \cline{2-6}
                            &
 $\ve$ [km/s]               &
 274.8 & $274.8 \~ ^{+ 8.8}_{- 9.9} \~ (^{+17.6}_{-22.0})$ & [264.9, 283.6] & [252.8, 292.4] \\
 \hline
 \multirow {2} {*}
  {\begin{minipage} {3.25 cm}
    \begin{center}
       231.99       \\ \vspace{0.65 ex}
      (\PeriodCc)
    \end{center}
   \end{minipage}}          &
 $v_0$ [km/s]               &
 185.0 & $185.0    \pm  9.0         \~ (^{+19.0}_{-16.0})$ & [176.0, 194.0] & [169.0, 204.0] \\
 \cline{2-6}
                            &
 $\ve$ [km/s]               &
 262.7 & $262.7    \pm  9.9         \~ (^{+18.7}_{-22.0})$ & [252.8, 272.6] & [240.7, 281.4] \\
 \hline
 \multirow {2} {*}
  {\begin{minipage} {3.25 cm}
    \begin{center}
       323.24       \\ \vspace{0.65 ex}
      (\PeriodCd)
    \end{center}
   \end{minipage}}          &
 $v_0$ [km/s]               &
 186.0 & $185.0 \~ ^{+10.0}_{-~9.0} \~ (^{+21.0}_{-17.0})$ & [176.0, 195.0] & [168.0, 206.0] \\
 \cline{2-6}
                            &
 $\ve$ [km/s]               &
 248.4 & $249.5 \~ ^{+~9.9}_{-11.0} \~ (^{+18.7}_{-26.4})$ & [238.5, 259.4] & [223.1, 268.2] \\
 }
{500}
{parameters $v_0$ and $\ve$ as well as}
{advanced}
 \def \Fittingfv             {Dv}
 \def \FittingPara           {v_0-D_v}
 \InsertPlotNvBayesianAnnual
  {As in Figs.~\ref{fig:N_v-Bayesian-Eq-Dv-050-04949},
   reconstructed with the modified velocity distribution
   $f_{1, \sh, \Delta v}(v; v_0, \Delta v)$,
   except that
   500 total events on average
   in each 60-day observation period of the advanced seasons
   have been simulated.
   }
\InsertResultsTableNvBayesianAnnual
{modified velocity distribution
 $f_{1, \sh, \Delta v}(v; v_0, \Delta v)$}
{\multirow {2} {*}
  {\begin{minipage} {3.25 cm}
    \begin{center}
        49.49       \\ \vspace{0.65 ex}
      (\PeriodCa)
    \end{center}
   \end{minipage}}          &
 $v_0$ [km/s]               &
 185.0 & $185.0    \pm  9.0         \~ (^{+20.0}_{-16.0})$ & [176.0, 194.0] & [169.0, 205.0] \\
 \cline{2-6}
                            &
 $\Delta v$ [km/s]          &
  78.6 & $ 78.6 \~ ^{+15.3}_{-18.7} \~ (^{+30.6}_{-42.5})$ &  [59.9,  93.9] &  [36.1, 109.2] \\
 \hline
 \multirow {2} {*}
  {\begin{minipage} {3.25 cm}
    \begin{center}
       140.74       \\ \vspace{0.65 ex}
      (\PeriodCb)
    \end{center}
   \end{minipage}}          &
 $v_0$ [km/s]               &
 186.0 & $185.0 \~ ^{+ 9.0}_{- 8.0} \~ (^{+19.0}_{-16.0})$ & [177.0, 194.0] & [169.0, 204.0] \\
 \cline{2-6}
                            &
 $\Delta v$ [km/s]          &
  88.8 & $ 90.5 \~ ^{+15.3}_{-18.7} \~ (^{+28.9}_{-39.1})$ &  [71.8, 105.8] &  [51.4, 119.4] \\
 \hline
 \multirow {2} {*}
  {\begin{minipage} {3.25 cm}
    \begin{center}
       231.99       \\ \vspace{0.65 ex}
      (\PeriodCc)
    \end{center}
   \end{minipage}}          &
 $v_0$ [km/s]               &
 185.0 & $185.0    \pm  9.0         \~ (^{+19.0}_{-16.0})$ & [176.0, 194.0] & [169.0, 204.0] \\
 \cline{2-6}
                            &
 $\Delta v$ [km/s]          &
  78.6 & $ 78.6 \~ ^{+17.0}_{-18.7} \~ (^{+30.6}_{-40.8})$ &  [59.9,  95.6] &  [37.8, 109.2] \\
 \hline
 \multirow {2} {*}
  {\begin{minipage} {3.25 cm}
    \begin{center}
       323.24       \\ \vspace{0.65 ex}
      (\PeriodCd)
    \end{center}
   \end{minipage}}          &
 $v_0$ [km/s]               &
 185.0 & $185.0 \~ ^{+10.0}_{-~9.0} \~ (^{+21.0}_{-17.0})$ & [176.0, 195.0] & [168.0, 206.0] \\
 \cline{2-6}
                            &
 $\Delta v$ [km/s]          &
  65.0 & $ 65.0 \~ ^{+17.0}_{-20.4} \~ (^{+32.3}_{-45.9})$ &  [44.6,  82.0] &  [19.1,  97.3] \\
 }
{500}
{parameters $v_0$ and $\Delta v$ as well as}
{advanced}
\subsubsection{Fitted radial WIMP velocity distribution
               in the Galactic frame}
\label{sec:N_v-Bayesian-G-500}

 Finally,
 we present here also
 our Bayesian reconstruction result of
 the radial distribution of the 3-D WIMP velocity
 in the Galactic coordinate system
 with 500 total events on average
 recorded in one entire year.
 Note that
 the scanning range of the fitting parameter
 has been shrunk to 190 km/s $\le v_0 \le$ 250 km/s.

 In Figs.~\ref{fig:N_v-Bayesian-G-v0-500-00000},
 we show the reconstructed radial distributions of
 the 3-D WIMP velocity
 and the 1(2)$\sigma$ statistical uncertainty bands
 by using the simple Maxwellian
 velocity distribution
 $f_{1, \Gau}(v)$
 given by Eq.~(\ref{eqn:f1v_Gau_vesc}).
 As in Sec.~\ref{sec:N_v-Bayesian-G-050},
 our best--fit reconstruction distributions
 could match the input
 simple Maxwellian velocity distribution perfectly
 and the fitting parameter $v_0$
 could also be pinned down very precisely
 with the $\sim$ 1/3 reduced statistical uncertainties.

 In Table \ref{tab:N_v-Bayesian-G-500-00000},
 we summarize
 the reconstructed results of the fitting parameter $v_0$
 and their 1(2)$\sigma$ statistical uncertainty ranges
 of the median values
 by using the simple Maxwellian velocity distribution
 with 500 total events on average in one entire year
 shown in Figs.~\ref{fig:N_v-Bayesian-G-v0-500-00000}.

 \def \ShortFrame       {G}
 \def \Fittingfv        {v0}
 \InsertPlotNvBayesianD [b!] {v_0}
  {As in Figs.~\ref{fig:N_v-Bayesian-G-v0-050-00000},
   reconstructed with the simple Maxwellian velocity distribution
   $f_{1, \Gau}(v)$,
   except that
   500 total events on average in one entire year
   have been simulated.
   Note that
   the scanning range of the fitting parameter $v_0$
   is shrunk to between 190 km/s and 250 km/s.
   }
 \def \PlotNumber             {\PlotNumberAa}
\InsertResultsTableNvBayesian [b!]
{{$f_{1, \Gau}(v)$} &
  $v_0$ [km/s]      &
  220.0 & $220.0   \pm  4.2         \~ (^{+ 9.0}_{- 8.4})$ & [215.8, 224.2] & [211.6, 229.0] \\
 }
{500 total events on average in the observation period of \PeriodA\ day
 }
{The summary of
 the reconstructed results of the fitting parameter $v_0$
 and their 1(2)$\sigma$ statistical uncertainty ranges
 of the median values
 by using the simple Maxwellian velocity distribution
 with 500 total events on average in one entire year
 shown in Figs.~\ref{fig:N_v-Bayesian-G-v0-500-00000}.
 }
\section{Summary and conclusions}

 In this paper,
 as a preparation for our future study
 aiming to develop data analysis procedures
 using and/or combining 3-dimensional information
 offered by directional Dark Matter detection experiments,
 to, e.g., reconstruct the 3-dimensional WIMP velocity distribution,
 we simulated 3-D velocity information
 (magnitude, direction and measuring time)
 of WIMP events in the Galactic coordinate system,
 transformed them to different celestial coordinate systems,
 and then
 investigated the angular distribution patterns
 of the 3-dimensional WIMP velocity
 as well as
 reconstructed its radial distribution (magnitude)
 by applying the Bayesian fitting technique.

 Our simulations indicate that,
 first,
 with $\cal O$(50) total WIMP events
 recorded in one entire year,
 the angular distribution patterns of the 3-D WIMP velocity
 transformed to the Ecliptic and Equatorial coordinate systems
 would already show clear anisotropy.
 And,
 more precisely,
 the most frequent directions of the 3-D WIMP velocity
 would be close to,
 but somehow deviate from
 the theoretically predicted direction of the WIMP wind,
 i.e.,
 the opposite direction of the Solar movement
 in our Galaxy.
 Once $\cal O$(500) total events could be accumulated
 and thus a higher analysis resolution would be used,
 the systematic bias could be reduced,
 but a small deviation might still exist.

 Moreover,
 for demonstrating
 the ``annual'' modulation of
 the main direction of the WIMP wind
 due to the Earth's orbital motion around the Sun,
 we considered two sets of observation periods:
 four normal seasons
 and four advanced seasons,
 $\cal O$(50) total recorded events in 60 observation days
 ($\pm$ 30 days from the central date)
 had been considered for each season.
 It has been found that,
 with only $\cal O$(50) total WIMP events
 recorded in one season,
 the pattern of the angular distribution
 of the 3-D WIMP velocity
 in the Equatorial coordinate system
 would already vary slightly but clearly
 with a clockwise rotation
 around the theoretical main direction of the WIMP wind.
 This would be,
 besides the pure anisotropy of the WIMP velocity,
 a second (important) characteristic
 for identifying directional WIMP signals
 and discriminating from any (unexpected) backgrounds
 with some specified incoming directions.
 Meanwhile,
 the angular distribution pattern
 in our Earth coordinate system
 in both of four normal and four advanced seasons
 would also show clear anisotropy
 with an annually rotation.

 In addition,
 in our laboratory--dependent horizontal coordinate system,
 as expected,
 our simulation results
 with $\cal O$(50) total events
 recorded in each (60-day) season
 show clearly reversed distribution patterns
 for laboratories located
 in the Northern or the Southern Hemisphere.
 The simulations for different underground laboratories
 show also that,
 with more WIMP events
 and a higher analysis resolution,
 we might even be able to demonstrate
 a more detailed latitude--dependent pattern.

 Regarding the original purpose
 of directional Dark Matter detection experiments,
 we demonstrated also
 the laboratory--dependent diurnal modulation of
 the angular WIMP velocity distribution pattern
 in our laboratory coordinate system
 with $\cal O$(50) total WIMP events
 recorded in each 4-hour daily shift
 (in the 60-day observation period).
 Not surprisingly,
 such a laboratory--dependent diurnal modulation
 shows a similar variation pattern
 to the annual modulation
 in our horizontal coordinate system,
 requires however much more recorded events.

 On the other hand,
 we found unexpectedly that,
 with only $\cal O$(50) total WIMP events
 recorded in one entire year,
 the radial component of
 the transformed 3-D WIMP velocity
 would already show a distribution shape,
 which is a little bit more concentrated
 than the theoretically derived
 shifted Maxwellian velocity distribution
 in the range around the average WIMP velocity,
 but a little bit reduced
 in the lower and higher velocity ranges.
 By using the Bayesian fitting technique,
 we obtained the slightly--smaller best--fit values of
 the fitting parameter $v_0$
 (under the constraint of $\ve = 1.05 \~ v_0$)
 and $\ve$
 (under the constraint of $v_0 = 220$ km/s),
 although
 the reconstructed velocity distributions could match
 the theoretical velocity distribution very well.

 However,
 once we release the constraints on $v_0$ and $\ve$
 and treat them as
 two independent fitting parameters,
 the reconstructed velocity distributions
 would have a clearly different shape
 from the theoretical prediction,
 although,
 with only $\cal O$(50) total events
 recorded in one entire year
 and thus a relatively large statistical uncertainties,
 the 1$\sigma$ statistical uncertainty bands
 of the reconstructed velocity distributions
 could still cover
 the theoretical distribution
 in most part of the velocity range.
 The best--fit values of
 the fitting parameters $v_0$ and $\ve$
 would also be \mbox{$\sim$ 16\%} smaller and $\sim$ 14\% larger
 than their theoretical values,
 respectively.

 Once we raise the event number to
 $\cal O$(500) total WIMP events,
 the best--fit velocity distributions
 would differ pretty strongly
 from the theoretical distribution
 and
 their narrower 1$\sigma$ statistical uncertainty bands
 could not cover the expected curve
 in most part of the velocity ranges;
 the fitting parameters $v_0$ and $\ve$
 would also be 3$\sigma$ -- 4$\sigma$
 smaller or larger
 than their theoretical values,
 respectively.
 Additionally,
 our reconstructed radial distributions
 as well as
 their 1(2)$\sigma$ statistical uncertainty bands
 seem not to fit
 the radial distribution of the simulated 3-D WIMP velocity
 in the middle velocity range very well.
 These would imply not only that
 the radial distribution of the 3-D WIMP velocity
 in the Equatorial as well as the in laboratory coordinate systems
 would differ from the theoretically derived analytic form,
 but also
 a requirement of the modification of the analytic form of
 the fitting velocity distribution function.

 Nevertheless,
 with $\cal O$(50) total WIMP events
 recorded in each (60-day) season,
 the annual modulation of
 the radial distribution of the 3-D WIMP velocity
 could already be identified very clearly
 by observing
 the $\sim$ 5\% annual variation of the best--fit values of
 the second fitting parameter $\ve$.
 In contrast and interestingly,
 the best--fit values of the first fitting parameter $v_0$
 in all observation periods stay fixed,
 although it is also unconstraint.

 In our simulations presented in this paper,
 50 and 500 total WIMP events on average
 in (one daily shift of) one observation period
 (365 days/year,
  60 days/season,
  or 4 hours/shift $\times$ 60 days)
 have been considered.
 In this and probably the next decades,
 even only $\cal O$(50) total WIMP signals (per year) to observe
 would be a strong challenge
 for all (non)directional
 direct Dark Matter detection experiments.
 Fortunately,
 the recorded 3-D velocity information of WIMP signals
 offered by different underground laboratories
 could (in principle) be
 transformed to all laboratory--independent
 (Galactic,
  Ecliptic,
  Equatorial,
  and Earth)
 coordinate systems.
 Hence,
 data offered by more than one directional DM detection experiment
 could be combined for:
\begin{itemize}
\item
 the identification of
 the anisotropy of
 the angular velocity distribution
 of incident WIMPs;
\item
 the demonstration of
 the annual modulation of
 the angular velocity distribution pattern
 of incident WIMPs;
\item
 the Bayesian reconstruction of
 the radial distribution of the 3-D WIMP velocity;
\item
 the demonstration of
 the annual modulation of
 the (Bayesian reconstructed)
 radial distribution of the 3-D WIMP velocity.
\end{itemize}
 These analyses could be applied for reconstructing
 the 3-dimensional WIMP velocity distribution
 in both of the Equatorial and Galactic coordinate systems,
 for checking our theoretical models of the Galactic Dark Matter halo.

 In the long term,
 with enough WIMP events
 observed in one underground laboratory,
 the angular distribution patterns of the 3-D WIMP velocity
 in the laboratory--dependent (horizontal and laboratory)
 coordinate systems
 could be used to demonstrate:
\begin{itemize}
\item
 the diurnal modulation of
 the angular velocity distribution pattern
 of incident WIMPs;
\item
 the latitude--dependence of
 the angular velocity distribution pattern
 of incident WIMPs.
\end{itemize}

 Regarding the observation periods
 considered in our simulations
 presented in this paper,
 we used several approximations
 about the Earth's orbital motion
 in the Solar system.
 First,
 the Earth's orbit around the Sun is perfectly circular
 on the Ecliptic plane
 and the orbital speed is thus a constant.
 Second,
 the date of the vernal equinox is fixed exactly
 at the end of the 79th day (the May 20th) of
 a 365-day year
 and the few extra hours
 in an actual Solar year
 have been neglected.
 Nevertheless,
 considering the very low WIMP scattering event rate
 and thus only a few tens of total (combined) WIMP events
 required in at least a few tens of days
 (an overall event rate of $\lsim \~ 1$ event/day)
 for the first--phase analyses,
 these approximations should be acceptable.

 For our simulations presented in this paper,
 we also assumed implicitly that
 3-D information on the velocity of incident WIMPs
 (magnitudes and directions/angles)
 can be measured
 by using data
 from directional detection experiments.
 However,
 so far we could only measure,
 besides recoil energies,
 recoil tracks and in turn recoil angles
 of the recoiled nuclei.
 A way to use the experimental measurable data
 (recoil energies and recoil tracks/angles
  of the recoiled nuclei)
 to reconstruct 3-D WIMP velocities
 requires more investigations.

 Furthermore,
 for generating the 3-D WIMP velocity
 we considered in this paper only
 the simplest isotropic Maxwellian velocity distribution.
 Effects of
 other theoretically predicted structures of
 the Galactic Dark Matter halo,
 including some possible streams
 and/or the bulk rotation
 discussed in literature,
 and the possibility of distinguishing these models
 would also be interesting to investigate.

 In summary,
 we have begun to extend our earlier works and to study
 how to use information provided by
 directional Dark Matter detection experiments
 to understand
 the 3-dimensional (velocity) distribution of
 Galactic DM particles.
 Hopefully,
 this and more works achieved in the future
 could offer useful information
 on the structure of our Galaxy
 as well as
 for data analyses
 in indirect Dark Matter detection experiments.

\subsubsection*{Acknowledgments}

 The author appreciates
 N.~Bozorgnia and P.~Gondolo
 for useful discussions
 about the transformations between the celestial coordinate systems.
 The author would also like to thank
 the friendly hospitality of
 the Gran Sasso Science Institute,
 the School of Physics
 at the University of New South Wales,
 the School of Physics and Astronomy
 at the Monash University,
 and
 the Department of Nuclear Physics
 at the Australian National University
 during the completion and finalization of this work.

\appendix
\setcounter{equation}{0}
\setcounter{figure}  {0}
\setcounter{table}   {0}
\renewcommand{\theequation}{A\arabic{equation}}
\renewcommand{\thefigure}  {A\arabic{figure}}
\renewcommand{\thetable}   {A\arabic{table}}
%
%
% Appendix A
%
% 7/10
 \renewcommand{\arraystretch}{1.5}
 %
%
% Appendix A
%
\section{Definitions of and transformations between the celestial coordinate systems}
\label{appx:XYZ}

 In this section,
 we describe the definitions of
 the celestial coordinate systems
 used in our simulations and data analyses
 presented in this paper
 as well as
 the transformation matrices
 between these coordinate systems
 in detail.
 In order to avoid possible confusion,
 the conventional astronomical definitions
 of these coordinate systems
 will also be mentioned.

\subsection{Definitions of the horizontal and the laboratory coordinate systems}
\label{appx:XYZ_H-Lab}

 Firstly,
 in order to connect
 the celestial (laboratory--independent)
 (Galactic,
  Ecliptic,
  and Equatorial)
 coordinate systems
 with
 the geographic (laboratory--dependent)
 (horizontal and laboratory)
 coordinate systems,
 we defined particularly
 the Earth coordinate system
 as described in Sec.~\ref{sec:XYZ_E}
 and sketched in Fig.~\ref{fig:E-Eq-S}
 \cite{Bandyopadhyay10}.

 Then we discuss here
 two coordinate systems
 depending on the geographic location of
 the underground laboratory of interest
 as well as
 the measuring time of WIMP scattering events.

\subsubsection{Conventional definition}
\label{appx:XYZ_H-Lab-convention}
\InsertSKPPlotS
 {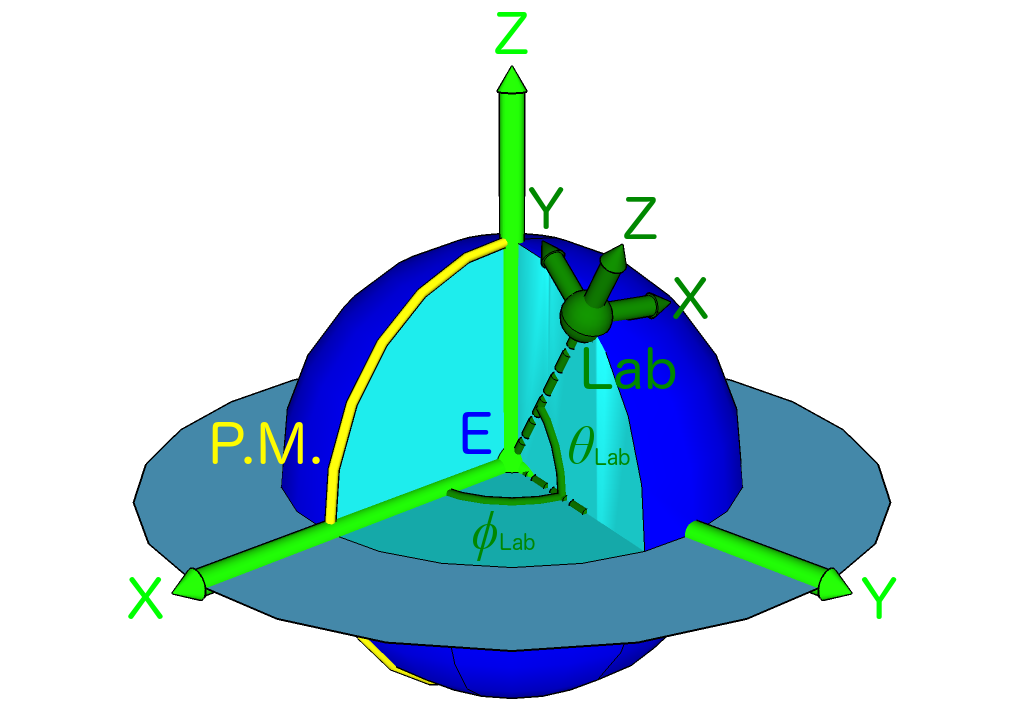}
 {The sketch of
  the conventional definition of
  the (dark--green) horizontal/laboratory coordinate system:
  it is basically the same as our definition of
  the horizontal coordinate system
  described in Sec.~\ref{sec:XYZ_H}
  and sketched in Fig.~\ref{fig:E-H-north},
  except that
  the primary direction (the ${\bf X}_{\rm H/Lab}$--axis)
  points here towards east.
  Our (light--green)
  Earth coordinate system
  is also given here.
  }

 Conventionally,
 as shown in Fig.~\ref{fig:E-H-east},
 the horizontal/laboratory coordinate system
 is defined with
 the origin at the geographic location
 of the laboratory of interest
 at, e.g., 12 midnight (the beginning) of
 (the UTC time of)
 each single Solar day,
 the primary direction (the ${\bf X}_{\rm H/Lab}$--axis)
 and the ${\bf Z}_{\rm H/Lab}$--axis
 pointing towards east
 and the zenith,
 respectively,
 and the right--handed convention
 for defining the ${\bf Y}_{\rm H/Lab}$--axis.

 In Fig.~\ref{fig:E-H-east},
 we sketch
 the conventionally defined
 horizontal/laboratory coordinate system
 with our Earth coordinate system
 together.
 It can easily find out that,
 by rotating the conventional horizontal/laboratory
 coordinate system
 at first $- \abrac{\pi / 2 - \thetaLab}$
 around the ${\bf X}_{\rm H/Lab}$--axis
 and then $- \abrac{\pi / 2 + \phiLab}$
 around the ${\bf Z}_{\rm H/Lab}$--axis,
 one can obtain our Earth coordinate system.
 Hence,
 the transformation matrix
 from the conventional horizontal/laboratory coordinate system
 to our Earth coordinate system
 can be expressed by
\cheqnXa{A}
\beqn
     \MaHEconv(\phiLab, \thetaLab)
 \=  \left[\begin{array}{c c c}
             \cos\bbrac{-\abrac{\pi / 2 + \phiLab}} ~&~
             \sin\bbrac{-\abrac{\pi / 2 + \phiLab}} ~&~
              0                                      \\
            -\sin\bbrac{-\abrac{\pi / 2 + \phiLab}} ~&~
             \cos\bbrac{-\abrac{\pi / 2 + \phiLab}} ~&~
              0                                      \\
              0                                     ~&~
              0                                     ~&~
              1                                      \\
           \end{array}\right]
     \non\\
 \conti ~~~~ ~~ \times % 6
     \left[\begin{array}{c c c}
              1                                       ~&~
              0                                       ~&~
              0                                        \\
              0                                       ~&~
             \cos\bbrac{-\abrac{\pi / 2 - \thetaLab}} ~&~
             \sin\bbrac{-\abrac{\pi / 2 - \thetaLab}}  \\
              0                                       ~&~
            -\sin\bbrac{-\abrac{\pi / 2 - \thetaLab}} ~&~
             \cos\bbrac{-\abrac{\pi / 2 - \thetaLab}}  \\
           \end{array}\right]
     \non\\
 \=  \left[\begin{array}{c c c}
            -\sin\phiLab               ~&~
            -\cos\phiLab \sin\thetaLab ~&~
             \cos\phiLab \cos\thetaLab  \\
             \cos\phiLab               ~&~
            -\sin\phiLab \sin\thetaLab ~&~
             \sin\phiLab \cos\thetaLab  \\
              0                        ~&~
                         \cos\thetaLab ~&~
                         \sin\thetaLab  \\
           \end{array}\right]
\~,
\label{eqn:Ma_H_E-convention}
\eeqn
 where
 $\phiLab$ and $\thetaLab$
 are the longitude and the latitude of
 the location of the laboratory,
 respectively.
 Conversely,
 the transformation matrix
 from our Earth coordinate system
 to the conventional horizontal/laboratory coordinate system
 can be given directly as
\cheqnXb{A}
\beqn
     \MaEHconv(\phiLab, \thetaLab)
 \=  \MaHEconv^{\rm T}(\phiLab, \thetaLab)
     \non\\
 \=  \left[\begin{array}{c c c}
            -\sin\phiLab               ~&~
             \cos\phiLab               ~&~
              0                         \\
            -\cos\phiLab \sin\thetaLab ~&~
            -\sin\phiLab \sin\thetaLab ~&~
                         \cos\thetaLab  \\
             \cos\phiLab \cos\thetaLab ~&~
             \sin\phiLab \cos\thetaLab ~&~
                         \sin\thetaLab  \\
           \end{array}\right]
\~.
\label{eqn:Ma_E_H-convention}
\eeqn
\cheqnX{A}
\subsubsection{Our definitions}
\label{appx:XYZ_H-Lab-AMIDAS-2D}

 In order to understand and compare
 the angular distribution patterns
 of the 3-D WIMP velocity
 offered by different underground laboratories
 more easily,
 as sketched in Fig.~\ref{fig:E-H-north},
 for our simulations presented in this paper
 we defined the horizontal coordinate system
 by rotating the conventional coordinates 90$^{\circ}$
 around the ${\bf Z}_{\rm H(/Lab)}$--axis,
 so that
 the $\xH$--axis points now towards north.
 Remind that,
 as our Earth coordinate system,
 for each single Solar day,
 our horizontal coordinate system
 is fixed at 12 midnight (the beginning) of
 (the UTC time of)
 the day.

 By this definition,
 the transformation
 from our horizontal coordinate system
 to the Earth coordinate system
 can be done by rotating
 at first $\pi / 2 - \thetaLab$ around the $\yH$--axis
 and then $\pi     - \phiLab$   around the $\zH$--axis.
 Thus
 the transformation matrix
 can be given by
\cheqnXa{A}
\beqn
     \MaHE(\phiLab, \thetaLab)
 \=  \left[\begin{array}{c c c}
             \cos\abrac{\pi - \phiLab} ~&~
             \sin\abrac{\pi - \phiLab} ~&~
              0                         \\
            -\sin\abrac{\pi - \phiLab} ~&~
             \cos\abrac{\pi - \phiLab} ~&~
              0                         \\
              0                        ~&~
              0                        ~&~
              1                         \\
           \end{array}\right]
     \non\\
 \conti ~~~~ ~~ \times % 6
     \left[\begin{array}{c c c}
            \cos\abrac{\pi / 2 - \thetaLab} ~&~ 0 ~&~ -\sin\abrac{\pi / 2 - \thetaLab} \\
             0                              ~&~ 1 ~&~   0                              \\
            \sin\abrac{\pi / 2 - \thetaLab} ~&~ 0 ~&~  \cos\abrac{\pi / 2 - \thetaLab} \\
           \end{array}\right]
     \non\\
 \=  \left[\begin{array}{c c c}
            -\cos\phiLab \sin\thetaLab ~&~
             \sin\phiLab               ~&~
             \cos\phiLab \cos\thetaLab  \\
            -\sin\phiLab \sin\thetaLab ~&~
            -\cos\phiLab               ~&~
             \sin\phiLab \cos\thetaLab  \\
                         \cos\thetaLab ~&~
              0                        ~&~
                         \sin\thetaLab  \\
           \end{array}\right]
\~,
\label{eqn:Ma_H_E}
\eeqn
 and then,
 conversely,
 we have
\cheqnXb{A}
\beqn
     \MaEH(\phiLab, \thetaLab)
 \=  \MaHE^{\rm T}(\phiLab, \thetaLab)
     \non\\
 \=  \left[\begin{array}{c c c}
            -\cos\phiLab \sin\thetaLab &
            -\sin\phiLab \sin\thetaLab &
                         \cos\thetaLab \\
             \sin\phiLab               &
            -\cos\phiLab               &
              0                        \\
             \cos\phiLab \cos\thetaLab &
             \sin\phiLab \cos\thetaLab &
                         \sin\thetaLab \\
           \end{array}\right]
\~.
\label{eqn:Ma_E_H}
\eeqn
\cheqnX{A}

 On the other hand,
 considering the long running time of
 (directional) direct Dark Matter detection experiments
 as well as
 for identifying the diurnal modulation of
 the angular WIMP velocity distribution,
 we also defined
 the laboratory coordinate system
 by taking into account
 the instantaneous measuring time of
 each recorded WIMP scattering event.
 As sketched in Fig.~\ref{fig:E-Lab},
 it is defined
 by rotating our horizontal coordinate system
 of the laboratory of interest
 around the Earth's north polar ($\zEq/\zE$--)axis instantaneously
 by the angle of $\omega \tPM$,
 where we define
\cheqnref{eqn:omega}
\beq
         \omega
 \equiv  \frac{2 \pi}{1~{\rm day}}
\~,
\eeq
\cheqnXN{A}{-1}
 and $\tPM$ indicates
 the fractional part of
 the measuring UTC time $t$ of each recorded WIMP event
 in unit of day.
 Hence,
 by adding the longitude of the laboratory of interest
 in the transformation matrices
 (\ref{eqn:Ma_H_E}) and (\ref{eqn:Ma_E_H})
 with this extra time--dependent term:
\beq
       \phiLab
 \lto  \phiLab + \omega \tPM
\~,
\label{eqn:phi_omega_tPM}
\eeq
 we can obtain the transformation matrices
 between the laboratory and the Earth coordinate systems
 straightforwardly as
\cheqnXa{A}
\beqn
 \conti
     \MaLabE(t, \phiLab, \thetaLab)
     \non\\
 \=  \left[\begin{array}{c c c}
            -\cos\abrac{\phiLab + \omega \tPM} \sin\thetaLab ~&~
             \sin\abrac{\phiLab + \omega \tPM}               ~&~
             \cos\abrac{\phiLab + \omega \tPM} \cos\thetaLab  \\
            -\sin\abrac{\phiLab + \omega \tPM} \sin\thetaLab ~&~
            -\cos\abrac{\phiLab + \omega \tPM}               ~&~
             \sin\abrac{\phiLab + \omega \tPM} \cos\thetaLab  \\
                                               \cos\thetaLab ~&~
              0                                              ~&~
                                               \sin\thetaLab  \\
           \end{array}\right]
\~,
     \non\\
\label{eqn:Ma_Lab_E}
\eeqn
 and
\cheqnXb{A}
\beqn
 \conti
     \MaELab(t, \phiLab, \thetaLab)
     \non\\
 \=  \left[\begin{array}{c c c}
            -\cos\abrac{\phiLab + \omega \tPM} \sin\thetaLab ~&~
            -\sin\abrac{\phiLab + \omega \tPM} \sin\thetaLab ~&~
                                               \cos\thetaLab  \\
             \sin\abrac{\phiLab + \omega \tPM}               ~&~
            -\cos\abrac{\phiLab + \omega \tPM}               ~&~
              0                                               \\
             \cos\abrac{\phiLab + \omega \tPM} \cos\thetaLab ~&~
             \sin\abrac{\phiLab + \omega \tPM} \cos\thetaLab ~&~
                                               \sin\thetaLab  \\
           \end{array}\right]
\~.
\label{eqn:Ma_E_Lab}
\eeqn
\cheqnX{A}%
 Moreover,
 by combining Eq.~(\ref{eqn:Ma_E_Lab})
 with Eq.~(\ref{eqn:Ma_H_E}),
 the transformation matrices
 between the horizontal and the laboratory coordinate systems
 can be found as
\cheqnXa{A}
\beqn
 \conti
     \MaHLab(t, \phiLab, \thetaLab)
     \non\\
 \=  \MaELab(t, \phiLab, \thetaLab) \~
     \MaHE(\phiLab, \thetaLab)
     \non\\
 \=  \left[\begin{array}{c c c}
                       \cos \omega \tPM  \sin^2\thetaLab + \cos^2\thetaLab ~&~
                       \sin \omega \tPM  \sin  \thetaLab                   ~&~
            \abrac{1 - \cos \omega \tPM} \sin  \thetaLab   \cos  \thetaLab  \\
                     - \sin \omega \tPM  \sin  \thetaLab                   ~&~
                       \cos \omega \tPM                                    ~&~
                       \sin \omega \tPM                    \cos  \thetaLab  \\
            \abrac{1 - \cos \omega \tPM} \sin  \thetaLab   \cos  \thetaLab ~&~
                     - \sin \omega \tPM                    \cos  \thetaLab ~&~
                       \cos \omega \tPM  \cos^2\thetaLab + \sin^2\thetaLab  \\
           \end{array}\right]
\~,
     \non\\
\label{eqn:Ma_Lab_H}
\eeqn
 and
\cheqnXb{A}
\beqn
 \conti
     \MaLabH(t, \phiLab, \thetaLab)
     \non\\
 \=  \MaHLab^{\rm T}(t, \phiLab, \thetaLab)
     \non\\
 \=  \left[\begin{array}{c c c}
                       \cos \omega \tPM  \sin^2\thetaLab + \cos^2\thetaLab ~&~
                     - \sin \omega \tPM  \sin  \thetaLab                   ~&~
            \abrac{1 - \cos \omega \tPM} \sin  \thetaLab   \cos  \thetaLab  \\
                       \sin \omega \tPM  \sin  \thetaLab                   ~&~
                       \cos \omega \tPM                                    ~&~
                     - \sin \omega \tPM                    \cos  \thetaLab  \\
            \abrac{1 - \cos \omega \tPM} \sin  \thetaLab   \cos  \thetaLab ~&~
                       \sin \omega \tPM                    \cos  \thetaLab ~&~
                       \cos \omega \tPM  \cos^2\thetaLab + \sin^2\thetaLab  \\
           \end{array}\right]
\~.
     \non\\
\label{eqn:Ma_H_Lab}
\eeqn
\cheqnX{A}%
 Note that
 the transformations
 between the horizontal and the laboratory coordinate systems
 are longitude ($\phiLab$) independent and
 depend in fact only on
 the latitude of the laboratory location $\thetaLab$
 (and the measuring time $t$).

\subsection{Definitions of the Ecliptic and the Equatorial coordinate systems}
\label{appx:XYZ_S-Eq}

 Now we turn to define
 the most important coordinate systems
 in our data analysis procedures:
 the Ecliptic and the Equatorial coordinate systems.
 The transformation matrices
 between the Earth and these two coordinate systems
 will be derived in detail here.

\subsubsection{Conventional definitions}
\label{appx:XYZ_S-Eq-convention}
\InsertSKPPlotS [12]
 {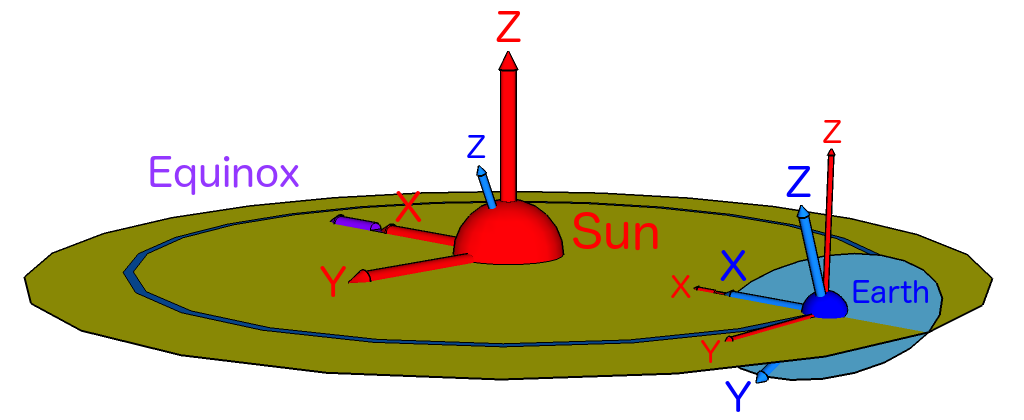}
 {The sketch of
  the astronomical definitions of
  the (red) Ecliptic and
  the (blue) Equatorial coordinate systems:
  they are basically the same as our definitions
  described in Sec.~\ref{sec:XYZ_S-Eq}
  and sketched in Fig.~\ref{fig:Eq-S}(a),
  except that
  the common primary direction (the $\xS$/$\xEq$--axis)
  is now the direction
  pointing from the Earth's center to that of the Sun
  at 12 noon of the date of the vernal equinox
  (i.e.,
   pointing towards the celestial Equinox),
  and their $\yS$-- and $\yEq$--axes
  are then defined as usual by
  the right--handed convention.
  }

 As shown in Fig.~\ref{fig:Eq-S-convention},
 in astronomy
 the Ecliptic and the Equatorial coordinate systems
 are conventionally defined as follows:
 their origins are located at
 the center of the Sun
 and that of the Earth,
 respectively,
 the common primary direction (the $\xS$/$\xEq$--axis)
 is now the direction
 pointing from the Earth's center to that of the Sun
 at 12 noon of the date of the vernal equinox
 (i.e.,
  pointing towards the celestial Equinox),
 the fundamental ($\xS - \yS$ and $\xEq - \yEq$) planes
 are the Ecliptic plane
 and the Equatorial plane,
 respectively,
 the $\zS$-- and $\zEq$--axes
 are perpendicular to the Ecliptic
 or the Equatorial plane,
 respectively,
 and their $\yS$-- and $\yEq$--axes
 are then defined as usual by
 the right--handed convention.

\subsubsection{Our definitions}
\label{appx:XYZ_S-Eq-AMIDAS-2D}

 Our definitions of
 the Ecliptic and
 the Equatorial coordinate systems
 are described in Sec.~\ref{sec:XYZ_S-Eq}
 and sketched in Fig.~\ref{fig:Eq-S}(a).
 Note only that,
 firstly,
 the Equatorial coordinate system
 is fixed and moves with the Earth,
 during the Earth's orbital motion around the Sun
 combined with
 the motion of the Solar system in the Galaxy.
 Secondly,
 since in our definitions
 the common primary direction (the $\xS$/$\xEq$--axis)
 point towards the opposite direction
 of the conventional astronomical defintion,
 in our calculations given in Appendix \ref{appx:XYZ_G}
 there is always a 180$^{\circ}$
 (or $-90^{\circ}$)
 difference from the conventional data
 provided by, e.g., Ref.~\cite{Wiki-Galactic}
 (and those used in Ref.~\cite{Bandyopadhyay10}).

\InsertSKPPlotS
 {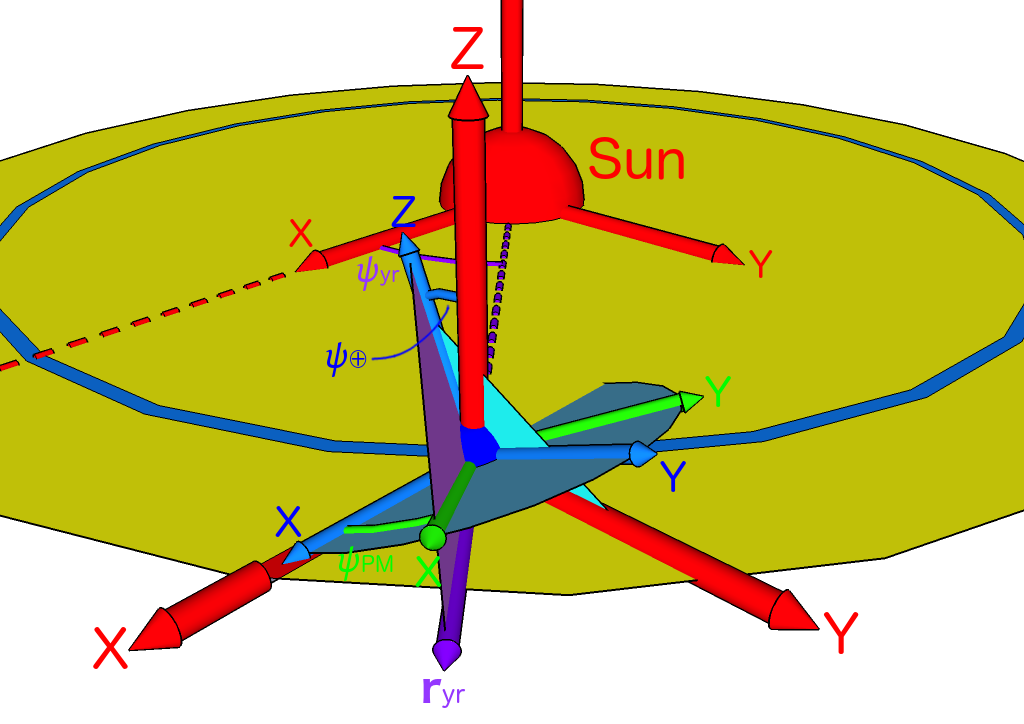}
 {The sketch of
  the relations between
  the (light--green) Earth,
  the (blue) Equatorial,
  and the (red) Ecliptic coordinate systems
  (cf.~Fig.~\ref{fig:E-Eq-S}).
  Here
  $\psiyear$
  indicates the angle swept by the connection
  between the Solar and the Earth's centers,
  $\ryear$,
  from the day of the vernal equinox,
  $\psiEarth = 23.4^{\circ}$
  is the Earth's obliquity,
  and $\psiPM$
  is the angle
  between the $\xEq$-- and the $\xE$--axes.
  Other notations are the same as
  in Fig.~\ref{fig:E-Eq-S}.
  }

 In Fig.~\ref{fig:E-Eq-S-transformation},
 we sketch
 the relations between
 the Earth,
 the Equatorial,
 and the Ecliptic coordinate systems
 (cf.~Fig.~\ref{fig:E-Eq-S}).
 Here
 $\psiyear$
 indicates the angle swept by the connection
 between the Solar and the Earth's centers,
 $\ryear$,
 from the day of the vernal equinox,
 $\psiEarth = 23.4^{\circ}$
 is the Earth's obliquity,
 and $\psiPM$
 is the angle
 between the $\xEq$-- and the $\xE$--axes.
 It can be found that,
 firstly,
 the Equatorial coordinate system
 can be obtained by simply rotating
 the Ecliptic coordinate system
 by the Earth's obliquity $\psiEarth$
 around the common $\xS$/$\xEq$--axis.
 Hence,
 the transformation matrix
 from the Ecliptic
 to the Equatorial coordinate system
 can be given directly as
\cheqnXa{A}
\beqn
     \MaSEq
  =  \left[\begin{array}{c c c}
             1 ~&~   0            ~&~  0            \\
             0 ~&~  \cos\psiEarth ~&~ \sin\psiEarth \\
             0 ~&~ -\sin\psiEarth ~&~ \cos\psiEarth \\
           \end{array}\right]
  =  \left[\begin{array}{c c c}
             1 ~&~  0       ~&~ 0       \\
             0 ~&~  0.91775 ~&~ 0.39715 \\
             0 ~&~ -0.39715 ~&~ 0.91775 \\
           \end{array}\right]
\~,
\label{eqn:Ma_S_Eq}
\eeqn
 and,
 conversely,
\cheqnXb{A}
\beqn
     \MaEqS
  =  \MaSEq^{\rm T}
  =  \left[\begin{array}{c c c}
             1 ~&~  0            ~&~   0            \\
             0 ~&~ \cos\psiEarth ~&~ -\sin\psiEarth \\
             0 ~&~ \sin\psiEarth ~&~  \cos\psiEarth \\
           \end{array}\right]
  =  \left[\begin{array}{c c c}
             1 ~&~ 0       ~&~  0       \\
             0 ~&~ 0.91775 ~&~ -0.39715 \\
             0 ~&~ 0.39715 ~&~  0.91775 \\
           \end{array}\right]
\~.
     \non\\
\label{eqn:Ma_Eq_S}
\eeqn
\cheqnX{A}

 On the other hand,
 following the calculations
 done by A.~Bandyopadhyay and D.~Majumdar
 in Ref.~\cite{Bandyopadhyay10},
 since $\ryear$ is the intersection vector
 of the (purple) $\xE - \zE$ plane
 and the (yellow) Ecliptic ($\xS - \yS$) plane,
 one has
\beqn
     \ryear
 \=  \cos\psiyear \~ \xS
   + \sin\psiyear \~ \yS
     \non\\
 \=  \cos\psiyear               \~ \xEq
   + \sin\psiyear \cos\psiEarth \~ \yEq
   - \sin\psiyear \sin\psiEarth \~ \zEq
\~,
\label{eqn:r_yr}
\eeqn
 and the vector perpendicular to
 the (purple) $\zE - \xE -\ryear$ plane
 (i.e.,
  paralle to the $\yE$--axis)
 can be obtained by
\beqn
     \zE \times \ryear
 \=  \zEq \times
     \aBig{  \cos\psiyear               \~ \xEq
           + \sin\psiyear \cos\psiEarth \~ \yEq
           - \sin\psiyear \sin\psiEarth \~ \zEq }
     \non\\
 \=  \cos\psiyear               \~ \yEq
   - \sin\psiyear \cos\psiEarth \~ \xEq
\~.
\eeqn
 Here
 the angle $\psiyear$ can be estimated
 by the measuring time $t$ of one WIMP event as
\beq
         \psiyear(t)
 \equiv  \frac{2 \pi}{365} \bBig{\abrac{t - \tPM} - 79.0}
\~.
\label{eqn:psi_yr}
\eeq
 Then
 the vector perpendicular to
 the $\yE - \zE$ plane
 (i.e.,
  paralle to the $\xE$--axis)
 can be given by
\beqn
     \aBig{\zE \times \ryear} \times \zE
 \=  \aBig{  \cos\psiyear               \~ \yEq
           - \sin\psiyear \cos\psiEarth \~ \xEq } \times
     \zEq
     \non\\
 \=  \cos\psiyear               \~ \xEq
   + \sin\psiyear \cos\psiEarth \~ \yEq
\~,
\eeqn
 and its length can thus be obtained as
\beq
     \vbrac{\aBig{\zE \times \ryear} \times \zE}
  =  \sqrt{  \cos^2\psiyear
           + \sin^2\psiyear \cos^2\psiEarth }
\~.
\eeq
 Hence,
 the unit vector $\xE$ of
 the Earth coordinate system
 can be expressed as
\cheqnXa{A}
\beqn
     \xE
 \=  \frac{       \abrac{\zE \times \ryear} \times \zE }
          {\vbrac{\abrac{\zE \times \ryear} \times \zE}}
     \non\\
 \=  \frac{  \cos\psiyear               \~ \xEq
           + \sin\psiyear \cos\psiEarth \~ \yEq }
          {\sqrt{  \cos^2\psiyear
                 + \sin^2\psiyear \cos^2\psiEarth }}
     \non\\
 \=  \gamma \cos\psiyear               \~ \xEq
   + \gamma \sin\psiyear \cos\psiEarth \~ \yEq
\~,
\label{eqn:X_E}
\eeqn
 and thus
\cheqnXb{A}
\beqn
     \yE
 \=  \zE \times \xE
     \non\\
 \=  \zEq \times
     \aBig{  \gamma \cos\psiyear               \~ \xEq
           + \gamma \sin\psiyear \cos\psiEarth \~ \yEq }
     \non\\
 \=  \gamma \cos\psiyear               \~ \yEq
   - \gamma \sin\psiyear \cos\psiEarth \~ \xEq
\~,
\label{eqn:Y_E}
\eeqn
\cheqnX{A}
 where we define
\beq
         \gamma
 \equiv  \frac{1}
              {\sqrt{  \cos^2\psiyear
                     + \sin^2\psiyear \~ \cos^2\psiEarth } }
\~.
\label{eqn:psi_yr_Earth}
\eeq
 Therefore,
 the transformation matrix
 from the Equatorial
 to the Earth coordinate system
 can be given by
\cheqnXa{A}
\beq
     \MaEqE(t)
  =  \left[\begin{array}{c c c}
             \gamma \cos\psiyear                  ~&~
             \gamma \sin\psiyear \~ \cos\psiEarth ~&~
              0                                    \\
            -\gamma \sin\psiyear \~ \cos\psiEarth ~&~
             \gamma \cos\psiyear                  ~&~
              0                                    \\
              0                                   ~&~
              0                                   ~&~
              1                                    \\
           \end{array}\right]
\~,
\label{eqn:Ma_Eq_E}
\eeq
 and,
 conversely,
\cheqnXb{A}
\beq
     \MaEEq(t)
  =  \MaEqE^{\rm T}(t)
  =  \left[\begin{array}{c c c}
             \gamma \cos\psiyear                  ~&~
            -\gamma \sin\psiyear \~ \cos\psiEarth ~&~
              0                                    \\
             \gamma \sin\psiyear \~ \cos\psiEarth ~&~
             \gamma \cos\psiyear                  ~&~
              0                                    \\
              0                                   ~&~
              0                                   ~&~
              1                                    \\
           \end{array}\right]
\~.
\label{eqn:Ma_E_Eq}
\eeq
\cheqnX{A}
 Finally,
 by combining the matrices
 given in Eqs.~(\ref{eqn:Ma_Eq_E}) and (\ref{eqn:Ma_S_Eq}),
 the transformation matrix
 from the Ecliptic
 to the Earth coordinate system
 can be obtained by
\cheqnXa{A}
\beqn
     \MaSE(t)
 \=  \MaEqE(t) \~ \MaSEq
     \non\\
 \=  \left[\begin{array}{c c c}
             \gamma \cos\psiyear                                     ~&~
             \gamma \sin\psiyear \~ \cos^2\psiEarth                  ~&~
             \gamma \sin\psiyear \~ \sin  \psiEarth \~ \cos\psiEarth  \\
            -\gamma \sin\psiyear \~ \cos  \psiEarth                  ~&~
             \gamma \cos\psiyear \~ \cos  \psiEarth                  ~&~
             \gamma \cos\psiyear \~ \sin  \psiEarth                   \\
              0                                                      ~&~
            -                       \sin  \psiEarth                  ~&~
                                    \cos  \psiEarth                   \\
           \end{array}\right]
\~,
\label{eqn:Ma_S_E}
\eeqn
 and,
 conversely,
\cheqnXb{A}
\beqn
     \MaES(t)
 \=  \MaSE^{\rm T}(t)
  =  \left[\begin{array}{c c c}
             \gamma \cos\psiyear                                     ~&~
            -\gamma \sin\psiyear \~ \cos  \psiEarth                  ~&~
              0                                                       \\
             \gamma \sin\psiyear \~ \cos^2\psiEarth                  ~&~
             \gamma \cos\psiyear \~ \cos  \psiEarth                  ~&~
            -                       \sin  \psiEarth                   \\
             \gamma \sin\psiyear \~ \sin  \psiEarth \~ \cos\psiEarth ~&~
             \gamma \cos\psiyear \~ \sin  \psiEarth                  ~&~
                                    \cos  \psiEarth                   \\
           \end{array}\right]
\~.
     \non\\
\label{eqn:Ma_E_S}
\eeqn
\cheqnX{A}
\subsubsection{Velocity of the Earth in the Ecliptic coordinate system}
\label{appx:v_Earth_S}

 The orbital velocity of the Earth's rotation around the Sun
 in the Ecliptic coordinate system
 can be expresses as
\beqn
     \VEarthS(t)
 \=  \vEarthS
     \aBig{- \sin\psiyear(t) \~ \xS
           + \cos\psiyear(t) \~ \yS }
  =  \vEarthS
     \left[\begin{array}{c}
            -\sin\psiyear(t) \\
             \cos\psiyear(t) \\
              0                    \\
           \end{array}\right]_{\rm S}
\~,
\label{eqn:V_Earth_S}
\eeqn
 where
 the Earth's orbital speed
 can be estimated by
 \cite{RPP18AP}
\beq
     \vEarthS
  =  \frac{2 \pi \times 1.495978707 \times 10^8~{\rm km}}
          {3.15569252 \times 10^7~{\rm s}}
  \simeq  29.79~{\rm km/s}
\~.
\label{eqn:v_Earth_S}
\eeq
 Note that
 the Earth's orbit around the Sun
 has been assumed here to be perfectly circular
 on the Ecliptic plane
 and the orbital speed is thus a constant.

\subsection{Definition of the Galactic coordinate system}
\label{appx:XYZ_G}

 Finally,
 we come to define the Galactic coordinate system.
 The Earth's velocity relative to the Dark Matter halo
 in the Ecliptic and the Equatorial coordinate systems
 and its annual modulation
 in four normal and four advanced seasons
 given in Table \ref{tab:period_year}
 will also be discussed here in detail.

\subsubsection{Conventional definition}
\label{appx:XYZ_G-convention}

\InsertSKPPlotS
 {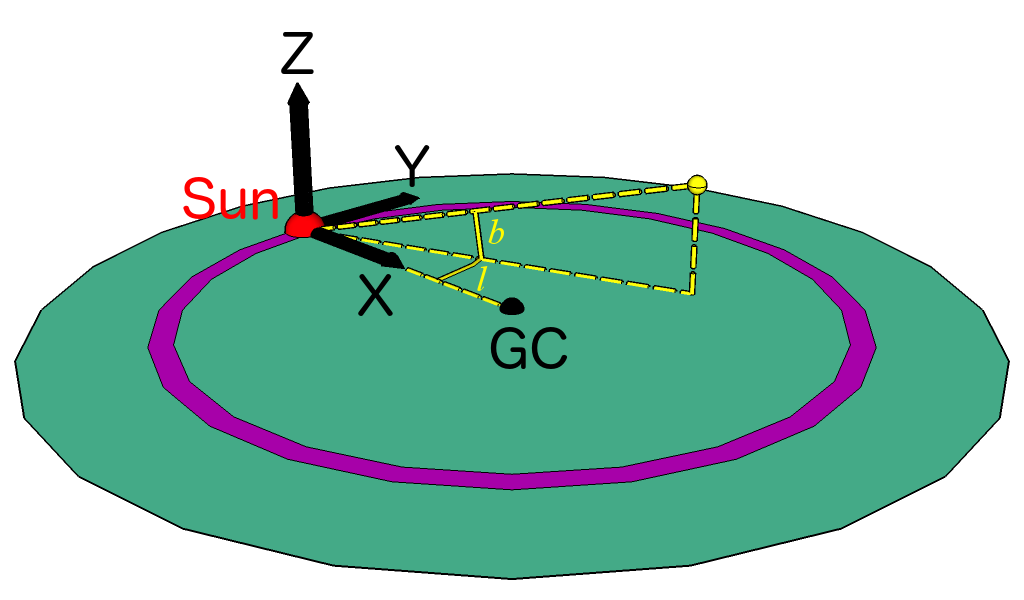}
 {The sketch of
  the astronomical definitions of
  the (black) Galactic coordinate system
  in both of the Cartesian and the spherical coordinates
  with
  the origin located at the center of the Sun
  (not the Galactic Center).
  The primary direction (the $\xG$--axis)
  points from the Solar center to
  the approximate center of the our Galaxy,
  the $\zG$--axis
  points to the Galactic North Pole,
  and
  the $\yG$--axis is then defined as usual
  by the right--handed convention
  \cite{Wiki-Galactic}.
  On the other hand,
  the Galactic longitude ($l$)
  is measured from the $\xG$--axis
  on the Galactic plane,
  while
  the Galactic latitude ($b$)
  measures the angle of the object
  above or below the approximate Galactic plane
  \cite{Wiki-Galactic}.
  }

 In Fig.~\ref{fig:l-b_G},
 we sketch
 the astronomical definitions of
 the Galactic coordinate system
 in both of the Cartesian and the spherical coordinates
 with
 the origin located at the center of the Sun
 (not the Galactic Center).
 The primary direction (the $\xG$--axis)
 points from the Solar center to
 the approximate center of the our Galaxy,
 the $\zG$--axis
 points to the Galactic North Pole,
 and
 the $\yG$--axis is then defined as usual
 by the right--handed convention
 \cite{Wiki-Galactic}.
 On the other hand,
 the Galactic longitude ($l$)
 is measured from the $\xG$--axis
 on the Galactic plane,
 while
 the Galactic latitude ($b$)
 measures the angle of the object
 above or below the approximate Galactic plane
 \cite{Wiki-Galactic}.

\subsubsection{Our definition}
\label{appx:XYZ_G-AMIDAS-2D}

 Our definition of
 the Galactic coordinate system
 is described in Sec.~\ref{sec:XYZ_G}
 and sketched in Fig.~\ref{fig:G-rotated}.
 Then,
 according to Ref.~\cite{Wiki-Galactic},
 the direction of the Galactic Center
 in the Equatorial coordinate system
 can be expressed by
\beqn
     \xGEq
 \=  \left[\begin{array}{c c c}
            \cos\thetaGC \cos\phiGC ~&~
            \cos\thetaGC \sin\phiGC ~&~
            \sin\thetaGC             \\
           \end{array}\right]_{\rm Eq}
     \non\\
 \eqncong
     \left[\begin{array}{c c c}
             0.05495 ~&~
             0.87340 ~&~
            -0.48389  \\
           \end{array}\right]_{\rm Eq}
\~,
\label{eqn:GC_Eq}
\eeqn
 where
 we have adopted the values
 provided by Ref.~\cite{Wiki-Galactic}%
\footnote{
 Remind that,
 in the Ecliptic and the Equatorial coordinate systems
 defined in Sec.~\ref{sec:XYZ_S-Eq},
 the common $\xS$/$\xEq$--axis
 points from the center of the Sun to that of the Earth,
 which is opposite to the $\bf X$ direction
 defined conventionally.
 Hence,
 the right ascensions of the Galactic Center
 and the Galactic North Pole
 in the Equatorial and the Ecliptic coordinate systems
 adopted in this section
 as well as
 those calculated later
 would differ from the conventional values
 by $12^{\rm h} = 180^{\circ}$.
}
\cheqnXa{A}
\beq
     \phiGC
  =  5^{\rm h} 45.6^{\rm m}
  =  86.40^{\circ}
\~,
\label{eqn:phi_GC_Eq}
\eeq
 and
\cheqnXb{A}
\beq
     \thetaGC
  = -28.94^{\circ}
\~,
\label{eqn:theta_GC_Eq}
\eeq
\cheqnX{A}
 as the right ascension and the declination of the Galactic Center
 in the Equatorial coordinate system
 (from the $\xEq$($= \xS$)--axis
  and the Earth's/celestial Equator ($\xEq - \yEq$) plane),
 respectively.
 Meanwhile,
 the direction of the Galactic North Pole
 in our Equatorial coordinate system
 can also be given as
 \cite{Wiki-Galactic}
\beqn
     \zGEq
 \=  \left[\begin{array}{c c c}
            \cos\thetaGNP \cos\phiGNP ~&~
            \cos\thetaGNP \sin\phiGNP ~&~
            \sin\thetaGNP              \\
           \end{array}\right]_{\rm Eq}
     \non\\
 \eqncong
      \left[\begin{array}{c c c}
            0.86769 ~&~
            0.19793 ~&~
            0.45601  \\
           \end{array}\right]_{\rm Eq}
\~,
\label{eqn:GNP_Eq}
\eeqn
 where we have
\cheqnXa{A}
\beq
     \phiGNP
  =  51.4^{\rm m}
  =  12.85^{\circ}
\~,
\label{eqn:phi_GNP_Eq}
\eeq
 and
\cheqnXb{A}
\beq
     \thetaGNP
  =  27.13^{\circ}
\~,
\label{eqn:theta_GNP_Eq}
\eeq
\cheqnX{A}
 as the right ascension and the declination of the Galactic North Pole
 in the Equatorial coordinate system,
 respectively.
 By combining Eqs.~(\ref{eqn:GC_Eq}) and (\ref{eqn:GNP_Eq}),
 the $\yG$--axis of the Galactic coordinate system
 in the Equatorial coordinate system
 can be calculated by
\beqn
     \yGEq
 \=  \zGEq \times \xGEq
     \non\\
 \=  \left[\begin{array}{c}
              \cos\thetaGNP \sin\phiGNP
              \sin\thetaGC
            - \sin\thetaGNP
              \cos\thetaGC  \sin\phiGC  \\
              \sin\thetaGNP
              \cos\thetaGC  \cos\phiGC
            - \cos\thetaGNP \cos\phiGNP
              \sin\thetaGC              \\
              \cos\thetaGNP \cos\thetaGC
              \abrac{\cos\phiGNP \sin\phiGC - \sin\phiGNP \cos\phiGC} \\
           \end{array}\right]_{\rm Eq}^{\rm T}
     \non\\
 \eqncong
      \left[\begin{array}{c c c}
            -0.49406 ~&~
             0.44492 ~&~
             0.74696  \\
           \end{array}\right]_{\rm Eq}
\~.
\label{eqn:Y_G_Eq}
\eeqn
 Then
 the transformation matrix
 from the Equatorial
 to the Galactic coordinate system
 can be given by
\cheqnXa{A}
\beq
     \MaEqG
  =  \left[\begin{array}{c}
            \xGEq \\
            \yGEq \\
            \zGEq \\
           \end{array}\right]
  =  \left[\begin{array}{c c c}
             0.05495 ~&~ 0.87340 ~&~ -0.48389 \\
            -0.49406 ~&~ 0.44492 ~&~  0.74696 \\
             0.86769 ~&~ 0.19793 ~&~  0.45601 \\
           \end{array}\right]
\~,
\label{eqn:Ma_Eq_G}
\eeq
 and,
 conversely,
 we have
\cheqnXb{A}
\beqn
     \MaGEq
 \=  \MaEqG^{\rm T}
  =  \left[\begin{array}{c c c}
             0.05495 ~&~ -0.49406 ~&~ 0.86769 \\
             0.87340 ~&~  0.44492 ~&~ 0.19793 \\
            -0.48389 ~&~  0.74696 ~&~ 0.45601 \\
           \end{array}\right]
\~.
\label{eqn:Ma_G_Eq}
\cheqnX{A}
\eeqn
 Furthermore,
 by using the transformation matrices
 between the Equatorial and the Ecliptic coordinate systems
 in Eqs.~(\ref{eqn:Ma_Eq_S}) and (\ref{eqn:Ma_S_Eq}),
 the transformation matrices
 between the Ecliptic and the Galactic coordinate systems
 can be given by
\cheqnXa{A}
\beqn
     \MaGS
 \=  \MaEqS \MaGEq
  =  \left[\begin{array}{c c c}
             0.05495 ~&~ -0.49406 ~&~ 0.86769 \\
             0.99374 ~&~  0.11168 ~&~ 0.00055 \\
            -0.09723 ~&~  0.86223 ~&~ 0.49711 \\
           \end{array}\right]
\~,
\label{eqn:Ma_G_S}
\eeqn
 and
\cheqnXb{A}
\beqn
     \MaSG
 \=  \MaGS^{\rm T}
  =  \left[\begin{array}{c c c}
             0.05495 ~&~ 0.99374 ~&~ -0.09723 \\
            -0.49406 ~&~ 0.11168 ~&~  0.86223 \\
             0.86769 ~&~ 0.00055 ~&~ 0.49711  \\
           \end{array}\right]
\~.
\label{eqn:Ma_S_G}
\eeqn
\cheqnX{A}%
 Finally,
 from the matrix $\MaGS$
 given in Eq.~(\ref{eqn:Ma_G_S}),
 one can obtain the $\xG$-- and $\zG$--axes
 in the Ecliptic coordinate system as
\beqn
     \xGS
 \=  \MaGS
     \left[\begin{array}{c c c}
             1 \\
             0 \\
             0 \\
           \end{array}\right]_{\rm S}
  =  \left[\begin{array}{c c c}
             0.05495 \\
             0.99374 \\
            -0.09723 \\
           \end{array}\right]_{\rm S}
\~,
\label{eqn:X_G_S}
\eeqn
 and
\beqn
     \zGS
 \=  \MaGS
     \left[\begin{array}{c c c}
             0 \\
             0 \\
             1 \\
           \end{array}\right]_{\rm S}
  =  \left[\begin{array}{c c c}
            0.86769 \\
            0.00055 \\
            0.49711 \\
           \end{array}\right]_{\rm S}
\~.
\label{eqn:Z_G_S}
\eeqn
 These give in turn
 the right ascensions and the declinations of
 the Galactic Center and the Galactic North Pole
 in the Ecliptic coordinate system as
\cheqnXa{A}
\beq
     \phiGC[S]
  =  86.84^{\circ}
  =   5.79^{\rm h}
\~,
\label{eqn:phi_GC_S}
\eeq
\cheqnXb{A}
\beq
     \thetaGC[S]
  = -5.58^{\circ}
\~,
\label{eqn:theta_GC_S}
\eeq
\cheqnX{A}
 and
\cheqnXa{A}
\beq
     \phiGNP[S]
  =  0.036^{\circ}
  =   2.17^{\rm m}
\~,
\label{eqn:phi_GNP_S}
\eeq
\cheqnXb{A}
\beq
     \thetaGNP[S]
  =  29.81^{\circ}
\~,
\label{eqn:theta_GNP_S}
\eeq
\cheqnX{A}
 respectively.

\subsubsection{Direction of the movement of the Solar system
               around the Galactic center}
\label{appx:v_Sun_G}
\InsertSKPPlotS
 {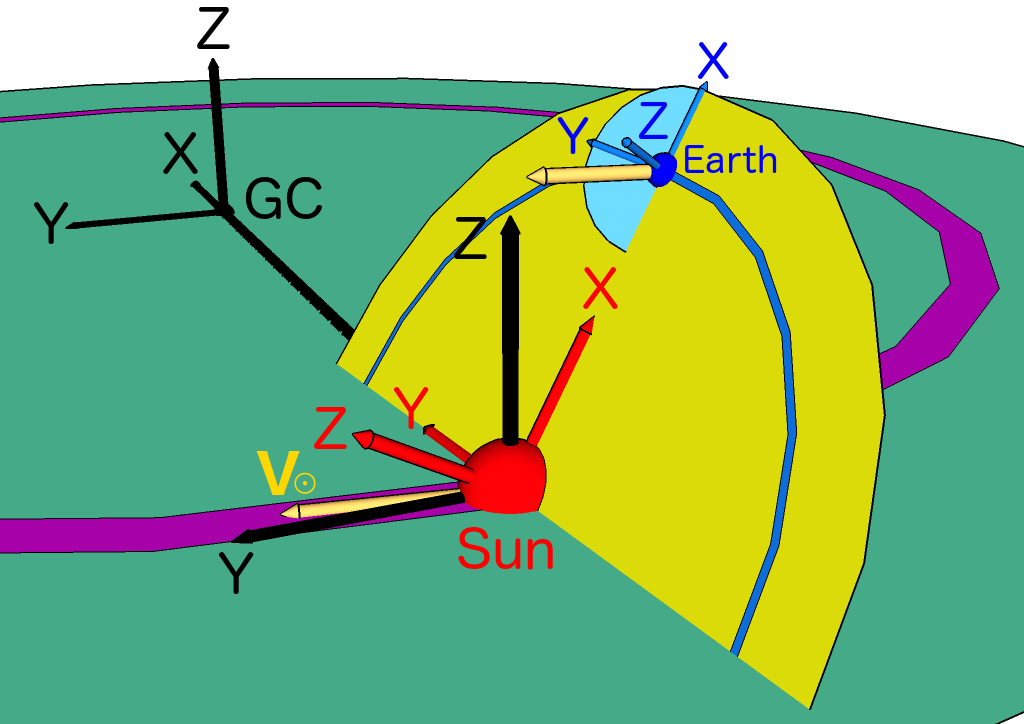}
 {The same as in Fig.~\ref{fig:Eq-S}(b),
  except that
  the additional (golden) arrows here
  indicate the direction of the movement of the Solar system
  around the Galactic center.
  Note that
  the moving direction of the Solar system
  is not parallel to,
  but only approximately along
  the $\yG$--axis
  (with an included angle of 8.87$^{\circ}$ = 35.48$^{\rm m}$),
  nor on the (approximate) $\xG - \yG$ (Galactic) plane
  (0.60$^{\circ}$ above).
  }

 In Fig.~\ref{fig:v_Sun_Eq-S-G-rotated},
 we add the (golden) arrow
 to indicate
 the direction of the movement of the Solar system
 around the Galactic center
 (cf.~%
  Fig.~\ref{fig:Eq-S}(b)).
 The Galactic movement of the Solar system
 pointing currently towards the CYGNUS constellation
 can be given
 in the Equatorial coordinate system as
 \cite{Bandyopadhyay10}
\beqn
     \VSunEq
 \=  \vSunG
     \left[\begin{array}{c}
            \cos\thetaCYGNUS \cos\phiCYGNUS \\
            \cos\thetaCYGNUS \sin\phiCYGNUS \\
            \sin\thetaCYGNUS                \\
           \end{array}\right]_{\rm Eq}
  =  \vSunG
     \left[\begin{array}{c}
            -0.47069 \\
             0.57508 \\
             0.66913 \\
           \end{array}\right]_{\rm Eq}
 \cong
     \left[\begin{array}{c}
            -103.55~{\rm km/s} \\
             126.52~{\rm km/s} \\
             147.21~{\rm km/s} \\
           \end{array}\right]_{\rm Eq}
\~.
     \non\\
\label{eqn:V_Sun_Eq}
\eeqn
 Here we have used
\beq
          \vSunG
  \simeq  220~{\rm km/s}
\~,
\label{eqn:v_Sun_G}
\eeq
 and
 the right ascension and the declination of
 the direction of the CYGNUS constellation
 in the Equatorial coordinate system are
 \cite{Bandyopadhyay10}
\cheqnXa{A}
\beq
     \phiCYGNUS
  =    8.62^{\rm h}
  =  129.30^{\circ}
\~,
\label{eqn:phi_CYGNUS_Eq}
\eeq
 and
\cheqnXb{A}
\beq
     \thetaCYGNUS
  =  42^{\circ}
\~.
\label{eqn:theta_CYGNUS_Eq}
\eeq
\cheqnX{A}
 Moreover,
 by using the transformation matrices
 $\MaEqS$ and $\MaEqG$
 given in Eqs.~(\ref{eqn:Ma_Eq_S}) and (\ref{eqn:Ma_Eq_G}),
 the moving direction of the Solar system
 in the Ecliptic and the Galactic coordinate systems
 can be obtained as
\beqn
     \VSunS
 \=  \MaEqS \~ \VSunEq
     \non\\
 \=  \vSunG
     \left[\begin{array}{c}
            \cos\thetaCYGNUS \cos\phiCYGNUS               \\
            \cos\psiEarth \cos\thetaCYGNUS \sin\phiCYGNUS
           -\sin\psiEarth \sin\thetaCYGNUS                \\
            \sin\psiEarth \cos\thetaCYGNUS \sin\phiCYGNUS
           +\cos\psiEarth \sin\thetaCYGNUS                \\
           \end{array}\right]_{\rm S}
     \non\\
 \=  \vSunG
     \left[\begin{array}{c}
            -0.47069 \\
             0.26203 \\
             0.84249 \\
           \end{array}\right]_{\rm S}
     \non\\
 \eqncong
     \left[\begin{array}{c}
            -103.55~{\rm km/s} \\
              57.65~{\rm km/s} \\
             185.35~{\rm km/s} \\
           \end{array}\right]_{\rm S}
\~,
\label{eqn:V_Sun_S}
\eeqn
 and
\beqn
     \VSunG
 \=  \MaEqG \~ \VSunEq
  =  \vSunG
     \left[\begin{array}{c}
            0.15262 \\
            0.98823 \\
            0.01054 \\
           \end{array}\right]_{\rm G}
 \cong
     \left[\begin{array}{c}
              33.58~{\rm km/s} \\
             217.41~{\rm km/s} \\
               2.32~{\rm km/s} \\
           \end{array}\right]_{\rm G}
\~.
\label{eqn:V_Sun_G}
\eeqn
 These give in turn
 the right ascension and the declination of
 the moving direction of the Solar system
 in the Ecliptic coordinate system
 as
\cheqnXa{A}
\beq
     \phiCYGNUS[S]
  =  150.90^{\circ}
  =   10.06^{\rm h}
\~,
\label{eqn:phi_CYGNUS_S}
\eeq
\cheqnXb{A}
\beq
     \thetaCYGNUS[S]
  =  57.40^{\circ}
\~.
\label{eqn:theta_CYGNUS_S}
\eeq
\cheqnX{A}
 And,
 in the Galactic coordinate system,
 one has
\cheqnXa{A}
\beq
     \phiCYGNUS[G]
  =  81.22^{\circ}
  =   5.41^{\rm h}
\~,
\label{eqn:phi_CYGNUS_G}
\eeq
\cheqnXb{A}
\beq
     \thetaCYGNUS[G]
  =  0.60^{\circ}
\~.
\label{eqn:theta_CYGNUS_G}
\eeq
\cheqnX{A}
\subsubsection{Annual modulation of the Earth's velocity
               in the Ecliptic coordinate system}
\label{appx:v_Earth_chi_S}

 In this and the next subsections,
 we discuss
 the annual modulation of the Earth's velocity
 relative to the Dark Matter halo
 in the Ecliptic and the Equatorial coordinate systems
 separately.
 Remind that
 our calculations are based on two assumptions.
 First,
 the Earth's orbit around the Sun is perfectly circular
 on the Ecliptic plane
 and the orbital speed is thus a constant.
 Second,
 the date of the vernal equinox is fixed exactly
 at the end of the 79th day (the May 20th) of a 365-day year
 and the few extra hours in an actual Solar year
 has been neglected.

 We begin with four normal seasons,
 in which
 the orbital velocity of the Earth's rotation around the Sun
 in the Ecliptic coordinate system
 on the central date of each season
 is along the $\xS$-- or the $\yS$--axis,
 corresponding to the Earth's positions
 shown in Fig.~\ref{fig:v_Earth_chi_S}(a).
 By setting $\psiyear = 0$, $\pi / 2$, $\pi$, and $3 \pi / 2$,
 or,
 equivalently,
 $t = 79.0$ (the {\em beginning} of the {\em 21st} of March),
 170.25 (June 20th),
 261.50 (September 19th),
 and 352.75 day (December 19th),
 respectively,
 into the expression for
 the Earth's orbital velocity around the Sun
 given in Eq.~(\ref{eqn:V_Earth_S}),
 we have the Earth's velocities in four normal seasons
 in the Ecliptic coordinate system as
\cheqnXa{A}
\beq
     \VEarthST{79.00}
  =  \vEarthS \yS
\~,
\label{eqn:V_Earth_S_07900}
\eeq
\cheqnXb{A}
\beq
     \VEarthST{170.25}
  = -\vEarthS \xS
\~,
\label{eqn:V_Earth_S_17025}
\eeq
\cheqnXc{A}
\beq
     \VEarthST{261.50}
  = -\vEarthS \yS
\~,
\label{eqn:V_Earth_S_26150}
\eeq
 and
\cheqnXd{A}
\beq
     \VEarthST{352.75}
  =  \vEarthS \xS
\~,
\label{eqn:V_Earth_S_35275}
\eeq
\cheqnX{A}
 respectively.
 By combining with the velocity of the Solar system
 given in Eq.~(\ref{eqn:V_Sun_S})
 and taking the Earth's orbital speed
 given in Eq.~(\ref{eqn:v_Earth_S}),
 one can obtain the Earth's velocities
 relative to the Dark Matter halo
 in four normal seasons
 in the Ecliptic coordinate system
 as
\cheqnXa{A}
\beqn
     \VEarthchiST{79.00}
 \=  \VSunS + \VEarthST{79.00}
  =  \left[\begin{array}{c}
            -103.55~{\rm km/s} \\
              87.43~{\rm km/s} \\
             185.35~{\rm km/s} \\
           \end{array}\right]_{\rm S}
\~,
\label{eqn:V_Earth_chi_S_07900}
\eeqn
\cheqnXb{A}
\beqn
     \VEarthchiST{170.25}
 \=  \VSunS + \VEarthST{170.25}
  =  \left[\begin{array}{c}
            -133.34~{\rm km/s} \\
              57.65~{\rm km/s} \\
             185.35~{\rm km/s} \\
           \end{array}\right]_{\rm S}
\~,
\label{eqn:V_Earth_chi_S_17025}
\eeqn
\cheqnXc{A}
\beqn
     \VEarthchiST{261.50}
 \=  \VSunS + \VEarthST{261.50}
  =  \left[\begin{array}{c}
            -103.55~{\rm km/s} \\
              27.86~{\rm km/s} \\
             185.35~{\rm km/s} \\
           \end{array}\right]_{\rm S}
\~,
\label{eqn:V_Earth_chi_S_26150}
\eeqn
 and
\cheqnXd{A}
\beqn
     \VEarthchiST{352.75}
 \=  \VSunS + \VEarthST{352.75}
  =  \left[\begin{array}{c}
            - 73.77~{\rm km/s} \\
              57.65~{\rm km/s} \\
             185.35~{\rm km/s} \\
           \end{array}\right]_{\rm S}
\~,
\label{eqn:V_Earth_chi_S_35275}
\eeqn
\cheqnX{A}%
 respectively.

 On the other hand,
 as shown in Fig.~\ref{fig:v_Earth_chi_S}(b),
 the Earth's velocity
 relative to the Dark Matter halo
 should be maximal (minimal),
 once the Earth's velocity
 is (anti--)parallel to the projection of
 the moving direction of the Solar system
 on the Ecliptic plane.
 This requires that,
 from Eq.~(\ref{eqn:phi_CYGNUS_S}),
\beq
     \psiyear
  =  60.90^{\circ}
  =  0.3383 \pi
  =  1.0628
\~,
\label{eqn:psi_yr_max}
\eeq
 or,
 equivalently,
\beq
     \tmax
  =  140.74~{\rm day}
\~,
\label{eqn:psi_yr_max_day}
\eeq
 i.e.,
 around the 21st of May.
 Then,
 by setting $\psiyear = 1.0628$, 4.2044, $-0.5080$, and 2.6336
 (corresponding to the Earth's positions
  in four advanced seasons
  shown in Fig.~\ref{fig:v_Earth_chi_S}(b)),
 or,
 equivalently,
 $t = 140.74$ (May 21st),
 323.24 (November 20th),
  49.49 (February 19th),
 and 231.99 day (August 20th),
 respectively,
 into the expression (\ref{eqn:V_Earth_S}),
 we have the Earth's velocities as
\cheqnXa{A}
\beq
     \VEarthST{140.74}
  =  \vEarthS
     \left[\begin{array}{c}
            -0.874 \\
             0.486 \\
             0     \\
           \end{array}\right]_{\rm S}
  =  \left[\begin{array}{c}
            -26.02~{\rm km/s} \\
             14.49~{\rm km/s} \\
             0                \\
           \end{array}\right]_{\rm S}
\~,
\label{eqn:V_Earth_S_14074}
\eeq
\cheqnXb{A}
\beq
     \VEarthST{323.24}
  =  \vEarthS
     \left[\begin{array}{c}
             0.874 \\
            -0.486 \\
             0     \\
           \end{array}\right]_{\rm S}
  =  \left[\begin{array}{c}
             26.02~{\rm km/s} \\
            -14.49~{\rm km/s} \\
             0                \\
           \end{array}\right]_{\rm S}
\~,
\label{eqn:V_Earth_S_32324}
\eeq
\cheqnXc{A}
\beq
     \VEarthST{49.49}
  =  \vEarthS
     \left[\begin{array}{c}
             0.486 \\
             0.874 \\
             0     \\
           \end{array}\right]_{\rm S}
  =  \left[\begin{array}{c}
             14.49~{\rm km/s} \\
             26.02~{\rm km/s} \\
             0                \\
           \end{array}\right]_{\rm S}
\~,
\label{eqn:V_Earth_S_04949}
\eeq
 and
\cheqnXd{A}
\beqn
     \VEarthST{231.99}
  =  \vEarthS
     \left[\begin{array}{c}
            -0.486 \\
            -0.874 \\
             0     \\
           \end{array}\right]_{\rm S}
  =  \left[\begin{array}{c}
            -14.49~{\rm km/s} \\
            -26.02~{\rm km/s} \\
             0                \\
           \end{array}\right]_{\rm S}
\~,
\label{eqn:V_Earth_S_23199}
\eeqn
\cheqnX{A}%
 respectively.
 Then
 the maximal, the minimal and the middle values of
 the Earth's velocities
 relative to the Dark Matter halo
 in four advanced seasons
 can be obtained as
\cheqnXa{A}
\beqn
     \VEarthchiST{140.74}
 \=  \VSunS + \VEarthST{140.74}
  =  \left[\begin{array}{c}
            -129.58~{\rm km/s} \\
              72.14~{\rm km/s} \\
             185.35~{\rm km/s} \\
           \end{array}\right]_{\rm S}
\~,
\label{eqn:V_Earth_chi_S_14074}
\eeqn
\cheqnXb{A}
\beqn
     \VEarthchiST{323.24}
 \=  \VSunS + \VEarthST{323.24}
  =  \left[\begin{array}{c}
            - 77.53~{\rm km/s} \\
              43.16~{\rm km/s} \\
             185.35~{\rm km/s} \\
           \end{array}\right]_{\rm S}
\~,
\label{eqn:V_Earth_chi_S_32324}
\eeqn
\cheqnXc{A}
\beqn
     \VEarthchiST{49.49}
 \=  \VSunS + \VEarthST{49.49}
  =  \left[\begin{array}{c}
            - 89.06~{\rm km/s} \\
              83.67~{\rm km/s} \\
             185.35~{\rm km/s} \\
           \end{array}\right]_{\rm S}
\~,
\label{eqn:V_Earth_chi_S_04949}
\eeqn
 and
\cheqnXd{A}
\beqn
     \VEarthchiST{231.99}
 \=  \VSunS + \VEarthST{231.99}
  =  \left[\begin{array}{c}
            -118.04~{\rm km/s} \\
              31.62~{\rm km/s} \\
             185.35~{\rm km/s} \\
           \end{array}\right]_{\rm S}
\~,
\label{eqn:V_Earth_chi_S_23199}
\eeqn
\cheqnX{A}%
 respectively.
 One can thus estimate the magnitudes and the directions
 (the right ascensions and the declinations)
 of the Earth's velocity
 relative to the Dark Matter halo
 in four normal and four advanced seasons
 straightforwardly,
 which
 we summarize
 in Table \ref{tab:vEarthchiST}
 for readers' reference.

\begin{table} [t!]
\small
\begin{center}
\renewcommand{\arraystretch}{1.5}
 \begin{tabular}{|| c || c | c | c ||}
\hline
\hline
 \makebox[2.5 cm][c]{Date                          (day)}  &
 \makebox[4   cm][c]{Magnitude       $\vEarthchiS$ (km/s)} &
 \makebox[4.5 cm][c]{Right ascension $\phichiS$}           &
 \makebox[3.5 cm][c]{Declination     $\thetachiS$}         \\
\hline
\hline
  79.0  & 229.61 & 139.82$^{\circ}$ = ~9.32$^{\rm h}$ & 53.83$^{\circ}$ \\
 170.25 & 235.49 & 156.62$^{\circ}$ = 10.44$^{\rm h}$ & 51.91$^{\circ}$ \\
 261.50 & 214.13 & 164.94$^{\circ}$ = 11.00$^{\rm h}$ & 59.95$^{\circ}$ \\
 352.75 & 207.65 & 141.99$^{\circ}$ = ~9.47$^{\rm h}$ & 63.20$^{\circ}$ \\
\hline
  49.49 & 222.01 & 136.79$^{\circ}$ = ~9.12$^{\rm h}$ & 56.60$^{\circ}$ \\
 140.74 & 237.38 & 150.90$^{\circ}$ = 10.06$^{\rm h}$ & 51.34$^{\circ}$ \\
 231.99 & 222.01 & 165.00$^{\circ}$ = 11.00$^{\rm h}$ & 56.60$^{\circ}$ \\
 323.24 & 205.49 & 150.90$^{\circ}$ = 10.06$^{\rm h}$ & 64.42$^{\circ}$ \\
\hline
\hline
\end{tabular}
\end{center}
\caption{
 The list of
 the magnitudes,
 the right ascensions and the declinations of
 the Earth's velocity
 relative to the Dark Matter halo
 in four normal and four advanced seasons
 in the Ecliptic coordinate system.
}
\label{tab:vEarthchiST}
\end{table}

\begin{figure} [p!]
\begin{center}
 \includegraphics [height = 10.5 cm] {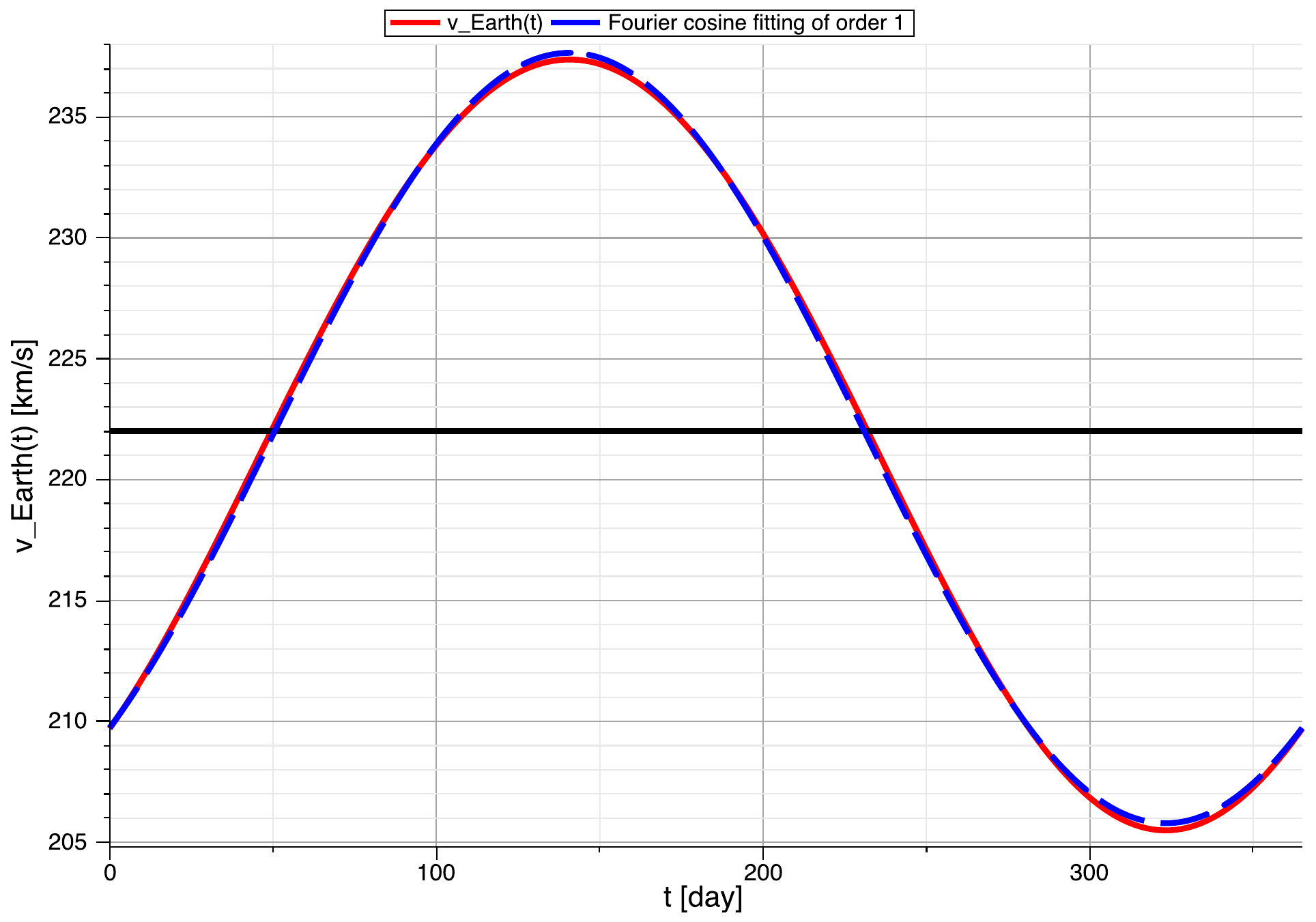} \\ \vspace{1.25 cm}
 \includegraphics [height = 10.5 cm] {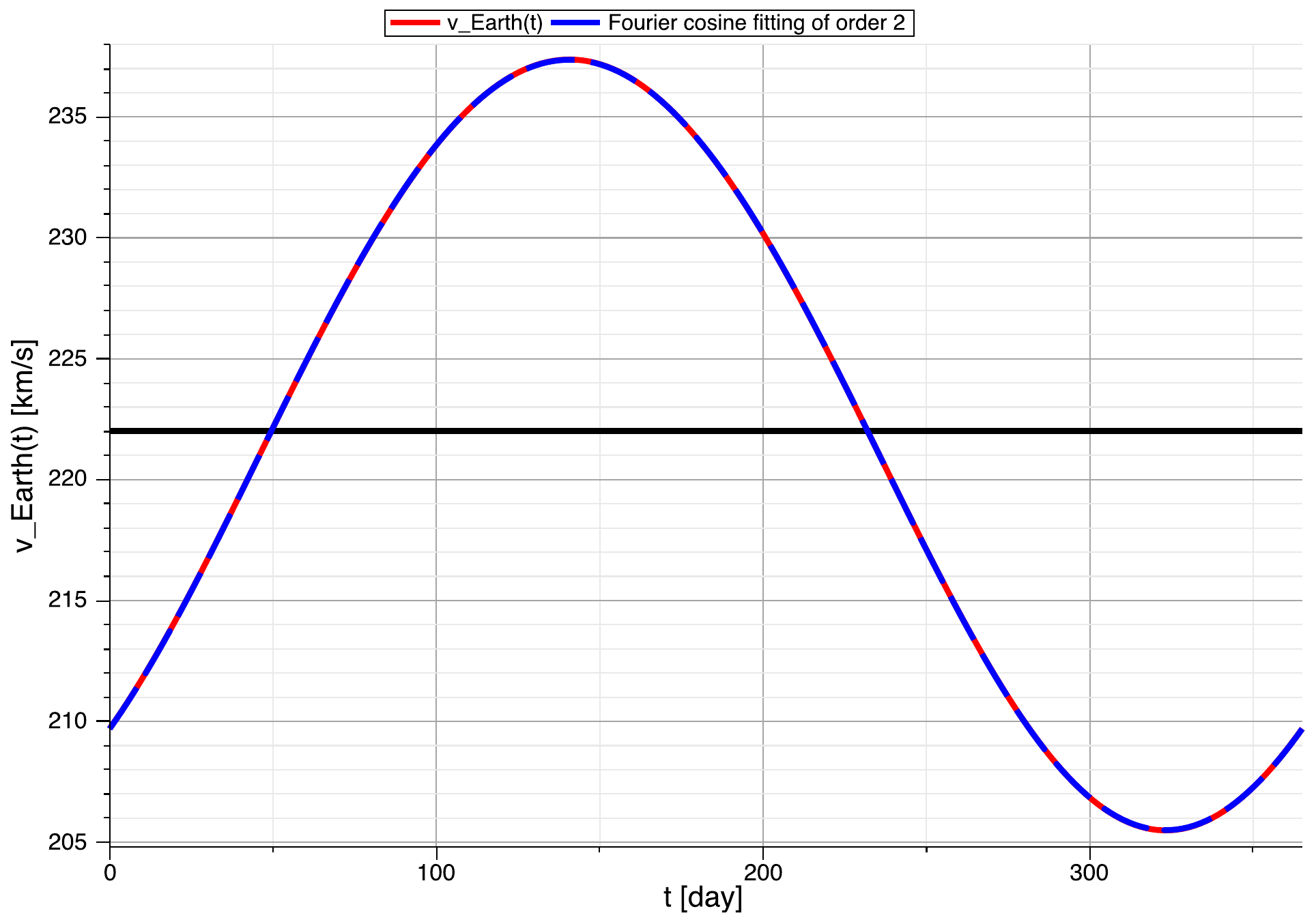}
\end{center}
\caption{
 Fourier cosine fitting of
 the time--dependent Earth's speed
 relative to the Dark Matter halo
 $\vEarthchiS(t)$
 of the order of 1 (upper)
 and 2 (below),
 respectively.
 The date of the maximal Earth's relative velocity is
 $\tmax = 140.74$ day.
}
\label{fig:v_Earth_chi_Sol_Fourier_cos}
\end{figure}

 Furthermore,
 one can in general have that
\beqn
     \VEarthchiS(t)
 \=  \VSunS
   + \VEarthS(t)
     \non\\
 \=  \vSunG
     \left[\begin{array}{c}
             \D
             \cos\thetaCYGNUS \cos\phiCYGNUS
            -\afrac{\vEarthS}{\vSunG}
             \sin\psiyear                                         \\
             \D
             \cos\psiEarth \cos\thetaCYGNUS \sin\phiCYGNUS
            -\sin\psiEarth \sin\thetaCYGNUS
            +\afrac{\vEarthS}{\vSunG}
             \cos\psiyear                                         \\
             \sin\psiEarth \cos\thetaCYGNUS \sin\phiCYGNUS
            +\cos\psiEarth \sin\thetaCYGNUS                       \\
           \end{array}\right]_{\rm S}
     \non\\
 \=  \vSunG
     \left[\begin{array}{c}
             \D
            -0.47069
            -\afrac{\vEarthS}{\vSunG}
             \sin\psiyear                                         \\
             \D
             0.26203
            +\afrac{\vEarthS}{\vSunG}
             \cos\psiyear                                         \\
             0.84249                                              \\
           \end{array}\right]_{\rm S}
\~.
\label{eqn:v_Earth_G}
\eeqn
 Hence,
 we can get that
\beqn
     \vEarthchiS(t)
 \=  \vSunG
     \bbrac{  1
            + \afrac{\vEarthS}{\vSunG}^2
            + 2
              \afrac{\vEarthS}{\vSunG}
              \aBig{  0.47069 \sin\psiyear
                    + 0.26203 \cos\psiyear}  }^{1 / 2}
     \non\\
 \=
     \vSunG
     \bbrac{
              \alpha_{v, 0}
            + \alpha_{v, 1}
              \cos \afrac{        2 \pi (t - \tmax)}{1~{\rm yr}}
            + \alpha_{v, 2}
              \cos \afrac{2 \cdot 2 \pi (t - \tmax)}{1~{\rm yr}}
            + \cdots  }
\~.
\label{eqn:v_Earth_G}
\eeqn
 Here,
 in the last line,
 we expand $\vEarthchiS(t)$
 by the Fourier cosine series
 around the date of the maximal Earth's relative velocity
 $\tmax = 140.74$ day.
 By some numerical calculations,
 one can find easily that
\cheqnXa{A}
\beq
     \alpha_{v, 0}
  =  1.0078
\~,
\label{eqn:alpha_v_0}
\eeq
\cheqnXb{A}
\beq
     \alpha_{v, 1}
  =  0.0724
\~,
\label{eqn:alpha_v_1}
\eeq
 and
\cheqnXc{A}
\beq
     \alpha_{v, 2}
  = -0.0013
\~.
\label{eqn:alpha_v_2}
\eeq
\cheqnX{A}
 Hence,
 the time--dependent Earth's speed
 relative to the Dark Matter halo
 can be expressed as
\beq
     \vEarthchiS(t)
  =  \vSunG
     \bbrac{
              1.0078
            + 0.0724
              \cos \afrac{        2 \pi (t - \tmax)}{1~{\rm yr}}
            - 0.0013
              \cos \afrac{2 \cdot 2 \pi (t - \tmax)}{1~{\rm yr}}  }
\~.
\label{eqn:v_Earth_G_Fourier_cos}
\eeq
 In Figs.~\ref{fig:v_Earth_chi_Sol_Fourier_cos},
 we show
 the fitted $\vEarthchiS(t)$
 of the order of 1 (upper)
 and 2 (below),
 respectively.

\subsubsection{Annual modulation of the Earth's velocity
               in the Equatorial coordinate system}
\label{appx:v_Earth_chi_Eq}

 First,
 by using the transformation matrix
 from the Ecliptic
 to the Equatorial coordinate system
 $\MaSEq$
 given in Eq.~(\ref{eqn:Ma_S_Eq}),
 the Earth's velocities
 relative to the Dark Matter halo
 in four normal seasons
 in the Equatorial coordinate system
 can be calculated directly as
\cheqnXa{A}
\beqn
     \VEarthchiEqT{79.00}
 \=  \MaSEq \~ \VEarthchiST{79.00}
  =  \left[\begin{array}{c}
            -103.55~{\rm km/s} \\
             153.85~{\rm km/s} \\
             135.38~{\rm km/s} \\
           \end{array}\right]_{\rm Eq}
\~,
\label{eqn:V_Earth_chi_Eq_07900}
\eeqn
\cheqnXb{A}
\beqn
     \VEarthchiEqT{170.25}
 \=  \MaSEq \~ \VEarthchiST{170.25}
  =  \left[\begin{array}{c}
            -133.34~{\rm km/s} \\
             126.52~{\rm km/s} \\
             147.21~{\rm km/s} \\
           \end{array}\right]_{\rm Eq}
\~,
\label{eqn:V_Earth_chi_Eq_17025}
\eeqn
\cheqnXc{A}
\beqn
     \VEarthchiEqT{261.50}
 \=  \MaSEq \~ \VEarthchiST{261.50}
  =  \left[\begin{array}{c}
            -103.55~{\rm km/s} \\
              99.18~{\rm km/s} \\
             159.04~{\rm km/s} \\
           \end{array}\right]_{\rm Eq}
\~,
\label{eqn:V_Earth_chi_Eq_26150}
\eeqn
 and
\cheqnXd{A}
\beqn
     \VEarthchiEqT{352.75}
 \=  \MaSEq \~ \VEarthchiST{352.75}
  =  \left[\begin{array}{c}
            - 73.77~{\rm km/s} \\
             126.52~{\rm km/s} \\
             147.21~{\rm km/s} \\
           \end{array}\right]_{\rm Eq}
\~,
\label{eqn:V_Earth_chi_Eq_35275}
\eeqn
\cheqnX{A}%
 respectively.
\begin{table} [t!]
\small
\begin{center}
\renewcommand{\arraystretch}{1.5}
 \begin{tabular}{|| c || c | c | c ||}
\hline
\hline
 \makebox[2.5 cm][c]{Date                           (day)}  &
 \makebox[4   cm][c]{Magnitude       $\vEarthchiEq$ (km/s)} &
 \makebox[4.5 cm][c]{Right ascension $\phichiEq$}           &
 \makebox[3.5 cm][c]{Declination     $\thetachiEq$}         \\
\hline
\hline
  79.0  & 229.61 & 123.94$^{\circ}$ =  8.26$^{\rm h}$ & 36.13$^{\circ}$ \\
 170.25 & 235.49 & 136.50$^{\circ}$ =  9.10$^{\rm h}$ & 38.69$^{\circ}$ \\
 261.50 & 214.13 & 136.24$^{\circ}$ =  9.08$^{\rm h}$ & 47.96$^{\circ}$ \\
 352.75 & 207.65 & 120.24$^{\circ}$ =  8.01$^{\rm h}$ & 45.15$^{\circ}$ \\
\hline
  49.49 & 222.01 & 120.63$^{\circ}$ =  8.04$^{\rm h}$ & 38.06$^{\circ}$ \\
 140.74 & 237.38 & 132.82$^{\circ}$ =  8.85$^{\rm h}$ & 36.58$^{\circ}$ \\
 231.99 & 222.01 & 138.99$^{\circ}$ =  9.27$^{\rm h}$ & 45.21$^{\circ}$ \\
 323.24 & 205.49 & 124.40$^{\circ}$ =  8.29$^{\rm h}$ & 48.11$^{\circ}$ \\
\hline
\hline
\end{tabular}
\end{center}
\caption{
 The list of
 the magnitudes,
 the right ascensions and the declinations of
 the Earth's velocity
 relative to the Dark Matter halo
 in four normal and four advanced seasons
 in the Equatorial coordinate system.
}
\label{tab:vEarthchiEqT}
\end{table}
 Similarly,
 for four advanced seasons,
 one has
\cheqnXa{A}
\beqn
     \VEarthchiEqT{140.74}
 \=  \MaSEq \~ \VEarthchiST{140.74}
  =  \left[\begin{array}{c}
            -129.58~{\rm km/s} \\
             139.81~{\rm km/s} \\
             141.45~{\rm km/s} \\
           \end{array}\right]_{\rm Eq}
\~,
\label{eqn:V_Earth_chi_Eq_14074}
\eeqn
\cheqnXb{A}
\beqn
     \VEarthchiEqT{323.24}
 \=  \MaSEq \~ \VEarthchiST{323.24}
  =  \left[\begin{array}{c}
            - 77.53~{\rm km/s} \\
             113.22~{\rm km/s} \\
             152.96~{\rm km/s} \\
           \end{array}\right]_{\rm Eq}
\~,
\label{eqn:V_Earth_chi_Eq_32324}
\eeqn
\cheqnXc{A}
\beqn
     \VEarthchiEqT{49.49}
 \=  \MaSEq \~ \VEarthchiST{49.49}
  =  \left[\begin{array}{c}
            - 89.06~{\rm km/s} \\
             150.40~{\rm km/s} \\
             136.87~{\rm km/s} \\
           \end{array}\right]_{\rm Eq}
\~,
\label{eqn:V_Earth_chi_Eq_04949}
\eeqn
 and
\cheqnXd{A}
\beqn
     \VEarthchiEqT{231.99}
 \=  \MaSEq \~ \VEarthchiST{231.99}
  =  \left[\begin{array}{c}
            -118.04~{\rm km/s} \\
             102.63~{\rm km/s} \\
             157.54~{\rm km/s} \\
           \end{array}\right]_{\rm Eq}
\~,
\label{eqn:V_Earth_chi_Eq_23199}
\eeqn
\cheqnX{A}%
 respectively.
 Then
 we can obtain the directions
 (the right ascensions and the declinations)
 of the Earth's velocity
 relative to the Dark Matter halo
 in four normal and four advanced seasons
 in the Equatorial coordinate system
 (i.e.,
  the blue--yellow points
  shown in Figs.~\ref{fig:N_phi_theta-Eq-050-07900}
  and \ref{fig:N_phi_theta-Eq-500-07900}
  as well as
  in Figs.~\ref{fig:N_phi_theta-Eq-050-04949}
  and \ref{fig:N_phi_theta-Eq-500-04949})
 summarized in Table \ref{tab:vEarthchiEqT}
 for readers' reference.

\subsubsection{Dates considered for demonstrating
               the diurnal modulation of the angular WIMP velocity distribution}
\label{appx:v_Earth_chi_S-20766}

 At the end of this section,
 we calculate the dates
 considered for demonstrating
 the diurnal modulation of
 the angular WIMP velocity distribution
 shown in Sec.~\ref{sec:N_phi_theta-Lab-050-20766}
 and Appendix \ref{appx:N_phi_theta-ULabs}.

 From Fig.~\ref{fig:E-Eq-S-transformation}
 and Eq.~(\ref{eqn:X_E}),
 one can have
\beqn
     \xE
 \=  \cos\psiPM \~ \xEq
   + \sin\psiPM \~ \yEq
     \non\\
 \=  \gamma \cos\psiyear               \~ \xEq
   + \gamma \sin\psiyear \cos\psiEarth \~ \yEq
\~.
\eeqn
 Then,
 by taking $\psiPM = \phiCYGNUS = 129.30^{\circ}$
 given in Eq.~(\ref{eqn:phi_CYGNUS_Eq}),
 the specified angle $\psiyear$ can be solved directly as
\beq
     \psiyear
  =  \tan^{-1} \afrac{\tan\psiPM}{\cos\psiEarth}
  =  126.9^{\circ}
\~.
\eeq
 This means in turn that
 the date
 on which
 the WIMP wind points straightly to the Prime Meridian
 in the night is%
\footnote{
 Remind that
 these dates are only pure theoretical estimations
 under some simplified assumptions.
 It is obviously that
 the Prime Meridian should be in the day (night)
 on the 207.66 (25.16) day,
 respectively.
}
\beq
     126.9^{\circ} \afrac{365~{\rm day}}{360^{\circ}} + 79.0~{\rm day}
  =  207.66~{\rm day}
\~.
\eeq
 And the date
 on which
 the WIMP wind points straightly to the Prime Meridian in the day
 is 25.16 day.

%

%
%
% Appendix B
%
% 8/10
 %
%
\section{Angular distributions of the 3-D WIMP velocity
         observed at the known underground laboratories}
\label{appx:N_phi_theta-ULabs}

 For readers' reference,
 we summarize
 in this section
 the angular distributions of the 3-D WIMP velocity
 in two laboratory--dependent
 (horizontal and laboratory)
 coordinate systems
 at the locations of several underground laboratories.
 Due to their geographical advantages,
 two under--constructed laboratories
 have also been considered:
 the Agua Negra Deep Experiment Site (ANDES)
 on the border between Argentina and Chile
 as the second underground laboratory
 in the Southern Hemisphere
 and
 the Callio Laboratory located
 at the Pyh\"asalmi Mine in Finnland,
 which will be
 the northernmost (close to the Arctic Circle) underground laboratory
 in the future
 (the Boulby Laboratory is so far the northernmost one).

 For each laboratory,
 we provide the following figures:
\begin{itemize}
\item
 the angular distributions of the 3-D WIMP velocity
 in the horizontal coordinate system
 in one entire year
 and in the 60-day observation periods
 of four advanced seasons on the central date of
  49.49 day,
 140.74 day,
 231.99 day,
 and
 323.24 day,
 respectively;
\item
 the angular distributions of the 3-D WIMP velocity
 in the laboratory coordinate system
 in one entire year
 and in the 60-day observation periods
 of four advanced seasons;
\item
 the angular distributions of the 3-D WIMP velocity
 in the laboratory coordinate system
 in four 4-hour daily shifts
 in two 60-day observation periods
 with the central dates
 of 207.66 day and 25.16 (= 390.16) day.
\end{itemize}

 Remind that,
 firstly,
 50 and 500 total WIMP events on average
 in (one daily shift of) one observation period
 (365 days/year,
  60 days/season,
  or 4 hours/shift $\times$ 60 days)
 have been simulated.
 Secondly,
 in the simulations of
 the diurnal modulation of
 the angular WIMP velocity distribution
 presented here,
 we have re--calculated the local time
 of each simulated event
 of the considered laboratory
 from the generated measuring UTC time.
 Moreover,
 by comparing the angular distribution patterns
 observed at different laboratories,
 it can be found that,
 with $\cal O$(500) total WIMP events
 and a higher analysis resolution,
 we might even be able to demonstrate
 a more detailed latitude--dependent distribution pattern.

 For the identification of
 the annual modulation of
 the angular distribution pattern
 with real experimental data
 in the future,
 as the first confirmation of
 the directionality of the WIMP wind,
 we also summarize
 the directions of the simulated 3-D WIMP velocity
 with the highest event numbers
 ($>$ 4 times of the all--sky average value)
 in the {\em horizontal} coordinate system
 in one entire year
 and in four advanced seasons
 in one table for each laboratory.
 Note that
 in these tables
 we mark the ranges with a double--underline
 to indicate that,
 with $\cal O$(500) total WIMP events
 and a higher analysis resolution,
 some extra bins would exceed the high--WIMP--flux areas
 identified with only $\cal O$(50) total events
 and a lower resolution.

 \def \PeriodA     {\PeriodAa}
 \def \Perioda     {\PeriodCa}
 \def \Periodb     {\PeriodCb}
 \def \Periodc     {\PeriodCc}
 \def \Periodd     {\PeriodCd}
 \def \PlotNumberA {\PlotNumberAa}
 \def \PlotNumbera {\PlotNumberCa}
 \def \PlotNumberb {\PlotNumberCb}
 \def \PlotNumberc {\PlotNumberCc}
 \def \PlotNumberd {\PlotNumberCd}
\newpage
\subsection{Agua Negra Deep Experiment Site (ANDES)}
\label{appx:N_phi_theta-ANDES}
 \def \LabName     {ANDES}
 \def \LabLocation {(30.19$^{\circ}$S, 69.82$^{\circ}$W)}
 \def \DirectionAaTens
  {\vspace{1.2 ex} --- }
 \def \DirectionCaTens
  {  0$^{\circ}$  --  60$^{\circ}$N   \\ \vspace{0.75 ex}
   120$^{\circ}$W --  60$^{\circ}$W}
 \def \DirectionCbTens
  {  0$^{\circ}$  --  60$^{\circ}$N   \\ \vspace{0.75 ex}
    60$^{\circ}$E -- 180$^{\circ}$ }
 \def \DirectionCcTens
  { 30$^{\circ}$S --  30$^{\circ}$N   \\ \vspace{0.75 ex}
    60$^{\circ}$E -- 180$^{\circ}$ }
 \def \DirectionCdTens
  { 30$^{\circ}$S --  60$^{\circ}$N   \\ \vspace{0.75 ex}
   180$^{\circ}$  -- 120$^{\circ}$W}
 \def \DirectionAaHundreds
  {\vspace{1.2 ex} --- }
 \def \DirectionCaHundreds
  {  0$^{\circ}$  --  60$^{\circ}$N,
   120$^{\circ}$W --  60$^{\circ}$W   \\ \vspace{0.75 ex}
   \Dunderline{60$^{\circ}$N --  75$^{\circ}$N,
   120$^{\circ}$E -- 180$^{\circ}$}}
 \def \DirectionCbHundreds
  {  0$^{\circ}$  --  60$^{\circ}$N   \\ \vspace{0.75 ex}
    60$^{\circ}$E -- 150$^{\circ}$E}
 \def \DirectionCcHundreds
  { 30$^{\circ}$S --  30$^{\circ}$N   \\ \vspace{0.75 ex}
    90$^{\circ}$E -- \Dunderline{150$^{\circ}$W}}
 \def \DirectionCdHundreds
  {\Dunderline{45$^{\circ}$S} --  60$^{\circ}$N   \\ \vspace{0.75 ex}
   180$^{\circ}$  -- 120$^{\circ}$W}
 \InsertResultsTableNphithetaLab

\noindent
 Remark:
 in Fig.~\ref{fig:N_phi_theta-H-500-04949-ANDES}(b)
 one can find that,
 with $\cal O$(500) total WIMP events
 and a higher analysis resolution,
 a second high--WIMP--flux bin (``hot--point'')
 from 60$^{\circ}$N to 75$^{\circ}$N
 and 120$^{\circ}$E to 180$^{\circ}$
 in the angular distribution pattern
 in the horizontal coordinate system
 would be observed at the ANDES laboratory.

 \InsertPlotNphithetaULab
\newpage
\subsection{Boulby Laboratory}
\label{appx:N_phi_theta-Boulby}
 \def \LabName     {Boulby}
 \def \LabLocation {(54.55$^{\circ}$N, 0.82$^{\circ}$W)}
 \def \DirectionAaTens
  {\vspace{1.2 ex} --- }
 \def \DirectionCaTens
  { 30$^{\circ}$S --  30$^{\circ}$N   \\ \vspace{0.75 ex}
   120$^{\circ}$E -- 180$^{\circ}$ }
 \def \DirectionCbTens
  { 60$^{\circ}$S --   0$^{\circ}$    \\ \vspace{0.75 ex}
    60$^{\circ}$E -- 120$^{\circ}$E}
 \def \DirectionCcTens
  { 60$^{\circ}$S --  30$^{\circ}$S   \\ \vspace{0.75 ex}
   180$^{\circ}$  -- 120$^{\circ}$W +
     0$^{\circ}$  --  60$^{\circ}$E}
 \def \DirectionCdTens
  { 30$^{\circ}$S --   0$^{\circ}$    \\ \vspace{0.75 ex}
   180$^{\circ}$  --  60$^{\circ}$W}
 \def \DirectionAaHundreds
  {\vspace{1.2 ex} --- }
 \def \DirectionCaHundreds
  {\Dunderline{45$^{\circ}$S} --  30$^{\circ}$N   \\ \vspace{0.75 ex}
   \Dunderline{90$^{\circ}$E} -- \Dunderline{150$^{\circ}$W}}
 \def \DirectionCbHundreds
  { 60$^{\circ}$S -- \Dunderline{15$^{\circ}$N}   \\ \vspace{0.75 ex}
    60$^{\circ}$E -- \Dunderline{150$^{\circ}$E}}
 \def \DirectionCcHundreds
  { 60$^{\circ}$S --  30$^{\circ}$S   \\ \vspace{0.75 ex}
   150$^{\circ}$W -- \Dunderline{90$^{\circ}$W} +
   \Dunderline{30$^{\circ}$W} --  60$^{\circ}$E}
 \def \DirectionCdHundreds
  {\Dunderline{45$^{\circ}$S} --   0$^{\circ}$    \\ \vspace{0.75 ex}
   \Dunderline{150$^{\circ}$E} --  60$^{\circ}$W}
 \InsertResultsTableNphithetaLab

\noindent
 Remark:
 Fig.~\ref{fig:N_phi_theta-H-050-04949-Boulby}(d)
 shows that,
 with only $\cal O$(50) total WIMP events,
 one could already observe
 two separate high--WIMP--flux bins
 in the angular distribution pattern
 in the horizontal coordinate system
 of the Boulby laboratory;
 with $\cal O$(500) total events
 and a higher analysis resolution,
 these two hot--points
 could be identified more clearly
 (see Fig.~\ref{fig:N_phi_theta-H-500-04949-Boulby}(d)).

 \InsertPlotNphithetaULab
\newpage
\subsection{Callio Laboratory}
\label{appx:N_phi_theta-Callio}
 \def \LabName     {Callio}
 \def \LabLocation {(63.66$^{\circ}$N, 26.04$^{\circ}$E)}
 \def \DirectionAaTens
  {\vspace{1.2 ex} --- }
 \def \DirectionCaTens
  { 60$^{\circ}$S --  30$^{\circ}$N   \\ \vspace{0.75 ex}
    60$^{\circ}$E -- 180$^{\circ}$ }
 \def \DirectionCbTens
  { 60$^{\circ}$S --   0$^{\circ}$    \\ \vspace{0.75 ex}
    60$^{\circ}$E -- 120$^{\circ}$E}
 \def \DirectionCcTens
  { 60$^{\circ}$S --  30$^{\circ}$S   \\ \vspace{0.75 ex}
   180$^{\circ}$  --   0$^{\circ}$ (western)}
 \def \DirectionCdTens
  { 30$^{\circ}$S --   0$^{\circ}$    \\ \vspace{0.75 ex}
   120$^{\circ}$E --  60$^{\circ}$W}
 \def \DirectionAaHundreds
  {\vspace{1.2 ex} --- }
 \def \DirectionCaHundreds
  { 60$^{\circ}$S --  30$^{\circ}$N   \\ \vspace{0.75 ex}
    60$^{\circ}$E -- 180$^{\circ}$ }
 \def \DirectionCbHundreds
  { 60$^{\circ}$S --   0$^{\circ}$    \\ \vspace{0.75 ex}
   \Dunderline{30$^{\circ}$E} -- 120$^{\circ}$E}
 \def \DirectionCcHundreds
  { 60$^{\circ}$S --  30$^{\circ}$S,
   150$^{\circ}$W --  90$^{\circ}$W  \\ \vspace{0.75 ex}
    60$^{\circ}$S -- \Dunderline{15$^{\circ}$S},
    60$^{\circ}$W -- \Dunderline{30$^{\circ}$E}}
 \def \DirectionCdHundreds
  {\Dunderline{45$^{\circ}$S} -- \Dunderline{15$^{\circ}$N}   \\ \vspace{0.75 ex}
   150$^{\circ}$E --  90$^{\circ}$W}
 \InsertResultsTableNphithetaLab

\noindent
 Remark:
 By comparing Fig.~\ref{fig:N_phi_theta-H-500-04949-Callio}(d)
 with Fig.~\ref{fig:N_phi_theta-H-050-04949-Callio}(d),
 one can find that,
 with $\cal O$(500) total WIMP events
 and a higher analysis resolution,
 the high--WIMP--flux bins
 in the angular distribution pattern
 in the horizontal coordinate system
 of the Callio laboratory
 could be distinguished into two separate areas.

 \InsertPlotNphithetaULab
\newpage
\subsection{China Jinping Underground Laboratory (CJPL)}
\label{appx:N_phi_theta-CJPL}
 \def \LabName     {CJPL}
 \def \LabLocation {(28.15$^{\circ}$N, 101.71$^{\circ}$E)}
 \def \DirectionAaTens
  {\vspace{1.2 ex} --- }
 \def \DirectionCaTens
  { 60$^{\circ}$S --   0$^{\circ}$    \\ \vspace{0.75 ex}
    60$^{\circ}$E -- 120$^{\circ}$E}
 \def \DirectionCbTens
  { 60$^{\circ}$S --   0$^{\circ}$    \\ \vspace{0.75 ex}
   180$^{\circ}$  -- 120$^{\circ}$W}
 \def \DirectionCcTens
  { 30$^{\circ}$S --  30$^{\circ}$N   \\ \vspace{0.75 ex}
   180$^{\circ}$  --  60$^{\circ}$W}
 \def \DirectionCdTens
  { 30$^{\circ}$S --  30$^{\circ}$N   \\ \vspace{0.75 ex}
   120$^{\circ}$E -- 180$^{\circ}$ }
 \def \DirectionAaHundreds
  {\vspace{1.2 ex} --- }
 \def \DirectionCaHundreds
  {\Dunderline{75$^{\circ}$S} -- \Dunderline{15$^{\circ}$N}   \\ \vspace{0.75 ex}
    60$^{\circ}$E -- \Dunderline{150$^{\circ}$W}}
 \def \DirectionCbHundreds
  { 60$^{\circ}$S --   0$^{\circ}$    \\ \vspace{0.75 ex}
   150$^{\circ}$W -- \Dunderline{90$^{\circ}$W}}
 \def \DirectionCcHundreds
  { 30$^{\circ}$S --  30$^{\circ}$N   \\ \vspace{0.75 ex}
   \Dunderline{150$^{\circ}$E} --  90$^{\circ}$W}
 \def \DirectionCdHundreds
  {\Dunderline{45$^{\circ}$S} -- \Dunderline{45$^{\circ}$N}   \\ \vspace{0.75 ex}
   120$^{\circ}$E -- \Dunderline{150$^{\circ}$W}}
 \InsertResultsTableNphithetaLab

\noindent
 Remark:
 Fig.~\ref{fig:N_phi_theta-H-500-04949-CJPL}(b)
 shows that,
 with $\cal O$(500) total WIMP events
 and a higher analysis resolution,
 the high--WIMP--flux bins
 in the angular distribution pattern
 in the horizontal coordinate system
 of the CJPL laboratory
 could spread towards the southeast pretty widely.

 \InsertPlotNphithetaULab
\newpage
\subsection{Deep Underground Science and Engineering Laboratory (DUSEL)}
\label{appx:N_phi_theta-DUSEL}
 \def \LabName     {DUSEL}
 \def \LabLocation {(44.35$^{\circ}$N, 103.75$^{\circ}$W)}
 \def \DirectionAaTens
  {\vspace{1.2 ex} --- }
 \def \DirectionCaTens
  { 30$^{\circ}$S --  30$^{\circ}$N   \\ \vspace{0.75 ex}
   180$^{\circ}$  --  60$^{\circ}$W}
 \def \DirectionCbTens
  { 30$^{\circ}$S --  30$^{\circ}$N   \\ \vspace{0.75 ex}
   120$^{\circ}$E -- 120$^{\circ}$W}
 \def \DirectionCcTens
  { 60$^{\circ}$S --   0$^{\circ}$    \\ \vspace{0.75 ex}
    60$^{\circ}$E -- 180$^{\circ}$ }
 \def \DirectionCdTens
  { 60$^{\circ}$S --  30$^{\circ}$S   \\ \vspace{0.75 ex}
   180$^{\circ}$  -- 120$^{\circ}$W}
 \def \DirectionAaHundreds
  { 15$^{\circ}$S --   0$^{\circ}$    \\ \vspace{0.75 ex}
   150$^{\circ}$E -- 150$^{\circ}$W}
 \def \DirectionCaHundreds
  {\Dunderline{45$^{\circ}$S} --  15$^{\circ}$N   \\ \vspace{0.75 ex}
   180$^{\circ}$  --  60$^{\circ}$W}
 \def \DirectionCbHundreds
  { 30$^{\circ}$S -- \Dunderline{45$^{\circ}$N}   \\ \vspace{0.75 ex}
   120$^{\circ}$E -- 120$^{\circ}$W}
 \def \DirectionCcHundreds
  {\Dunderline{75$^{\circ}$S} -- \Dunderline{30$^{\circ}$N}   \\ \vspace{0.75 ex}
    90$^{\circ}$E -- 150$^{\circ}$E}
 \def \DirectionCdHundreds
  { 60$^{\circ}$S -- \Dunderline{15$^{\circ}$S}   \\ \vspace{0.75 ex}
   150$^{\circ}$W -- 120$^{\circ}$W}
 \InsertResultsTableNphithetaLab

\noindent
 Remark:
 although
 due to the Earth's orbital motion around the Sun
 and thus the horizontal coordinate system
 rotates daily,
 by using data
 recorded in one entire year,
 the angular distribution of the 3-D WIMP velocity
 spreads out latitudinally
 and the anisotropy
 of the incident direction of Galactic WIMPs
 of the main direction of the WIMP wind
 would be in principle averaged out,
 in Fig.~\ref{fig:N_phi_theta-H-500-04949-DUSEL}(a)
 one can (unexpectedly) find that,
 with $\cal O$(500) total WIMP events
 and a higher analysis resolution,
 clear high--WIMP--flux bins
 from 15$^{\circ}$S to 0$^{\circ}$
 and 150$^{\circ}$E to 150$^{\circ}$W
 could be identified
 in the angular distribution pattern
 in the horizontal coordinate system
 of the DUSEL laboratory.

 \InsertPlotNphithetaULab
\newpage
\subsection{Kamioka Observatory}
\label{appx:N_phi_theta-Kamioka}
 \def \LabName     {Kamioka}
 \def \LabLocation {(36.43$^{\circ}$N, 137.31$^{\circ}$E)}
 \def \DirectionAaTens
  {\vspace{1.2 ex} --- }
 \def \DirectionCaTens
  { 60$^{\circ}$S --  30$^{\circ}$S   \\ \vspace{0.75 ex}
    60$^{\circ}$E -- 120$^{\circ}$E}
 \def \DirectionCbTens
  { 60$^{\circ}$S --   0$^{\circ}$    \\ \vspace{0.75 ex}
   180$^{\circ}$  --  60$^{\circ}$W}
 \def \DirectionCcTens
  { 30$^{\circ}$S --  30$^{\circ}$N   \\ \vspace{0.75 ex}
   120$^{\circ}$E -- 120$^{\circ}$W}
 \def \DirectionCdTens
  { 60$^{\circ}$S --  30$^{\circ}$N   \\ \vspace{0.75 ex}
   120$^{\circ}$E -- 180$^{\circ}$ }
 \def \DirectionAaHundreds
  {\vspace{1.2 ex} --- }
 \def \DirectionCaHundreds
  {\Dunderline{75$^{\circ}$S --  30$^{\circ}$S,
   150$^{\circ}$W --  90$^{\circ}$W}  \\ \vspace{0.75 ex}
    60$^{\circ}$S -- \Dunderline{15$^{\circ}$S},
   \Dunderline{30$^{\circ}$E} --  90$^{\circ}$E}
 \def \DirectionCbHundreds
  { 60$^{\circ}$S --   0$^{\circ}$    \\ \vspace{0.75 ex}
   180$^{\circ}$  -- \Dunderline{30$^{\circ}$W}}
 \def \DirectionCcHundreds
  { 30$^{\circ}$S --  30$^{\circ}$N   \\ \vspace{0.75 ex}
   150$^{\circ}$E -- \Dunderline{90$^{\circ}$W}}
 \def \DirectionCdHundreds
  { 60$^{\circ}$S -- \Dunderline{45$^{\circ}$N}   \\ \vspace{0.75 ex}
   120$^{\circ}$E -- 180$^{\circ}$ }
 \InsertResultsTableNphithetaLab

\noindent
 Remark:
 as at the ANDES laboratory,
 in Fig.~\ref{fig:N_phi_theta-H-500-04949-Kamioka}(b)
 one can find clearly that,
 with $\cal O$(500) total WIMP events
 and a higher analysis resolution,
 a second ``hot--point''
 from 75$^{\circ}$S to 30$^{\circ}$S,
 and 150$^{\circ}$W to 90$^{\circ}$W
 in the angular distribution pattern
 in the horizontal coordinate system
 would also be observed
 at the Kamioka laboratory.

 \InsertPlotNphithetaULab
\newpage
\subsection{Laboratori Nazionali del Gran Sasso (LNGS)}
\label{appx:N_phi_theta-LNGS}
 \def \LabName     {LNGS}
 \def \LabLocation {(42.45$^{\circ}$N, 13.58$^{\circ}$E)}
 \def \DirectionAaTens
  {\vspace{1.2 ex} --- }
 \def \DirectionCaTens
  { 30$^{\circ}$S --  30$^{\circ}$N   \\ \vspace{0.75 ex}
   120$^{\circ}$E -- 180$^{\circ}$ }
 \def \DirectionCbTens
  { 60$^{\circ}$S --   0$^{\circ}$    \\ \vspace{0.75 ex}
    60$^{\circ}$E -- 120$^{\circ}$E}
 \def \DirectionCcTens
  { 60$^{\circ}$S --  30$^{\circ}$S   \\ \vspace{0.75 ex}
   180$^{\circ}$  -- 120$^{\circ}$W}
 \def \DirectionCdTens
  { 30$^{\circ}$S --  30$^{\circ}$N   \\ \vspace{0.75 ex}
   120$^{\circ}$E --  60$^{\circ}$W}
 \def \DirectionAaHundreds
  {\vspace{1.2 ex} --- }
 \def \DirectionCaHundreds
  {\Dunderline{45$^{\circ}$S} -- \Dunderline{45$^{\circ}$N}   \\ \vspace{0.75 ex}
   \Dunderline{90$^{\circ}$E} -- 180$^{\circ}$ }
 \def \DirectionCbHundreds
  { 60$^{\circ}$S -- \Dunderline{15$^{\circ}$N}   \\ \vspace{0.75 ex}
    60$^{\circ}$E -- 120$^{\circ}$E}
 \def \DirectionCcHundreds
  { 60$^{\circ}$S -- \Dunderline{15$^{\circ}$S},
   150$^{\circ}$W -- 120$^{\circ}$W  \\ \vspace{0.75 ex}
   \Dunderline{60$^{\circ}$S --  45$^{\circ}$S,
    30$^{\circ}$W --  30$^{\circ}$E}}
 \def \DirectionCdHundreds
  { 30$^{\circ}$S --  15$^{\circ}$N   \\ \vspace{0.75 ex}
   150$^{\circ}$E --  90$^{\circ}$W}
 \InsertResultsTableNphithetaLab

\noindent
 Remark:
 as at the ANDES and the Kamioka laboratories,
 in Fig.~\ref{fig:N_phi_theta-H-500-04949-LNGS}(d)
 one could also find clearly
 a second ``hot--point''
 from 60$^{\circ}$S to 45$^{\circ}$S,
 and 30$^{\circ}$W to 30$^{\circ}$E
 in the angular distribution pattern
 in the horizontal coordinate system
 of the LNGS laboratory.

 \InsertPlotNphithetaULab
\newpage
\subsection{Laboratoire Souterrain Canfranc (LSC)}
\label{appx:N_phi_theta-LSC}
 \def \LabName     {LSC}
 \def \LabLocation {(42.81$^{\circ}$N, 0.56$^{\circ}$W)}
 \def \DirectionAaTens
  {\vspace{1.2 ex} --- }
 \def \DirectionCaTens
  { 30$^{\circ}$S --  30$^{\circ}$N   \\ \vspace{0.75 ex}
   120$^{\circ}$E -- 180$^{\circ}$ }
 \def \DirectionCbTens
  { 60$^{\circ}$S --   0$^{\circ}$    \\ \vspace{0.75 ex}
    60$^{\circ}$E -- 120$^{\circ}$E}
 \def \DirectionCcTens
  { 60$^{\circ}$S --  30$^{\circ}$S   \\ \vspace{0.75 ex}
   180$^{\circ}$  -- 120$^{\circ}$W}
 \def \DirectionCdTens
  { 30$^{\circ}$S --   0$^{\circ}$    \\ \vspace{0.75 ex}
   180$^{\circ}$  --  60$^{\circ}$W}
 \def \DirectionAaHundreds
  {\vspace{1.2 ex} --- }
 \def \DirectionCaHundreds
  {\Dunderline{45$^{\circ}$S} -- \Dunderline{45$^{\circ}$N}   \\ \vspace{0.75 ex}
   120$^{\circ}$E -- \Dunderline{150$^{\circ}$W}}
 \def \DirectionCbHundreds
  { 60$^{\circ}$S -- \Dunderline{60$^{\circ}$N}   \\ \vspace{0.75 ex}
    90$^{\circ}$E -- \Dunderline{150$^{\circ}$E}}
 \def \DirectionCcHundreds
  { 60$^{\circ}$S -- \Dunderline{15$^{\circ}$S},
   150$^{\circ}$W -- 120$^{\circ}$W  \\ \vspace{0.75 ex}
   \Dunderline{60$^{\circ}$S --  45$^{\circ}$S,
     0$^{\circ}$  --  60$^{\circ}$E}}
 \def \DirectionCdHundreds
  {\Dunderline{45$^{\circ}$S} -- \Dunderline{15$^{\circ}$N}   \\ \vspace{0.75 ex}
   \Dunderline{150$^{\circ}$E} --  60$^{\circ}$W}
 \InsertResultsTableNphithetaLab

\noindent
 Remark:
 as at the ANDES,
 the Kamioka,
 and the LNGS laboratories,
 in Fig.~\ref{fig:N_phi_theta-H-500-04949-LSC}(d)
 one could also find clearly
 a second ``hot--point''
 from 60$^{\circ}$S to 45$^{\circ}$S,
 and 0$^{\circ}$ to 60$^{\circ}$E
 in the angular distribution pattern
 in the horizontal coordinate system
 of the LSC laboratory.

 \InsertPlotNphithetaULab
\newpage
\subsection{Laboratoire Souterrain de Modane (LSM)}
\label{appx:N_phi_theta-LSM}
 \def \LabName     {LSM}
 \def \LabLocation {(45.14$^{\circ}$N, 6.70$^{\circ}$E)}
 \def \DirectionAaTens
  {\vspace{1.2 ex} --- }
 \def \DirectionCaTens
  { 30$^{\circ}$S --  30$^{\circ}$N   \\ \vspace{0.75 ex}
   120$^{\circ}$E -- 180$^{\circ}$ }
 \def \DirectionCbTens
  { 60$^{\circ}$S --   0$^{\circ}$    \\ \vspace{0.75 ex}
    60$^{\circ}$E -- 120$^{\circ}$E}
 \def \DirectionCcTens
  { 60$^{\circ}$S --  30$^{\circ}$S   \\ \vspace{0.75 ex}
   180$^{\circ}$  -- 120$^{\circ}$W}
 \def \DirectionCdTens
  { 30$^{\circ}$S --   0$^{\circ}$    \\ \vspace{0.75 ex}
   120$^{\circ}$E --  60$^{\circ}$W}
 \def \DirectionAaHundreds
  {\vspace{1.2 ex} --- }
 \def \DirectionCaHundreds
  {\Dunderline{45$^{\circ}$S} -- \Dunderline{45$^{\circ}$N}   \\ \vspace{0.75 ex}
   \Dunderline{90$^{\circ}$E} -- \Dunderline{150$^{\circ}$W}}
 \def \DirectionCbHundreds
  {\Dunderline{75$^{\circ}$S} -- \Dunderline{15$^{\circ}$N}   \\ \vspace{0.75 ex}
    60$^{\circ}$E -- 120$^{\circ}$E}
 \def \DirectionCcHundreds
  { 60$^{\circ}$S -- \Dunderline{15$^{\circ}$S},
   150$^{\circ}$W -- 120$^{\circ}$W  \\ \vspace{0.75 ex}
   \Dunderline{60$^{\circ}$S --  45$^{\circ}$S,
    30$^{\circ}$W --  30$^{\circ}$E}}
 \def \DirectionCdHundreds
  { 30$^{\circ}$S -- \Dunderline{15$^{\circ}$N}   \\ \vspace{0.75 ex}
   150$^{\circ}$E --  60$^{\circ}$W}
 \InsertResultsTableNphithetaLab

\noindent
 Remark:
 as at the ANDES,
 the Kamioka,
 the LNGS,
 and the LSC laboratories,
 in Fig.~\ref{fig:N_phi_theta-H-500-04949-LSM}(d)
 one could also find clearly
 a second ``hot--point''
 from 60$^{\circ}$S to 45$^{\circ}$S,
 and 30$^{\circ}$W to 30$^{\circ}$E
 in the angular distribution pattern
 in the horizontal coordinate system
 of the LSM laboratory.

 \InsertPlotNphithetaULab
\newpage
\subsection{Sudbury Neutrino Observatory (SNOLAB)}
\label{appx:N_phi_theta-SNOLAB}
 \def \LabName     {SNOLAB}
 \def \LabLocation {(46.47$^{\circ}$N, 81.19$^{\circ}$W)}
 \def \DirectionAaTens
  {\vspace{1.2 ex} --- }
 \def \DirectionCaTens
  { 30$^{\circ}$S --  30$^{\circ}$N   \\ \vspace{0.75 ex}
   180$^{\circ}$  --  60$^{\circ}$W}
 \def \DirectionCbTens
  { 30$^{\circ}$S --  30$^{\circ}$N   \\ \vspace{0.75 ex}
   120$^{\circ}$E -- 180$^{\circ}$ }
 \def \DirectionCcTens
  { 30$^{\circ}$S --   0$^{\circ}$    \\ \vspace{0.75 ex}
    60$^{\circ}$E -- 120$^{\circ}$E}
 \def \DirectionCdTens
  { 60$^{\circ}$S --   0$^{\circ}$    \\ \vspace{0.75 ex}
   180$^{\circ}$  -- 120$^{\circ}$W}
 \def \DirectionAaHundreds
  { 15$^{\circ}$S --   0$^{\circ}$    \\ \vspace{0.75 ex}
   150$^{\circ}$E -- 150$^{\circ}$W}
 \def \DirectionCaHundreds
  { 30$^{\circ}$S --  15$^{\circ}$N   \\ \vspace{0.75 ex}
   \Dunderline{150$^{\circ}$E} --  90$^{\circ}$W}
 \def \DirectionCbHundreds
  {\Dunderline{45$^{\circ}$S} -- \Dunderline{45$^{\circ}$N}   \\ \vspace{0.75 ex}
   120$^{\circ}$E -- \Dunderline{150$^{\circ}$W}}
 \def \DirectionCcHundreds
  {\Dunderline{60$^{\circ}$S} -- \Dunderline{15$^{\circ}$N}   \\ \vspace{0.75 ex}
    90$^{\circ}$E -- \Dunderline{150$^{\circ}$E}}
 \def \DirectionCdHundreds
  { 60$^{\circ}$S --  15$^{\circ}$S   \\ \vspace{0.75 ex}
   150$^{\circ}$W -- 120$^{\circ}$W}
 \InsertResultsTableNphithetaLab

\noindent
 Remark:
 as at the DUSEL laboratory,
 with $\cal O$(500) total WIMP events
 recorded in one entire year
 and a higher analysis resolution,
 in the angular distribution pattern
 in the horizontal coordinate system
 of the SNOLAB laboratory
 shown in Fig.~\ref{fig:N_phi_theta-H-500-04949-SNOLAB}(a),
 clear high--WIMP--flux bins
 from 15$^{\circ}$S to 0$^{\circ}$
 and 150$^{\circ}$E to 150$^{\circ}$W
 could be identified.

 \InsertPlotNphithetaULab
\newpage
\subsection{Stawell Underground Physics Laboratory (SUPL)}
\label{appx:N_phi_theta-SUPL}
 \def \LabName     {SUPL}
 \def \LabLocation {(37.07$^{\circ}$S, 142.81$^{\circ}$E)}
 \def \DirectionAaTens
  {\vspace{1.2 ex} --- }
 \def \DirectionCaTens
  { 30$^{\circ}$S --  30$^{\circ}$N   \\ \vspace{0.75 ex}
   120$^{\circ}$E -- 120$^{\circ}$W}
 \def \DirectionCbTens
  { 30$^{\circ}$S --  60$^{\circ}$N   \\ \vspace{0.75 ex}
   180$^{\circ}$  -- 120$^{\circ}$W}
 \def \DirectionCcTens
  { 30$^{\circ}$N --  60$^{\circ}$N   \\ \vspace{0.75 ex}
   120$^{\circ}$W --  60$^{\circ}$W +
   120$^{\circ}$E -- 180$^{\circ}$ }
 \def \DirectionCdTens
  {  0$^{\circ}$  --  60$^{\circ}$N   \\ \vspace{0.75 ex}
    60$^{\circ}$E -- 180$^{\circ}$ }
 \def \DirectionAaHundreds
  {\vspace{1.2 ex} --- }
 \def \DirectionCaHundreds
  {\Dunderline{45$^{\circ}$S} --  30$^{\circ}$N   \\ \vspace{0.75 ex}
   120$^{\circ}$E -- 120$^{\circ}$W}
 \def \DirectionCbHundreds
  {\Dunderline{45$^{\circ}$S} --  60$^{\circ}$N   \\ \vspace{0.75 ex}
   150$^{\circ}$W -- \Dunderline{90$^{\circ}$W}}
 \def \DirectionCcHundreds
  {\Dunderline{15$^{\circ}$N} --  60$^{\circ}$N,
   120$^{\circ}$W --  60$^{\circ}$W  \\ \vspace{0.75 ex}
    30$^{\circ}$N --  60$^{\circ}$N,
   120$^{\circ}$E -- 150$^{\circ}$E}
 \def \DirectionCdHundreds
  {\Dunderline{15$^{\circ}$S} --  45$^{\circ}$N   \\ \vspace{0.75 ex}
    60$^{\circ}$E -- 180$^{\circ}$ }
 \InsertResultsTableNphithetaLab

\noindent
 Remark:
 as at the Boulby laboratory,
 with only $\cal O$(50) total WIMP events
 one could already observe
 two separate high--WIMP--flux bins
 in the angular distribution pattern
 in the horizontal coordinate system
 of the SUPL laboratory
 shown in Fig.~\ref{fig:N_phi_theta-H-050-04949-SUPL}(d);
 with $\cal O$(500) total events
 and a higher analysis resolution,
 these two hot--points
 could be identified more clearly
 (see Fig.~\ref{fig:N_phi_theta-H-500-04949-SUPL}(d)).

 \InsertPlotNphithetaULab
 \newpage
%

%

%
%
% Appendix C
%
% 9/10
%
 %
%
% Appendix C
%
\section{Annual modulation of the radial WIMP velocity distribution
         in four normal seasons}
\label{appx:N_v-Bayesian-Eq-07900}

 For the sake of completeness and readers' reference,
 in this section,
 we provide the reconstruction results of
 the annual modulation of
 the radial distribution of
 the 3-D WIMP velocity
 in the Equatorial coordinate system
 in the observation periods of four normal seasons.
 50 and 500 total events on average
 in each 60-day observation period
 have been simulated.

\subsection[With the one--parameter velocity distribution
            $f_{1, \sh, v_0}(v; v_0)$]
           {\boldmath
            With the one--parameter velocity distribution
            $f_{1, \sh, v_0}(v; v_0)$}
\label{sec:N_v-Bayesian-Eq-v0-07900}

 As in Sec.~\ref{sec:N_v-Bayesian-Eq-v0-050-04949},
 we start our reconstruction
 by using the one--parameter velocity distribution
 $f_{1, \sh, v_0}(v; v_0)$
 given by Eq.~(\ref{eqn:f1v_sh_v0})
 with $v_0$ as the fitting parameter
 scanned in the ranges of
 \mbox{160 km/s} $\le v_0 \le$ 270 km/s (50 events) and
 190 km/s $\le v_0 \le$ 240 km/s (500 events)
 as well as
 the constraint that
 $\ve = 1.05 \~ v_0$.

 \def \ShortFrame      {Eq}
 \def \EventNumber     {050}
 \def \Perioda         {\PeriodBa}
 \def \Periodb         {\PeriodBb}
 \def \Periodc         {\PeriodBc}
 \def \Periodd         {\PeriodBd}
 \def \PlotNumbera     {\PlotNumberBa}
 \def \PlotNumberb     {\PlotNumberBb}
 \def \PlotNumberc     {\PlotNumberBc}
 \def \PlotNumberd     {\PlotNumberBd}
 \def \Fittingfv       {v0}
 \def \FittingPara     {v_0}
 \InsertPlotNvBayesianAnnual
  {As in Figs.~\ref{fig:N_v-Bayesian-Eq-v0-050-04949},
   reconstructed with the one--parameter velocity distribution
   $f_{1, \sh, v_0}(v; v_0)$,
   except that
   four 60-day observation periods of the normal seasons
   have been considered.
   }
 \def \EventNumber     {500}
 \InsertPlotNvBayesianAnnual
  {As in Figs.~\ref{fig:N_v-Bayesian-Eq-v0-050-07900},
   except that
   500 total events on average
   in one 60-day observation period
   have been simulated.
   Remind that
   the scanning range of the fitting parameter $v_0$
   is shrunk to between 190 km/s and 240 km/s.
   }
 \def \PlotNumber                    {\PlotNumberBa}
\InsertResultsTableNvBayesianAnnualD
{one--parameter velocity distribution
 $f_{1, \sh, v_0}(v; v_0)$}
{\begin{minipage} {3.25 cm}
  \begin{center}
     ~          \\ \vspace{1.1  ex}
      79.0      \\ \vspace{0.65 ex}
    (\PeriodBa) \\ \vspace{1.1  ex}
  \end{center}
 \end{minipage}             &
 $v_0$ [km/s]               &
 218.3 & $217.2 \~ ^{+11.0}_{-~9.9} \~ (^{+23.1}_{-20.9})$ & [207.3, 228.2] & [196.3, 240.3] \\
 \hline
 \begin{minipage} {3.25 cm}
  \begin{center}
     ~          \\ \vspace{1.1  ex}
     170.25     \\ \vspace{0.65 ex}
    (\PeriodBb) \\ \vspace{1.1  ex}
  \end{center}
 \end{minipage}             &
 $v_0$ [km/s]               &
 220.5 & $219.4 \~ ^{+11.0}_{-~9.9} \~ (^{+23.1}_{-20.9})$ & [209.5, 230.4] & [198.5, 242.5] \\
 \hline
 \begin{minipage} {3.25 cm}
  \begin{center}
     ~          \\ \vspace{1.1  ex}
     261.5      \\ \vspace{0.65 ex}
    (\PeriodBc) \\ \vspace{1.1  ex}
  \end{center}
 \end{minipage}             &
 $v_0$ [km/s]               &
 212.8 & $211.7 \~ ^{+11.0}_{-~9.9} \~ (^{+22.0}_{-20.9})$ & [201.8, 222.7] & [190.8, 233.7] \\
 \hline
 \begin{minipage} {3.25 cm}
  \begin{center}
     ~          \\ \vspace{1.1  ex}
     352.75     \\ \vspace{0.65 ex}
    (\PeriodBd) \\ \vspace{1.1  ex}
  \end{center}
 \end{minipage}             &
 $v_0$ [km/s]               &
 209.5 & $209.5    \pm 11.0         \~ (^{+22.0}_{-20.9})$ & [198.5, 220.5] & [188.6, 231.5] \\
 }
{50}
{\begin{minipage} {3.25 cm}
  \begin{center}
     ~          \\ \vspace{1.1  ex}
      79.0      \\ \vspace{0.65 ex}
    (\PeriodBa) \\ \vspace{1.1  ex}
  \end{center}
 \end{minipage}             &
 $v_0$ [km/s]               &
 217.5 & $217.5 \~ ^{+ 3.0}_{- 3.5} \~ (^{+ 6.5}_{- 7.0})$ & [214.0, 220.5] & [210.5, 224.0] \\
 \hline
 \begin{minipage} {3.25 cm}
  \begin{center}
     ~          \\ \vspace{1.1  ex}
     170.25     \\ \vspace{0.65 ex}
    (\PeriodBb) \\ \vspace{1.1  ex}
  \end{center}
 \end{minipage}             &
 $v_0$ [km/s]               &
 220.0 & $220.0 \~ ^{+ 3.0}_{- 3.5} \~ (^{+ 6.5}_{- 7.0})$ & [216.5, 223.0] & [213.0, 226.5] \\
 \hline
 \begin{minipage} {3.25 cm}
  \begin{center}
     ~          \\ \vspace{1.1  ex}
     261.5      \\ \vspace{0.65 ex}
    (\PeriodBc) \\ \vspace{1.1  ex}
  \end{center}
 \end{minipage}             &
 $v_0$ [km/s]               &
 211.5 & $211.5 \~ ^{+ 3.5}_{- 3.0} \~ (^{+ 7.0}_{- 6.5})$ & [208.5, 215.0] & [205.0, 218.5] \\
 \hline
 \begin{minipage} {3.25 cm}
  \begin{center}
     ~          \\ \vspace{1.1  ex}
     352.75     \\ \vspace{0.65 ex}
    (\PeriodBd) \\ \vspace{1.1  ex}
  \end{center}
 \end{minipage}             &
 $v_0$ [km/s]               &
 209.5 & $209.5 \~ ^{+ 3.0}_{- 3.5} \~ (\pm \~  6.5)$      & [206.0, 212.5] & [203.0, 216.0] \\
 }
{500}
{parameter $v_0$ and}
{normal}

 In Figs.~\ref{fig:N_v-Bayesian-Eq-v0-050-07900}
 and \ref{fig:N_v-Bayesian-Eq-v0-500-07900},
 we show the reconstructed radial distributions of
 the 3-D WIMP velocity
 and the 1(2)$\sigma$ statistical uncertainty bands
 by using $f_{1, \sh, v_0}(v; v_0)$
 as well as
 the distributions of the fitting parameter $v_0$
 in all simulated experiments
 with 50 and 500 total events on average
 in each 60-day observation period of four normal seasons,
 respectively.
 Comparing with our results of the advanced seasons
 shown in Figs.~\ref{fig:N_v-Bayesian-Eq-v0-050-04949}
 and \ref{fig:N_v-Bayesian-Eq-v0-500-04949},
 one can see here,
 as expected,
 an {\em asymmetric} periodic variation of
 the best--fit values of the fitting parameter $v_0$
 (see the summary of
  the reconstructed results
  and the 1(2)$\sigma$ statistical uncertainty ranges
  of the median values
  in Table \ref{tab:N_v-Bayesian-Eq-v0-07900}).

\subsection[With the $v_0$--fixed velocity distribution
            $f_{1, \sh, \ve}(v; \ve)$]
           {\boldmath
            With the $v_0$--fixed velocity distribution
            $f_{1, \sh, \ve}(v; \ve)$}
\label{sec:N_v-Bayesian-Eq-ve-07900}

 As in Sec.~\ref{sec:N_v-Bayesian-Eq-ve-050-04949},
 we now consider the $v_0$--fixed velocity distribution
 $f_{1, \sh, \ve}(v; \ve)$
 given by Eq.~(\ref{eqn:f1v_sh_ve})
 with $\ve$ as the fitting parameter
 scanned in the ranges of
  90 km/s $\le \ve \le$ 330 km/s (50 events) and
 180 km/s $\le \ve \le$ 270 km/s (500 events)
 as well as
 the input condition that
 $v_0 = 220$ km/s.

 \def \EventNumber     {050}
 \def \Fittingfv       {ve}
 \def \FittingPara     {v_e}
 \InsertPlotNvBayesianAnnual
  {As in Figs.~\ref{fig:N_v-Bayesian-Eq-ve-050-04949},
   reconstructed with the $v_0$--fixed velocity distribution
   $f_{1, \sh, \ve}(v; \ve)$,
   except that
   four 60-day observation periods of the normal seasons
   have been considered.
   \vspace{0.5 cm}
   }
 \def \EventNumber     {500}
 \InsertPlotNvBayesianAnnual
  {As in Figs.~\ref{fig:N_v-Bayesian-Eq-ve-050-07900},
   except that
   500 total events on average
   in one 60-day observation period
   have been simulated.
   Remind that
   the scanning range of the fitting parameter $\ve$
   is shrunk to between 180 km/s and 270 km/s.
   }
\InsertResultsTableNvBayesianAnnualD
{$v_0$--fixed velocity distribution
 $f_{1, \sh, \ve}(v; \ve)$}
{\begin{minipage} {3.25 cm}
  \begin{center}
     ~          \\ \vspace{1.1  ex}
      79.0      \\ \vspace{0.65 ex}
    (\PeriodBa) \\ \vspace{1.1  ex}
  \end{center}
 \end{minipage}             &
 $\ve$ [km/s]               &
 231.6 & $231.6    \pm 26.4           \~ (^{+50.4}_{-55.2})$ & [205.2, 258.0] & [176.4, 282.0] \\
 \hline
 \begin{minipage} {3.25 cm}
  \begin{center}
     ~          \\ \vspace{1.1  ex}
     170.25     \\ \vspace{0.65 ex}
    (\PeriodBb) \\ \vspace{1.1  ex}
  \end{center}
 \end{minipage}             &
 $\ve$ [km/s]               &
 238.8 & $238.8 \~ ^{+24.0}_{-28.8} \~ (^{+50.4}_{-55.2})$ & [210.0, 262.8] & [183.6, 289.2] \\
 \hline
 \begin{minipage} {3.25 cm}
  \begin{center}
     ~          \\ \vspace{1.1  ex}
     261.5      \\ \vspace{0.65 ex}
    (\PeriodBc) \\ \vspace{1.1  ex}
  \end{center}
 \end{minipage}             &
 $\ve$ [km/s]               &
 217.2 & $217.2 \~ ^{+26.4}_{-28.8} \~ (^{+50.4}_{-60.0})$ & [188.4, 243.6] & [157.2, 267.6] \\
 \hline
 \begin{minipage} {3.25 cm}
  \begin{center}
     ~          \\ \vspace{1.1  ex}
     352.75     \\ \vspace{0.65 ex}
    (\PeriodBd) \\ \vspace{1.1  ex}
  \end{center}
 \end{minipage}             &
 $\ve$ [km/s]               &
 210.0 & $210.0 \~ ^{+26.4}_{-28.8} \~ (^{+52.8}_{-62.4})$ & [181.2, 236.4] & [147.6, 262.8] \\
 }
{50}
{\begin{minipage} {3.25 cm}
  \begin{center}
     ~          \\ \vspace{1.1  ex}
      79.0      \\ \vspace{0.65 ex}
    (\PeriodBa) \\ \vspace{1.1  ex}
  \end{center}
 \end{minipage}             &
 $\ve$ [km/s]               &
 232.2 & $232.2    \pm 8.1        \~ (^{+16.2}_{-17.1})$ & [224.1, 240.3] & [215.1, 248.4] \\
 \hline
 \begin{minipage} {3.25 cm}
  \begin{center}
     ~          \\ \vspace{1.1  ex}
     170.25     \\ \vspace{0.65 ex}
    (\PeriodBb) \\ \vspace{1.1  ex}
  \end{center}
 \end{minipage}             &
 $\ve$ [km/s]               &
 238.5 & $238.5 \~ ^{+7.2}_{-9.0} \~ (^{+15.3}_{-17.1})$ & [229.5, 245.7] & [221.4, 253.8] \\
 \hline
 \begin{minipage} {3.25 cm}
  \begin{center}
     ~          \\ \vspace{1.1  ex}
     261.5      \\ \vspace{0.65 ex}
    (\PeriodBc) \\ \vspace{1.1  ex}
  \end{center}
 \end{minipage}             &
 $\ve$ [km/s]               &
 216.9 & $216.9    \pm 8.1        \~ (\pm \~ 17.1)$      & [208.8, 225.0] & [199.8, 234.0] \\
 \hline
 \begin{minipage} {3.25 cm}
  \begin{center}
     ~          \\ \vspace{1.1  ex}
     352.75     \\ \vspace{0.65 ex}
    (\PeriodBd) \\ \vspace{1.1  ex}
  \end{center}
 \end{minipage}             &
 $\ve$ [km/s]               &
 210.6 & $210.6 \~ ^{+8.1}_{-9.0} \~ (^{+16.2}_{-18..0})$ & [201.6, 218.7] & [192.6, 226.8] \\
 }
{500}
{parameter $\ve$ and}
{normal}

 In Figs.~\ref{fig:N_v-Bayesian-Eq-ve-050-07900}
 and \ref{fig:N_v-Bayesian-Eq-ve-500-07900},
 we show the reconstructed radial distributions of
 the 3-D WIMP velocity
 and the 1(2)$\sigma$ statistical uncertainty bands
 by using $f_{1, \sh, \ve}(v; \ve)$
 as well as
 the distributions of the fitting parameter $\ve$
 in all simulated experiments
 with 50 and 500 total events on average
 in each 60-day observation period of four normal seasons,
 respectively.
 As in Sec.~\ref{sec:N_v-Bayesian-Eq-v0-07900},
 by comparing with our results of the advanced seasons
 shown in Figs.~\ref{fig:N_v-Bayesian-Eq-ve-050-04949}
 and \ref{fig:N_v-Bayesian-Eq-ve-500-04949},
 we can also find here
 an asymmetric periodic variation of
 the best--fit values of the fitting parameter $\ve$
 (see the summary of
  the reconstructed results
  and the 1(2)$\sigma$ statistical uncertainty ranges
  of the median values
  in Table \ref{tab:N_v-Bayesian-Eq-ve-07900}).

\subsection[With the simplified velocity distribution
            $f_{1, \sh}(v; v_0, \ve)$]
           {\boldmath
            With the simplified velocity distribution
            $f_{1, \sh}(v; v_0, \ve)$}
\label{sec:N_v-Bayesian-Eq-sh-07900}

 As in Sec.~\ref{sec:N_v-Bayesian-Eq-sh-050-04949},
 we release now the constraints on $v_0$ and $\ve$
 and consider the simplified velocity distribution
 $f_{1, \sh}(v; v_0, \ve)$
 given by Eq.~(\ref{eqn:f1v_sh})
 with $v_0$ and $\ve$ as two free fitting parameters
 scanned in the ranges of
  80 km/s $\le v_0 \le$ 340 km/s (50 events) and
 140 km/s $\le v_0 \le$ 240 km/s (500 events)
 as well as
   0      $\le \ve \le$ 380 km/s (50 events) and
 200 km/s $\le \ve \le$ 310 km/s (500 events),
 respectively.

 \def \EventNumber     {050}
 \def \Fittingfv       {sh}
 \def \FittingPara     {v_0-v_e}
 \InsertPlotNvBayesianAnnual
  {As in Figs.~\ref{fig:N_v-Bayesian-Eq-sh-050-04949},
   reconstructed with the simplified velocity distribution
   $f_{1, \sh}(v; v_0, \ve)$,
   except that
   four 60-day observation periods of the normal seasons
   have been considered.
   \vspace{0.5 cm}
   }
 \def \EventNumber     {500}
 \InsertPlotNvBayesianAnnual
  {As in Figs.~\ref{fig:N_v-Bayesian-Eq-sh-050-07900},
   except that
   500 total events on average
   in one 60-day observation period
   have been simulated.
   Remind that
   the scanning ranges of the fitting parameters $v_0$ and $\ve$
   are shrunk to between 140 km/s and 240 km/s
   and between 200 km/s and 310 km/s,
   respectively.
   }
\InsertResultsTableNvBayesianAnnualD
{simplified velocity distribution
 $f_{1, \sh}(v; v_0, \ve)$}
{\multirow {2} {*}
  {\begin{minipage} {3.25 cm}
    \begin{center}
%      ~            \\ \vspace{1.1  ex}
        79.0        \\ \vspace{0.65 ex}
      (\PeriodBa) % \\ \vspace{1.1  ex}
    \end{center}
   \end{minipage}}          &
 $v_0$ [km/s]               &
 191.8 & $186.6 \~ ^{+33.8}_{-23.4} \~ (^{+ 98.8}_{- 46.8})$ & [163.2, 220.4] & [139.8, 285.4] \\
 \cline{2-6}
                            &
 $\ve$ [km/s]               &
 262.2 & $266.0 \~ ^{+26.6}_{-41.8} \~ (^{+~53.2}_{-239.4})$ & [224.2, 292.6] &  [26.6, 319.2] \\
 \hline
 \multirow {2} {*}
  {\begin{minipage} {3.25 cm}
    \begin{center}
%      ~            \\ \vspace{1.1  ex}
       170.25       \\ \vspace{0.65 ex}
      (\PeriodBb) % \\ \vspace{1.1  ex}
    \end{center}
   \end{minipage}}          &
 $v_0$ [km/s]               &
 189.2 & $186.6 \~ ^{+33.8}_{-23.4} \~ (^{+101.4}_{-~46.8})$ & [163.2, 220.4] & [139.8, 288.0] \\
 \cline{2-6}
                            &
 $\ve$ [km/s]               &
 269.8 & $269.8 \~ ^{+26.6}_{-41.8} \~ (^{+~53.2}_{-239.4})$ & [228.0, 296.4] &  [30.4, 323.0] \\
 \hline
 \multirow {2} {*}
  {\begin{minipage} {3.25 cm}
    \begin{center}
%      ~            \\ \vspace{1.1  ex}
       261.5        \\ \vspace{0.65 ex}
      (\PeriodBc) % \\ \vspace{1.1  ex}
    \end{center}
   \end{minipage}}          &
 $v_0$ [km/s]               &
 191.8 & $186.6 \~ ^{+33.8}_{-26.0} \~ (^{+ 96.2}_{- 49.4})$ & [160.6, 220.4] & [137.2, 282.8] \\
 \cline{2-6}
                            &
 $\ve$ [km/s]               &
 250.8 & $254.6 \~ ^{+30.4}_{-45.6} \~ (^{+~53.2}_{-231.8})$ & [209.0, 285.0] &  [22.8, 307.8] \\
 \hline
 \multirow {2} {*}
  {\begin{minipage} {3.25 cm}
    \begin{center}
%      ~            \\ \vspace{1.1  ex}
       352.75       \\ \vspace{0.65 ex}
      (\PeriodBd) % \\ \vspace{1.1  ex}
    \end{center}
   \end{minipage}}          &
 $v_0$ [km/s]               &
 189.2 & $186.6 \~ ^{+39.0}_{-26.0} \~ (^{+ 96.2}_{- 49.4})$ & [160.6, 225.6] & [137.2, 282.8] \\
 \cline{2-6}
                            &
 $\ve$ [km/s]               &
 247.0 & $250.8 \~ ^{+26.6}_{-53.2} \~ (^{+~49.4}_{-231.8})$ & [197.6, 277.4] &  [19.0, 300.2] \\
 }
{50}
{\multirow {2} {*}
  {\begin{minipage} {3.25 cm}
    \begin{center}
%      ~            \\ \vspace{1.1  ex}
        79.0        \\ \vspace{0.65 ex}
      (\PeriodBa) % \\ \vspace{1.1  ex}
    \end{center}
   \end{minipage}}          &
 $v_0$ [km/s]               &
 185.0 & $185.0 \~ ^{+ 9.0}_{- 8.0} \~ (^{+19.3}_{-16.0})$ & [177.0, 194.0] & [169.0, 204.3] \\
 \cline{2-6}
                            &
 $\ve$ [km/s]               &
 269.3 & $269.3 \~ ^{+~8.8}_{-11.0} \~ (^{+17.6}_{-23.1})$ & [258.3, 278.1] & [246.2, 286.9] \\
 \hline
 \multirow {2} {*}
  {\begin{minipage} {3.25 cm}
    \begin{center}
%      ~            \\ \vspace{1.1  ex}
       170.25       \\ \vspace{0.65 ex}
      (\PeriodBb) % \\ \vspace{1.1  ex}
    \end{center}
   \end{minipage}}          &
 $v_0$ [km/s]               &
 186.0 & $186.0 \~ ^{+ 8.0}_{- 9.0} \~ (^{+18.0}_{-16.0})$ & [177.0, 194.0] & [170.0, 204.0] \\
 \cline{2-6}
                            &
 $\ve$ [km/s]               &
 272.6 & $273.7 \~ ^{+ 8.8}_{- 9.9} \~ (^{+16.8}_{-22.0})$ & [263.8, 282.5] & [251.7, 290.5] \\
 \hline
 \multirow {2} {*}
  {\begin{minipage} {3.25 cm}
    \begin{center}
%      ~            \\ \vspace{1.1  ex}
       261.5        \\ \vspace{0.65 ex}
      (\PeriodBc) % \\ \vspace{1.1  ex}
    \end{center}
   \end{minipage}}          &
 $v_0$ [km/s]               &
 185.0 & $185.0    \pm  9.0         \~ (^{+19.0}_{-17.0})$ & [176.0, 194.0] & [168.0, 204.0] \\
 \cline{2-6}
                            &
 $\ve$ [km/s]               &
 257.2 & $257.2 \~ ^{+~9.9}_{-11.0} \~ (^{+18.7}_{-24.2})$ & [246.2, 267.1] & [233.0, 275.9] \\
 \hline
 \multirow {2} {*}
  {\begin{minipage} {3.25 cm}
    \begin{center}
%      ~            \\ \vspace{1.1  ex}
       352.75       \\ \vspace{0.65 ex}
      (\PeriodBd) % \\ \vspace{1.1  ex}
    \end{center}
   \end{minipage}}          &
 $v_0$ [km/s]               &
 186.0 & $186.0    \pm  9.0         \~ (^{+21.0}_{-17.0})$ & [177.0, 195.0] & [169.0, 207.0] \\
 \cline{2-6}
                            &
 $\ve$ [km/s]               &
 250.6 & $250.6 \~ ^{+~9.9}_{-11.0} \~ (^{+19.8}_{-25.3})$ & [239.6, 260.5] & [225.3, 270.4] \\
 }
{500}
{parameters $v_0$ and $\ve$ as well as}
{normal}

 In Figs.~\ref{fig:N_v-Bayesian-Eq-sh-050-07900}
 and \ref{fig:N_v-Bayesian-Eq-sh-500-07900},
 we show the reconstructed radial distributions of
 the 3-D WIMP velocity
 and the 1(2)$\sigma$ statistical uncertainty bands
 by using $f_{1, \sh}(v; v_0, \ve)$
 as well as
 the distributions of the fitting parameters $v_0$ and $\ve$
 in all simulated experiments on the $v_0 - \ve$ plane
 with 50 and 500 total events on average
 in each 60-day observation period of four normal seasons,
 respectively.
 Besides that
 the asymmetric periodic variation of
 the best--fit values of the fitting parameter $\ve$
 can be seen clearly here,
 the {\em invariability} of the best--fit values of $v_0$
 as its annual average value
 can also be confirmed
 (see the summary of
  the reconstructed results
  and the 1(2)$\sigma$ statistical uncertainty ranges
  of the median values
  in Table \ref{tab:N_v-Bayesian-Eq-sh-07900}).

\subsection[With the modified velocity distribution
            $f_{1, \sh, \Delta v}(v; v_0, \Delta v)$]
           {\boldmath
            With the modified velocity distribution
            $f_{1, \sh, \Delta v}(v; v_0, \Delta v)$}
\label{sec:N_v-Bayesian-Eq-Dv-07900}

 Finally,
 as in Sec.~\ref{sec:N_v-Bayesian-Eq-Dv-050-04949},
 we consider the modified velocity distribution
 $f_{1, \sh, \Delta v}(v; v_0, \Delta v)$
 given by Eq.~(\ref{eqn:f1v_sh_Dv})
 with $v_0$ and $\Delta v$ as two free fitting parameters
 scanned in the ranges of
  80 km/s $\le v_0 \le$ 340 km/s (50 events) and
 140 km/s $\le v_0 \le$ 240 km/s (500 events)
 as well as
 $-190$ km/s $\le \Delta v \le$ 230 km/s (50 events) and
  $-20$ km/s $\le \Delta v \le$ 150 km/s (500 events),
 respectively.

 \def \EventNumber     {050}
 \def \Fittingfv       {Dv}
 \def \FittingPara     {v_0-D_v}
 \InsertPlotNvBayesianAnnual
  {As in Figs.~\ref{fig:N_v-Bayesian-Eq-Dv-050-04949},
   reconstructed with the modified velocity distribution
   $f_{1, \sh, \Delta v}(v; v_0, \Delta v)$,
   except that
   four 60-day observation periods of the normal seasons
   have been considered.
   }
 \def \EventNumber     {500}
 \InsertPlotNvBayesianAnnual
  {As in Figs.~\ref{fig:N_v-Bayesian-Eq-Dv-050-07900},
   except that
   500 total events on average
   in one 60-day observation period
   have been simulated.
   Remind that
   the scanning ranges of the fitting parameters $v_0$ and $\ve$
   are shrunk to between 140 km/s and 240 km/s
   and between $-20$ km/s and 150 km/s,
   respectively.
   }
\InsertResultsTableNvBayesianAnnualD
{modified velocity distribution
 $f_{1, \sh, \Delta v}(v; v_0, \Delta v)$}
{\multirow {2} {*}
  {\begin{minipage} {3.25 cm}
    \begin{center}
%      ~            \\ \vspace{1.1  ex}
        79.0        \\ \vspace{0.65 ex}
      (\PeriodBa) % \\ \vspace{1.1  ex}
    \end{center}
   \end{minipage}}          &
 $v_0$ [km/s]               &
 191.8 & $186.6 \~ ^{+36.4}_{-26.0} \~ (^{+ 88.4}_{- 46.8})$ & [163.2, 220.4] & [139.8, 278.3] \\
 \cline{2-6}
                            &
 $\Delta v$ [km/s]          &
  70.4 & $ 83.0 \~ ^{+46.2}_{-75.6} \~ (^{+~84.0}_{-273.0})$ &   [7.4, 129.2] & [$-190.0$, 167.0] \\
 \hline
 \multirow {2} {*}
  {\begin{minipage} {3.25 cm}
    \begin{center}
%      ~            \\ \vspace{1.1  ex}
       170.25       \\ \vspace{0.65 ex}
      (\PeriodBb) % \\ \vspace{1.1  ex}
    \end{center}
   \end{minipage}}          &
 $v_0$ [km/s]               &
 191.8 & $186.6 \~ ^{+33.8}_{-23.4} \~ (^{+ 93.6}_{- 46.8})$ & [163.2, 220.4] & [139.8, 280.2] \\
 \cline{2-6}
                            &
 $\Delta v$ [km/s]          &
  74.6 & $ 87.2 \~ ^{+42.0}_{-71.4} \~ (^{+~84.0}_{-274.1})$ &  [15.8, 129.2] & [$-186.9$, 171.2] \\
 \hline
 \multirow {2} {*}
  {\begin{minipage} {3.25 cm}
    \begin{center}
%      ~            \\ \vspace{1.1  ex}
       261.5        \\ \vspace{0.65 ex}
      (\PeriodBc) % \\ \vspace{1.1  ex}
    \end{center}
   \end{minipage}}          &
 $v_0$ [km/s]               &
 191.8 & $186.6 \~ ^{+33.8}_{-26.0} \~ (^{+ 89.1}_{- 49.4})$ & [160.6, 220.4] & [137.2, 275.7] \\
 \cline{2-6}
                            &
 $\Delta v$ [km/s]          &
  57.8 & $ 70.4 \~ ^{+46.2}_{-79.8} \~ (^{+~88.2}_{-260.4})$ & [$-9.4$, 116.6] & [$-190.0$, 158.6] \\
 \hline
 \multirow {2} {*}
  {\begin{minipage} {3.25 cm}
    \begin{center}
%      ~            \\ \vspace{1.1  ex}
       352.75       \\ \vspace{0.65 ex}
      (\PeriodBd) % \\ \vspace{1.1  ex}
    \end{center}
   \end{minipage}}          &
 $v_0$ [km/s]               &
 191.8 & $186.6 \~ ^{+39.0}_{-26.0} \~ (^{+ 88.4}_{- 49.4})$ & [160.6, 225.6] & [137.2, 275.0] \\
 \cline{2-6}
                            &
 $\Delta v$ [km/s]          &
  53.6 & $ 66.2 \~ ^{+46.2}_{-88.2} \~ (^{+~84.0}_{-256.2})$ & [$-22.0$, 112.4] & [$-190.0$, 150.2] \\
 }
{50}
{\multirow {2} {*}
  {\begin{minipage} {3.25 cm}
    \begin{center}
%      ~            \\ \vspace{1.1  ex}
        79.0        \\ \vspace{0.65 ex}
      (\PeriodBa) % \\ \vspace{1.1  ex}
    \end{center}
   \end{minipage}}          &
 $v_0$ [km/s]               &
 185.0 & $185.0 \~ ^{+ 9.0}_{- 8.0} \~ (^{+19.0}_{-16.0})$ & [177.0, 194.0] & [169.0, 204.0] \\
 \cline{2-6}
                            &
 $\Delta v$ [km/s]          &
  83.7 & $ 83.7    \pm 17           \~ (^{+32.3}_{-40.8})$ &  [66.7, 100.7] &  [42.9, 116.0] \\
 \hline
 \multirow {2} {*}
  {\begin{minipage} {3.25 cm}
    \begin{center}
%      ~            \\ \vspace{1.1  ex}
       170.25       \\ \vspace{0.65 ex}
      (\PeriodBb) % \\ \vspace{1.1  ex}
    \end{center}
   \end{minipage}}          &
 $v_0$ [km/s]               &
 186.0 & $186.0 \~ ^{+ 8.0}_{- 9.0} \~ (^{+18.0}_{-17.0})$ & [177.0, 194.0] & [169.0, 204.0] \\
 \cline{2-6}
                            &
 $\Delta v$ [km/s]          &
  87.1 & $ 88.8 \~ ^{+15.3}_{-18.7} \~ (^{+28.9}_{-39.1})$ &  [70.1, 104.1] &  [49.7, 117.7] \\
 \hline
 \multirow {2} {*}
  {\begin{minipage} {3.25 cm}
    \begin{center}
%      ~            \\ \vspace{1.1  ex}
       261.5        \\ \vspace{0.65 ex}
      (\PeriodBc) % \\ \vspace{1.1  ex}
    \end{center}
   \end{minipage}}          &
 $v_0$ [km/s]               &
 185.0 & $185.0    \pm  9.0         \~ (^{+19.0}_{-17.0})$ & [176.0, 194.0] & [168.0, 204.0] \\
 \cline{2-6}
                            &
 $\Delta v$ [km/s]          &
  71.8 & $ 73.5 \~ ^{+17.0}_{-20.4} \~ (^{+30.6}_{-42.5})$ &  [53.1,  90.5] &  [31.0, 104.1] \\
 \hline
 \multirow {2} {*}
  {\begin{minipage} {3.25 cm}
    \begin{center}
%      ~            \\ \vspace{1.1  ex}
       352.75       \\ \vspace{0.65 ex}
      (\PeriodBd) % \\ \vspace{1.1  ex}
    \end{center}
   \end{minipage}}          &
 $v_0$ [km/s]               &
 186.0 & $186.0    \pm  9.0         \~ (^{+21.0}_{-17.0})$ & [177.0, 195.0] & [169.0, 207.0] \\
 \cline{2-6}
                            &
 $\Delta v$ [km/s]          &
  65.0 & $ 65.0 \~ ^{+17.0}_{-18.7} \~ (^{+34.0}_{-45.9})$ &  [46.3,  82.0] &  [19.1,  99.0] \\
 }
{500}
{parameters $v_0$ and $\Delta v$ as well as}
{normal}

 In Figs.~\ref{fig:N_v-Bayesian-Eq-Dv-050-07900}
 and \ref{fig:N_v-Bayesian-Eq-Dv-500-07900},
 we show the reconstructed radial distributions of
 the 3-D WIMP velocity
 and the 1(2)$\sigma$ statistical uncertainty bands
 by using $f_{1, \sh, \Delta v}(v; v_0, \Delta v)$
 as well as
 the distributions of the fitting parameters $v_0$ and $\Delta v$
 in all simulated experiments on the $v_0 - \Delta v$ plane
 with 50 and 500 total events on average
 in each 60-day observation period of four normal seasons,
 respectively.
 Not surprisingly,
 one can find
 the asymmetric periodic variation of
 the best--fit values of the fitting parameter $\Delta v$
 as well as
 the invariability of the best--fit values of $v_0$
 as its annual average value
 (see the summary of
  the reconstructed results
  and the 1(2)$\sigma$ statistical uncertainty ranges
  of the median values
  given in Table \ref{tab:N_v-Bayesian-Eq-Dv-07900}).

%

%
%
% Appendix D
%
\section{Analytic forms of the shift Maxwellian velocity distribution}
\label{appx:f1v_sh_vesc}

\begin{figure} [h!]
\vspace{-0.25 cm}
\begin{center}
 \includegraphics [height = 10.5 cm] {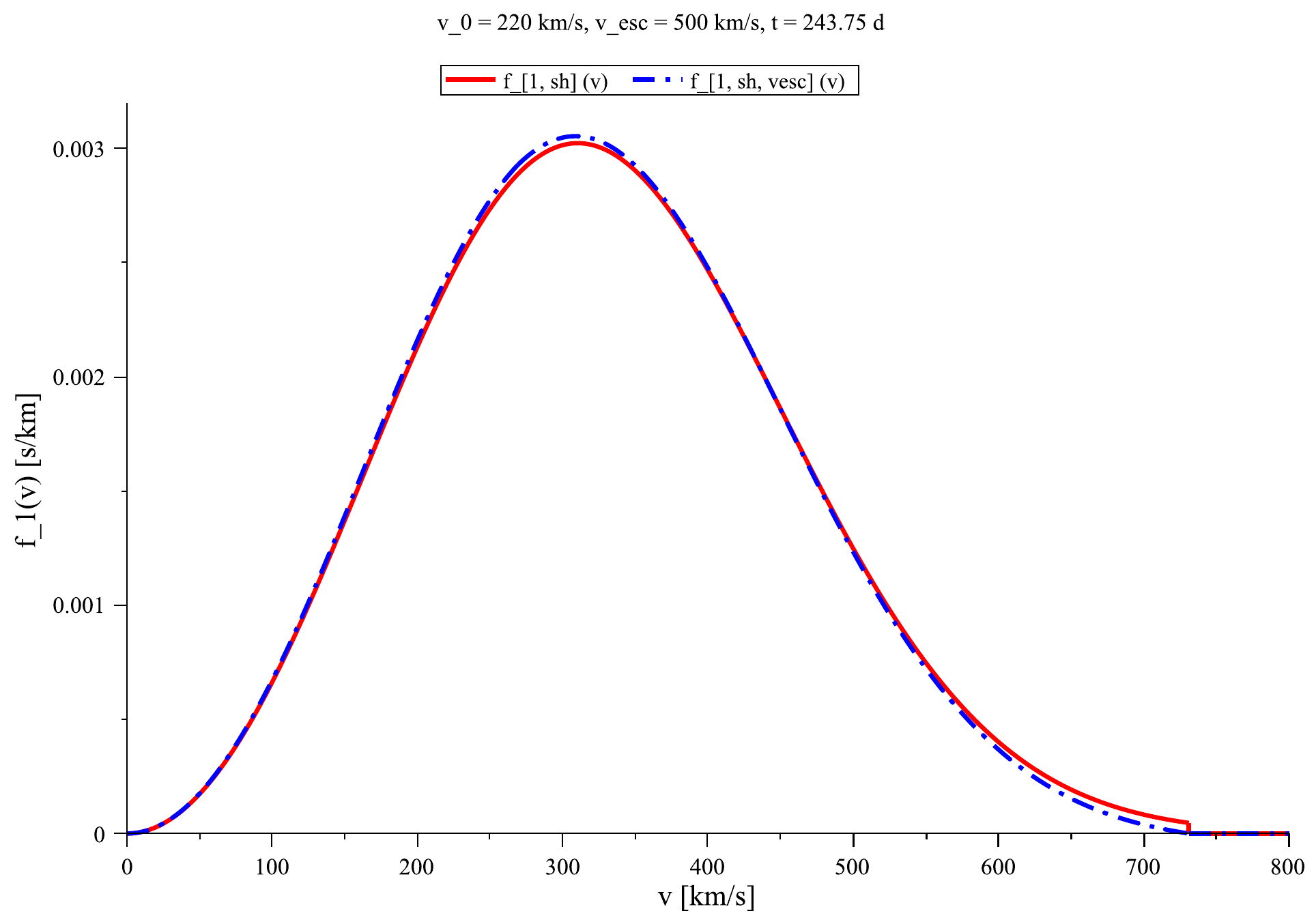}    \\ \vspace{ 0.75 cm}
 \includegraphics [height = 10.5 cm] {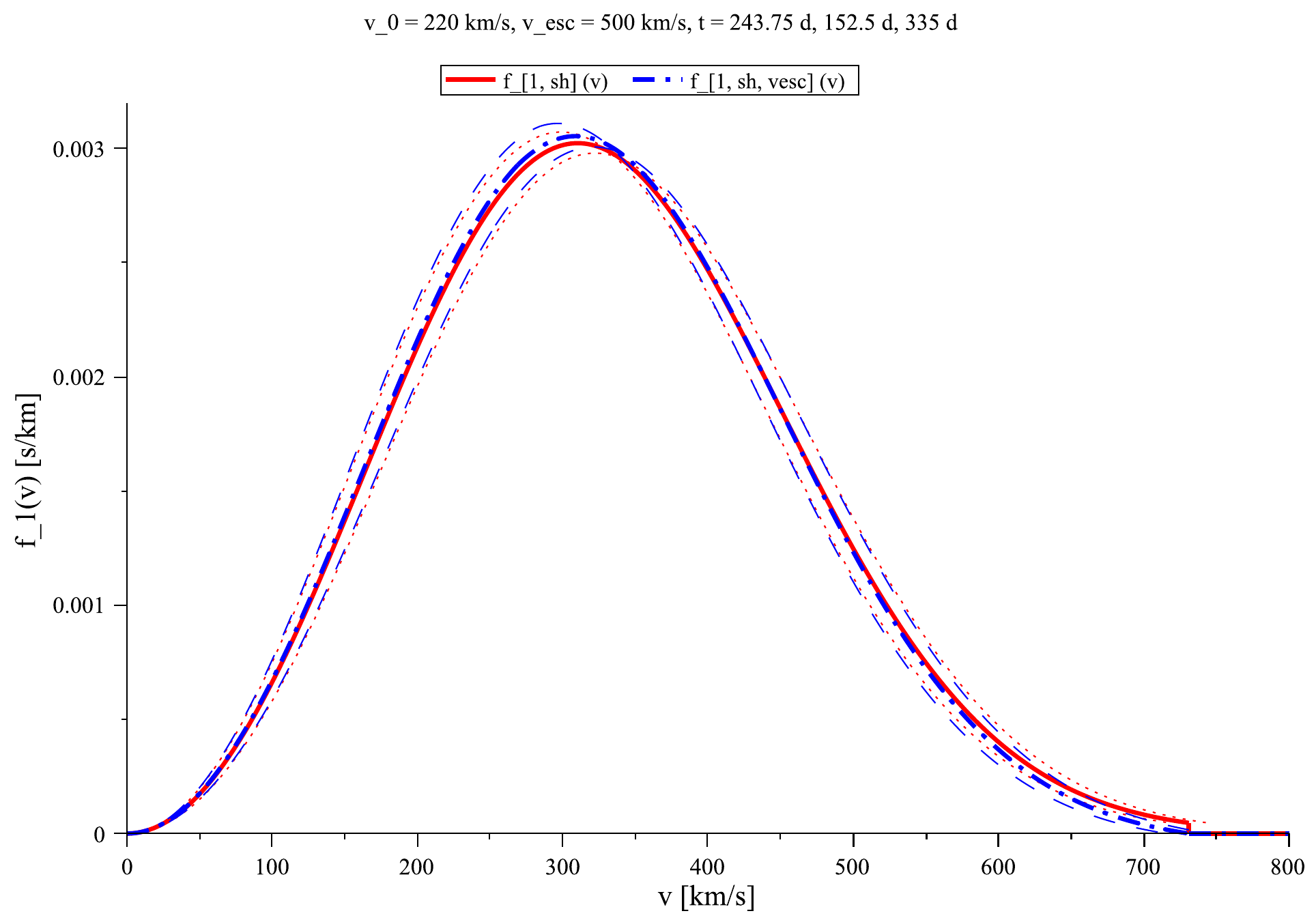} \\ \vspace{-0.25 cm}
\end{center}
\caption{
 The exact analytic form of
 the shift Maxwellian velocity distribution
 $f_{1, \sh, {\rm vesc}}(v)$
 given in Eq.~(\ref{eqn:f1v_sh_vesc})
 (dash--dotted blue)
 and the simplified form
 $f_{1, \sh}(v)$
 given in Eq.~(\ref{eqn:f1v_sh_vmax})
 (solid red)
 with
 the Galactic escape velocity of $\vesc = 500$ km/s.
 In the lower frame,
 the boundaries of two velocity distributions
 due to the annual modulation of
 the Earth's Galactic velocity $\ve(t)$
 given by Eq.~(\ref{eqn:ve})
 have been additionally given
 as the dashed blue and the dotted red curves
 for the exact and the simplified expressions,
 respectively.
\vspace{-1.5 cm}
}
\label{fig:f1v_sh_vesc-500}
\end{figure}

\begin{figure} [h!]
\vspace{-0.25 cm}
\begin{center}
 \includegraphics [height = 10.5 cm] {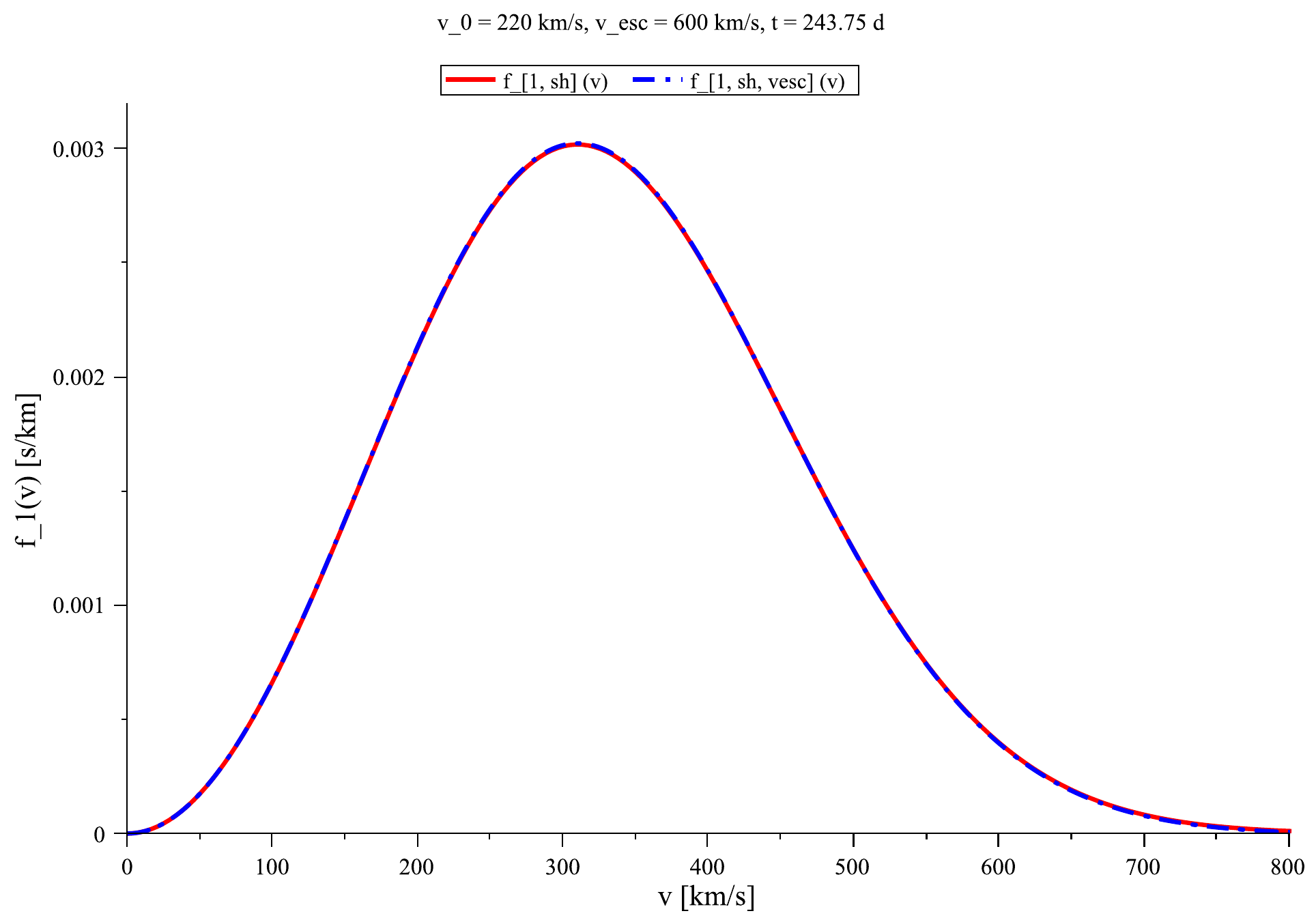}    \\ \vspace{ 0.75 cm}
 \includegraphics [height = 10.5 cm] {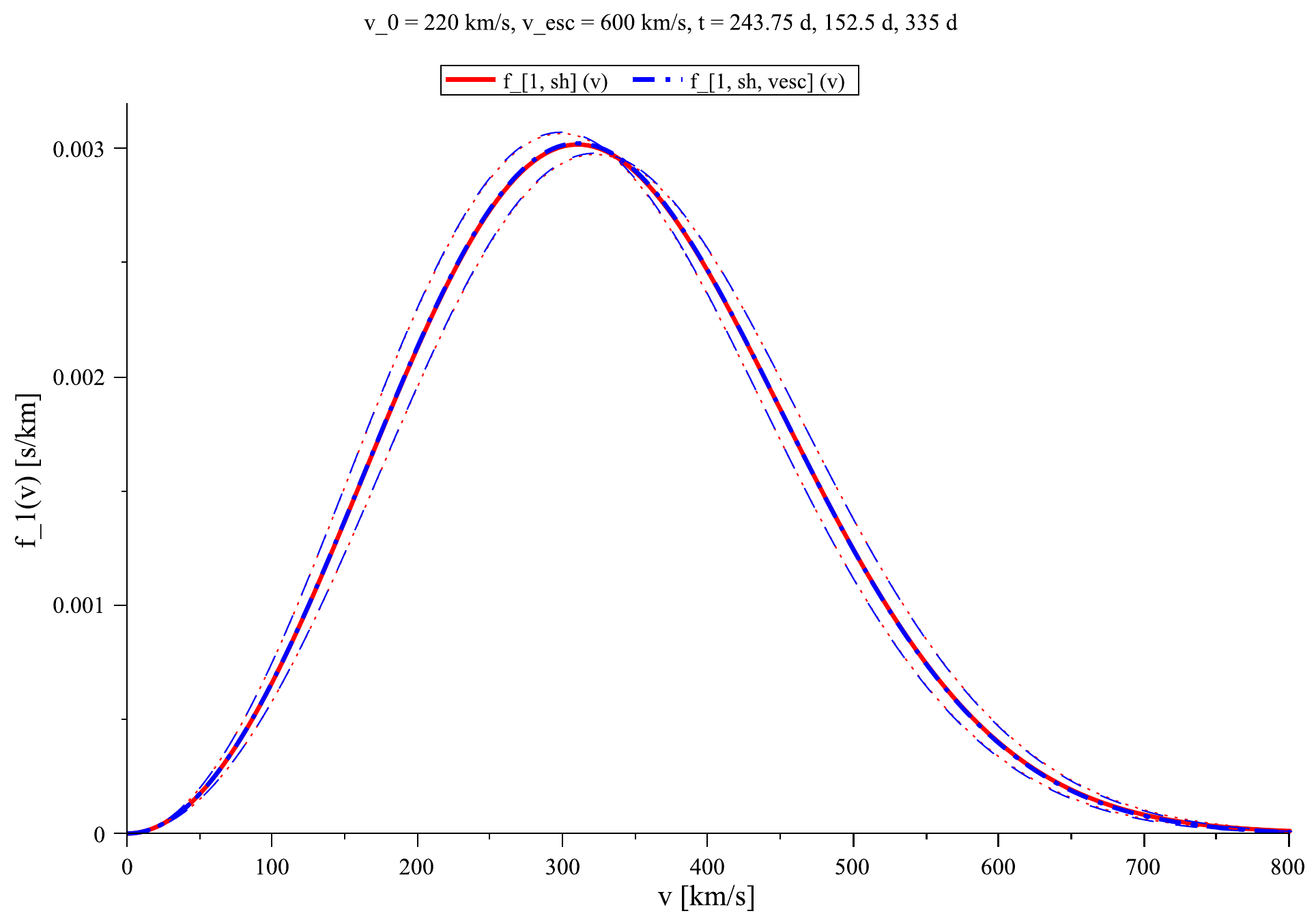} \\ \vspace{-0.25 cm}
\end{center}
\caption{
 As in Figs.~\ref{fig:f1v_sh_vesc-500},
 except that
 the Galactic escape velocity
 has been raised to \mbox{$\vesc = 600$ km/s}.
\vspace{0.5 cm}
}
\label{fig:f1v_sh_vesc-600}
\end{figure}

 In this section,
 we show the plots of
 the exact analytic expression of
 the shift Maxwellian velocity distribution
 $f_{1, \sh, {\rm vesc}}(v)$
 given in Eq.~(\ref{eqn:f1v_sh_vesc})
 and the simplified form
 $f_{1, \sh}(v)$
 given in Eq.~(\ref{eqn:f1v_sh_vmax}).
 The Galactic escape velocity
 has been
 set as $\vesc = 500$ km/s
 in Figs.~\ref{fig:f1v_sh_vesc-500}
 and
 raised to \mbox{$\vesc = 600$ km/s}
 in Figs.~\ref{fig:f1v_sh_vesc-600}
 \cite{RPP18AP}.
 Additionally,
 as a comparison of the shape difference
 between two expressions,
 their boundaries
 due to the annual modulation of
 the Earth's Galactic velocity $\ve(t)$
 given by Eq.~(\ref{eqn:ve})
 have also been given.

 It can be seen clearly that,
 with the Galactic escape velocity of $\vesc = 500$ km/s,
 the shape difference between
 the exact analytic expression of
 the shift Maxwellian velocity distribution
 $f_{1, \sh, {\rm vesc}}(v)$
 and the simplified form $f_{1, \sh}(v)$
 is much smaller than
 the annual variation of two expressions.
 This tiny difference
 can also be neglected
 compared to
 the much larger statistical uncertainties
 on the recorded WIMP events
 (shown in e.g., Fig.~\ref{fig:N_v-Eq-050-00000}).
 Moreover,
 once the Galactic escape velocity
 would be as large as $\vesc = 600$ km/s,
 the tiny shape difference
 between two expressions
 could even vanish.

%

%
%
% References
%
% 10/10
 %
%

%
%

%

%
%
%
\end{document}